\documentclass[headers,manuscript]{ucalgmthesis}
\fullpagethesisBind
\setsecnumdepth{subsection}
  
\usepackage[hidelinks]{hyperref} % for hyperlinks, the option is to remove the red boxes

\usepackage{authblk}

\usepackage{physics}

\usepackage[utf8]{inputenc}
\usepackage{amsmath}
\usepackage{amsthm}
\usepackage{amssymb}
\usepackage{mathrsfs}

\usepackage{subfiles}

\usepackage{bbm} %for mathbbm blackboard bold for number
\usepackage{soul} %for strikingthrough text use \st 

\usepackage[T1]{fontenc}
\usepackage{bm} %for bold greek letter

\usepackage{graphicx}

\usepackage[dvipsnames]{xcolor}% http://ctan.org/pkg/xcolor
\usepackage{caption}
\usepackage{subcaption}
\usepackage{enumitem}  %For enumeration item styles

\usepackage{xparse}
\usepackage{mathtools}
\usepackage{setspace}
\usepackage{tikz}
\usepackage{qcircuit}

\usepackage[numbered]{bookmark}

\usepackage{Acorn,lettrine}

\usepackage{algorithm}
\usepackage{algpseudocode}
\tikzset{algpxIndentLine/.style={draw=blue,dashed}:}

\usepackage[most]{tcolorbox}
\newtcolorbox{defbox}[2][]{
  colback=purple!7,
  frame hidden,
  left = 10pt,
  arc=0mm,
  halign       = flush left,
  fonttitle    = \bfseries\sffamily,
  colbacktitle = purple!50,
  title        = #2,
  enhanced,
  #1
}
\newtcolorbox{myt}[2][]{%
  attach boxed title to top center
               = {yshift=-4pt},
  colback      = blue!5!white,
  colframe     = blue!75!black,
  halign       = flush left,
  fonttitle    = \bfseries\sffamily,
  colbacktitle = blue!65!black,
  title        = #2,#1,
  enhanced,
}
\newtcolorbox{propbox}[2][]{
  colback=RedOrange!7,
  frame hidden,
  left = 10pt,
  arc=0mm,
  halign       = flush left,
  fonttitle    = \bfseries\sffamily,
  colbacktitle = BurntOrange,
  title        = #2,
  enhanced,
}

\newtcolorbox{probbox}[2][]{
  colback=White,
  left = 10pt,
  arc=0mm,
  halign       = flush left,
  fonttitle    = \bfseries\sffamily,
  coltitle = Black,
  colbacktitle = White,
  title        = #2,
  enhanced,
}

\NewTcbTheorem[number within=section]{theorembox}{Theorem}% 
{ colback=blue!3,
  colframe=blue!35!black,
  #1,
  fonttitle=\bfseries}{th}

\newtheoremstyle{romantheorem}% name
  {}% Space above
  {}% Space below
  {\upshape}% Body font
  {}% Indent amount
  {\bfseries}% Theorem head font
  {.}% Punctuation after theorem head
  {.5em}% Space after theorem head
  {}% Theorem head spec (can be left empty, meaning 'normal')

\theoremstyle{romantheorem}
\newtheorem{theorem}{Theorem}[section]

\newtheorem{example}[theorem]{Example}
\newtheorem{corollary}[theorem]{Corollary}
\newtheorem{lemma}[theorem]{Lemma}

\newtheorem{proposition}[theorem]{Proposition}

\theoremstyle{definition}
\newtheorem{definition}[theorem]{Definition}

\theoremstyle{remark}
\newtheorem*{remark}{Remark}

\def\*#1{\mathbf{#1}} % Tried of writing \mathbf{} everyday?
\def\rm#1{\mathrm{#1}}
\def\ssf#1{\mathsf{#1}}
\def\scr#1{\mathscr{#1}}
\def\bb#1{\mathbb{#1}}
\def\cal#1{\mathcal{#1}}
\newcommand{\fk}[1]{\mathfrak{#1}}

\newcommand{\knorm}[2]{\left\Vert#1\right\Vert_{(#2)}}
\def\magenta#1{\textcolor{magenta}{#1}}

\newcommand{\parm}{\mathord{\color{black!33}\bullet}}%

\def\point#1{\noindent\textbf{\underline{#1}}\newline}
\newcommand*{\rom}[1]{\expandafter\@slowromancap\romannumeral #1@}

\newcommand{\eqdef}{\coloneqq}
\newcommand{\da}{\downarrow}

\newcommand{\mbR}{\mathbb{R}}
\newcommand{\prob}{\mathrm{Prob}}
\newcommand{\stoch}{\rm{Stoch}}
\newcommand{\dstoch}{\rm{DStoch}}
\newcommand{\perm}{\mathrm{Perm}}
\newcommand{\cptp}{\mathrm{CPTP}}
\newcommand{\clch}{\mathsf{ClassicalChannels}}
\newcommand{\super}{\mathrm{SUPER}}

\newcommand{\CMO}{\mathrm{CMO}}
\newcommand{\CDS}{\mathrm{CDS}}

\newcommand{\conv}{\mathrm{Conv}}

\newcommand{\predict}{\mathbf{P}}
\newcommand{\pr}{\mathbf{Pr}} %Winning chance of a game

\newcommand{\rH}{H_\alpha}

\newcommand{\Hmin}{\underline{\bb{H}}}
\newcommand{\Hmax}{\overline{\bb{H}}}

\newcommand{\id}{\mathsf{id}}

\def\s{\mathbf{s}}
\def\p{\mathbf{p}}
\def\q{\mathbf{q}}

\def\e{\mathbf{e}}
\def\x{\mathbf{x}}
\def\y{\mathbf{y}}
\def\z{\mathbf{z}}
\def\t{\mathbf{t}}
\def\u{\mathbf{u}}
\def\w{\mathbf{w}}
\def\0{\mathbf{0}}

\def\b{\mathbf{b}}

\newcommand{\vctr}{\mathbf{r}}

\newcommand{\vctp}{\mathbf{p}}
\newcommand{\vctq}{\mathbf{q}}

\newcommand{\vctu}{\mathbf{u}}
\newcommand{\vctv}{\mathbf{v}}

\newcommand{\knot}{\mathbf{k}_0}
\newcommand{\pstar}{p^\star}

\newcommand{\mD}{\mathcal{D}}
\newcommand{\mE}{\mathcal{E}}
\newcommand{\mF}{\mathcal{F}}

\newcommand{\mI}{\mathcal{I}}
\newcommand{\mJ}{\mathcal{J}}

\newcommand{\mM}{\mathcal{M}}
\newcommand{\mN}{\mathcal{N}}

\newcommand{\mR}{\mathcal{R}}
\newcommand{\mS}{\mathcal{S}}

\def\addsymbol #1: #2{$#1$ \> \ \ \ \ \ \ \ \ \ \ \ \ \ \ \ \ \ \ \ \ \ \ \ \ \ \ \ \ \ \ \ \ \ \ \ \   \parbox{5in}{#2}\\}
\newcommand{\thesisbibliography}{  
  \if@unbold\addtocontents{toc}{\let\string\bfdefault\string\mddefault}\fi
  \addcontentsline{toc}{chapter}{Bibliography}
  }

\usepackage[style=numeric-comp,sorting=none, url=false, eprint=true,giveninits=true]{biblatex}
\DeclareNameAlias{author}{family-given}
\DeclareNameAlias{editor}{family-given}
\begin{document}

%%%%%%%%%%%%%%%%%%%%%%%%%%%%%%%%%%%%%%%%%%%%%%%%%%%%%%%%%%%%%%%%%%%%%%%%
%%                                                                    %%
%% Title page                                                         %%
%%                                                                    %%
%%%%%%%%%%%%%%%%%%%%%%%%%%%%%%%%%%%%%%%%%%%%%%%%%%%%%%%%%%%%%%%%%%%%%%%%

%%%%%%%%%%%%%%%%%%%%%%%%%%%%%%%%%%%%%%%%%%%%%%%%%%%%%%%%%%%%%%%%%%%%%%%%
%% Instructions for title page information:                           %%
%%                                                                    %%
%%  Fill in the following fields with the required information:       %%
%%   - \title{...}        title of the thesis                         %%
%%   - \author{...}       your full name                              %%
%%   - \dept{...}         full name of the graduate department        %%
%%                          or degree program                         %%
%%   - \degree{...}       full name of the degree obtained            %%
%%                          (i.e. Doctor of Philosophy)               %%
%%   - \gradyear{...}     year of submission                          %%
%%   - \monthname{...}    month of submission                         %%
%%%%%%%%%%%%%%%%%%%%%%%%%%%%%%%%%%%%%%%%%%%%%%%%%%%%%%%%%%%%%%%%%%%%%%%%
  \title{
   Uncertainty and entropies of classical channels
   \\ \bigskip
  %  [Thesis Title Second Line (optional)]
   }
  \author{Takla Nateeboon}
  \dept{Graduate Program in Mathematics and Statistics}
  \degree{Master of Science} % or, e.g., Master of Science
  \gradyear{2025}
  \monthname{July}
  %\thesis{Thesis}              
  % may change `Thesis' to `Dissertation' depending on type of work

%%%%%%%%%%%%%%%%%%%%%%%%%%%%%%%%%%%%%%%%%%%%%%%%%%%  
%% Make the thesis title page.
%%%%%%%%%%%%%%%%%%%%%%%%%%%%%%%%%%%%%%%%%%%%%%%%%%%
  \frontmatter%% Don't remove this line.
  \makethesistitle%% Don't remove this line.

%%%%%%%%%%%%%%%%%%%%%%%%%%%%%%%%%%%%%%%%%%%%%%%%%%%%%%%%%%%%%%%%%%%%%%%%
%%                                                                    %%
%% Prefatory pages                                                    %%
%%                                                                    %%
%%%%%%%%%%%%%%%%%%%%%%%%%%%%%%%%%%%%%%%%%%%%%%%%%%%%%%%%%%%%%%%%%%%%%%%%
%% The following sections are in the correct order    
%% as specified by the April 2014 thesis guidelines
%% set by the University of Calgary.
%%
%% You may remove the optional sections, but do not
%% change their order.
%%%%%%%%%%%%%%%%%%%%%%%%%%%%%%%%%%%%%%%%%%%%%%%%%%%

%%%%%%%%%%%%%%%%%%%%%%%%%%%%%%%%%%%%%%%%%%%%%%%%%%%
%%
%% Abstract page (required)
%%%%%%%%%%%%%%%%%%%%%%%%%%%%%%%%%%%%%%%%%%%%%%%%%%%
  % \begin{thesisabstract}
  \section*{Abstract}
    In this thesis, I studied a mathematical development to define and quantify the uncertainty inherent in classical channels. 
    % \lettrine{I}{n} this thesis, I studied a mathematical development to define and quantify the uncertainty inherent in classical channels. 
    This thesis starts with the introduction and background on how to formally think about uncertainty in the domain of classical states. The concept of probability vector majorization and its variants, relative majorization and conditional majorization, are reviewed. This thesis introduces three conceptually distinct approaches to formalize the notion of uncertainty inherent in classical channels. These three approaches define the same preordering on the domain of classical channels, leading to characterizations from many perspectives. With the solid foundation of uncertainty comparison, classical channel entropy is then defined to be an additive monotone with respect to the majorization relation. The well-known entropies in the domain of classical states are uniquely extended to the domain of channels via the optimal extensions, providing not only a solid foundation but also the quantifiers of uncertainty inherent in classical channels.
  % \end{thesisabstract}

%%%%%%%%%%%%%%%%%%%%%%%%%%%%%%%%%%%%%%%%%%%%%%%%%%%
%%
%% Preface page (optional)
%%%%%%%%%%%%%%%%%%%%%%%%%%%%%%%%%%%%%%%%%%%%%%%%%%%

%% This section is required if the work presented 
%% in the thesis is done as part of a collaboration.
%% In this case, the preface must state which parts
%% of the thesis are the author's original work.
%% Otherwise this section is optional.

\chapter{Preface}
The new results in this thesis are partly from the author's collaboration with Gilad Gour, Doyeong Kim, Guy Shemesh, and Goni Yoeli. We published together an article under the name ``Inevitable negativity: quantum additivity commands negative quantum channel entropy''~\cite{GKNSY2025}. Some figures have previously appeared or have been adapted from the published manuscript.

%%%%%%%%%%%%%%%%%%%%%%%%%%%%%%%%%%%%%%%%%%%%%%%%%%%
%%
%% Acknowledgements page (required)
%%%%%%%%%%%%%%%%%%%%%%%%%%%%%%%%%%%%%%%%%%%%%%%%%%%

  \chapter{Acknowledgements}  
I want to express my gratitude to both of my supervisors, Carlo Maria Scandolo and Gilad Gour. Their support, guidance, and insightful feedback throughout my master's program have been essential in shaping this thesis and my academic growth. Their kindness proved especially valuable during my first few months as I adjusted to this new environment. With their help, my first experience studying abroad proved to be both rewarding and enjoyable.

I want to extend my thanks to my colleagues, Doyeong Kim, Samuel Steakley, Guy Shemesh, and Goni Yoeli, for their insightful academic discussions. Specifically to Doyeong and Sam, you two have been good company both in terms of our academic journey, working through the courses, and someone to hang out with. Thank you very much.

I would like to acknowledge the supportive environment the University of Calgary has provided me through financial support, scholarships, and academic resources.

I am incredibly grateful to my parents for their unwavering love and support; their warmth and assurance have been the cornerstone of my success throughout this journey. I would also like to express my heartfelt appreciation to Pattarawadee Srestabunjong. Her compassion and unwavering presence have been invaluable, as she has always listened and supported me through all my ups and downs.

Thank you all for your contributions to this important milestone in my academic life.
    
%%%%%%%%%%%%%%%%%%%%%%%%%%%%%%%%%%%%%%%%%%%%%%%%%%
%%
%% Dedication page (this section is optional)
%%%%%%%%%%%%%%%%%%%%%%%%%%%%%%%%%%%%%%%%%%%%%%%%%%

  \chapter[Dedication]{}
  \begin{center}
    To my parents
    % \vspace{10pt}
    % \includegraphics[width=2in]{../figures/Dedication.pdf}
  \end{center}

%%%%%%%%%%%%%%%%%%%%%%%%%%%%%%%%%%%%%%%%%%%%%%%%%%
%%
%% Table of Contents (required)
%%%%%%%%%%%%%%%%%%%%%%%%%%%%%%%%%%%%%%%%%%%%%%%%%%

  \tableofcontents

%%%%%%%%%%%%%%%%%%%%%%%%%%%%%%%%%%%%%%%%%%%%%%%%%%
%%
%% Various lists
%%%%%%%%%%%%%%%%%%%%%%%%%%%%%%%%%%%%%%%%%%%%%%%%%%
%%%%%%%%%%%%%%%%%%%%%%%%%%%%%%%%%%%%%%%
%% List of figures (required, if any)
  \listoffigures
  
%%%%%%%%%%%%%%%%%%%%%%%%%%%%%%%%%%%%%%%
%% List of tables (required, if any)
  % \listoftables
  
%%%%%%%%%%%%%%%%%%%%%%%%%%%%%%%%%%%%%%%
%% List of Symbols, abbreviations, and 
%% nomenclature (required, if any)   
%%
%% (Note: You may use your own format for your list of symbols, 
%%  abbreviations, etc. This format is just a guideline)
%%
  \chapter{List of Symbols, Abbreviations, and Nomenclature}      
  \begin{tabbing}
    % Symbol~~~~~\= \ \ \ \ \ \ \ \ \ \ \ \ \ \ \ \ \ \ \ \ \ \ \ \ \ \ \ \ \ \ \ \ \ \ \ \  \parbox{5in}{Definition}\\
    Symbol~~~~~\= \ \ \ \ \ \ \ \ \ \ \ \ \ \ \ \ \ \ \ \ \ \ \ \ \ \ \ \ \ \ \ \ \ \ \ \  \parbox{5in}{Definition}\\

    \addsymbol\mbox{$A, B, C, \ldots$}: {Hilbert spaces corresponding to or labels for physical systems}
    \addsymbol\mbox{$X, Y, Z, W$}: {Hilbert spaces corresponding to or labels for classical physical systems}
    \addsymbol\mbox{$\cptp(A\to B)$}: {The set all quantum channels from the system $A$ to the system $B$}
    \addsymbol\mbox{$\clch$}: {The set of all classical channels}
    \addsymbol\mbox{$\stoch(n)$}: {The set of column stochastic matrices of dimension $n\times n$}
    \addsymbol\mbox{$\prob(n)$}: {The set of probability vectors of dimension $n$}
    \addsymbol\mbox{$\super((A\to B) \to (A'\to B'))$}: {The set of all superchannels from $(A\to B)$ to $(A'\to B')$}
    % \addsymbol\mbox{$\cmo((A\to B))$}
    \addsymbol\mbox{$\mbR$}: {The set of all real numbers}
    \addsymbol\mbox{$\mbR^n$}: {The set of all real n-dimensional vectors}
    \addsymbol\mbox{$\mbR_+$}: {The set of all non-negative real numbers}
    \addsymbol\mbox{$\mbR_{++}$}: {The set of all positive real numbers}
    \addsymbol\mbox{$\id^{A}$}: {An identity superchannel on $\mathfrak{L}(A)$ to itself.}
    \addsymbol\mbox{$\id^{(A\to B)}$}: {An identity superchannel on $\cptp(A\to B)$ to itself.}
    %
    % Add more symbols here...
    % . 
  % .KP{O(·············¨‚∏Ø¨∏∏∏∏∏∏∏∏∏∏∏∏∏—‚¨r5)
    % .
    %
    % ALWAYS KEEP THE FOLLOWING LINE IF YOU USE THIS LIST OF SYMBOLS
  \end{tabbing}

%%%%%%%%%%%%%%%%%%%%%%%%%%%%%%%%%%%%%%%%%%%%%%%%%%%%%%%%%%%%%%%%%%%%%%%%
%%                                                                    %%
%% Main matter                                                        %%
%%                                                                    %%
%%%%%%%%%%%%%%%%%%%%%%%%%%%%%%%%%%%%%%%%%%%%%%%%%%%%%%%%%%%%%%%%%%%%%%%%

  %% Do not remove this line
  \mainmatter
    
%%%%%%%%%%%%%%%%%%%%%%%%%%%%%%%%%%%%%%%%%%%%%%%%%%
%%
%% Chapters
%%%%%%%%%%%%%%%%%%%%%%%%%%%%%%%%%%%%%%%%%%%%%%%%%%

\setcounter{chapter}{-1}
\chapter[Introduction]{Introduction}

%Big picture; nature of information theory; irrelevance of physical systems; abstraction of the notion of information.
Throughout this thesis, we shall concern ourselves with the study which stems from the development of information theory, where in this thesis we refer to it as the \emph{classical} information theory. Unlike any physical theories, classical information theory distances itself from a specific physical system by abstracting the information as a currency of ability to perform tasks, for example, sending a message, encoding, and decoding. Centered on this field study is entropy, especially Shannon entropy, which is proposed to reflect how much information one can obtain from a random variable~\cite{Shannon1948}. The quantity of entropy itself depends on the probability distribution of the random variable of interest. The theory has been successful so far in describing and giving insight into communication, cryptography, statistical testing, and more disciplines~\cite{Wyner1974,DHW2008}. 

%Shift to quantum
Quantum information theory has emerged as a natural progression from classical information theory. Paradoxically, the development of the physical theory, known as quantum theory, has led to a paradigm shift in the physical-independent theory of information theory. Random variables and their probability distributions, once considered the most fundamental objects, have been replaced by quantum states, or in certain contexts, referred to as density operators. This shift has profound implications. A notable example is quantum entanglement, a property where two physical systems cannot be fully described individually, despite appearing to be physically separated. Entanglement enables the formation of correlations beyond the capabilities of any classical correlation, leading to the development of new capabilities. 

%The challenges : what is studied in many quantum information theory
Despite the powerful capability inherent in quantum properties, many implementations are hard to realize as there are physical constraints. In the notable example of entanglement, the creation and maintaining of such properties have been challenging. This frames entanglement as a valuable resource, which one cannot easily create, hence it requires optimally utilizing the available resource. Similar situations also appear in coherence~\cite{DBG2015,WY2015,CG2016,ZMCFV2017}, purity~\cite{SKWGB2018}, and athermality~\cite{BHNOW2015,GMVRN2015} as well. These constraints lead to the study of multiple quantum resource theories. 

%Classical<->Quantum information theory interaction
Although there is a shift in the nature of information from classical to quantum, concepts in classical information theory extend to the quantum regime. The von Neumann entropy, which was previously devised to put quantum mechanics into a thermodynamic perspective~\cite{Neumann1927,Petz2001}, turns out to be applicable in quantum information theory. The von Neumann entropy extends Shannon entropy from the domain of classical random variables, where we will refer to them as classical states, to the domain of quantum states. The entropy not only quantifies information contained in a quantum state but also provides a measure of resource in multiple resource theories, including quantum resource of entanglement~\cite{BBPS1996,Nielsen1999,Vidal2000,NV2001}, purity~\cite{SKWGB2018}, and coherence~\cite{Aberg2006,ZMCFV2017}. 

%What is an entropy (<- let’s think about how to approach this section later…)
Shannon entropy and von Neumann entropy are said to be measures of uncertainty. But what makes entropy a measure of uncertainty? Can we have a more varied set of entropies? The more general approach is to define uncertainty by a preorder, namely, majorization. Independently of information theory and physics, the majorization itself is a concept developed to quantify how much the wealth distribution spread out~\cite{HLP1934}. The key characterization of majorization with a convertibility via \emph{noisy} operation leads to the motivation to ground the definition of entropy with the majorization~\cite{Birkhoff1946}. The development of the field in the past decades, 2010s, has shown that not only can we define entropy as an additive monotone (in more strict term, an additive antitone) of majorization, but also we can do with relative entropies and conditional entropies~\cite{GGH+2018,BGWG2021,VSGC2022}, for both classical and quantum domains. Majorization also serves as a quintessential tool in quantum information theory. For example, majorization is used in describing the uncertainty relation between non-commuting observables~\cite{Partovi2011,PRZ2013}, classical and quantum thermodynamics~\cite{HO2013,BHNOW2015,GMVRN2015}, and many quantum resource theories~\cite{Renes2016,ZMCFV2017}. Most prominently, Nielsen's majorization theorem~\cite{Nielsen1999,NV2001} gives a sufficient and necessary condition of converting quantum state via local operation and classical communication by majorization relation.  
% \magenta{Also}~\cite{LQ2020, ZLXW+2023}.

%Resource theories beyond state: dynamical resource theories — Channel as a resource by itself
The recognition of what constitutes a resource in quantum theories extends beyond static resources of quantum states. The operations on quantum states, quantum channels, have been recognized as resources~\cite{Devetak2005,HW2010,RR2011,CMW2016,HS2018,GS2020a,GS2021}, leading to the development of dynamical resource theories~\cite{LY2020,TSP2020,GS2020b,RT2021,RT2021b}. Recently, a dynamical resource has been experimentally measured in~\cite{HWXLG2022}. Taking an example from the entanglement theories, one can consume a maximally entangled state, which is a static resource, to perform a teleportation~\cite{BBCJ+1993} or dense coding protocols~\cite{BW1992}. These operations can be viewed as a creation of a teleportation channel or a dense coding channel with an entanglement pair, rendering such channels as a resourceful object. On the contrary, if one has access a bipartite channel that can perform non-LOCC operation, then one can use such a channel instead of consuming an entanglement pair in every operation. This is an example that put quantum channels as a resourceful object in its own right.

%Can we have an entropy for quantum channel?
Similar questions occurred in the dynamical resource theories. Can quantum channels have entropy? The assumption was that they can. The earlier attempts~\cite{Yuan2019,LBL2020,GW2021} to define quantum channel entropy are done with divergences or relative entropies of quantum channels, which measure a degree of distinguishability between any two quantum channels. The quantum channel entropy is then defined to be a measure of the distance of a quantum channel from the totally mixed channel, a channel that always produces a maximally mixed state regardless of the input. Questions still remain. What does it mean for one quantum channel to have more entropy compared to another channel? What is the uncertainty of a quantum channel? What is the more fundamental rationale for using such a definition as the entropy of a quantum channel?

%Classical <-> Quantum theory interaction in the level of channel %%% --- Problem statement for the research --- %%% ,,,,, Problem statement ,,,,
To answer these questions, this thesis approaches from and focuses on the classical channel domain, a domain of quantum channels where the output is incoherent and completely independent of the coherent part of the quantum state. We will attempt to define the uncertainty of classical channels via a preorder. This preorder should be devised from convertibility using \emph{noisy} operations, which will provide a foundational and operational interpretation of what uncertainty is in the channel domain. Then, we will define an entropy, similar to quantum and classical states’ entropies, to be an additive monotone of classical channel majorization.

%Organization of the thesis 
To do this, we firstly begin with the review of necessary mathematical concepts and tools in Chapter~1. Starting with the framework of quantum information theory: Hilbert spaces, operators, quantum states, quantum channels, and quantum superchannels. 
In Chapter~2, we provide a review of entropy measures on classical states and its foundational preorder, including majorization, relative majorization, and the conditional majorization.
Our original contribution begins in Chapter~3, where we provide our answer to the main question of this thesis: What is the uncertainty of a classical channel? Along this, we characterize the uncertainty via multiple mathematical perspectives. In Chapter~4, we define the entropy of a classical channel as a function reflecting the uncertainty of a classical channel. We provide examples of entropy functions and provide a discussion on the uniqueness of functions extending entropies. Lastly, we conclude our answer and provide a look ahead in Chapter~5.

\subsection*{Organization}
Throughout this thesis, we display the important definitions and propositions with color boxes: \textcolor{BurntOrange}{orange} boxes are for theorems, lemmas, or propositions and \textcolor{purple!50}{pink} boxes are for definitions. The complementary definitions and propositions are presented in the usual definition, lemma, or theorem environment without a box.

% %%% --- Problem statement for the research --- %%%
% Majorization, a fundamental concept in linear algebra, is extensively explored in Marshall and Olkin's seminal text. Its applicability extends to various fields, such as economics, thermodynamics, and quantum information theory, showcasing its broad relevance and significance. In quantum information theory, majorization plays a pivotal role in characterizing and manipulating quantum states. Nielsen's majorization theorem, as highlighted in~\cite{Nielsen1999,NV2001}, establishes that the interconvertibility among pure bipartite quantum states via local operations and classical communication (LOCC) is governed by a majorization relation between their Schmidt coefficients. These insights not only provide a foundational understanding of quantum entanglement but also have practical applications in entanglement distillation, quantum state discrimination, and quantum key distribution. Over time, the theory of majorization has evolved into an indispensable toolkit in quantum information science, as discussed in~\cite{Partovi2011, PRZ2013, HO2013, GMNS+2015, Renes2016, LQ2020, ZLXW+2023}.

% %How the thesis is organised 

% Reference %%%%% for editing purpose only %%%%%%%%
% \bibliography{QuantumEntropy}
% \bibliographystyle{unsrt}
% \printbibliography
%%%%%%%%%%%%%%%%%%%%%%%%%%%%%%

\chapter{Mathematical framework for quantum information theory}

\section{Mathematical path to quantum theory}
Quantum theory is a physical theory that describes a physical system. In quantum theory, a physical system, denoted by $A$, is described by a \emph{Hilbert space} $\scr{H}_A$. One of the fundamental objects of interest is a quantum state, which is analogue to position and momentum in classical mechanics. A quantum state is built on a mathematical object called \emph{wavefunction}. A wavefunction $\psi$ belongs to the Hilbert space $\scr{H}_A$. For convenience, we will use $A$ to denote both the physical system and the associated Hilbert space. We denote a set of complex or real numbers with $\bb{F}$.

To begin with, a Hilbert space is a complete inner product space. An inner product space is defined as follows.
\begin{definition}[Inner Product Space]
 An inner product space is a vector space $V$ over the field $\bb{F}$ equipped with a binary map $\braket{\parm}{\parm}: V\cross V \to \bb{F}$ such that $\forall \phi,\psi,\chi \in V$ and $a,b\in\bb{F}$ the followings holds.
    \begin{enumerate}
        \item{Conjugate Symmetry:} $\braket{\phi}{\psi} = \overline{\braket{\psi}{\phi}}$
        \item{Linearity in the second argument:} $\braket{\phi}{a\psi+b\chi} = a\braket{\phi}{\psi}+b\braket{\phi}{\chi}$
        \item{Positive-definiteness:} $\braket{\phi}{\phi}>0$ except at $\phi=0$, $\braket{\phi}{\phi}=0$
    \end{enumerate}
\end{definition}
An inner product space $(V,\braket{\parm}{\parm})$ is said to be complete if any Cauchy sequence in $V$ converges to an element in $V$. The Cauchy sequence is defined with respect to metric $d$, which is induced from a norm, $d(x,y) = \norm{x-y}$.
\begin{definition}[Norm]
 A norm is a mapping from a vector space $V$ to a real number, denoted by $\norm{\parm}: V\to \bb{R}$, that satisfies the following properties for all $a\in\bb{F}$ and $\psi,\phi\in V$.
    \begin{enumerate}
        \item $\norm{\psi}>0$ except when $\psi =0$ we have $\norm{\psi}=0$.
        \item $\norm{a\psi}=\abs{a}\norm{\psi}$.
        \item Triangle inequality: $\norm{\psi+\phi}\leq\norm{\psi}+\norm{\phi}$.
    \end{enumerate}
\end{definition}
Now, we can define Hilbert space 
\begin{definition}
 A Hilbert space is a vector space $\mathscr{H}$ over a field $\bb{F}$ endowed with an inner product which is complete as a metric space.
\end{definition}

In this thesis, we work exclusively on the finite dimensional Hilbert space. A dimension of a Hilbert space is defined from the cardinality of its basis. 

\subsection{Dirac notation}
\begin{definition}
 Given a vector space $\scr{H}$ over a field $\bb{F}$. Its dual space $\scr{H}^*$ is a set of all linear functionals $f:V\to\bb{F}$,
    \begin{equation*}
 \scr{H}^* = \qty{f:\scr{H}\to \bb{F}~\text{linear}}
    \end{equation*}
\end{definition}

In our particular case of quantum theory, we are interested in finite-dimensional Hilbert spaces. These spaces come with a special property that all linear functionals are representable with matrices, hence bounded and continuous. By the Riesz representation theorem, any elements in the dual $f \in \scr{H}^*$ can be identified uniquely with an element $\psi_f \in \scr{H}$, in exactly the following form 
\begin{equation*}
 f(\phi) = \braket{\psi_f}{\phi}.
\end{equation*}
Indeed, any $\psi \in \scr{H}$ can be identified with a linear function $f_\psi (\phi) = \braket{\psi}{\phi}$, establishing an isomorphism between the original and the dual. This motivates the bra-ket notation to be used. For a wavefunction $\phi\in A$, the ket of $\phi$, $\ket{\phi}$, denotes $\phi\in A$, which is the wavefunction itself. On the other hand, the bra of $\phi$, $\bra{\phi}$, denotes the dual of $\phi$.

Within the dirac notation, since $\ket{\phi}$ denotes an element in the original function, the expression $\braket{\phi}{M\psi}$ is often expressed as $\bra{\phi} M \ket{\psi}$, and equivalently $\braket{M^* \phi}{\psi}$. Note that in $\braket{\phi}{M\psi}$ and $\bra{\phi} M \ket{\psi}$ the inner product are between two elements in $B$ while $\braket{M^* \phi}{\psi}$ is between two elements in $A$.
% Note that when written $\bra{M^* \phi}$, the operator $M^*:\mathfrak{L}(A,B)$ is applied to the dual $\bra{\phi}$ and the elements $M^*$. 

\subsection{Multipartite systems}
One of the key tasks in quantum information theory is to include or discard physical systems from consideration. The inclusion of a physical system into consideration is represented by taking the tensor product of two Hilbert spaces, e.g. a Hilbert space $A$ and a Hilbert space $B$. The product space is denoted by $A\otimes B$ or simply by their concatenation $AB$. 

In the finite dimensional case, which this thesis is based on, we have that the Kronecker product is the corresponding product between vectors from two Hilbert spaces. 

On the other hand, the discarding of a physical system is represented by taking a trace or partial trace. Formally, it is defined as
\begin{equation*}
 \Tr _{A} \bqty{\rho^{AB}} = \sum_{x\in[n]} (\bra{x} \otimes I^{B}) \rho^{AB} (\ket{x}\otimes I^{B}).
\end{equation*}

\subsection{Operators on Hilbert space}
% A linear operator is a mapping $M$.
A set of all linear operators from $A$ to $B$ is denoted with $\mathfrak{L}(A,B)$. A set of all linear operators onto itself, e.g., from $A$ to $A$ is denoted with $\mathfrak{L}(A)$. In quantum theory, there are two important classes of operators: Unitary operators and Hermitian operators. Both are defined on their relations with their adjoint.
\begin{definition}[Adjoint of an operator]
 Suppose $M\in \mathfrak{L}(A,B)$. An operator $M^* \in \mathfrak{L}(B,A)$ is an adjoint of $M$ if $\braket{\phi}{M \psi} = \braket{M^* \phi}{\psi}$ for all $\phi\in B$ and all $\psi \in A$.
\end{definition}

\begin{lemma}
 Given an operator $M\in\mathfrak{L}(A,B)$, its adjoint is unique. 
\end{lemma}
\begin{proof}
 Suppose $N$ and $M^*$ are adjoints of $M$, not necessarily distinct. We have the following equalities for all $\phi\in B$ and $\psi \in A$,
    \begin{equation*}
 \braket{\phi}{M\psi} = \braket{M^* \phi}{\psi} = \braket{N \phi}{\psi}.
    \end{equation*}
 That is $\braket{M^* \phi - N \phi}{\psi} = 0$ for all $\psi\in A$. This implies $M^* \phi - N \phi = 0$. Since this is for all $\phi \in B$, we have that $N = M^*$, proving the uniqueness.
\end{proof}

Suppose $A$ is a finite-dimensional Hilbert space, and each operator $V\in \mathfrak{L}(A,B)$ has a matrix representation. To create a matrix representation, we prescribe bases in which we would like to write this matrix in. For example, suppose $\qty{\phi_x}$ be an orthonormal basis of $A$ and $\qty{\psi_x}$ be an orthonormal basis of $B$. A matrix $M$ represents $V$ in this basis is 
\begin{equation}\label{eq:matrix-rep}
 M = \sum_{x,x'} \bra{\psi_x} V \ket{\phi_{x'}} \ketbra{\psi_x}{\phi_{x'}}.
\end{equation}

\subsection{Important classes of operators on Hilbert space and their properties}

Now, we define the three important classes of operators.
\begin{definition}[Unitary operator]
 An operator $U\in\mathfrak{L}(A)$ is unitary if its adjoint is its inverse. In symbol, $U$ is unitary if 
    \begin{equation*}
 U^* = U^{-1}.
    \end{equation*}
\end{definition}
Every unitary matrix can be written as 
\begin{equation*}
 U = \sum_{x\in[n]} \ketbra{\phi_x}{\psi_x}
\end{equation*}
where $\qty{\phi_x:x\in[n]}$ and $\qty{\psi_x : x \in[n]}$ each is an orthonormal basis of $A$ and $B$ respectively.

A class of unitary operators is contained within a class of isometry operators. An isometry is an operator that preserves an inner product between two vectors.
\begin{definition}
 An operator $V\in\mathfrak{L}(A,B)$ is an isometry if for all $\phi,\psi\in A$
    \begin{equation*}
 \braket{ V \phi}{V \psi} = \braket{\phi}{\psi}.
    \end{equation*}
\end{definition}
Moreover, if $V$ is an isometry, we have that $V^* V = I$. This follows from $\braket{\phi}{\psi} = \braket{V\phi}{V\psi} = \bra{\phi}V^* V\ket{\psi}$ for all $\phi, \psi \in A$. Applying a matrix representation in the Equation~\eqref{eq:matrix-rep} will show that an identity matrix is a matrix representing $V^* V$.

Similar to unitary operator, for any isometry $V\in\mathfrak{L}\pqty{A,B}$,
\begin{equation*}
 V = \sum_{x\in[n]} \ketbra{\phi_x}{\psi_x}
\end{equation*} 
where $\qty{\phi_x}$ is an orthonormal set, need not span $B$, i.e. may not be a basis.

\begin{definition}[Hermitian matrix]
 An operator $H\in\mathfrak{L}\pqty{A,B}$ is Hermitian, or in more modern terms self-adjoint, if its adjoint is itself. In symbols, $H$ is Hermitian if 
    \begin{equation*}
 H^* = H.
    \end{equation*}
\end{definition}

For a Hermitian matrix, we have that $H$ has a spectral decomposition 
\begin{equation*}
 H = \sum_{x\in[n]} \lambda \ketbra{\psi_\lambda}{\psi_\lambda}
\end{equation*}
where $\qty{\psi_\lambda}$ form an orthonormal basis of $\scr{H}$. That is, every Hermitian matrix is diagonalizable with an orthonormal basis. 

% \section{Quantum information theory}
\section{Quantum state}
A quantum state is represented by a density matrix. It describes the property of a quantum object, e.g. an electron and possible energy levels, positions, or momenta.
% A quantum state, also called density matrix, is representation of quantum object which is in a particular state, e.g. a energy level, position, or momentum.
\begin{definition}
 A \emph{quantum state} $\rho$ on the Hilbert space $A$ is a Hermitian operator with the following properties,
    \begin{itemize}
        \item Positive semidefinite: $\forall \psi \in A$ we have that $\bra{\psi}\rho\ket{\psi}\geq 0$
        \item Trace one: $\Tr[\rho] =1$
    \end{itemize}
 We denote the set of all density matrices on a system $A$ with $\fk{D}(A)$.
\end{definition}

We denote a quantum state $\rho$ on system $A$ as $\rho^A$. This is to signify its corresponding Hilbert space.

Any quantum state $\rho$ has a spectral decomposition of the form
\begin{equation}
 \rho = \sum_{i\in[m]} p_i \ketbra{\psi_i}{\psi_i}; \quad\text{$\sum_{i\in[m]} p_i = 1$;$\quad \forall i\in[m]$, $p_i \geq 0$ }
\end{equation}
where $m$ is the dimension of the Hilbert space $A$. This leads to an interpretation that the quantum state is a probability mixture of wavefunctions. In particular, we call
\begin{equation*}
\qty{(p_x, \phi_x) : \phi_x \in \rm{PURE}(A), ~p_x \in[0,1], ~\sum_{x\in[m]} p_x = 1}
\end{equation*}
a probabilistic ensemble of wavefunctions. With the ensemble, there is a unique density matrix related to it. However, it is not true in general that given a density matrix, a wavefunction ensemble is unique.

\subsection{Pure states, mixed states, and purifications}
A pure state describes perfect knowledge of what quantum state the system is in. A mixed state, however, describes an ensemble of possible quantum states. 
\begin{definition}
 A quantum state $\rho\in\frak{D}$ is \emph{pure} if $\Tr[\rho^2]=1$. Otherwise, $\Tr[\rho^2] < 1$ and the quantum state $\rho$ is said to be \emph{mixed}.
\end{definition}

However, given a mixed state $\rho^{A}$ one can always find a pure state $\psi^{AB}$ such that its partial trace $\Tr_B [\psi^{AB}] = \rho^A$. Such a pure state is said to be a \emph{purification} of $\rho^{A}$. An example of the purification of $\rho^{A}$ is  
\begin{equation}\label{eq:purification}
 \ket{\psi}^{\tilde{A}A} = \sqrt{\rho}^{\tilde{A}} \otimes I^{{A}} (\ket{\Phi}^{\tilde{A}A})
\end{equation}
where $\ket{\Phi}^{\tilde{A}A}$ is called maximally entangled state and has the form 
\begin{equation*}
 \ket{\Phi}^{\tilde{A}A} = \sum_{x\in[\abs{A}]} \ket{x}^{\tilde{A}} \otimes \ket{x}^{A}.
\end{equation*}
In fact, any two purifications of the same quantum state are related through an isometry channel, a type of linear map between quantum states which we will introduce in the next section. To see that this is a purification, tracing out $\tilde{A}$ will lead to 
\begin{align*}
 \Tr_{\tilde{A}} [\ketbra{\psi}{\psi}^{\tilde{A}A}] &= \Tr_{\tilde{A}} \bqty{(\sqrt{\rho^T}^{\tilde{A}} \otimes I^{{A}})\ket{\Phi}^{\tilde{A}{A}} \bra{\Phi}^{\tilde{A}A}(\sqrt{\rho^T}^{\tilde{A}} \otimes I^{{A}})}\\
    &= \Tr_{\tilde{A}} \bqty{(\sqrt{\rho^T}^{\tilde{A}} \otimes I^{{A}})\pqty{\sum_{x,y\in[n]} \ketbra{x}{y}^{\tilde{A}} \otimes\ketbra{x}{y}^{A}}(\sqrt{\rho^T}^{\tilde{A}} \otimes I^{{A}})}\\
    &= \sum_{x,y \in[n]}\Tr \bqty{\sqrt{\rho^T}\ketbra{x}{y}\sqrt{\rho^T}~}\ketbra{x}{y}^A\\
    &= \sum_{x,y \in[n]}\bra{y}\rho^T\ket{x} \ketbra{x}{y}^A = \sum_{x,y\in[n]} \rho_{xy} \ketbra{x}{y}^A = \rho ^A.
\end{align*}

Similar to Equation~\eqref{eq:purification}, any bipartite state $\ket{\psi}^{AB}$ can be written as a product of the maximally entangled state $\ket{\Phi}^{\tilde{B}B}$ with some matrix $M\otimes I$ where $M\in \mathfrak{L}(\tilde{B},A)$. Suppose $\abs{A} = n $ and $\abs{B} = m$. Suppose that 
\begin{equation*}
 \ket{\psi}^{AB} = \sum_{\substack{x\in[n]\\y\in[m]}} c_{xy} \ket{x}^{A}\otimes \ket{y}^{B}.
\end{equation*}
Define $M$ to be 
\begin{equation*}
 M = \sum_{\substack{x\in[n]\\ y'\in[m]}} c_{xy'} \ketbra{x}{y'}.
\end{equation*}
Then, it is straightforward to verify that 
\begin{equation}\label{eq:bipartite-max-ent}
 \ket{\psi}^{AB} = (M\otimes I)\ket{\Phi}^{\tilde{B}B}.
\end{equation}

\subsection{Probability distribution from quantum state}
A probability distribution can be obtained from a quantum state via \emph{measurement}. Mathematically, measurement is defined as a set of matrices satisfying a condition. 
\begin{definition}[Measurement]
 A set of matrices $\qty{M_x:x\in[n]}\subset\bb{C}^{\ell\times m}$ is a \emph{measurement} if 
    \begin{equation*}
 \sum_{x\in[n]} M^* _x M_x = I^{(m)}.
    \end{equation*}
\end{definition}

A measurement maps a quantum state $\rho$ to the measurement outcome $\sigma_x$ 
\begin{equation*}
 \sigma_x = \frac{1}{p_x} M_x \rho M^* _x \quad \rm{where}~p_x = \Tr[M_x \rho M^* _x].
\end{equation*}
Then, it is easy to see that $\sum_{x\in[n]} \Tr [M^* _x M_x \rho ] =\Tr \bqty{\sum_{x\in[n]} M^* _x M_x \rho} = \Tr[\rho] = 1$. Since $M^* _x M_x$ is Hermitian, its eigenvalues are real numbers. Moreover, for any $\rho\in\frak{D}(A)$, $\Tr[M^* M\rho]$ is positive since it is a product of $M$ and $M^*$. For each $x\in[n]$, we interpret $p_x$ to be a probability in which a quantum state after the measurement is $\sigma_x$.

\section{Classical state}
A notion of quantum state extends a classical state. In terms of density operators, a classical state is a quantum state that is diagonal with respect to a particular orthonormal basis. This particular orthonormal basis is often referred to as the classical basis, which is picked depending on the physical systems and physical laws describing the system. 

For example, a quantum communication using photonic qubit encodes information in 2 distinct states, says the states $\ket{0}$ and $\ket{1}$. The measurement device in this setup measures a quantum state sent in the $\qty{\ket{0},\ket{1}}$ basis. Notice that a diagonal matrix in this basis may not be diagonal in other bases. For example, $\ketbra{0}{0}$ is diagonal in this basis but not in $\qty{\ket{+},\ket{-}}$ basis, where $\ket{\pm} = \frac{1}{\sqrt{2}}\pqty{\ket{0}\pm\ket{1}}$. 

\subsection{Probability vectors}
Since any classical state $\rho\in \mathfrak{D}(X)$ can be written as a diagonal matrix, it can be put in a probability vector of dimension equal to $\abs{X}$. For example, a classical state $\rho^{X} = \sum_{x\in[n]} p_x \ketbra{x}{x}^{X}$ where $n = \abs{X}$. Then, a probability vector associated with a classical state is
\begin{equation*}
    \*{p} = \sum_{x\in[n]} p_x \*{e}_x.
\end{equation*}
We denoted by $\prob(n)$ a set of all probabilities of dimension $n$. A probability vector is a non-negative vector with the sum of all entries equal to one.

\section{Quantum channel}
A quantum channel is a linear map with extra properties that always maps a quantum state to a quantum state. These properties will be introduced shortly in a formal definition. The linearity requirement comes from the preservation of the probabilistic interpretation of the quantum state. Suppose that one has an ensemble of pure quantum states,
\begin{equation*}
 \qty{(p_x, \psi_x) :x\in[n], p_x \in[0,1],\sum_{x\in[n]} p_x = 1, \psi_x \in \rm{PURE}(A)}. 
\end{equation*}
Mapping the quantum state $\rho = \sum_{x\in[n]} p_x \psi_x$ associated with this ensemble, by a channel $\cal{E}$ must result identically to mapping the pure quantum state in the ensemble individually. We denote $\fk{L}(\fk{L}(A), \fk{L}(B))$ with $\fk{L}(A\to B)$.
\begin{definition}[Quantum channel]
 Quantum channel $\cal{E}^{A\to B}$ is a linear map $\fk{L}(A\to B)$, which is
    \begin{itemize}
        \item completely-positive (CP): for any Hilbert space $R$, an action of the channel on a positive operator $\rho^{RA} \geq 0$ results in a positive operator. That is 
        \begin{equation*}
 \id^{R} \otimes \cal{E}^{A\to B} (\rho^{RA}) \geq 0.
        \end{equation*}
        \item trace-preserving (TP): for any operator $\sigma^A\in\fk{L} (A)$, we have that 
        \begin{equation*}
 \Tr [\cal{E}^{A\to B} (\sigma^A)] = \Tr [\sigma^A]
        \end{equation*}
    \end{itemize}
\end{definition}
The set of all quantum channels in $\fk{L}(\fk{L}(A), \fk{L}(B))$ is denoted by $\cptp(A\to B)$. The question is how one could verify these properties? One of the approaches is to characterize it with the Choi-Jamiołkowski isomorphism, which is an isomorphism between maps of quantum channel types and matrices.

\subsection{Choi-Jamiołkowski isomorphism between channel and state}

\begin{definition}[A Choi matrix]
 Suppose $\cal{E}\in\cptp(A\to B)$ is a quantum channel. Suppose $\abs{A} = n$ and $\abs{B} = m$. A \emph{Choi matrix} denoted with $J_\cal{E} \in \bb{C}^{mn \times mn}$ is defined to be an application of the channel $\cal{E}$ on the unnormalized maximally entangled state, $\Phi ^{A\tilde{A}} = \sum_{i,j\in[n]} \ketbra{i,i}{j,j}$.
    \begin{equation*}
 J_\cal{E} ^{AB} = \id^{A\to A}\otimes\cal{E}^{\tilde{A}\to B}(\Phi^{A\tilde{A}}),
    \end{equation*}
 where $\id^{A\to A}$ is an identity channel.
\end{definition}

One can compute an action of channel $\cal{E}^{A\to B}$ using its Choi matrix $J_\cal{E}$, 
\begin{align*}
 \cal{E}^{A\to B}(\rho) &= \Tr_{A} [J_\cal{E} (\rho^{T}\otimes I)].
\end{align*}
To see this, consider the right side of the equation
\begin{align*}
 \Tr_{A} [J_\cal{E} (\rho^{T}\otimes I)] &=\Tr_A \bqty{\id^{A} \otimes \cal{E}^{\tilde{A}\to B} (\Phi^{A\tilde{A}}) (\rho^T \otimes I)}\\
    &=\Tr_A \bqty{\sum_{x,y} \rho^{T} \ketbra{x}{y}^A \otimes \cal{E}^{\tilde{A}\to B}(\ketbra{x}{y}^{\tilde{A}})}\\
    &=\sum_{x,y} \Tr[\rho^{T}\ketbra{x}{y}^A] \cal{E}^{\tilde{A}\to B}(\ketbra{x}{y}^{\tilde{A}})\\
    &=\sum_{x,y} \rho_{xy} \cal{E}(\ketbra{x}{y}) = \cal{E}(\rho).
\end{align*}
% \begin{lemma}
%     A Choi-Jamiołkowski isomorphism, denote with $\mathfrak{C}:\fk{L}(A\to B)\to \fk{L}(AB)$ is an isomorphism.
% \end{lemma}

\subsubsection{Characterizations of quantum channels}
\begin{theorem}\label{thm:choi-cp}
 A linear map $\cal{E}\in\mathfrak{L}(A\to B)$ is completely positive if and only if its Choi matrix $J_\cal{E} ^{AB}$ is positive semidefinite.
\end{theorem}
\begin{proof}
 Suppose that $\cal{E}$ is completely positive. Its Choi matrix is 
    \begin{align*}
 J^{AB} _\cal{E} &= \id^{A\to A}\otimes\cal{E}^{\tilde{A}\to B}(\Phi^{A\tilde{A}})\geq 0 
    \end{align*}
 where the positive semidefiniteness follows directly from the definition of completely positive.

 Conversely, if $J^{AB} _{\cal{E}} \geq 0$. Suppose $R$ is an auxiliary system and $\ket{\psi}^{RA}$ is a bipartite wavefunction on the product Hilbert space $RA$. Suppose $M\in\mathfrak{L}(\tilde{A},R)$ is a matrix such that 
    \begin{equation*}
 \ket{\psi}^{RA} = (M\otimes I) \ket{\Phi}^{\tilde{A}A}.
    \end{equation*}
 Furthermore, we define a linear map $\cal{M} \in \mathfrak{L}(A\to R)$ to be $\cal{M}(\rho^A) = M\rho M^*$.
 To check complete positivity, consider 
    \begin{align*}
 \id^{R\to R} \otimes \cal{E} (\ketbra{\psi}{\psi}^{RA}) &= (\id^{R\to R} \otimes \cal{E}) \circ (\cal{M}\otimes \id^{A\to A})(\Phi^{\tilde{A}A})\\
        &=  (\cal{M}\otimes \id^{B\to B}) \circ (\id^{R\to R} \otimes \cal{E}) (\Phi^{\tilde{A}A}) \\
        &= (\cal{M}\otimes \id^{B\to B}) J^{AB} _{\cal{E}}.
    \end{align*}
 Next is to show that $(\cal{M}\otimes \id^{B\to B}) J^{AB} _{\cal{E}}$ is positive semidefinite. For any $\psi \in RB$ consider 
    \begin{align*}
 \bra{\psi}(\cal{M}\otimes \id^{B\to B}) J^{AB} _{\cal{E}}\ket{\psi} &= \bra{\psi}(M\otimes I)J^{AB} _{\cal{E}}(M^*\otimes I)\ket{\psi} \\
        &= \bra{(M^*\otimes I) \psi} J^{AB} _{\cal{E}} \ket{(M^*\otimes I)\psi} \geq 0,
    \end{align*}
 since $ J^{AB} _{\cal{E}} \geq 0$ and $(M^*\otimes I) \psi^{RA} \in AB$.
\end{proof}

\begin{theorem}
 A linear map $\cal{E}\in\mathfrak{L}(A\to B)$ is trace-preserving if and only if the marginal of its Choi matrix $J_\cal{E} ^{A} = \Tr_B [J_\cal{E}]$ is an identity matrix on $A$. 
\end{theorem}
\begin{proof}
 Suppose that $\cal{E}$ is trace-preserving. Then its Choi matrix has a partial trace on $B$ as 
    \begin{align*}
 \Tr_{B} [J^{AB} _\cal{E}] &= \Tr_B [\id^{A\to A} \otimes \cal{E}^{\tilde{A} \to B} (\Phi^{A\tilde{A}})]\\
        &= \Tr_{B} [J^{AB} _\cal{E}] = \sum_{x,y\in[n]} \Tr[ \cal{E}^{\tilde{A}\to B} (\ketbra{x}{y})] \ketbra{x}{y} \\ 
        &= \sum_{x,y\in[n]} \Tr[\ketbra{x}{y}] \ketbra{x}{y} = \sum_{x,y\in[n]} \delta_{xy} \ketbra{x}{y} = I^{A},
    \end{align*}
 where $\delta_{xy}$ is equal to one if $x=y$ and is equal to zero otherwise and $\abs{A} = n$.

 Conversely, suppose $J^A _\cal{E} = I^A$ and $m = \abs{B}$. The trace of $\cal{E}(\rho)$ is 
    \begin{align*}
 \Tr[\cal{E}(\rho)] &= \Tr_B [\Tr_A[J^{AB} _\cal{E} (\rho^T \otimes I^B)]]\\
        &= \Tr[ \sum_{y\in[m]} (I^A \otimes \bra{y}^B) J^{AB} _\cal{E} (I^A \otimes \ket{y}^B)(\rho^{T})]\\ 
        &= \Tr[ \Tr_B[J^{AB} _\cal{E}] (\rho^T)] = \Tr [I^A \rho^T] = \Tr[ \rho],
    \end{align*}
 showing trace-preserving property of $\cal{E}$.
\end{proof}

\subsection{Notable examples of quantum channels}
\subsubsection{Isometry channel}
Suppose $V:\mathfrak{L}(A,B)$ is an isometry matrix. An isometry channel $\cal{V}$ is defined as 
\begin{equation*}
 \cal{V}(\rho^A) = V\rho^{A} V^*.
\end{equation*}
The map $\cal{V}$ is completely positive and trace-preserving. For complete positivity, notice that an isometry channel is a specific case of the channel $\cal{M}$ defined in the proof of Theorem~\ref{thm:choi-cp}.
For trace-preserving, it is straightforwardly
\begin{align*}
 \Tr [\cal{V}(\rho)] = \Tr[ V\rho V^*] = \Tr[ V^* V \rho] = \Tr[\rho].
\end{align*}

\subsubsection{Discarding a system}
Tracing a system out is completely positive and trace-preserving. An act of tracing out a specific system, $\Tr_A$ is a linear map in $\mathfrak{L}(A\to \bb{C})$. Its Choi matrix is 
\begin{equation*}
 J_{\rm{Tr}} ^A = \Tr_{\tilde{A}} [ \Phi^{A\tilde{A}}] = I^A \geq 0,
\end{equation*}
showing that $\Tr_{A}$ is completely-positive. For trace-preserving, it is trivially followed from $\Tr[\rho]$ giving exactly the trace of the operator $\rho$.

For the partial trace, its complete-positivity and trace-preserving follow directly from the complete-positivity and trace-preserving of the trace.

Any quantum channel $\cal{E}$ can be implemented with an isometry and the discarding of a physical system. This is due to the well-known Stinespring dilation theorem~\cite{Stinespring1955}.
\begin{theorem}[Stinespring dilation]\label{th:stinespring}
 A channel $\cal{E}\in\cptp(A\to B)$ can be expressed as a composition of an isometry channel $\cal{V}\in\cptp(A\to BR)$ and the discarding of an auxiliary system $R$.
\end{theorem}

We will follow the proof that is laid out in~\cite{Gour2024a}. This start with introducing an operator sum representation (Kraus representation) of a quantum channel~\cite{Kraus1983}.
\begin{theorem}[Krause representation]
 A quantum channel $ \cal{E}\in\cptp(A\to B) $ can be expressed as 
   \begin{equation*}
   \cal{E}(\rho) = \sum_{x\in[n]} M_x \rho M^* _x
   \end{equation*}
where $n = \abs{AB}$ and $\qty{M_x :M_x \in \mathfrak{L}(A\to B), x\in[n]}$ is a set of matrices.
\end{theorem}
\begin{proof}
 Since $\cal{E}$ is a quantum channel, its Choi matrix is positive semidefinite, hence Hermitian and diagonalizable, 
    \begin{equation*}
 J^{AB} _{\cal{E}} = \sum_{x \in [n]} \lambda_x \ketbra{x}{x}
    \end{equation*}
 where $\qty{x}$ is an orthonormal basis of $AB$ where the Choi matrix is diagonal. Absorbing the factor $\lambda_x$ into both vector $\ket{x}$ and $\bra{x}$, rename it $\ket{\phi_x}$ we have, 
    \begin{equation*}
 J^{AB} _{\cal{E}} = \sum_{x \in [n]} \ketbra{\phi_x}{\phi_x}.
    \end{equation*}
 Similarly to the case of a bipartite wavefunction, we can rewrite $\ket{\phi_x}$ as 
    \begin{equation*}
 \ket{\phi_x} = M_x \otimes I \ket{\Phi}^{A\tilde{A}}.
    \end{equation*}
 The action of the channel $\cal{E}$ is expressed in terms of the Choi matrix as 
    \begin{align*}
 \cal{E}(\rho) &= \sum_{x\in[n]} \Tr_A [ (I^A \otimes M_x) \Phi^{A\tilde{A}} (I^A \otimes M_x ^*) (\rho^{T} \otimes I)]
        \\
        &= \sum_{x\in[n]} \Tr_A [ (I^A \otimes M_x) \Phi^{A\tilde{A}} (\rho^{T} \otimes I) (I^A \otimes M_x ^*)] \\ 
        &= \sum_{x\in[n]} M_x \Tr_A [ \Phi^{A\tilde{A}} (I\otimes \rho)] M^* _x  = \sum_{x\in[n]} M_x \rho M_x ^*.
    \end{align*}  
\end{proof}
\begin{corollary}
 A linear map $\cal{E}\in\mathfrak{L}(A\to B)$ defined as 
    \begin{equation*}
 \cal{E}(\rho) = \sum_{x\in[n]} M_x \rho M^* _x
    \end{equation*}
 is trace-preserving if and only if $\sum_{x\in[n]} M_x ^* M_x = I$
\end{corollary}
\begin{proof}
 The map $\cal{E}$ is trace-preserving if and only if its Choi has a marginal be $I^A$ on $A$. That is,
    \begin{align*}
 I^A &= \Tr_B [J^{AB} _\cal{E}] = \Tr_B [\sum_{x\in[n]} (I\otimes M_x) \Phi^{A\tilde{A}} (I \otimes M^* _x)] \\ 
            &= \Tr_B [\sum_{x\in[n]} (M_x ^T \otimes  I) \Phi^{\tilde{B}B} ((M^* _x)^T \otimes I)] = \sum_{x\in[n]} M_x ^T (M^* _x) ^T = \sum_{x\in[n]} M_x ^* M_x.
    \end{align*}
\end{proof}

\begin{proof}[Proof of Theorem~\ref{th:stinespring}]
 Such isometry is 
    \begin{equation*}
 V = \sum_{x\in[n]} M_x \otimes \ket{x}^R 
    \end{equation*}
 where $\qty{\ket{x}^R}$ is an orthonormal basis for $R$. Notice that $R$ can be set to have dimension $\abs{AB}$. This is an isometry, because 
    \begin{align*}
 V^* V = \sum_{x,y\in[n]} M^* _y M_x \braket{y}{x} = \sum_{x\in[n]} M_x ^* M_x  = I.
    \end{align*}
    
 To see that this representation emulates the channel $\cal{E}$, consider
    \begin{align*}
 \Tr_R &\bqty{\pqty{\sum_{x\in[n]} M_x \otimes \ket{x}^R} \rho \pqty{\sum_{y\in[n]} M_y \otimes \ket{y}^R}} \\
        &= \sum_{x\in[n]} \Tr_R \bqty{M_x \rho M^* _y \otimes \ketbra{x}{y}^R}\\
        &= \sum_{x,y\in[n]} M_x \rho M^* _y \Tr[\ketbra{x}{y}] = \sum_{x\in[n]} M_x \rho M^* _x = \cal{E}(\rho).
    \end{align*}
\end{proof}

\subsubsection{Completely dephasing channel}
A completely dephasing channel $\cal{D}\in\cptp(A\to A)$ is a channel such that for all $\rho\in\mathfrak{D}(A)$,
\begin{equation*}
 \cal{D}(\rho) = \sum_{x\in[n]} \ketbra{x}{x} \bra{x}\rho\ket{x},
\end{equation*}
with respect to an orthogonal basis $\qty{\ket{x}:x\in[n]}$ and $n = \abs{A}$.
This channel, with respect to the basis, removes all off-diagonal elements of the input density matrix while keeping the diagonal elements intact.

\subsubsection{Quantum measurement}
A quantum measurement can be modeled with a quantum channel. Suppose that we have a physical system $A$ and once we perform a measurement, it puts a quantum state in the physical system $B$ and records the outcome in a classical memory $X$. For this process, a quantum channel performing this measurement is $\cal{E}\in\cptp(A\to BX)$, which has the form, 
\begin{equation*}
\cal{E}^{A\to BX} (\rho) = \sum_{x\in[n]} \cal{N}_x ^{A \to B} (\rho) \otimes \ketbra{x}{x}^X
\end{equation*}
where $\qty{\cal{N}_x ^A: x\in[n]}$ is a set of completely positive trace-non-increasing operations that sum to a map in $\cptp(A\to B)$.

\section{Classical channel}
A classical channel is a channel $\cal{N}$ that always output a classical state and treat an input classically. Specifically, a classical channel is invariant under the pre- and post-compositions with a totally dephasing channel $\cal{D}$ on a specific basis.
\begin{defbox}{Classical channel}
    \begin{definition}
 A channel $\cal{N}\in\rm{CPTP}(X\to Y)$ is a \emph{classical channel} if
        \begin{equation}
 \cal{N}=\cal{D}^Y  \circ \cal{N}\circ \cal{D}^X
        \end{equation}
    \end{definition}
\end{defbox}
As this thesis discusses mainly classical channels, we use probability vectors to represent classical states and transition matrices to represent classical channels. A vector in the standard basis $\*{e}_x = \ketbra{x}{x}$ represents a state of the definite outcome labeled with $x$. For example, a channel $\cal{N}^{X\to Y}$ is classical with respect to the choice of basis $\qty{\ket{\psi_i}^X}_i$ and $\qty{\ket{\phi_j}^Y}_j$. This channel has an associated transition matrix $N$ whose elements are defined from
\begin{equation}
N_{ij} = \Tr \bqty{\ketbra{\phi_j}{\phi_j} \cal{N}\pqty{\ketbra{\psi_i}{\psi_i}}}.
\end{equation}

\section{Quantum superchannel}
\begin{defbox}{A superchannel}
    \begin{definition}
 A \emph{superchannel}, $\Theta^{(A\to B)\to (A'\to B')}$, is a linear map between $\mathfrak{L}(\mathfrak{L}(A),\mathfrak{L}(B))$ to $\mathfrak{L}(\mathfrak{L}(A'),\mathfrak{L}(B'))$, denoted as $\mathfrak{L}((A\to B)\to (A'\to B'))$, with the following properties
        \begin{enumerate}
            \item 
 Completely completely-positive preserving: given any complete positive map $\cal{E}^{AR_0 \to BR_1}\in\rm{CP}(AR_0 \to BR_1)$, the map
            \begin{equation}
 \Theta^{(A\to B)\to (A'\to B')} \otimes \mathrm{1}^{(R_0\to R_1)} \qty[\cal{E}^{AR_0 \to BR_1}] 
            \end{equation}
 is completely positive, where $\mathrm{1}^{(R_0\to R_1)}$ is an identity supermap in $\mathfrak{L}((R_0\to R_1))$.
            \item Trace-preserving preserving: given any trace-preserving map $\cal{E}\in \rm{TP}(A\to B)$, the map
            \begin{equation}
 \Theta^{(A\to B)\to (A'\to B')}\bqty{\cal{E}^{AR_0 \to BR_1}}
            \end{equation}
 is trace-preserving.
        \end{enumerate}
    \end{definition}
\end{defbox}

A superchannel can be realized with quantum channels and quantum memory~\cite{CDP2008}. To see this, we first have to introduce the Choi isomorphism between superchannel and quantum channel. The proof of the realization theorem follows the previously appeared in~\cite{Gour2019}.

\subsection{Choi isomorphism and superchannel}
A Choi map associated with $\Theta$ is $\mJ_{\Theta}$ defined as 
\begin{equation}
 \mJ_{\Theta} = \Theta^{(C\to D)\to (A'\to B')}[\cal{Q}^{AC\to BD}]
\end{equation}
where $C\to D$ is a replica of $A\to B$ and $\cal{Q}$ is a maximally entangled maps defined as follows.
\begin{equation}
 \cal{Q}^{AC\to BD} = \sum_{\mu\in\fk{I}} \cal{E}^{A\to B} _\mu \otimes \cal{E}^{C\to D}_\mu
\end{equation}
where $\cal{E}_\mu ^{A\to B}$ is defined as 
\begin{equation}
\cal{E}^{A\to B} _\mu \pqty{\rho ^A} := \Tr\bqty{\rho^{A} \ketbra{x'}{x}^{A}}\ketbra{y}{y'}^B
\end{equation}
and $\mu = (x,x',y,y')$.

\begin{propbox}{Realization theorem}
\begin{theorem}\label{th:super-realization}
 Any superchannel $\Theta\in\rm{SUPER}((A\to B)\to (A'\to B'))$ can be realized by pre-processing channel, post-processing channel, and a quantum memory $R$. That is for any $\cal{N}\in\rm{CPTP}(A\to B)$, the action of the superchannel $\Theta$ is equal to
    \begin{equation}\label{eq:th-super-realization}
 \Theta[\cal{N}] = \cal{E}^{BR \to B'}\circ \cal{N}^{A\to B}\circ \cal{V}^{A' \to AR}
    \end{equation}
 where $\cal{E}$ and $\cal{F}$ are quantum channels.
\end{theorem}
\end{propbox}
To prove this theorem, we make use of the lemma, 
\begin{lemma}\label{lm:choi-super-F}
 A superchannel that has the form as in Equation~\eqref{eq:th-super-realization}. A map $\cal{F}\in\rm{CPTP}(A'B \to AB')$ defined as 
    \begin{equation}
 \cal{F}^{A'B \to AB'} = \cal{E}^{RB\to B'} \circ\cal{V}^{A' \to RA}
    \end{equation}
 has the same Choi matrix as the superchannel.
\end{lemma}
\begin{proof}
 A Choi map associated with $\Theta\in\super((A\to B) \to (A'\to B'))$ is $\mJ_{\Theta}\in\cptp(AA'\to BB')$ defined as 
    \begin{equation}
 \Theta^{(C\to D)\to (A'\to B')}[\cal{Q}^{AC\to BD}] = \cal{E}^{DR\to B'} \circ \cal{Q}^{AC\to BD}\circ \cal{V}^{A' \to CR}.
    \end{equation}
 Its Choi matrix is defined as 
    \begin{equation}
 J_\Theta ^{ABA'B'} = \cal{J}^{\tilde{A}\tilde{A'}\to BB'}_\Theta \pqty{\Omega^{A\tilde{A}}\otimes \Omega^{A'\tilde{A'}}}.
    \end{equation}
 Consider the action of the isometry channel $\cal{V}^{A'\to CR}$ on the maximally entangled state,
    \begin{equation}
 \cal{V}^{\tilde{A}'\to CR} \pqty{\Omega^{A\tilde{A}}\otimes\Omega^{A'\tilde{A'}}} = \cal{V}^{\tilde{C}\tilde{R}\to A'} _t \pqty{\Omega^{A\tilde{A}}\otimes\Omega^{C\tilde{C}}\otimes\Omega^{R\tilde{R}}}.
    \end{equation}
 Compose with $\cal{Q}^{\tilde{A}C\to BD}$,
    \begin{align}
 \cal{Q}^{\tilde{A}C\to BD} \circ \cal{V}^{A'\to CR} \pqty{\Omega^{A\tilde{A}}\otimes\Omega^{A'\tilde{A'}}} &= \cal{Q}^{\tilde{A}C\to BD}\circ \cal{V}^{\tilde{C}\tilde{R}\to A'} _t \pqty{\Omega^{A\tilde{A}}\otimes\Omega^{C\tilde{C}}\otimes\Omega^{R\tilde{R}}}\\
 \magenta{\pqty{\text{Choi of } \cal{Q}}\longrightarrow}&= \cal{V}^{\tilde{C}\tilde{R}\to A'} _t \pqty{\Omega^{BD}\otimes\Omega^{A\tilde{C}}\otimes\Omega^{R\tilde{R}}}\\
        &= \cal{V}^{\tilde{A'}\to CR}\pqty{\Omega^{AC}\otimes\Omega^{A'\tilde{A'}}}\\
 \cal{E}^{RD\to B'}\circ\cal{Q}^{\tilde{A}C\to BD} \circ \cal{V}^{A'\to CR} \pqty{\Omega^{A\tilde{A}}\otimes\Omega^{A'\tilde{A'}}} &= \cal{E}^{RD\to B'}\circ \cal{V}^{\tilde{A'}\to CR}\pqty{\Omega^{AC}\otimes\Omega^{A'\tilde{A'}}}
    \end{align}
 The last equality shows that the Choi of superchannel and $\cal{F}$ are identical. Since Choi's isomorphism is an isomorphism.
\end{proof}

Now we can prove the theorem.
\begin{proof}[Proof of the Theorem~\ref{th:super-realization}]
 Suppose that we have a Choi matrix $\*{J}^\bb{AB}_{\Theta}$ of the superchannel $\Theta$, we want to shows that we can define $\cal{E}\in\rm{CPTP}(RA_1 \to B_1)$ and $\cal{V}\in\rm{CPTP}(B_0 \to RA_0)$ such that $\cal{F}:= \cal{E}\circ \cal{V}$ has the same Choi matrix as $\Theta$. This is so that Lemma~\ref{lm:choi-super-F} implies $\Theta$ is in the form in Equation~\ref{eq:th-super-realization}. 

 To begin with, we define $\omega^{\bb{AB}}:= \*{J}^\bb{AB}_{\Theta}/\abs{A_0 A_1 B_0 B_1}$ to be a normalization of the Choi matrix of the superchannel. We pick the following
    \begin{enumerate}
    \item a system $E$ such that it allows us to define $\psi^{\bb{AB}E}$ to be a purification of $\omega^{\bb{AB}}$.
    \item a system $R$ such that it allows us to define $\phi^{A_0 B_0 R}$ to be a purification of the marginal omega $\omega^{A_0B_0}$.
    \end{enumerate} 
 where $\omega^{A_0B_0}:= \Tr_{A_1B_1} [\omega^{\bb{AB}}]$ is one of its marginals.  With the condition that $\Theta$ is a superchannel, then it is TP-preserving, which is equivalent to 
    \begin{equation}\label{eq:TPP-equiv}
 \Tr_{B_0}[\*{J}^{\bb{A}B_0}_\Theta] = \*{J}^{A_0 B_0}_\Theta \otimes \*{u}^{A_1} \quad \rm{and}\quad \Tr_{A_0 A_1 B_1} [\*{J}^{\bb{AB}}_{\Theta}] = \abs{A_1} I^{B_0}.
    \end{equation}
 We make use of the first equality and use purification with $R$, 
    \begin{align}
 \omega^{\bb{A}B_0} = \omega^{A_0 B_0}\otimes \*{u}^{A_1} \xrightarrow{\rm{Purification}} \phi^{A_0 B_0 R}\otimes \Phi^{A_1 \tilde{A}_1}.
    \end{align}
 Another purification of $\omega^{\bb{AB}}$ is the state $\psi^{\bb{AB}E}$. The two purifications relate to one another by an isometry $\cal{U}$,
    \begin{equation}
 \psi^{\bb{AB}E} = ~\cal{U}^{R\tilde{A}_1\to EB_1} \pqty{\phi^{A_0B_0R}\otimes \Phi^{A_1 \tilde{A}_1}}.
    \end{equation}
 From the above equation, tracing out the auxiliary system $E$ results in 
    \begin{equation}
 \omega^{\bb{AB}} = \underbrace{\Tr_E \circ~\cal{U}^{R\tilde{A}_1\to EB_1}}_{=:\cal{E}^{R\tilde{A}_1\to B_1}} \pqty{\phi^{A_0B_0R}\otimes \Phi^{A_1 \tilde{A}_1}}
    \end{equation}
 By the second equality in Equation~\eqref{eq:TPP-equiv}, we have that tracing out $\phi^{A_0B_0R}$ over $A_0 R$ results in a uniform vector $\*{u}$. Purification of the state $\*{u}$ with a copy of $B_0$ is a maximally entangled state, which related to $\phi^{A_0B_0R}$ by an isometry$\cal{V}^{\tilde{B}_0\to A_0 R}$
    \begin{equation}
 \phi^{A_0B_0R} = \cal{V}^{\tilde{B}_0 \to A_0 R} (\Phi^{B_0\tilde{B}_0}).
    \end{equation}
 All of these together, we have 
    \begin{equation}
 \omega^{\bb{AB}} = \cal{E}^{R\tilde{A}_1\to B_1}\circ \cal{V}^{\tilde{B}_0 \to A_0 R} \pqty{\Phi^{B_0\tilde{B}_0}\otimes \Phi^{A_1 \tilde{A}_1}}.
    \end{equation}
 The right-hand side is the Choi matrix of the superchannel of the form as in Equation~\eqref{eq:th-super-realization}, by Lemma~\ref{lm:choi-super-F}. As the Choi isomorphism is an isomorphism, the superchannel $\Theta$ can be realized as such.
\end{proof}

% \tableofcontents
\setcounter{chapter}{1}
\chapter{Uncertainty and majorizations of classical states}
In this chapter, we provide a review of a mathematical conceptualization of the uncertainty inherence in a classical state, which later on will be applicable to a quantum state. The chapter begin with the main tool termed \emph{Majorization} which developed initially in the early 20th century by Hardy, Littlewood and P\'olya \cite{HLP1934} to study inequality of wealth distribution. Later on, the concept found its way into the study of quantum resource theories as a comparison tools of uncertainty.

In this chapter, we provide an introduction to three variants of majorization relation on the domain of probability vectors: majorization, relative majorization, and conditional majorization. The majorization, or the standard relation, defines concretely what is an uncertainty inherent in a random variable $X$ with a specific probability distribution. The conditional majorization compares and defines the notion of uncertainty of a random variable $X$ similarly to the standard probability vector majorization, but with a conditioning on accessing information of $Y$, a potentially non-independent random variable.  Lastly, the relative majorization is a majorization comparing distinguishability between probability distributions. That is, how likely one can tell two probability distributions apart by just learning their outcomes.

\section{Majorization}
Majorization compares two distributions, e.g. probability distributions or wealth distribution of two nations. We can represent a distribution among the finite number of parties, e.g. random variable outcomes or individuals in a nation, with a vector.
An intuitive way to compare the concentration of the distribution is to compare the sums of the first $k$ the largest weight of each distribution. For example, in case of a nation wealth, it could be a sum of the top $k\%$. Mathematically, the sum of the first $k$ absolutely the largest elements of a vector is a norm.
\begin{defbox}{The Ky Fan $k$-norm for vectors}
\begin{definition}\label{def:KyFanNorm}
    Given a vector $\vctr\in\bb{R}^n$ and $k$ is a positive integer no greater than $n$. The \emph{Ky Fan $k$-norm} of the vector $\vctr$ is
    \begin{equation}
        \knorm{\vctr}{k} = \sum_{i=0} ^{k} \abs{r_i^\da},
    \end{equation}
    where $\vctr^\da = \pqty{r_1 ^\da, r_2 ^\da, \ldots, r_n ^\da}^T$ is a rearrangement of the elements of $\vctr$ such that $\abs{r_1 ^\da} \geq \abs{r_2 ^\da} \geq \ldots \geq \abs{r_n ^\da}$.
\end{definition}
\end{defbox}

% \begin{propbox}{}
\begin{theorem}
    The Ky Fan $k$-norm is a norm. For any $\alpha \in \bb{R}$, $\vctr,\s\in\bb{R}^n$, and $n$ is a positive integer the followings hold.
    \begin{enumerate}
        \item $\knorm{\vctr}{k}>0$ except when $\vctr = 0$, $\knorm{\vctr}{k}=0$.
        \item $\knorm{\alpha\vctr}{k} = \abs{\alpha} \knorm{\vctr}{k}$.
        \item $\knorm{\vctr + \s}{k} \leq \knorm{\vctr}{k}+\knorm{\s}{k}$
    \end{enumerate}
\end{theorem}
% \end{propbox}
\begin{proof}
    1. and 2. follow directly from the definition. To see the inequality in 3., consider 
    \begin{align*}
        \knorm{\vctr + \s}{k} &= \sum_{x\in[k]} \abs{(\vctr+\s)^\da _x} = \max \qty{\abs{\vctr+\s}\cdot\t~\big\vert~\t\in\qty{0,1}^n, ~ \norm{\*{t}}_1 = k} \\
        \magenta{\pqty{\abs{\vctr+\s}\leq\abs{\vctr}+\abs{\s}}\rightarrow}&\leq \max \qty{\abs{\vctr}\cdot\t_1+\abs{\s}\cdot\t_2~\big\vert~\t_1=\t_2 \in\qty{0,1}^n, ~ \norm{\t_1}_1=\norm{\t_2}_1 = k} \\
        \magenta{\pqty{\text{being~superset}}\rightarrow}&\leq \max \qty{\abs{\vctr}\cdot\t_1+\abs{\s}\cdot\t_2~\big\vert~\t_1,\t_2\in\qty{0,1}^n, ~ \norm{\t_1}_1=\norm{\t_2}_1 = k} \\
        &= \max \qty{\abs{\vctr}\cdot \t_1 ~\big\vert~\t_1 \in\qty{0,1}^n, ~ \norm{\t_1}_1=k}\\ 
        &\qquad +\max \qty{\abs{\s}\cdot \t_2 ~\big\vert~\t_2 \in\qty{0,1}^n, ~ \norm{\t_2}_1=k}\\
        &= \knorm{\vctr}{k} +\knorm{\s}{k},
    \end{align*}
    showing that we have triangle inequality. 
    %The second to last inequality is from the maximum of the sum of two positive numbers is the sum of the maximum of the two positive numbers.
\end{proof}

Majorization $(\succ)$ is a preorder on $\bb{R}^n$ defined from the $k^\rm{th}$ Ky Fan norm of the vector. 
\begin{defbox}{Majorization}
    \begin{definition}
        Suppose $\*{r},\*{s}\in \bb{R}^{n}$. we say $\vctr$ \emph{majorizes} $\s$ and write $\*{r} \succ \*{s}$ if 
        \begin{equation}\label{eq:def-major-knorm}
            \norm{\*{r}}_{(k)} \geq \norm{\*{s}}_{(k)} \quad \forall k\in [n-1]
        \end{equation}
        with the equality at $k=n$. Moreover, if $\vctr\succ\*{s}$ and $\*{s}\succ\vctr$ we said that both vectors are equivalent under majorization relation and write $\*{r}\sim\*{s}$.
    \end{definition}
\end{defbox}

\begin{propbox}{}
    \begin{theorem}
        Majorization, $\succ$, is a preorder. It satisfies
        \begin{itemize}
            \item Transitivity: for any $\vctr ,\s ,\t\in\bb{R}^n$, we have $\vctr \succ \s$ and $\s \succ \t$ imply $\vctr \succ \t$.
            \item Reflexivity: $\vctr \succ \vctr$ for any $\vctr \in \bb{R}^n$.
        \end{itemize}
    \end{theorem}
\end{propbox}
\begin{remark}
    Majorization is not a total preorder, i.e. there are non-majorizing pairs of vectors. For example $\vctr = \pqty{70,15,15}^T$ and $\s = \pqty{50,45,5}^T$, both vectors have equal Ky Fan $3$-norm but $\knorm{\vctr}{1}>\knorm{\s}{1}$ and $\knorm{\vctr}{2}<\knorm{\s}{2}$.
\end{remark}
\begin{proof}
For transitivity, $\vctr\succ\s$ implies $\knorm{\vctr}{k} \geq \knorm{\s}{k}$ for all $k$. Similarly, with the pair $\s$ and $\t$. By transitivity of the $\geq$ total order, we have $\knorm{\vctr}{k}\geq \knorm{\s}{k}$ for any $k\in[n]$ as well. Reflexivity follows from $\knorm{\vctr}{k}$ equals to itself.
\end{proof}

\section{Probability vector majorization}
The majorization relation is defined on the domain of any real vectors. In this thesis, we focus specifically on the probability vectors, as it is associated with a probability distribution. Given a probability distribution of a random variable $X: \Omega \to \bb{R}$, if the image of the sample space $\Omega$ under $X$ is a finite set, then we can assign each outcome a label by number $1,2,\ldots,n$ and arrange them into a vector. Mathematically, probability vector is defined as follows. 
\begin{defbox}{Probability vector}
    \begin{definition}
        A vector $\p \in\bb{R}^n$ is said to be an \emph{$n$-dimensional probability vector} and write $\p\in\prob(n)$ if its elements are non-negative, and they sum to 1. That is 
        \begin{equation*}
            \p\in\prob(n) \quad :\iff \quad \sum_{i\in[n]} p_i =1 \quad\text{and}\quad p_i \geq 0 \quad \forall i \in[n].
        \end{equation*}
    \end{definition}
\end{defbox}
Restricting to the subset $\rm{Prob}(n) \subset \bb{R}^n$ defines a preorder on probability vectors.
The following is an example of two vectors that majorized one another.
\begin{equation*}
    \*{p} = \pmqty{1 & 0 & 0}^T, \quad \*{q} = \frac{1}{3} \pmqty{1 & 1 & 1}^T, \quad \*{p} \succ \*{q}
\end{equation*}
To motivate why the definition of majorization captures uncertainty inherent in the probability distribution, we follow similar characterizations in~\cite{FGG2013} and~\cite{BGG2022}. The majorization between probability vectors can be characterized from three conceptually distinct perspectives: constructive, axiomatic, and operational.

\subsection{Characterizations of majorization with uncertainty}
From constructive and axiomatic perspectives, we want to characterize majorization as an ability to transform a probability distribution without decreasing its uncertainty. That is, we will motivate two classes of transformation $M$ to be classes of uncertainty non-increasing operations: one in axiomatic and another in constructive ways. As we will see shortly, these two approaches reach the same class of stochastic matrices—giving a solid notion of uncertainty comparison.

In defining a set of mixing operations in both constructive and axiomatic characterization, the mixing operations need to be linear, i.e.
\begin{equation}\label{eq:prob-interpretation-vector}
    M\qty(\sum_{x\in[m]} t_x \*{p}_x) = \sum_{x\in[m]} t_x  M(\*{p}_x)
\end{equation}
where $\*{t} = (t_x) \in \rm{Prob}(m)$. The requirement of linearity is from a probabilistic interpretation of probability distribution. The left-hand side of the above equation is application of the mixing $M$ to a probabilistic ensemble of probability distributions. The linearity minimally implies that applying $M$ to the ensemble has to be equivalent to applying $M$ to each individual distribution and mixed into probabilistic ensemble. In addition, $M$ must map a probability vector to another probability vector, this requires $M$ to be a column stochastic matrix, $M \in \rm{Stoch}(n,n)$. Later on whenever we say stochastic matrix we mean a column stochastic matrix. 

\textbf{Constructive characterization.} Majorization relation between probability vector can be characterized as an ability to convert probability vector with a random permutation matrix. Consider a stochastic matrix $M$ representing an operation of randomly permute the outcomes. The intuition behind this is that relabelling of the outcomes do not change the level of uncertainty. However, by randomly relabelling, one increase the uncertainty of the distribution. That is $M$ is a convex combination of a permutation matrix $\Pi_y$ where $y$ denotes the permutation scheme,
\begin{equation}
    M = \sum_{y\in[n!]} s_y \Pi_y
\end{equation}
for some $\*{s} \in \rm{Prob}(n!)$. Suppose $\p$ and $\q$ are probability vectors. If $\p$ can be randomly permuted into $\q$, then the distribution associated with $\q$ is at least as uncertain as $\p$.

\textbf{Axiomatic characterization.}
Two majorization-comparable probability vectors can be characterized with convertibility via a doubly stochastic matrix. One would expect an uncertainty non-decreasing operation to preserve the most uncertain distributions for a dimension $n$, which is a uniform distribution $\*{u}^{(n)} \in \rm{prob}(n)$. 
From this perspective, $M$ is defined to be a column stochastic matrix which preserve $\*{u}^{(n)} \in \rm{prob}(n)$ an $n$-dimension uniform vector. This set of matrix is known as a doubly-stochastic matrix.
\begin{lemma}
    A stochastic matrix $M\in\stoch(n,n)$ satisfies $M\u^{(n)} =\u^{(n)}$ if and only if it is a doubly stochastic matrix.
\end{lemma}
\begin{proof}
    A doubly stochastic matrix is a stochastic matrix which the sum of elements in each row is equal to one. That is we want to show that $\sum_{j\in[n]} M_{ij} = 1$ for any column $i\in[n]$.
    \begin{align*}
        M\u^{(n)} = \u^{(n)} &\implies  \sum_{i\in[n]} \sum_{j\in[m]} M_{ij} \frac{1}{n} \*{e}_i = \sum_{i\in[n]} \frac{1}{n} \*{e}_i\\
        &\implies \sum_{j\in[n]} M_{ij} = 1 \implies M_{ij} \in \dstoch(n).
    \end{align*}
\end{proof}

\textbf{Operational interpretation.}
Given two probability vectors $\*{p}$ and $\*{q}$ representing two probability distributions, a $k$-game is a game of chance where a player gives $k$ guesses to a sample draw from a probability distribution. The player of this game has full knowledge of the probability distributions. The winning strategy is to guess $k$ most likely outcomes. The winning chance with a distribution $\*{p}$ is 
\begin{equation}
    \rm{Pr}_{k} (\*{p}) = \knorm{\*{p}}{k}
\end{equation}
If a probability vector $\p$ has a higher chance of winning a $k$-game for any $k\in[n]$, then for any $k\in[n]$, $\rm{Pr}_{k} (\*{p}) \geq \rm{Pr}_{k} (\*{q})$ or equivalently, $\knorm{\*{p}}{k}\geq \knorm{\*{q}}{k}$, 
which is exact the same as Equation~\eqref{eq:def-major-knorm}. 

\begin{propbox}{Characterization of majorization with uncertainty}
    \begin{theorem}
        The following statements are equivalent
        \begin{enumerate}
            \item There exists $M$ a random permutation matrix transforming $\p$ to $\q$
            \item There exists $M$ a doubly stochastic matrix transforming $\p$ to $\q$.
            \item For any $k$, $\rm{Pr}_{k} (\p) \geq \rm{Pr}_{k} \pqty{\q}$.
        \end{enumerate}
    \end{theorem}
\end{propbox}
The direction $1. \implies 2.$ is given by the Birkhoff--von Neumann theorem~\cite{Birkhoff1946}. The direction $2. \implies 3.$, is given by a stochastic matrix being a contraction on the Ky Fan $k$-norms. 
\begin{proof}[Proof of 2. $\implies$ 3.]
    Suppose that $\*{q} = M \*{p}$ where $M$ is a doubly stochastic matrix, we want to show $\knorm{\*{p}}{k} \geq \knorm{\*{q}}{k}$ for all $k\in[n]$. We make use of the Birkhoff--von Neumann theorem to express $M$ as a convex sum of permutation matrices,
    \begin{equation}
        M = \sum_{w\in[n!]} t_w \Pi_w, \quad \sum_{w\in[n!]} t_w = 1, \textrm{ and} \quad t_w \geq 1 \quad \forall w\in[n!].
    \end{equation}
    Now consider the following with any $k\in[n!]$,
    \begin{align}
        \rm{Pr}_k (\q) = \knorm{\*{q}}{k} &= \knorm{\sum_{w\in[n!]} t_w \Pi_w \*{p}}{k} \leq \sum_{w\in[n!]} t_w  \knorm{\Pi_w \*{p} }{k} = \knorm{\*{p}}{k} = \rm{Pr}_k (\p).
    \end{align}
\end{proof}
We will use $T$-transform to prove the direction $3. \implies 1.$.
\begin{defbox}{$T$-transform}
\begin{definition}
    A square matrix $T\in\bb{R}^{n\times n}$ is said to be a \emph{$T$-transform} if it can be put in the form 
    \begin{equation}
    T = t I_n + (1-t) \Pi
    \end{equation}
    where $t\in[0,1]$ and $\Pi$ is a permutation matrix swapping only two elements.
\end{definition}
\end{defbox}
There are two remarks. First, any action of a $T$-transform on a probability vector $\p$ will result in a new probability vector, denoted with $\vctr$. The vector $\vctr$ has elements $r_z$, and it always exists two distinct indices $x,y$ such that
\begin{equation}
r_z = \begin{cases}
    p_z &(z\neq x \text{ and } z\neq y)\\
    t p_x + (1-t) p_y  &(z=x)\\
    t p_y + (1-t) p_x  &(z=y)
\end{cases}.
\end{equation}
Secondly, the resultant vector is always majorized by the original vector. That is $\vctr= T\p$ implies $\p\succ \vctr$. This can be verified by expressing the $T$, Ky Fan norm, and exploiting the triangle inequality.
\begin{align}
    \forall k\in[n]\qquad\knorm{\vctr}{k} = \knorm{t\p + (1-t)\Pi\p}{k} \leq t\knorm{\p}{k} + (1-t)\knorm{\Pi \p}{k} = \knorm{\p}{k}.
\end{align}

\begin{lemma}\label{lm:T-transform_majorization}
$\*{p}\succ\*{q}$ if and only if there exists finite number of $T$-transforms such that $\q = T_n \ldots T_2 T_1 \p$.
\end{lemma}
\begin{proof}
In the only-if direction, since $\vctr= T\p$ implies $\p\succ \vctr$ for any $\p\in\prob(n)$, therefore the result of applying $T_n \ldots T_2 T_1$ on $\p$ is majorized by $\p$.

Conversely, suppose $\p\succ\q$. Assume $\p$ and $\q$ are in non-increasing order. If it is not, then first apply $T$-transform to permute $\p$ to its non-increasing order $\p^{\da}$ and similarly with $\q$. Furthermore, assume $\p\neq \q$, otherwise $T$-transform is an identity and the proof is trivial. We define each $T$-transform such that the resultant vector $\vctr$ satisfies the followings.
\begin{enumerate}
    \item $\p\succ \vctr\succ\q$
    \item There are distinct $x$ and $y$ such that $r_x\neq p_x$ and $r_y\neq p_y$.
    \item $r_x = q_x$ or $r_y = q_y$.
\end{enumerate}
Such $T$-transform can be defined by $T = tI + (1-t) \Pi$ when $\Pi$ swap $x,y$ and $t$, $x$, and $y$ are defined as follows. $x$ is defined to be the largest index such that $p_x > q_x$. Since $\p \succ \q$ and $\p\neq\q$, such $x$ always exists. $y$ is defined to be the smallest index such that $p_y < q_y$. Such $y$ always exists since the probability vector must have a unit 1-norm and there exists $p_x > q_x$. Lastly, $t$ is defined as 
\begin{align*}
    t &= 1 - \frac{\varepsilon}{p_x - p_y} \\
    \varepsilon &= \min \qty{p_x - q_x, q_y - p_y}.
\end{align*}
To show that $x,y,$ and $t$ result in $\vctr$ that satisfies 1., 2., and 3., consider the elements $r_x$ and $r_y$ 
\begin{align*}
r_x &= tp_x + (1-t) p_y = p_y + t(p_x -p_y)\\
r_y &= tp_y + (1-t) p_x = p_x + t(p_y -p_x).
\end{align*}
If $\varepsilon = p_x - q_x$, then $t = 1 - \frac{p_x - q_x}{p_x - p_y} = \frac{q_x - p_y}{p_x - p_y}$ and 
$$r_x = p_y +\pqty{\frac{q_x - p_y}{p_x - p_y}} \pqty{p_x - p_y} = q_x.$$ 
Otherwise, $\varepsilon = p_y - q_y$, then 
$$r_y = p_x +\pqty{\frac{p_x - q_y}{p_x - p_y}} \pqty{p_y - p_x} = q_y.$$
Both of the two cases satisfy 3. To show 2., since $p_x > q_x$ and $ p_y < q_y$, either $r_x = q_x$ or $r_y = q_y$, and the $T$-transform changes two elements, therefore $p_x\neq r_x$ and $p_y \neq r_y$. Note here that $x>y$ as implied by the majorization relation. For the majorization condition of $\vctr\succ \q$, consider 
\begin{align*}
    &\text{For $1\leq k < x$}, \quad \knorm{\vctr}{k} = \knorm{\vctr}{k} \geq \knorm{\q}{k}.\\
    &\text{For $x\leq k < y$}, \quad \knorm{\vctr}{k} \geq \knorm{\p}{x-1} + \underbrace{r_x}_{r_x \geq q_x} + \sum_{w=x+1} ^{k} \underbrace{p_w}_{=q_w} \geq \knorm{\q}{k},
    % &\text{For $x\leq k < y$}, \quad \knorm{\vctr}{k} \geq \knorm{\p}{x-1} + \underbrace{r_x}_{r_x \geq q_x} + \sum_{w=x+1} ^{k} \underbrace{p_w}_{\substack{=q_w\\\forall w\in\qty{x+1, \ldots, y-1, y}}} \geq \knorm{\q}{k},
\end{align*}
for $y \leq k$; use $r_x + r_y = p_x + p_y$, we have that $\knorm{\vctr}{k} = \knorm{\p}{k}$. 

The requirement number 3. implies that after each application of the $T$-transform, one will have a new probability vector $r$ which has one more elements identical to the element $q_x$. Repeating this process over a certain finite iterations results in the vector $\q$. This completes the proof.
\end{proof}
\begin{proof}[Proof of 3. $\implies$ 1.]
    By Lemma~\ref{lm:T-transform_majorization} and $T$-transform being a doubly-stochastic matrix, we have that $M = T_n \ldots T_2 T_1 \in\dstoch(n)$.
\end{proof}
\subsection{Comparison between probability vector of different dimensions}
\begin{definition}
    Suppose $\p\in\prob(n)$ and $\q\in\prob(m)$ where $n\leq m$. Define $\ell = m-n$, we say $\p\succ \q$ if $\p \oplus \*{0}^{(\ell)} \succ \q$. Conversely, we say $\q \succ \p$ if $\q \succ \p \oplus \*{0}^{(\ell)}$.
\end{definition}
\begin{remark}
    Since we would like to capture uncertainty inherit in a distribution, the comparison should be invariant under the removal of events of zero probability. With this extension, we have that for any $n,\ell\in\bb{N}$ and any $\p\in\prob(n)$,
    \begin{equation}
        \p \sim \p \oplus \*{0}^{(\ell)}.
    \end{equation} 
    That is, adding or removing impossible outcomes into consideration do not affect the uncertainty as captured by majorization relation. 
    % A reader with resource theories background may wonder why the extension of $\p$ into a higher dimension is via appending zeroes instead of tensoring with a state $\*{u}^{(n)}$ for some $n\in\bb{N}$. The definition of majorization compares the Ky Fan $k$-norm of two probability vectors. This extension give us 
    % \begin{equation*}
    %     \knorm{\p}{k} = \knorm{\p\oplus\*{0}^{(\ell)} }{k} \quad \forall k\in[n]
    % \end{equation*}
    % while the similar equality for $\knorm{\p\otimes \*{u}^{(n')}}{k}$ does not hold in general,
    % \begin{equation}
    %     \knorm{\p}{k}  \neq \knorm{\p\otimes \*{u}^{(n')}}{k} \quad \forall k\in[n].
    % \end{equation}
\end{remark}

\subsection{Majorization and entanglement theory}
Majorization can be extended to the domain of quantum states and provide a useful characterization of entanglement resources. Suppose $\rho,\sigma\in\frak{D}(A)$, the majorization relation compares the vectors form by eigenvalues of $\rho$ and $\sigma$, denoted by $\lambda_\rho$ and $\lambda_\sigma$ respectively.
\begin{definition}
    Suppose $\rho,\sigma\in\frak{D}(A)$, we say that the state $\rho$ majorizes the state $\sigma$ and write 
    $\rho \succ \sigma$ if $\lambda_\rho \succ \lambda_\sigma$ where $\lambda_\rho$ and $\lambda_\sigma$ are vectors of eigenvalues of $\rho$ and $\sigma$ respectively.
\end{definition}
When $\rho$ and $\sigma$ are classical, the majorizations, classical and quantum, coincide. This definition is equivalent to the existence of a \emph{unital} channel transforming one quantum state to another.
% \begin{propbox}{Equivalent statement for quantum majorization}
    \begin{lemma}
        Given $\rho,\sigma\in\fk{D}(A)$, $\rho \succ \sigma$ if and only if there exists a unital channel $\cal{E}\in\rm{CPTP}(A\to A)$ such that $\sigma = \cal{E}(\rho)$.
    \end{lemma}
% \end{propbox}
\begin{proof}
    Suppose $\rho\succ\sigma$. Then, there exists a doubly stochastic matrix $D=(d_{i,j})_{i,j}$ such that $\lambda_\sigma = D \lambda_\rho$. Suppose $\rho$ and $\sigma$ are diagonal in the basis $\qty{\psi_i : i \in[n]}$  and $\qty{\phi_i : i \in[n]}$ respectively. We define a matrix $W$ to be 
    \begin{equation*}
        W = \sum_{i,j} d_{ij} \ketbra{\phi_i}{\psi_j}.
    \end{equation*}
    A channel $\cal{U}$ defined as 
    \begin{equation*}
        \cal{U}(\rho) = W \rho W^*
    \end{equation*}
    is a unital channel transforming $\rho \mapsto \sigma$.

    Conversely, suppose that there exists a unital channel $\cal{U}$ mapping $\rho$ to $\sigma$. Define
    \begin{equation*}
        D = \sum_{i,j\in[n]} \bra{\phi_j} \cal{U}(\ketbra{\psi_i}{\psi_i}) \ket{\phi_j} \*{e}_j\*{e}_i ^T .
    \end{equation*}
    The matrix $D$ is doubly stochastic and maps $\lambda_\rho \mapsto \lambda_\sigma$.
\end{proof}

\begin{remark}
    There is no constructive approach for quantum state majorization. Not all unital channel can be written as a convex sum of unitary channels.
\end{remark}
One of the most famous application of majorization relation in the field of quantum information theorem is Nielsen majorization theorem~\cite{Nielsen1999, Gour2024a}. The theorem pertains to convertibility relation of a bipartite quantum state using only \emph{local operations and classical communications} (LOCC). The LOCC condition captures a physical constraint where a quantum state is spatially separated, e.g. two photons are in two different labs. The only allowed operation is the operation where the experimentalists perform operation on one side of the system and communicating with another lab with only classical communications, e.g. electrical signal, text messages, phone calls.
\begin{propbox}{Nielsen majorization theorem}
    \begin{theorem}
        Let $\psi, \phi \in \operatorname{Pure}(A B)$ be two bipartite quantum states, and let $\rho^A:=\rm{Tr}_B\left[\psi^{A B}\right]$ and $\sigma^A:=\operatorname{Tr}_B\left[\phi^{A B}\right]$ be their corresponding reduced density matrices. Then, $\psi$ is convertible to $\phi$ via LOCC if and only if $\sigma^A \succ \rho^{B}$.
    \end{theorem}
\end{propbox}

\section{Monotones of probability vector majorization}
In this section, we discuss the conditions and characteristics of a family of function which behave monotonically under majorization relation. First, we recite the definition of Schur-convex function on a subset $A$ of $\bb{R}^n$~\cite{MOA2011}.
\begin{defbox}{Schur-convex function}
    \begin{definition}
        A function $f:A\subseteq \bb{R}^n \to \bb{R}$ is \emph{Schur-convex} on $A$ if $\p\succ\q$ implies $f(\p)\geq f(\q)$ for any pairs $\p,\q\in A$. A function $f$ is Schur-concave if its negative $-f$ is Schur-convex.
    \end{definition}
\end{defbox}
The merit of this definition is in its generality. The function need not be defined all on $\bb{R}^n$ but only on a subset $A$. In the interest of this thesis, Schur-convex functions are usually defined on a set of non-negative $n$-dimensional vectors $\bb{R}^n _{+}$.

One of the immediate necessary condition of $f$ to be Schur-convex is that $f$ need to be invariant under permutation of its inputs. That is if $A$ is symmetric set, which mean invariant under permutation, then $f(\p) = f(\Pi \p)$ for all $\p\in A$ and $\Pi\in\perm(n)$. This is the case because $\p\succ\Pi\p$ and $\Pi\p\succ\p$ as well. This together with another condition give us a sufficient condition for $f$ to be Schur-convex.

\begin{propbox}{}
\begin{theorem}
    If $f$ is symmetric and convex then it is Schur-convex.
\end{theorem}
\end{propbox}
\begin{proof}
    Suppose $\p\succ\q$. Then, there exists a doubly stochastic matrix $D \in\dstoch(n)$ such that $\q = D\p$. Therefore, we have the following chain of inequalities,
    \begin{align*}
    f(\q) &= f(D\q) = f\pqty{\sum_{i\in[n!]} t_i \Pi_i \p} \quad (t_i \geq 0, \sum_{i\in[n!]} t_i = 1)\\
        &\leq \sum_{i\in[n!]} t_i f\pqty{\Pi_i \p} \\
        &=  \sum_{i\in[n!]} t_i f(\p) = f(\p),
    \end{align*}
    showing that $f$ is Schur-convex.
\end{proof}

Note that the converse is not true. If we have Schur-convexity, need not the function be convex. One of the examples is min-entropy, which will be introduced in the next section.

The sufficient and necessary condition of a continuously differentiable function to be Schur-convex is succinctly stated by the following theorem called Schur test.
\begin{propbox}{Schur test}
\begin{theorem}
    Suppose $f:A\subseteq \bb{R}^n \to \bb{R}$ where $\prob(n)\subseteq A \subseteq \bb{R}^n$ and the relative interior of $\prob(n)$ is in the interior of $A$. Suppose $f$ is continuous on $A$ and continuously differentiable on the interior of $A$. $f$ is Schur-convex on $\prob(n)$ if and only if 
    \begin{enumerate}
        \item $f$ is symmetric.  
        \item $f$ satisfies \emph{Schur condition}. That is for any $\p$ in the relative interior of $\prob(n)$ and $i,j\in[n]$
        \begin{equation}
            \pqty{p_i - p_j}\pqty{\pdv{f(\p)}{p_i} -\pdv{f(\p)}{p_j} } \geq 0.
        \end{equation}
    \end{enumerate}
\end{theorem}
\end{propbox}
\begin{remark}
    The condition of $f$ and the domain $A$ are to guarantee existence of the derivative on $\prob(n)$.
\end{remark}
\begin{proof}
Suppose $f$ is Schur-convex on $A$. Suppose $\p\in\prob(n)$ and $p_1 > p_2$. Otherwise, if $p_1 < p_2$, applied a permutation matrix to swap them, or if $p_1= p_2$ the proof is trivial. 

First, $f$ is symmetric since for any permutation matrices $\Pi\in\perm(n)$ $\p\succ\Pi\p$ implies $f(\p) \geq f(\Pi\p)$ and $\Pi\p\succ \p$ implies $f(\Pi\p) \geq f(\p)$, therefore $f(\p) = f(\Pi\p)$ for any permutation matrix $\Pi$. 

To show the Schur condition, consider a $T$-transform $T = t I_n + (1-t)\Pi$, where $t\in[1/2,1]$ and $\Pi$ exchanges the first and second elements.   Define $\vctv = (-1,1,0,\ldots,0)^T$ and $\p_\varepsilon := \p +\varepsilon \*{v}$ for any $\varepsilon \in [0, \frac{1}{2}(p_1 - p_2)]$. The transformation $T$ maps $\p$ to $\p_\varepsilon$ with $\varepsilon = (1-t)(p_1 - p_2)$. This means $f(\p_\varepsilon) \leq f(\p)$. By taking the limit $t \to 1/2 ^+$ or equivalently $\varepsilon \to 0^+$, we have that
\begin{align}
    0 &\geq \lim_{\varepsilon\to 0 ^+} \frac{1}{\varepsilon}\pqty{f(\p_\varepsilon) - f(\p)} = \lim_{\varepsilon\to 0^+} \frac{1}{\varepsilon} \pqty{f(\p + \varepsilon \vctv) + f(\p)} \\
    &=  D_\vctv f (\p) = \nabla f\vctv =  - \partial_{1} f (\p) + \partial_{2} f(\p).
\end{align}
Since $p_1> p_2$ and for any other pair of $p_i, p_j$ we can permute them to the first and second elements, thus we have the satisfaction of the Schur condition. 

Conversely, we want to show that if $f$ satisfies both conditions then $f(\p) \geq f(\q)$ whenever $\p \succ \q$. Notice that if $\p\succ\q$, then there exists a doubly stochastic matrix such that it transform $\p$ into $\q$ and any doubly-stochastic can be written as finite applications of $T$-transforms. Therefore, it is enough to show that $f(\p) \geq f(T\p)$ for any $T$ being a $T$-transform.

Suppose $T = t I + (1-t) \Pi$. Since $f$ is symmetric, we can assume the following without the loss of generality.
\begin{enumerate}
    \item $T$ swap the first and second elements Otherwise, apply $\Pi'$ to permute the wanted elements to 1 and 2.
    \item The parameter $t$ is in the interval $t\in[\frac{1}{2},1]$. If $t\in[0,\frac{1}{2})$, apply $\Pi$ to $T\p$. We have $f(\Pi T\p) = f(T\p)$.
    \item $p_1 > p_2$, otherwise switch the ordering or if $p_1 = p_2$ the proof is trivial.
\end{enumerate}
Define $\p_\varepsilon$ to be a probability vector $T\p$ and $\varepsilon = (1-t)(p_1 - p_2)$. The parameter $\varepsilon \in [0,\frac{1}{2}(p_1 - p_2)]$. The vector $\p_\varepsilon$ can be written as 
\begin{equation}
    \p_\varepsilon = \pqty{p_1 - \varepsilon, p_2 + \varepsilon, p_3 , p_4, \ldots, p_n}^T.
\end{equation}
Therefore,
\begin{align*}
    \dv{\varepsilon} f(\p_\varepsilon) &= \pdv{p_1} f(\p_\varepsilon) \dv{p_1}{\varepsilon} + \pdv{p_2} f(\p_\varepsilon) \dv{p_2}{\varepsilon} \\ 
    &= - \varepsilon \pqty{\pdv{p_1} f(\p_\varepsilon) -\pdv{p_2} f(\p_\varepsilon)} \leq 0 
\end{align*}
The inequality is given by the Schur condition. Define $g(\varepsilon)=f(\p_\varepsilon)$.  Since $\varepsilon > 0$ we have that ${\pdv{p_1} f(\p_\varepsilon) -\pdv{p_2} f(\p_\varepsilon)} \geq 0$. The inequality shows that the function $g$ is non-increasing with $\varepsilon$ on the interval $\bqty{0,\frac{1}{2}(p_1 - p_2)}$. Therefore, $f(\p) \geq f(T\p)$.
\end{proof}

\subsection{Entropies}
Entropy in information theory was first purposed by Claude Shannon in 1948~\cite{Shannon1948}. Later in 1961, Alfr\'ed R\'enyi axiomatize and purpose a family of entropies generalizing Shannon entropy. The family is later known as R\'enyi entropies~\cite{Renyi1961}. There are multiples axiomatization of Shannon entropy~\cite{Khinchin1957,Renyi1961,AFN1974,NPVV1992}. However, a recent axiomatic definition of ground the definition of entropy by a majorization~\cite{GT2021}. As we have seen in the earlier section, the majorization is closely related with uncertainty. We present the definition in the following.
\begin{defbox}{Entropy}
\begin{definition}
    A function $f:\bigcup_{n\in\bb{N}} \prob(n) \times \prob(n) \to \bb{R}$ is said to be an \emph{entropy} if it satisfies the following conditions.
    \begin{enumerate}
        \item It is Schur-concave (it is an antitone of majorization).
        \item It is additive under Kronecker tensor product of probability vector, i.e.
        \begin{equation*}
            f(\p\otimes \q) = f(\p) + f(\q)
        \end{equation*}
        for any $\p,\q\in\prob(n)$ and any $n\in\bb{N}$.
    \end{enumerate}
\end{definition} 
\end{defbox}
The additivity condition together with being Schur-concave imply non-negativity of entropy on the domain of probability vector. To see this, consider an entropy of 
\begin{equation*}
 f(p) = f(\p\otimes 1) = f(\p) + f(1) \implies f(1) = 0.
\end{equation*}
The first equality comes from $\p = \p\otimes 1$. Since $1\succ \q$ for any $\q\in\prob(m)$ and any $m\in\bb{N}$. Therefore, $f(1) = 0 \leq f(\q)$ for any $\q\in\prob(m)$ and for any $m\in\bb{N}$.

\subsection{R\'enyi entropies}
R\'enyi entropies are parametrized with a parameter $\alpha\in\bb{R}_+ \cup \qty{\infty}$, an $\alpha$-R\'enyi entropy of a probability vector $\p\in\prob(n)$ is defined as
\begin{equation*}
    H_\alpha (\p) = \frac{1}{1-\alpha} \log \pqty{\sum_{x\in[n]} p_x ^\alpha}.
\end{equation*}
Notice that in the above equation at $\alpha = 0, 1,$ and $\infty$, $H_\alpha$ are not-well defined. 
At those $\alpha$'s, the R\'enyi entropy is defined as the limit of $\alpha\to 0, 1,$ or $\infty$.

For $\alpha =0$, then entropy is called max-entropy and Hartley entropy.
\begin{align}
    H_0 (\p) &:= \lim_{\alpha\to 0^+} \frac{1}{1-\alpha} \log\pqty{\sum_{x\in[n]} p_x ^\alpha} = \lim_{\alpha \to 0 ^+} \log(\sum_{x\in[n]} p_x ^\alpha) \\
    % \magenta{\pqty{\text{continuity of log and sum}}\rightarrow}
    &= \log \pqty{\sum_{x\in[n]} \lim_{\alpha \to 0 ^+}  p_x ^\alpha}
\end{align}
The limit $\alpha\to 0^+$ of $p^\alpha$ is 1 if $p \neq 0$ and is $0$ if $p=0$. This result in 
\begin{equation}
    H_0 (\p) = \log\abs{\rm{supp}(\p)}
\end{equation}
where $\rm{supp}(\p) = \qty{x: p_x \neq 0}$.

For $\alpha =1$, the entropy is the famous Shannon entropy, denoted with $H$,
\begin{equation}
H = H_1 := \lim_{\alpha \to 1} \frac{1}{1-\alpha} \log\pqty{\sum_{x\in[n]} p_i ^\alpha}.
\end{equation}
Applying l'Hopital's rule results in
\begin{align}
    H_1 &= \lim_{\alpha \to 1} \frac{-1}{\sum_{x\in[n]} p_x ^\alpha} \sum_{x\in[n]} p_x ^\alpha \log(p_x) = \sum_{x\in[m]} -p_x \log p_x,
\end{align}
where we use the convention $0 \log (0) = 0$.

For $\alpha = \infty$, the entropy is called min-entropy. We apply l'Hopital's rule in the following derivation,
\begin{align}
H_\infty (\p) &= \lim_{\alpha\to\infty} \frac{1}{1-\alpha} \log \pqty{\sum_{x\in[n]} p_x ^\alpha}\\
&= \lim_{\alpha \to \infty} \frac{-1}{\sum_{x\in[n]} p_x ^\alpha} \sum_{x\in[n]} p_x ^\alpha \log(p_x) 
\end{align}
Consider the numerator terms and suppose $p_m = \max_{x\in[n]} p_x$
\begin{align*}
    - \frac{1}{p_m ^\alpha} \lim_{\alpha \to \infty} \sum_{x\in[n]} p_x ^\alpha \log(p_x) &= - \sum_{x\in[n]} \lim_{\alpha \to \infty}  \pqty{\frac{p_x}{p_m}}^\alpha\log (p_x)\\
    &= -n\log(p_m), 
\end{align*}
where $n$ is the number of $x$'s that have $p_x = p_m$. The last equality is because of each term in the sum being equal to $0$ if $p_x < p_m$ and being equal to $\log(p_x)$ if $p_x = p_m$. Similarly, for the denominator, 
\begin{align*}
    \frac{1}{p_m ^\alpha} \lim_{\alpha\to\infty} \sum_{x\in[n]} p_x ^\alpha = \lim_{\alpha\to\infty} \sum_{x\in[n]} \pqty{\frac{p_x}{p_m}}^\alpha = n.
\end{align*}
The min entropy is 
\begin{equation}
H_\infty (\p) = -\log(p_m).
\end{equation}

\subsubsection{Properties of R\'enyi entropies}

\begin{proposition}
    For any $\alpha \in[0,\infty]$, an $\alpha$-R\'enyi entropy is quasi concave. Moreover, for any $\alpha \in[0,1]$ the entropy is concave.
\end{proposition}
\begin{proof}
    Suppose that $\p,\q\in\prob(n)$ and $\rH(\p) \geq \rH(\q)$, that is $\rH(\q) = \min \qty{\rH(\p),\rH(\q)}.$ 

    For $0<\alpha<1$, we have that
    \begin{equation}
        \sum_{x\in[n]} p_x ^\alpha \geq \sum_{x\in[n]} q_x ^\alpha.
    \end{equation}
    On the other hand, for $\alpha> 1$, we have that 
    \begin{equation}
        \sum_{x\in[n]} p_x ^\alpha \leq \sum_{x\in[n]} q_x ^\alpha 
    \end{equation}
    Notice that the inequality sign flip because $\frac{1}{1-\alpha}$ is negative when $\alpha > 1$.
    Consider an $\alpha$-R\'enyi entropy on a convex combination of $\p$ and $\q$, denoted with $\vctr$, 
    \begin{align*}
        \rH (\vctr) &= \frac{1}{1-\alpha} \log \pqty{\sum_{x\in[n]} \pqty{t p_x + (1-t) q_x}^\alpha}\\
        &\geq \frac{1}{1-\alpha}\log\pqty{\sum_{x\in[n]} t p_x ^\alpha + (1-t) q_x ^\alpha}\\
        &\geq \frac{1}{1-\alpha}\log\pqty{\sum_{x\in[n]} t q_x ^\alpha + (1-t) q_x ^\alpha}= \frac{1}{1-\alpha}\log\pqty{\sum_{x\in[n]} q_x ^\alpha} = \rH(\q).
    \end{align*}
    The last inequality holds in $\alpha \in (0,1)$ due to $\frac{1}{1-\alpha}$ is positive, substitution by a smaller term leads to a smaller output of $\log$ function. For $\alpha \in(0,1)$, $\frac{1}{1-\alpha}$ is negative, substitution by a bigger term leads to smaller output of $\log$ function.

    For $\alpha = 0,1,$ and $\infty$, take a sequence of $\alpha$ to each limit point. Since $\rH$ converges to point-wise, the quasi concavity follows.

    For the concavity statement, consider $\alpha \in(0,1)$ we have that
    \begin{align*}
        \rH (\vctr) &\geq \frac{1}{1-\alpha}\log\pqty{\sum_{x\in[n]} t q_x ^\alpha + (1-t) q_x ^\alpha} \\ 
        &= \frac{1}{1-\alpha}\log\pqty{t \sum_{x\in[n]} p_x ^\alpha + (1-t) \sum_{w\in[n]} q_w ^\alpha} \\
        &\geq \frac{t}{1-\alpha} \log\pqty{ \sum_{x\in[n]} p_x ^\alpha} + \frac{1-t}{1-\alpha} \log\pqty{ \sum_{w\in[n]} q_w ^\alpha}  = t \rH(\p) + (1-t) \rH(\q).
    \end{align*}
\end{proof}

\subsection{The min-entropy is Schur-concave but not concave.}
It is enough to show that there is $t\in[0,1]$ and $\p,\q\in\prob(n)$
\begin{equation*}
    H_\rm{min} (t \p + (1-t) \q) < t H_\rm{min} (\p) + (1-t) H_\rm{min} (\q),
\end{equation*}
to conclude that $H_\rm{min}$ is Schur-concave but not concave. Consider $t = 1/2$, $\p = \*{e}_1$, and $\q = \frac{1}{2} (1, 1)^T$. The min entropy for each probability vector is 
\begin{align*}
H_\rm{min} (\p) &=  - \log \max{1,0} = - \log(1) = 0,\\
H_\rm{min} (\q) &=  - \log \max{1/2,1/2} = - \log(1/2) = 1,\\
H_\rm{min} (0.5\p + 0.5\q) &=  - \log \max{3/4,1/4} = - \log(3/4) = \log(4) - \log(3) \approx 0.415.
\end{align*}
Notice that $0.5H_\rm{min} (\p) + 0.5H_\rm{min} (\q) = 0.5> 0.415$.

\subsection{Incompleteness of R\'enyi entropies}
Is the condition $H_\alpha (\p) \geq H_\alpha(\q)$ for any $\alpha \in [0,\infty]$ sufficient to conclude that $\p\succ\q$? If the mentioned condition is sufficient, we would say that a family of R\'enyi entropies is a complete family of majorization monotones.
As it turns out, R\'enyi entropies do not give a sufficient condition for majorization. One of the simplest examples is from~\cite{Hasselbarth1986}, and we present it below. 
\begin{propbox}{R\'enyi entropies form an incomplete family of majorization monotones}
    \begin{proposition}
        There is a pair of probability vectors such that 
        \begin{equation}
            H_\alpha (\p) \leq H_\alpha (\q) \quad \forall \alpha \in [0,\infty] 
        \end{equation}
        but $\p\not\succ\q$ and $\q\not\succ\p$.
    \end{proposition}
\end{propbox}
\begin{proof}
To prove this proposition, it is enough to give an example. Consider $\p,\q\in\prob(8)$, 
\begin{align*}
    \p &= \frac{1}{36} \pmqty{16 & 4 & 4 & 4 & 4 & 4 & 0 & 0}\\
    \q &= \frac{1}{36} \pmqty{8 & 8 & 8 & 8 & 1 & 1 & 1 & 1}
\end{align*}
$\q\not\succ \p$ because $\knorm{\q}{1}<\knorm{\p}{1}$ and $\p\not\succ\q$ because $\knorm{\p}{4} < \knorm{\q}{4}$. 

Next, we will show that $H_\alpha (\p) \leq H_\alpha (\q)$ for any $\alpha\in[0,\infty]$.
First, consider at $\alpha  = 0$, 
\begin{equation*}
    H_0 (\p) = \log \pqty{\sum_{x:p_x \neq 0} 1} = \log(6) \qquad H_0 (\q) = \log \pqty{\sum_{x:q_x \neq 0} 1} = \log(8).
\end{equation*}
We have that $H_0(\p) < H_0 (\q)$. For $H$, Shannon entropy,
\begin{align*}
    H (\p) = - \sum_{x\in[8]} p_x \log \pqty{p_x} = - \frac{16}{36} \log\pqty{\frac{16}{36}}  - 5\times \frac{4}{36} \log\pqty{\frac{4}{36}} =  \log(9) - \frac{8}{9} \log(2) \\
    H (\q) = - \sum_{x\in[8]} q_x \log \pqty{q_x} = -4 \times \frac{8}{36}\log\pqty{\frac{8}{36}} - 4 \times \frac{1}{36} \log\pqty{\frac{1}{36}}= \log(9) - \frac{6}{9} \log(2) 
\end{align*}
We have that $H (\p) < H(\q)$. For $H_\infty$,
\begin{equation*}
    H_\infty (\p) = - \log \pqty{\max_{x\in[8]} p_x} = \log (36) - \log (16) \qquad
    H_\infty (\q) = - \log \pqty{\max_{x\in[8]} q_x} = \log (36) - \log (8)
\end{equation*}
We have that $H_\infty (\p) < H_\infty (\q)$. 

For any other $\alpha \neq 0, 1,$ or $\infty$, if there is an $\alpha'$ such that $H_{\alpha'} (\p) > H_{\alpha'} (\q)$, then there is $\alpha$ such that $H_{\alpha} (\p) = H_{\alpha} (\q)$, by continuity of the R\'enyi entropy on the parameter $\alpha$. We will show that there is no such $\alpha\in(0,\infty)\setminus\qty{1}$. For any $\alpha \in (0,\infty)\setminus\qty{1}$, 
\begin{align*}
    H_\alpha (\p) = \frac{1}{1-\alpha} \log \pqty{\sum_{x\in[8]} p_x ^\alpha}\\ 
    H_\alpha (\q) = \frac{1}{1-\alpha} \log \pqty{\sum_{x\in[8]} q_x ^\alpha}.
\end{align*}
Enforcing that both entropies are equal, we have that 
\begin{align*}
    \sum_{x\in[8]} p_x ^\alpha &= \sum_{x\in[8]} q_x ^\alpha \\
    16^\alpha + 5 \times 4^\alpha &= 4\times 8^\alpha + 4 \\
    2^{4\alpha} + 5 \times 2^{2\alpha} &= 4 \times 2^{3\alpha} +4\\ 
    \magenta{ \pqty{a:=2^\alpha}\longrightarrow}\quad  a^4 -4a^3 + 5 a^2 - 4 &= 0 \\
    (a-2)(a(a-1)^2 +2) &= 0.
\end{align*}
If $a = 2$, then $\alpha = 1$ and it is outside the consideration as we suppose $\alpha \neq 1$. For all $a>0$, the term $a(a-1)^2 +2 > 0$. Therefore, for any $\alpha\in[0,\infty]$, $H_{\alpha'} (\p) < H_{\alpha'} (\q)$, completing the example.
\end{proof}

\section{Relative majorization and distinguishability of distributions}
Majorization can be viewed as a special case of \emph{relative majorization}, which is defined as follows.
\begin{defbox}{Relative majorization}
    \begin{definition}
        Suppose $\p,\q \in \prob(n)$ and $\p', \q' \in \prob(n')$. We say that the pair of probability vectors $\p,\q$ \emph{relatively majorizes} the pair $\p',\q'$ and write $(\p,\q) \succ (\p',\q')$ if there exists a stochastic matrix $E\in \stoch(n',n)$ such that $(\p',\q') = (E\p,E\q)$.
    \end{definition}
\end{defbox}
Relative majorization relation captures the distinguishability between pairs of objects. One of the implications of relative majorization is that 
\begin{equation*}
    \p' = E \p \quad \textrm{and} \quad \q' = E \q \implies \norm{\p - \q}_1 \geq \norm{\p' - \q'}_1.
\end{equation*}
That is the distance, measured by the 1-norm, between the pair $\p$ and $\q$ are greater than the distance between $\p'$ and $\q'$. However, the converse direction of the implication is not true. There are pairs of probability vectors which are not majorizing one another but the 1-norm of their pairwise difference are greater than one another.

\subsection{Properties of relative majorization}
For a pair with a rational vector in the second component, they are equivalent to $\pqty{r,\u^{(n)}}$.
\begin{propbox}{A pair with a rational vector.}
    \begin{lemma}\label{lm:rel-major-rational-vector}
        Suppose $\p,\q \in\prob(n)$ and $\q$ has rational components. Define $k$ and $\qty{k_i: i\in[n]}$ to be non-negative integers such that 
        \begin{equation}
            \q = \pmqty{\frac{k_1}{k},\frac{k_2}{k},\dots,\frac{k_n}{k}} \quad \rm{and}\quad k = \sum_{i\in[n]} k_i. 
        \end{equation} 
        The probability vector $\vctr$ defined as 
        \begin{equation*}
            \vctr= \bigoplus_{x\in[n]} p_x \u^{(k_x)} = \pqty{\underbrace{\frac{p_1}{k_1},\ldots,\frac{p_1}{k_1}}_{k_1~\text{repeats}},\underbrace{\frac{p_2}{k_2},\ldots,\frac{p_2}{k_2}}_{k_2~\text{repeats}},\ldots,\underbrace{\frac{p_n}{k_n},\ldots,\frac{p_n}{k_n}}_{k_n~\text{repeats}}}^T.
        \end{equation*}
        We have that $\pqty{\p,\q}\sim\pqty{\vctr,\u^{(k)}}$.
    \end{lemma}
\end{propbox}
\begin{proof}
    \point{$(\p,\q)\succ (\vctr,\u^{(k)})$}
    We want to show that there exists $\exists E \in \stoch(k,n)$ such that $E\p = \vctr$ and $E\q = \*{u}$.
    Define $E^{(x)}\in\bb{R}_+ ^{k_x \times n}$ to be a matrix with $\u^{(k_x)}$ in the $x^{\text{th}}$ column and zero elsewhere,
    \begin{equation}
        E^{(x)} = \pqty{ \*{0}^{(k_x)}, \ldots, \*{0}^{(k_x)}, \underbrace{\*{u}^{(k_x)}}_{x^{\rm{th}}-column},\*{0}^{(k_x)},\ldots, \*{0}^{(k_x)}}.
    \end{equation}
    With this, 
    \begin{align*}
        E^{(1)} \p &= p_1 \*{u}^{(k_1)} = \frac{p_1}{k_1} \*{1}^{(k_1)}\\
        E^{(1)} \q &= q_1 \*{u}^{(k_1)} = \frac{k_1}{k} \times \frac{1}{k_1} \*{1}^{(k_1)} = \frac{1}{k} \*{1}^{(k_1)}.
    \end{align*}
    Concatenation of each submatrix $E^{(x)}$ gives the desired matrix $E$,
    \begin{equation}
        E = \pmqty{E_1 \\ E_2 \\ \vdots \\ E_n}.
    \end{equation}
    \point{$ (\vctr,\u^{(k)})  \succ (\p,\q)$}
    Similarly, we will construct a matrix $F$ such that $F\vctr= \p$ and $F\q = \*{u}^{(k)}$. Define $F^{(x)}$ as 
    \begin{equation*}
        F^{(x)} = \pmqty{ \*{0}^{(k_x)~T}\\ \*{0}^{(k_x)~T}\\ \vdots \\*{1}^{(k_x)~T}\\ \*{0}^{(k_x)~T}\\ \vdots\\ \*{0}^{(k_x)~T} }.
    \end{equation*}
    With a concatenation, we define $F$ 
    \begin{equation*}
        F = \pmqty{ F^{(1)} & F^{(2)} & \ldots & F^{(n)}}.
    \end{equation*}
    Application of $F$ on the vector $\vctr$ will take a sum $\sum_{i\in[k_{x}]} \frac{p_x}{k_x} = p_x$ and assign the result to the $x$-th row resulting in the vector $\p$. Similarly, application of $F$ on the vector $\*{u}^{(k)}$ will sum $\frac{1}{k}$ for $k_x$ times and assign the result to the $x$-th row resulting in the vector $\q$.
\end{proof}

\subsection{Hypothesis testing and testing region}
Specifically, relative majorization is related to the operational task called hypothesis testing. The hypothesis testing is a task characterized with the following conditions and constraints.
\begin{probbox}{Hypothesis testing}
    \textbf{Given:} a set of samples $\qty{x_1, x_2, \ldots, x_n}$ drawn independently with an unknown source.\\
    \textbf{Promise:} the distribution of this source is either $\p$ or $\q$.\\
    \textbf{Goal:} determine which distribution the samples are drawn from.
\end{probbox}
Notice that the goal of this problem is to determine the source of the sample, a common task in the field of statistics. The optimal solution to this problem is generally not by {a perfect accuracy} but with a bounded error. Suppose that the distribution $\p$ is our null hypothesis and $\q$ is an alternative hypothesis. There are two possible types of errors.
\begin{enumerate}
    \item \textbf{Type-I error:} a rejection of the null hypothesis when the null hypothesis is true. That is inferring $\q$ as the source when $\p$ is actually the source.
    \item \textbf{Type-II error:} a failure to reject the null hypothesis when the null hypothesis is false. That is inferring $\p$ as the source when $\q$ is actually the source.
\end{enumerate}

We can model a decision made based on a length-$n$ string of numbers $\x = x_1x_2\ldots x_n$ with a decision function $f:[m]^n \to \qty{0,1}$. If $f(\x) = 0$, then we deterministically infer that the source is $\p$. On the other hand, if $f(\x) = 1$, then we deterministically infer that the source is $\q$, rejecting the null hypothesis. Now, a probabilistic decision-making, where we have probabilistically apply one of $\ell$ decision functions $\qty{f_j:j\in[\ell]}$ with probability $t_j$. Define a function $g : [m]^n \to [0,1]$ where 
\begin{equation*}
    g(\x) = \sum_{j\in[\ell]} t_j f_j (\x).
\end{equation*}
The function $g(\x)$ evaluate to 0 or 1 means deterministically infer $\p$ or $\q$. The values between $0$ and $1$ tell the probability of rejecting the null hypothesis. The function $g$ is linear; it can be rewritten as 
\begin{equation*}
    g(\x) = \*{e}_{\x} \cdot \*{t}
\end{equation*}
for some vector $\t \in [0,1]^n$, where $\*{e}_{\x}$ is a standard basis vector with the label $\x$ in base-$m$. That is each probabilistic decision-making can be represented with a unique vector $\t$.

A probability of type-I error, denoted with $\alpha(\t)$, conditioned on the source being $\p$ is 
\begin{equation}
    \alpha(\t) = \sum_{\x':t_{\x'} = 1} \rm{Prob}(\x = \x' \vert \text{ source is } \p) = 1 - \t \cdot \p^{\otimes n}.
\end{equation}
On the other hand, a probability of type-II error, denoted with $\beta(\t)$, conditioned on the source being $\q$ is 
\begin{equation}
    \beta(\t) = \sum_{\x':t_{\x'} = 0} \rm{Prob}(\x = \x' \vert \text{ source is } \q) = \t \cdot \q^{\otimes n}.
\end{equation}

There are many strategies in optimally determined the source. The strategy that related to relative majorization is an asymmetric method, where the type-II error is minimized under a bounded type-I error. Suppose that the error bound of type-I error is $\varepsilon$, the optimal type-II error $\beta_{n} ^* (\varepsilon)$ given $n$-sample in this approach is 
\begin{equation}
    \beta_{n} ^* (\varepsilon) = \min \qty{\t \cdot \q^{\otimes n} : \t \in [0,1]^n \text{ and } 1- \t \cdot \p^{\otimes n} \leq \varepsilon}.
\end{equation}

In the scenario where there is only one sample drawn, relative majorization compares the optimal type-II error $\beta^* (\varepsilon)$ between two cases of probability vector pairs for any $\varepsilon\in[0,1]$. Notice that the subscript $n$ is dropped for simplicity.
\begin{propbox}{Relative majorization and hypothesis testing}
    \begin{proposition}
        Suppose $\p,\q\in\prob(n)$ and $\p',\q' \in \prob(n')$. We have that $(\p,\q) \succ (\p',\q')$ if and only if $\beta^* (\varepsilon) (\p,\q) < \beta^* (\varepsilon) (\p',\q')$ for any $\varepsilon\in[0,1]$.
    \end{proposition}
\end{propbox}
The proof of this proposition is giving by the Theorem~\ref{th:Blackwell-theorem}. All possible value of $\alpha(\t)$ and $\beta(\t)$ in a single-shot regime for a pair $(\p,\q)$ form a \emph{testing region}.
\begin{defbox}{Testing region}
    \begin{definition}
        Suppose $\p,\q\in\prob(n)$. The \emph{testing region} $\fk{T}(\p,\q)$ is a subset of $\bb{R}^2$ defined as 
        \begin{equation}
            \fk{T}(\p,\q) = \qty{(\t\cdot\p,\t\cdot\q): \t \in [0,1]^n }.
        \end{equation}
    \end{definition}
\end{defbox}
% \magenta{points in testing regions tell us a rate of error type-I type-II}

A testing region of a pair $\p,\q$ is convex, closed, and symmetric in such a way that $(x,y)\in\fk{T}(\p,\q)\implies (1-x,1-y)\in\fk{T}(\p,\q)$. 
\begin{proposition}
    A testing region is convex and rotational symmetric around $(0.5,0.5)$.
\end{proposition}
\begin{proof}
    Suppose $\p,\q\in\prob(n)$ and $r = (r_1, r_2)$ and $s = (s_1, s_2)$ are in the testing region $\fk{T}(\p,\q)$. 
    For convexity, a convex sum of the two points is $t r + (1-t) s = (t r_1 - (1-t)s_1, t r_2 - (1-t)s_2)$. Using the fact that both points are in $\fk{T}$, we have that for some $\t_1, \t_2 \in [0,1]^n$
    \begin{align*}
        t r + (1-t) s &= \big(t (\p\cdot \t_1) - (1-t)(\p\cdot \t_2)~,~t (\q\cdot\t_1) - (1-t)(\q\cdot\t_2)\big)\\
        &= \big( \p \cdot(t\t_1+(1-t)\t_2)~,~\q \cdot(t\t_1+(1-t)\t_2) \big).
    \end{align*}
    Since $(t\t_1+(1-t)\t_2) \in [0,1]^n$, the convex sum of two points are in $\fk{T}(\p,\q)$, showing convexity of the testing region.

    For the symmetry, since $(r_1, r_2) = (\t\cdot\p,\t\cdot\q)$, testing with vector $(\*{1}-\t)$ gives $(1-r_1, 1-r_2)$.
\end{proof}

It's important to introduce the following form which make the subsequent definition and theorem easier to state.
\begin{defbox}{The standard form of relative majorization}
    \begin{definition}
        A pair of probability vectors $\p,\q\in\prob(n)$, $(\p,\q)$, is in the \emph{standard form} if 
        \begin{enumerate}
            \item there is no $x\in[n]$ such that $p_x = q_x = 0$, and 
            \item both vectors are arranged such that the ratios $p_x/q_x$ are non-increasing orders, i.e.
            $$\frac{p_1}{q_1}\geq \frac{p_2}{q_2} \geq \ldots \geq \frac{p_n}{q_n}.$$
            The convention used here is $p_x/q_x =\infty$ if $p_x > 0$ and $q_x =0$.
        \end{enumerate}
    \end{definition}
\end{defbox}
Any pair of probability vectors $(\p,\q)$ can be put in the standard form by permuting elements and remove any entries where $p_x$ and $q_x$ are both zero.

\begin{figure}[h]
    \centering
    \begin{tikzpicture}
        %Frame for 1x1 square
        \draw [color = black!40, dashed] (0,6) -- (6,6);
        \draw [color = black!40, dashed] (6,0) -- (6,6);
        
        %fill
        \fill[color=BurntOrange!40] (0,0) -- (5.4,0.6) -- (6,5.4) -- (6,6) -- (0.6,5.4) -- (0,0.6);
        %Lower lorenz curves
        % (0.9,0.1)
        \draw [color = black!40, dashed] (5.4,0) -- (5.4,0.6);
        \draw [color = black!40, dashed] (0,0.6) -- (5.4,0.6);
        \draw [thick, color = BurntOrange] (0,0) -- (5.4,0.6);
        \draw (5.4,-0.05) -- (5.4,0.05);
        \draw (-0.05,0.6) -- (0.05,0.6);
        \node[left] at (0, 0.6) {$0.1$};
        \node[below] at (5.4, 0) {$0.9$};
        
        % (1,0.9)
        \draw [color = black!40, dashed] (6,5.4) -- (6,5.4);
        \draw [color = black!40, dashed] (0,5.4) -- (6,5.4);
        \draw [thick, color = BurntOrange] (5.4,0.6) -- (6,5.4);
        \draw [thick, color = BurntOrange] (6,5.4) -- (6,6);
        \node[left] at (0, 5.4) {$0.9$};
        % (0,0.1)
        \draw [thick, color = BurntOrange] (0,0) -- (0,0.6);
        %(0.1,0.9)
        \draw [thick, color = BurntOrange] (0,0.6) -- (0.6,5.4);
        %(1,1)
        \draw [color = black!40, dashed] (0.6,0) -- (0.6,5.4);
        \draw (5.4,-0.05) -- (5.4,0.05);
        \draw [thick, color = BurntOrange] (0.6,5.4) -- (6,6);
        \node[below] at (0.6,0) {$0.1$};

        \draw [color = black] (6,-0.05) -- (6,0.05);
        \draw [color = black] (-0.05,6) -- (0.05,6);
        \node[below] at (6,0) {$1$};
        \node[left] at (0,6) {$1$};
        \draw [thick, ->] (0,0) -- (8,0);
        \draw [thick, ->] (0,0) -- (0,7);
    \end{tikzpicture}
    \caption{The testing region of the pair $\p = (0.9, 0.1,0)^T, \q = (0.1,0.8,0.1)^T$.}
    \label{fig:lowerLorenzCurve-Example}
\end{figure}
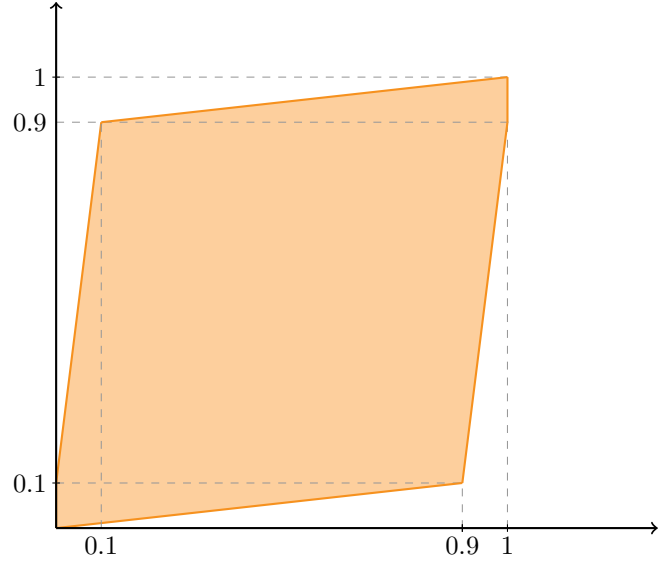

The first application of the standard form is in describing \emph{Lorenz curves}, a boundary of a testing region. A Lorenz curve can be described by its extreme points. Because of the symmetry, it is enough to gives a lower boundary. 
\begin{propbox}{Lower Lorenz curve}
    \begin{theorem}
        Suppose $\p,\q\in\prob(n)$. A lower Lorenz curve of the testing region $\fk{T}(\p,\q)$ is given by a linear interpolation of points $\qty{(a_k,b_k): k\in[n]}$ where 
        \begin{equation}
            (a_k,b_k) = \pqty{\sum_{x\in[k]} p_x, \sum_{x\in[k]} q_x, }
        \end{equation}
        and $\pqty{\p,\q}$ is in the standard form.
    \end{theorem}
\end{propbox}
\begin{proof}
    Suppose $f:[0,1]\to [0,1]$ is a lower bound for a testing region $\fk{T}(\p,\q)$. The lower bound $f(r)$ is a minimization of $\q\cdot\t$ over a valid testing vector $\t$ subjected to the condition $\p\cdot\t = r$, 
    \begin{equation}
        f(r) = \min\qty{\q\cdot \t \vert ~t\in[0,1]^n , \p \cdot \t = r}.
    \end{equation}
    If there exists a vector $\t\in[0,1]^n$ such that $\p \cdot \t > r$ then there is a vector $\t'$ that satisfies the equality as well. If $\p \cdot \t = r/a$ for some $a \in(0,1)$, then $\p\cdot a\t = r$ and $a\t\in[0,1]^n$.
    % \begin{equation*}
    % \exists a \in (0,1)\quad \p \cdot \t = a r \implies \p\cdot \t / a = r,~ \t / a \in [0,1]^n, \text{ and } \q \cdot \t/a \leq \q \cdot \t.
    % \end{equation*}
    Therefore, the condition $\t \in[0,1]^n$ can be replaced with $\*{e}_i \cdot \t \leq 1$ and $ \t \geq 0$. This put $f(r)$ to be an optimal value from a linear program,
    \begin{align*}
        \min \q\cdot \t &=: \alpha \\ 
        \text{subjected to}\quad \p\cdot \t &\geq r \\
        -\*{e}_i \cdot \t &\geq -1 \quad \forall i\in[n]\\ 
        \text{and }\quad \t &\geq 0 .
    \end{align*}
    Define $A = (\p, -\*{e}_1, -\*{e}_2, \ldots, -\*{e}_n )^T$ and $\b = \pmqty{r\\ -\*{1}^{(n)}}$. The dual of this linear program is 
    \begin{align*}
        \max \b \cdot \y &=: \beta \\
        \text{subjected to}\quad A^T \y &\leq \q\\
                \text{and}\quad \y&\geq 0.
    \end{align*}
    Define $\y = \pmqty{s \\ \vctv}$ where $s\in\bb{R}_+$, $\vctv\in\bb{R}^n _{++}$. Then, 
    \begin{equation*}
        \b\cdot \y = \pmqty{r \\ -\*{1}^{(n)}} \cdot \pmqty{s \\ \vctv} = rs - \vctv\cdot \*{1}^{(n)}
    \end{equation*}
    The constraint can be put as
    \begin{equation*}
        A^T \y = s\p - \vctv \leq \q \iff s\p - \q \leq \vctv.
    \end{equation*}
    For linear programming, the strong duality holds if the feasibility set of the primal program is not empty. Since $\t=\*{1}$ gives $\p\cdot\t= 1 \geq r$ for any $r\in[0,1]$, the feasibility set is not empty. Therefore, the strong duality holds, $\alpha = \beta$. 
    
    Next, we consider the dual problem. Notice that $rs- \*{1} \cdot \vctv$ is decreasing with $\norm{\vctv}_1$. For each $s$, the vector $\*{v}$ with the smallest $\norm{\*{v}}_1$ is the optimal $\*{v}$, i.e.
    \begin{equation*}
        \vctv = \pqty{s \p - \q}_+ 
    \end{equation*}
    where the $\pqty{\p}_+$ is the application of positive cut-off into each element of a vector $\p$. Define $s_x = q_x / p_x$ and the objective function becomes
    \begin{align*}
        f(r) = rs - \*{1}^{(n)} \cdot \vctv = rs - \sum_{x\in[n]} (sp_x - q_x)_+ = rs - \sum_{x\in[n]} p_x (s-s_x)
    \end{align*}
    For any $s\geq 0$, there exists an $\ell \in\qty{0,1,2,\ldots,n+1}$ such that $s_\ell \leq s < s_{\ell +1}$, where $s_0 = 0$ and $s_{n+1} = \infty$. The dual function becomes 
    \begin{equation*}
        f(r) = rs - \sum_{x\in[ell]} p_x (s-s_x) = rs- \sum_{x\in[\ell]} sp_x - q_x = s(r-a_\ell) - b_\ell.
    \end{equation*}
    
    Next, we will apply divide-and-conquer strategy to search for the optimal $s$ in each interval $[s_\ell, s_{\ell +1})$.
    \begin{equation}
        f(r) = \max_{\ell\in\qty{0,1,\dots,n}} \sup_{s\in[s_\ell, s_{\ell+1})} \qty{ s(r-a_\ell) + b_{\ell}}.
    \end{equation}
    If the optimal $\ell$ satisfies $\ell \leq k$ where $r\in[a_k, a_{k+1}]$ then $r- a_\ell \geq 0$, $M_\ell :=  s(r-a_\ell) + b_{\ell} = s_{\ell+1} (r-a_\ell) +b_\ell.$
    Moreover, $\ell = k$ gives the largest $M_\ell$ because $M_\ell$ is increasing with $\ell$. To see this increasing, consider 
    \begin{align*}
        M_{\ell} - M_{\ell-1} &= s_{\ell+1} (r-a_\ell) + b_\ell -s_\ell (r-a_{l-1})-b_{l-1} \\
            &= s_{\ell +1} (r-a_\ell) -s_\ell (r-a_{\ell-1}) + q_\ell \\
            \magenta{\pqty{a_{\ell-1} = a_\ell - p_\ell}\longrightarrow} &= s_{\ell+1} (r-a_{\ell}) - s_{\ell} (r - a_\ell + p_\ell) +q_\ell \\
            &= \underbrace{(s_{\ell+1} -s_\ell)}_{\geq 0}\underbrace{(r-a_\ell)}_{\geq 0} + \underbrace{q_\ell - s_\ell p_\ell}_{=0} \geq 0.
    \end{align*}
    On the other hand, if the optimal $\ell$ satisfies $\ell > k$, then $r- a_\ell <0$,
    \begin{equation*}
        \sup_{s\in[s_\ell, s_{\ell+1}]} \qty{s(r-a_\ell) + b_\ell} = s_\ell (r- a_\ell) + b_\ell 
    \end{equation*}
    $\ell = k+1$ gives the optimal, since $M_\ell$ is decreasing with $\ell$. The decreasing part can be seen by modifies the above chain of equalities. 
    Now, we compare candidate of $\ell$, $\ell = k,k+1$.
    \begin{align*}
        M_k &= s_{k+1} (r-a_k) + b_k \\ 
        M_{k+1} &= s_{k+1} (r -a_{k+1}) + b_{k+1}
    \end{align*}
    Apply $a_k = a_{k+1} - p_{k+1}$,
    \begin{equation*}
        M_k = s_{k+1} (r-a_k + p_{k+1}) +b_k = s_{k+1} (r-a_k) + q_{k+1} + b_k = M_{k+1}.
    \end{equation*}
    Therefore, the function $f(r)$ is 
    \begin{equation}
        f(r) = s_{k+1} (r-a_k) + b_k \quad \text{for } r\in[s_{k}, s_{k+1})
    \end{equation}
    which is a linear interpolation between points $\qty{(a_k, b_k): k\in\qty{0,1,2,\dots,n}}.$
\end{proof}
\begin{example}
    Suppose $\p = (0.9, 0.1,0)^T, \q = (0.1,0.8,0.1)^T$. The Lower Lorenz curve of $(\p,\q)$ is an interpolation of 
    \begin{align*}
        &\pqty{a_0,b_0} = (0,0) ;
        &\pqty{a_1,b_1} = (0.9,0.1) ;\\
        &\pqty{a_1,b_1} = (1,0.9) ;
        &\pqty{a_2,b_2} = (1,1). \\
    \end{align*}
    Once the lower Lorenz curve for the pair is known, one can compute the upper Lorenz curve of the pair by interpolation of $(\bar{a}_i,\bar{b}_i) = (1-a_i,1-b_i)$. See the Figure~\ref{fig:lowerLorenzCurve-Example} for an illustration of this example.
\end{example}

One useful consequence of the characterization of the lower Lorenz curve is the removal of a point when three or more points are collinear.
\begin{theorem}
    Suppose $\p$,$\q\in\prob(n)$. There is a pair of probability vectors $\pqty{\p_0, \q_0}$ equivalent with $\pqty{\p,\q}$ such that there is no repeating ratio of $p_{0,x} / q_{0,x}$ for any $x\in[n']$.
\end{theorem}

\begin{proof}
First, we will turn a pair $\pqty{\p,\q}$ into its equivalent pair $\pqty{\p_0, \q_0}$ where $\pqty{\p_0,\q_0}$ is a pair where $\fk{T}\pqty{\p_0, \q_0}=\fk{T}\pqty{\p, \q}$ but there is no $x$ such that the ratio between elements of $\p_0$ and $\q_0$ are not repeated, i.e. there is no $x\in[n]$ that makes $p_{0,x}/q_{0,x} = p_{0,x+1}/q_{0,x+1}$. Suppose $\pqty{\p,\q}$ has exactly one repeat ratio, $\frac{p_x}{q_x} = \frac{p_{x+1}}{q_{x+1}}$. Then, we set $E\in\stoch(n-1,n)$ to be a matrix representation of a linear map $g:\bb{R}^{n} \to \bb{R}^{n-1}$ such that 
    \begin{equation*}
        g(p_1, \ldots, p_x, p_{x+1}, \ldots, p_n) = (p_1, \ldots, p_x + p_{x+1}, \ldots, p_n).
    \end{equation*}
    To see that this operation is linear, consider $g$ on $\alpha \p + \beta\q$ for any $\alpha,\beta\in\bb{R}$,
    \begin{equation*}
        g(\alpha \p + \beta\q) = \pqty{\alpha p_1 + \beta q_1, \ldots,\alpha (p_x+p_{x+1}) + \beta (q_x - q_{x+1}), \ldots, \alpha p_n + \beta q_n} = \alpha g(\p) + \beta g(\q).
    \end{equation*}
    % \begin{equation}
    %     E = \pmqty{ 1 & 0       & \ldots  & 0 &  &  &  & 0 \\
    %                 0 & 1       & \ldots  & 0 &  &  &  & 0 \\
    %            \vdots & \vdots  & \ddots  & 0 &  &  &  & 0 \\ 
    %                 0 & \ldots  &    0    & 1 & 1 & 0  & \ldots &0\\
    %                   &         &         & 0 & 0 & 1  & \ldots &0\\
    %                   &         &         & 0 & 0 & 0  & \ddots &0\\
    %                   &         &         &   &   &    & 0 &1
    %                   }
    % \end{equation}
    The linear map $g$ preserves probability vectors. This operation has a left inverse, $h$, suppose $\vctr,\s\in\prob(n-1)$, 
    \begin{equation*}
        h(\vctr) = \pqty{r_1, \ldots, r_{x-1},r_x \frac{p_x}{p_{x}+p_{x+1}}, r_x \frac{p_{x+1}}{p_{x}+p_{x+1}},  r_{x+1},  \ldots, r_{n}}.
    \end{equation*}
    Image of $h$ on $f(\p)$ is $\p$ and the image on $f(\q)$ is $\q$. The later is less obvious, 
    \begin{equation}
        h(f(\q)) = \pqty{q_x + q_{x+1}} \pqty{\frac{p_x }{p_x + p_{x+1}}} = \pqty{\frac{q_x + q_{x+1}}{p_x + p_{x+1}}} p_x = \frac{q_x}{p_x} p_x = q_x.
    \end{equation}
    Since $h$ is linear, there exists a matrix representing $h$, call it $F$. This $F$ is stochastic since $h$ preserve probability vectors.
\end{proof}

The testing region plays important role in characterizing relative majorization with the hypothesis testing task via the Blackwell's theorem.

\begin{propbox}{Blackwell's theorem}
    \begin{theorem}\label{th:Blackwell-theorem}
        Suppose $\p,\q\in\prob(n)$ and $\p',\q' \in \prob(n')$. The following are equivalent
        \begin{enumerate}
            \item $(\p,\q) \succ (\p',\q')$
            \item for all $t\in\bb{R}$, $\norm{\p - t\q}_{1} \geq \norm{\p' - t\q'}_{1}$
            \item $\fk{T}(\p,\q) \supseteq \fk{T}(\p',\q')$
        \end{enumerate}
    \end{theorem}
\end{propbox}
\begin{proof}
    \point{$1. \implies 2.$} Suppose $\pqty{\p,\q} \succ \pqty{\p',\q'}$. For some $E\in\stoch(n,n')$, $\pqty{\p',\q'}=(E\p,E\q)$. For any $t\in\bb{R}$, we have that
    \begin{equation}
    \norm{\p'-t\q'}_1 = \norm{E\p-tE\q}_1 \geq \norm{\p - t \q}_1.
    \end{equation} 
    \point{$2. \implies 3.$} Recall the lower Lorenz curve $f(r)$ defined in previous proof. Specifically, $f_{(\p,\q)} (r)$ is a function that gives a lower Lorenz curves for the pair $(\p,\q)$, 
    \begin{equation*}
        f (r) = \max_{s\geq 0} \qty{sr - (s\p - \q)_+ \cdot \*{1}^{(n)}}
    \end{equation*}
    Define $t = 1/s$, 
    \begin{equation*}
        f (r) = \max_{t >  0} \qty{\frac{1}{t} \pqty{r - \pqty{ \p - t \q}_+ \cdot \*{1}^{(n)}}}.
    \end{equation*}
    Consider,
    \begin{align*}
        (\p - t \q)_+ \cdot \*{1}^{(n)} &= \frac{1}{2} \pqty{ \norm{\p - t \q}_1 + \pqty{\p - t\q}}\cdot\*{1}^{(n)} = \frac{1}{2} \norm{\p-t\q}_1 + \frac{1}{2} - \frac{1}{2}t.
    \end{align*}
    This 
    \begin{equation*}
        f (r) = \max_{t >  0} \qty{\frac{1}{t} \pqty{r - \frac{1}{2} \norm{\p-t\q}_1 + \frac{1}{2} - \frac{1}{2}t}}.
    \end{equation*}
    Suppose a lower Lorenz curve for the pair $(\p',\q')$ is given by $f_{(\p',\q')} (r)$. Since $\norm{\p-t\q}_1 \geq \norm{\p' - t \q'}_1 $ for all $t\in\geq 0$, therefore $f_{(\p,\q)} (r) \geq f_{(\p',\q')} (r)$, i.e. $T(\p,\q)\supseteq T(\p',\q')$.
    
    \point{$3. \implies 1.$}
    The proof of this will be postponed until we proved all required lemmas.
\end{proof}

\begin{lemma}
    Suppose $\q,\q'\in\prob(n)$ has a rational component. If $\fk{T}(\p,\q) \supseteq \fk{T}(\p',\q')$, then $\pqty{\p,\q}\succ \pqty{\p',\q'}$.
\end{lemma}    
\begin{proof}
    From Lemma~\ref{lm:rel-major-rational-vector}, suppose $\pqty{\vctr',\*{u}^{(k)}}\sim\pqty{\*{p},\*{q}}$ and $\pqty{\vctr',\*{u}^{(k)}}\sim \pqty{\p',\q'}$. Notice that we can make both $\*{r}$ and $\*{r'}$ to be the same dimension. If two pairs of probability vectors are equivalent, then their testing regions are necessarily the same. This can be verified quickly by applying a chain of implication in the previous proof $1.\implies 2. \implies 3.$ in both directions: $\succ$ and $\prec$. Then, the premise is equivalent to 
    \begin{equation}
        \fk{T}(\*{r},\*{u}^{(k)}) \supseteq \fk{T}(\*{r}',\*{u}^{(k)}).
    \end{equation}
    The above containment implies that $\*{r} \succ \*{r}'$ and consequently $(\*{r},\*{u}^{(k)})\succ (\*{r}',\*{u}^{(k)})$. Since we have, $\pqty{\p,\q} \succ (\*{r},\*{u}^{(k)}) \succ (\*{r}',\*{u}^{(k)}) \succ (\p',\q')$, we have $(\p,\q) \succ (\p',\q')$ by transitivity of the relative-majorization preorder.
\end{proof}

\begin{lemma}\label{lm:rel-maj-T-epsilon}
    Suppose $\fk{T}(\p,\q)\supseteq \fk{T}(\p',\q')$. For any $\varepsilon \in (0,1)$, there exist $\q^{\pqty{\varepsilon}} \in \prob(n)$ and $\q'^{\pqty{\varepsilon}} \in \prob(n')$ such that $\q^{(\varepsilon)} \approx_{\varepsilon} \q$, $\q'^{(\varepsilon)} \approx_{\varepsilon} \q'$, and $\fk{T}(\p,\q ^{(\varepsilon)})\supseteq \fk{T}(\p',\q'^{(\varepsilon)})$.
\end{lemma}
\begin{remark}
    $\p \approx_\varepsilon \q$ if and only if $\frac{1}{2} \norm{\p - \q}_1 \leq \varepsilon$.

    Which norm we are using doesn't matter
\end{remark}
\begin{proof}
    The idea of the proof is show that, to make component rationals, we can add small number $\varepsilon_x$ to the almost all elements of $\q$ and subtract the sum of all $\varepsilon_x$ from the last element. The subtraction from the last element ensures that the final vector is a probability vector. In doing this, the lower Lorenz curve will move up a bit. Similarly, we can also move a Lorenz curve down a little. To keep the proof simple, we want to preserve the ordering of $\qty{p_x/q_x}_{x\in[n]}$ after adding the small amount into the elements of $\q$. See Figure~\ref{fig:ratio_nudge} for visualization of this process.
    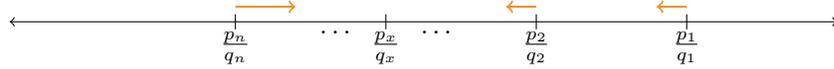
\begin{figure}[h]
        \centering
        \begin{tikzpicture}[scale=2]
            % Number line
            \draw[<->] (-1.5,0) -- (4,0) node[below] {};
            \foreach \x in {0,1,2,3} {\draw (\x,0.05) -- (\x,-0.05);}
            % Points and labels
            \node[below] at (3, 0) {\(\frac{p_1}{q_1}\)};
            \node[below] at (2, 0) {\(\frac{p_2}{q_2}\)};
            \node[below] at (0, 0) {\(\frac{p_n}{q_n}\)};
            \node[below left] at (1.5, 0) {$\ldots$};
            \node[below right] at (0.5, 0) {$\ldots$};
            \node[below] at (1, 0) {\(\frac{p_x}{q_x}\)};
            % Green arrow
            \draw[->,thick,BurntOrange] (0,0.1) -- (0.4,0.1);
            % Other arrows
            \draw[->,thick,BurntOrange] (2,0.1) -- (1.8,0.1);
            \draw[->,thick,BurntOrange] (3,0.1) -- (2.8,0.1);    
        \end{tikzpicture}
        \caption{Ratios between elements of $\p$ and $\q$ on a real line. The leftmost point is $p_n / q_n$. If $p_{n-1}/q_{n-1}= p_n / q_n$, then adding small number to $q_{n-1}$ will shift the fraction $p_{n-1}/q_{n-1}$ to the left while $p_{n}/q_{n}$ is shifted to the right changing the ordering of $\qty{p_x/q_x}_{x\in[n]}$.}
        \label{fig:ratio_nudge}
    \end{figure}

    % Now, we can suppose without loss of generality that a pair $\pqty{\p,\q}$ has no repeated ratio $\qty{p_x/q_x}_{x\in[n]}$. 
    Suppose that the pair $\pqty{\p',\q'}$ has that the last $n'$ elements with the identical ratio $\frac{p_n}{q_n}$.
    Pick $\vec{\varepsilon} = \pqty{\varepsilon_1, \ldots, \varepsilon_n}$ to be positive except at the last $n'$ elements which will be set to be $-\frac{1}{n'} \sum_{x\in[n-n']} \varepsilon_x$. By denseness of $\bb{Q}$, we can pick $\vec{\varepsilon}$ such that 
    \begin{enumerate}
        \item it shifts all $q'_x$ to be rational numbers, $q'_x + \varepsilon_x \in\bb{Q}_+$ for all $x\in[n-n']$, 
        \item its sum is not greater than the element $q_n$, i.e. $\sum_{x\in[n-1]} \varepsilon_x \leq q_n$,  
        \item the shift is smaller than $\varepsilon/2$, i.e. $\sum_{x\in[n-1]} \varepsilon_x < \varepsilon /2$, and
        \item preserve the ordering 
        \begin{equation}
            \frac{p_1}{q_1 '^{(\varepsilon)}} \leq \frac{p_2}{q_2 '^{(\varepsilon)}} \leq \ldots \leq \frac{p_n}{q_n '^{(\varepsilon)}}.
        \end{equation}
    \end{enumerate}
    We denote the shifted probability vector $\q'^{(\varepsilon)} = \q' + \vec{\varepsilon}$. The points on the lower Lorenz curve of $\fk{T}(\p',\q'^{(\varepsilon)})$ are 
    $$(a_k, b' _k) = \pqty{\sum_{x\in[k]} p' _x , \sum_{x\in[k]} q'_x + \varepsilon_x}.$$

    Next, we pick $\vec\nu$ to define an approximate $\q^{(\varepsilon)}$. Suppose similarly that the pair $\pqty{\p,\q}$ has last $m$ identical elements. Similar to $\vec{\varepsilon}$, by denseness of $\bb{Q}$ we can pick the vector $\vec\nu$ to be such that
    \begin{enumerate}
        \item it shifts all $q_x$ to be rational numbers, $q_x - \nu_x \in\bb{Q}_+$ for all $x\in[n-m]$, 
        \item its sum is not greater than the element $q_n$, i.e. $\sum_{x\in[n-m]} \nu_x \leq q_n$,  
        \item the shift is smaller than $\varepsilon/2$, i.e. $\sum_{x\in[n-m]} \nu_x < \varepsilon /2$, 
        \item preserve the ordering 
        \begin{equation}
            \frac{p_1}{q_1 ^{(\varepsilon)}} \leq \frac{p_2}{q_2 ^{(\varepsilon)}} \leq \ldots \leq \frac{p_n}{q_n ^{(\varepsilon)}},
        \end{equation}
    \end{enumerate}
    Since $LC(\p,\q)$ is under $LC(\p',\q')$, shifting one up and another one down will not make lower Lorenz curves cross, i.e. $LC(\p,\q^{(\varepsilon)})$ is under $LC(\p',\q'^{(\varepsilon)})$. Therefore, $\fk{T}(\p,\q ^{(\varepsilon)})\supseteq \fk{T}(\p',\q'^{(\varepsilon)})$
\end{proof}

Now we can proof the direction $3. \implies 1.$ 
\begin{proof}[Proof of $3. \implies 1.$]
    Set $\qty{\varepsilon_\ell : \varepsilon_\ell \in (0,1), \ell \in \bb{N}}$ converging to zero. By the Lemma~\ref{lm:rel-maj-T-epsilon}, there are probability vectors $\*{q}^{(\varepsilon_\ell)}$ and $\q'^{(\varepsilon_\ell)}$ such that $\fk{T}(\p,\q ^{(\varepsilon)})\supseteq \fk{T}(\p',\q'^{(\varepsilon)})$ for any $\ell\in\bb{N}$. By the Lemma~\ref{lm:rel-major-rational-vector},  $(\p,\q ^{(\varepsilon_\ell)})\succ (\p',\q'^{(\varepsilon_\ell)})$. Define $E^{(\ell)}\in\stoch(n',n)$ to be such that $(\p',\q'^{(\varepsilon_\ell)})=(E^{(\ell)}\p,E^{(\ell)}\q^{(\varepsilon_\ell)})$. Since a set of stochastic matrices is compact $\qty{E^{(\ell)}}_{\ell\in\bb{N}}$ has a converging subsequence. Set $E^{\pqty{\ell_j}}$ to be the subsequent and supose that it converges to $E$. Therefore,
    \begin{align*}
        E \p &= \lim_{j\to \infty} E^{\pqty{\ell_j}} \p = \lim_{j\to \infty }\p' = \p' \\
        E \q &= \pqty{\lim_{j\to \infty} E^{\pqty{\ell_j}}} \pqty{\lim_{k\to\infty} \q^{(\varepsilon_k)}} = \lim_{j\to \infty} E^{\pqty{\ell_j}} \q^{(\varepsilon_\ell)} = \q',
    \end{align*} 
    proving that $\pqty{\p,\q}\succ \pqty{\p',\q'}$. This completes the proof of the Blackwell theorem.
\end{proof}

% \subsection{d-majorization and thermo-majorization}
% \begin{defbox}{d-majorization}
% \begin{definition}
%     Suppose $n\in\bb{N}$ and $\p,\q,\*{d} \in\prob(n)$. We say that the vector $\p$ d-majorizes a vector $\q$ and write $\p \succ_d \q$ if $\pqty{\p,\*{d}} \succ \pqty{\q,\*{d}}$.
% \end{definition}
% \end{defbox}
% \begin{example}
%     $\*{d} = \*{u}^{(n)}$ we have the classical probability vector majorization 
% \end{example}
% \begin{example}
%     $\*{d} = \vec{\tau}$, a Gibbs state, we have \emph{thermo}-majorization.
% \end{example}

\section{Monotones of relative majorization}
\subsection{Divergences}
A divergence is a monotone for relative majorization. It quantifies how far apart one distribution is from another one. One of the most important divergences is called Kullback-Leibler divergence put forward by Solomon Kullback and Richard Leibler in 1951 to generalize Shannon entropy~\cite{KL1951}. Later, a definition of divergence grounded on two axioms is purposed by Gilad Gour and Marco Tomamichel in 2020~\cite{GT2020}.
\begin{defbox}{Divergence}
    \begin{definition}
        A function $\bb{D}: \bigcup_{n\in\bb{N}} \prob(n)\times\prob(n) \to \bb{R}\cup\qty{\infty}$ is a \emph{divergence} if 
        \begin{enumerate}
            \item (Data processing inequality: DPI) For any stochastic map $E\in\stoch(m,n)$ for any $m,n\in\bb{N}$ and $\*{p},\*{q} \in \prob(n)$, an inequality $\bb{D}(E\*{p}\Vert E \*{q}) \leq \bb{D}(\*{p} \Vert \*{q})$ holds.
            \item (Normalization) $\bb{D}(1\Vert 1) = 0$.
        \end{enumerate}
    \end{definition}
\end{defbox}
Notice that the data processing inequality is the condition that make a divergence a relative majorization monotone. 
% In some literature, a function $\bb{D}$ is a divergence if it satisfies data processing inequality (not necessary normalization).

One immediate result is non-negativity of the divergence. To see this, consider any probability pair $(\p,\q)$. The pair majorizes $(1,1)$. By the DPI,
\begin{equation}
(\p,\q) \succ (1,1) \implies D(\p\Vert \q) \geq D(1\Vert 1) = 0.
\end{equation}
with the equality if (but not only if) $\p=\q$. 
For example, 
\begin{equation*}
\bb{D}(\p\Vert\q) = \begin{cases}
    0 & \rm{supp}(\p) = \rm{supp}(\q)\\
    \infty & \rm{otherwise}
\end{cases}.
\end{equation*}
This is a divergence. Suppose that $\p$ and $\q$ are $n$-dimensional probability vector. We want to shows that $\bb{D}$ is a divergence, i.e. for any $E\in\stoch(m,n)$, any $m\in\bb{N}$, we have that $\bb{D}(\p\Vert\q) \geq \bb{D}(E\p,E\q)$. Notice that if $\rm{supp}(\p) \neq \rm{supp}(\q)$, then the proof of DPI is trivial. Suppose $\rm{supp}(\p) = \rm{supp}(\q)$. We claim that $\rm{supp}(E\p) = \rm{supp}(E\q)$ as well. This is because,
\begin{equation*}
    \rm{supp}\pqty{E\p} = \rm{supp}(\p) \cap \bigcup_{i\in[n]} \rm{supp}(E\e_i)
\end{equation*}
and similarly with $E\q$.

Any convex function can be extended to a divergence. Such divergence is called an $f$-divergence.
\begin{propbox}{$f$-divergence}
    \begin{theorem}
        If $f:\pqty{0,\infty}\to\bb{R}$ is a convex function such that $f(1)=0$, then a function $D_f: \bigcup_{n\in\bb{N}} \prob(n)\times\prob(n) \to \bb{R}$ defined by 
        \begin{equation}
            D_f (\*{p} \Vert \*{q}) = \sum_{x\in[n]} q_x f\pqty{\frac{p_x}{q_x}}
        \end{equation}
         is a divergence. The conventions are 
         \begin{equation*}
         f(0) = \lim_{r\to 0^+} f(r), \quad 0f\left(\frac{0}{0}\right)= \lim_{r\to 0^+} r f\pqty{\frac{r}{r}} = 0, \quad 0 f\pqty{\frac{a}{0}} = \lim_{r\to 0^+} r f\pqty{\frac{a}{r}} = a \lim_{s\to 0^+} s f\pqty{\frac{1}{s}}.
         \end{equation*}
    \end{theorem}
\end{propbox}
\begin{proof}
    Normalization follows directly from $f(1)=0$. To see that $D_f$ satisfies data processing inequality, suppose that $\p,\q\in\prob(n)$ and $E = \pqty{e_{x\vert y}:x\in[m],y\in[n]}\in\stoch(m,n)$ be any stochastic matrix. Denote $\pqty{\vctr,\s} = \pqty{E\p,E\q}$. The image of $D_f$ on the pair $(\vctr,\s)$ is 
    \begin{align}
        D_f (\vctr\Vert\s) &= \sum_{x\in[m]} s_x f\pqty{\frac{r_x}{s_x}}
    \end{align}
    Define $$t_{y\vert x} = \frac{e_{x\vert y} q_y}{s_x}.$$ Notice that $t_{y\vert x} \geq 0$ and $\sum_{y\in[n]} t_{y\vert x}= \sum_{y\in[n]} \frac{e_{x\vert y} q_y}{s_x}=1$. The fraction $r_x / s_x$ can be written as 
    \begin{equation*}
        \frac{r_x}{s_x} = \sum_{y\in[n]} \frac{e_{x\vert y} p_y}{s_x} =\sum_{y\in[n]} t_{y\vert x} \frac{p_y}{q_y}.
    \end{equation*}
    Using convexity of $f$, we have a chain of inequalities
    \begin{align*}
        D_f (\vctr\Vert \s) &= \sum_{x\in[m]}  s_x f\pqty{\sum_{y\in[m]} t_{y\vert x} \frac{p_y}{q_y}} \leq \sum_{\substack{x\in[m]\\y\in[n]}} s_{x} t_{y\vert x} f\pqty{\frac{p_y}{q_y}} \\ 
        &= \sum_{y\in[n]} \underbrace{\sum_{x\in[m]} e_{x\vert y}}_{=1} q_y f\pqty{\frac{p_y}{q_y}} = D_f (\p \Vert \q),
    \end{align*}
    showing that $D_f$ satisfies DPI.
\end{proof}
This theorem gives us a large class of divergences. The famous KL-divergence is in this class.
\begin{example}
    Kullback--Leibler divergence (KL divergence) is an $f$-divergence where
    \begin{equation}
        f(r) = r \log(r).
    \end{equation} 
    This gives a KL divergence of a pair of probability vectors $\p,\q\in\prob(n)$ to be 
    \begin{equation}
        D(\p\Vert\q) = \sum_{x\in[n]} q_x \frac{p_x}{q_x} \log\pqty{\frac{p_x}{q_x}} = \sum_{x\in[n]} p_x (\log\pqty{p_x} - \log\pqty{q_x}).
    \end{equation}
    This divergence is in some literature called the relative entropy.
\end{example}

% \begin{example}
% Trace distance
% \begin{equation}
%     \norm{\p - \q}_1
% \end{equation}
% \end{example}

\begin{example}
The chi-square distance 
\begin{equation}
    \chi^2 (\p\Vert\q)  = \sum_{x\in[n]} \frac{\pqty{p_x - q_x}^2}{q_x}
\end{equation}
is an $f$-divergence where 
\begin{equation}
    f(r) = \pqty{r-1}^2.
\end{equation}
\end{example}

\subsection{Relative entropies}
Similarly to the previous section on entropy. A monotone of relative majorization that is additive under tensor product and satisfies an additional normalization condition is called a relative entropy.
\begin{defbox}{Relative entropy}
    \begin{definition}
        A function $\bb{D}: \bigcup_{n\in\bb{N}} \prob(n)\times\prob(n) \to \bb{R}\cup\qty{\infty}$ is called a \emph{relative entropy} if it satisfies 
        \begin{enumerate}
            \item Data processing inequality: for any $n,m\in\bb{N}$, $\p,\q\in\prob(n)$ and $E\in\stoch(m,n)$ $$\bb{D}(\p\Vert \q) \geq \bb{D}(E \p\Vert E\q)$$
            \item Additivity under tensor product: for any $n,m\in\bb{N}$, $\p,\q\in\prob(n)$ and $\vctr,\s\in\prob(m)$ $$\bb{D}(\p\otimes\vctr\Vert \q\otimes\s) = \bb{D}(\p\Vert \q)+\bb{D}(\vctr\Vert \s)$$
            \item Normalized $\bb{D}(\*{e}_1\Vert \u^{(2)}) = 1$ where $\*{e}_1 = (1, 0)^T$ and $\u = \frac{1}{2}(1, 1)^T$.
        \end{enumerate}
    \end{definition}
\end{defbox}

\subsubsection{R\'enyi relative entropies}
For any $\alpha \in \bb{R}_+ \cup \qty{\infty}$, an $\alpha$-R\'enyi relative entropy, $D_\alpha$, of a pair of probability vectors $\p,\q\in\prob(n)$ is defined as 
\begin{equation*}
    D_\alpha (\p\Vert \q) = 
    \begin{cases}
        \frac{1}{\alpha -1} \log \pqty{\sum_{x\in[n]} p_x ^\alpha q_x ^{1-\alpha}} &\p\ll\q \text{ or } (\alpha\in[0,1) \text{ and } \p\cdot \q \neq 0) \\
        \infty &\text{otherwise}
    \end{cases}.
\end{equation*}
where $\p\ll\q$ denotes the condition $\rm{supp}(\p) \subseteq\rm{supp}(\q)$.

At $\alpha = 0, 1, \infty$, the $\alpha$-R\'enyi relative entropy is defined by taking limit of $\alpha$ to the value.

Particularly at $\alpha =1$, $D_1 := \lim_{\alpha \to 1} D_\alpha (\p\Vert \q)=  D(\p\Vert \q)$ is a KL divergence. 
Another two important R\'enyi relative entropies are $\alpha = 0$, termed min-relative entropy,
\begin{equation*}
    D_0 (\p\Vert \q)= D_{\rm{min}} (\p\Vert \q) = - \log \pqty{ \sum_{x\in\rm{supp}(\p)} q_x}
\end{equation*}
and $\alpha = \infty$, termed max-relative entropy,
\begin{equation*}
    D_\infty (\p\Vert \q)= D_{\rm{max}} (\p\Vert \q)= \log \max_{x\in[n]} \qty{\frac{p_x}{q_x}}.
\end{equation*}
These relative entropies are derived similarly to the case of $\alpha = 0, 1, \infty$ R\'enyi entropies.

The two min and max relative entropies give a sufficient condition for relative majorization.

\begin{propbox}{Sufficient condition for relative majorization}
    \begin{theorem}
        Suppose $\p,\q\in\prob(n)$ and $\p',\q'\in\prob(n')$. If $D_{\rm{max}} (\p' \Vert \q') \leq D_{\rm{min}} (\p\Vert \q)$, then $\pqty{\p,\q}\succ \pqty{\p',\q'}$.
    \end{theorem}
\end{propbox}
\begin{proof}
    Suppose that $D_\infty (\p'\Vert \q') \leq D_0 (\p\Vert\q)$. Suppose both pairs are in the standard form. The inequality can be put as 
    \begin{equation*}
        \log \max_{x\in[n]} \pqty{\frac{p' _x}{q'_x}} 
        = \log \pqty{\frac{p'_1}{q'_1}} 
        \leq - \log \sum_{x\in\rm{supp}(\p)} q_x
    \end{equation*}
    Exponentiate both side of the inequality leads to 
    \begin{equation*}
        \frac{q' _1}{p' _1} \geq \sum_{x\in\rm{supp}(\p)} q_x
    \end{equation*}
    The term $\frac{q' _1}{p' _1}$ gives a point $(a,b) = \pqty{p'_1, q'_1}$ on a lower Lorenz curve. Extrapolating the line segment linearly to intersect with the line $a = 1$, we have that the intersection is at $(1,\frac{q'}{p'})$. By convexity of the lower Lorenz curve, this extrapolation give the lowest possible Lorenz curve. On the other hand, $\sum_{x\in\rm{supp}(\p)} q_x$ gives the extreme point of a lower Lorenz curve
    \begin{equation*}
    (a, b) = \pqty{\sum_{x\in\rm{supp}(\p)} p_x,\sum_{x\in\rm{supp}(\p)} q_x} = \qty(1,\sum_{x\in\rm{supp}(\p)} q_x).
    \end{equation*}
    Notice that this point is on the line $a=1$ as well. 
    Intrapolating linearly between the origin and this yield a possible lower Lorenz curve for the pair $\pqty{\p,\q}$. This curve is the highest possible since the lower Lorenz curve must be convex.
    All of this information is illustrated in Figure~\ref{fig:lowerLorenzCurve-RelEntropy}. Since we have that the point $\pqty{1,\frac{q' _1}{p' _1}}$ is above $\sum_{x\in\rm{supp}(\p)} q_x$, the lower Lorenz curve of $\pqty{\p,\q}$ is under that of $\pqty{\p',\q'}$ showing the relative majorization relation $\pqty{\p,\q} \succ \pqty{\p',\q'}$. 
\end{proof}
    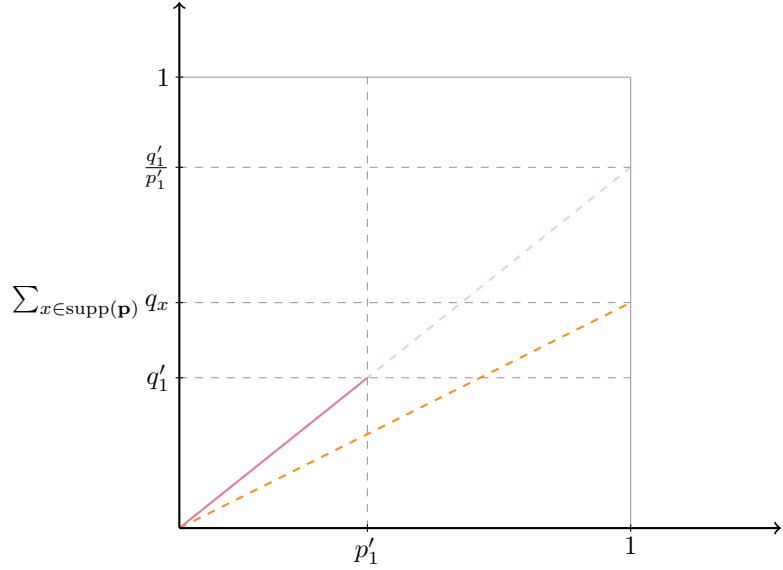
\begin{figure}[h]
        \centering
        \begin{tikzpicture}
            \draw [thick, dashed, color = BurntOrange] (0,0) -- (6,3);
            % \draw [thick, color = BurntOrange] (0,0) -- (1.4,0.7);
            \draw [color = black!40, dashed] (0,3) -- (6,3);
            \draw (-0.05,3) -- (0.05,3);
            \node[left] at (0, 3) {$\sum_{x\in\rm{supp}(\p)} q_x$};
            
            \draw [thick, dashed, color = purple!20] (0,0) -- (6,4.8);
            \draw [thick, color = purple!50] (0,0) -- (2.5,2);
            \draw [color = black!40, dashed] (0,4.8) -- (6,4.8);
            \draw [color = black!40, dashed] (2.5,0) -- (2.5,6);
            \draw [color = black, dashed] (2.5,-0.05) -- (2.5,0.05);
            \node[below] at (2.5,0) {$p' _1$};
            \draw [color = black!40, dashed] (0,2) -- (6,2);
            \draw [color = black, dashed] (-0.05,2) -- (0.05,2);
            \node[left] at (0,2) {$q' _1$};
            \draw (-0.05,4.8) -- (0.05,4.8);
            \node[left] at (0, 4.8) {$\frac{q_1 '}{p_1 '}$};
            
            \draw [thick, ->] (0,0) -- (8,0);
            \draw [thick, ->] (0,0) -- (0,7);
            \draw [color = black!40] (0,6) -- (6,6);
            \draw [color = black!40] (6,0) -- (6,6);
            \draw [color = black] (6,-0.05) -- (6,0.05);
            \draw [color = black] (-0.05,6) -- (0.05,6);
            \node[below] at (6,0) {$1$};
            \node[left] at (0,6) {$1$};
        \end{tikzpicture}
        \caption{Possible lower Lorenz curves for a pair $\pqty{\p,\q}$ (solid and dashed \textcolor{purple!80}{purple} line) and a pair $\pqty{\p',\q'}$ (dashed \textcolor{BurntOrange}{orange}). The dashed purple portion shows the lowest possible lower Lorenz curve for a given $D_\rm{max} (\p'\Vert \q') = p'_1 / q'_1$. The dashed orange shows the highest possible lower Lorenz curve for a given $D_\rm{min} (\p\Vert\q)=\sum_{x\in\rm{supp}(\p)}  q_x$.}
        \label{fig:lowerLorenzCurve-RelEntropy}
    \end{figure}
    \begin{remark}
        The converse of the above theorem is not true. There is a pair of probability vector pairs such that $\pqty{\p,\q} \succ \pqty{\p',\q'}$ but $D_\rm{max} (\p'\Vert\q') > D_\rm{min} (\p \Vert \q)$. For example, $\p = (0.9, 0.1,0)^T, \q = (0.1,0.8,0.1)^T$ and $\p' = \pqty{0.2,0.8}^T$, $\q' = \pqty{0.1,0.9}^T$.
    \end{remark}

\subsubsection{Mutual information}
A mutual information between two classical systems $X$ and $Y$, denoted with $I(X:Y)$, is defined as a divergence of a joint probability distribution from a product of the marginals,
\begin{equation}
    I(X:Y) = D(\p^{XY}\Vert \q^{X} \otimes \vctr^{Y})
\end{equation}
where $\p^{XY}$ joint probability vector and $\q^X$ and $\vctr^Y$ are marginal distribution.

\section{Conditional uncertainty of states}
Suppose we have two finite-dimensional classical systems $X$ and $Y$. A classical state on a bipartite system is described by $\p^{XY}$ a joint probability vectors. We can compare two bipartite probability vector with the majorization relation. In that case, we compare uncertainty when \emph{both} systems are inaccessible. On the other hand, the conditional majorization relation compares uncertainty relation when one specific system is accessible. In this thesis, we compare uncertainty of a system $X$ when a system $Y$ is accessible. 

There are three perspectives to define conditional majorization: constructive, axiomatic, and operational.  The conditional majorization was first introduced in~\cite{GGH+2018} with the constructive appoarch, where the author describe what operations are allowed to use and define conditional uncertainty comparison to be an ability to reconstruct one joint probability distribution into another. Later on, the development in axiomatic definition of quantum conditional majorization was introduced in~\cite{BGWG2021}, where the authors prescribe a properties of a general conditional uncertainty non-decreasing map and shows the coincident with the constructive approach. Moreover, they also purposed a family of games of chance that can fully characterize the majorization with operational tasks. The compilation of all three approaches in the classical regime was introduced in~\cite{Gour2024a}.

From constructive and axiomatic point of views, an unconditional uncertainty is defined via a convertibility relation of mixing operations and say $\p^{XY}$ conditionally majorizes $\q^{XY'}$ if there is a mixing operation transforming $\p^{XY}$ to $\q^{XY'}$. Similarly, to the probability vector majorization, this operation must be linear to preserve probabilistic interpretation. 

In the following section, we will introduce the axiomatic definition of conditional uncertainty comparison. Later, we will characterize this definition via constructive and operational perspectives.

\subsection{Axiomatic approach}
Suppose $\q^{XY'} = M \p^{XY}$. What should be the key properties of $M$ so that accessing $Y'$ when the distribution is $\q^{XY'}$ gives less certainty regarding $X$ compare to when accessing $Y$ and the distribution is $\p^{XY}$? 

Suppose $M$ sends an information of $X$ to $Y'$. Then accessing $Y'$ on $M\p^{XY}$  would be gain more information regarding $X$ compare to accessing $Y$, from $\p^{XY}$. This gives us the first requirement of the mapping $M$ must send no information from $X$ to $Y$, which is called \emph{$X\not\to Y$ no-signalling}.
Mathematically, this condition can be expressed as the following condition.
Suppose $M = \pqty{\mu_{x'y'\vert xy}}$. $M$ is $X\to Y$ no-signalling if 
\begin{equation}
    \sum_{x'} \mu_{x' y'\vert x y} = r_{y'\vert y}.
\end{equation}
That is an evolution of any classical joint distribution $\p^{XY}$ by $M$ is independent of $x$, making an agent on $Y,Y'$ cannot learn more about $X$ by applying $M$ even when they have access to $Y$ and $Y'$. Otherwise, consider a scenario where application of $M$ result in two different probability distributions on $Y$ depdending on what $x$ is. Then, one could with a hypothesis testing strategy to test if the samples $\qty{y_1, y_2, \ldots, y_k}$ is drawn from which source.

\begin{figure}
    \centering
    \includegraphics[height=3cm]{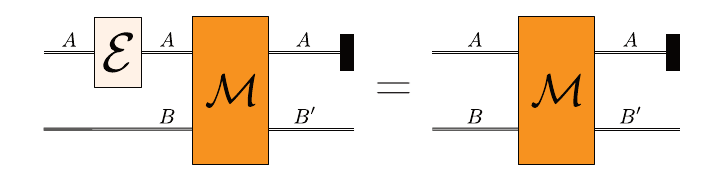}
    \caption{A diagramatic definition of semicausal channel.}
    \label{fig:semicausal}
\end{figure}

Secondly, such $M$ must preserves the most uncertain state on $X$ which is a uniform distribution. That is for any joint probability distribution in the form $\p^{XY} = \vctu ^X \otimes \p^{Y}$, we have that $M \p^{XY} = \vctu^X \otimes \q^{Y'}$ for some $\q^{Y'}$. We call such a matrix $M$ satisfying both conditions a \emph{conditionally mixing operation}.
\begin{defbox}{Conditionally mixing operation (CMO)}
\begin{definition}
    A stochastic matrix $M\in\stoch(nm',nm)$ is said to be a \emph{conditionally mixing operation} (CMO) conditioned on $Y$ if it satisfies
    \begin{enumerate}
        \item $M$ is $X\not\to Y$ no-signalling: for any $x\in[n]$, $y\in[m]$, and $y'\in[m']$, $\sum_{x'} \mu_{x' y'\vert x y} = r_{y'\vert y}$.
        \item $M$ preserves uniform distribution on $X$.
    \end{enumerate}
\end{definition}
\end{defbox}
Conditional majorization is defined to be a convertibility via this set of operations.
\begin{defbox}{Conditional majorization}
    \begin{definition}
        Suppose $\p^{XY} \in \prob(nm)$, $\q^{XY'} \in \prob(nm')$, $\abs{X} = n$, $\abs{Y} = m$, and $\abs{Y'} = m'$. We say that $\p^{XY}$ \emph{conditionally majorizes} $\q^{XY'}$ on $X$ and write $\p^{XY} \succ_{X} \q^{XY'}$ if there exists $M\in\CMO(nm',nm)$ such that $\q^{XY'} = M \p^{XY}$.
    \end{definition}
\end{defbox}
The no-signaling $X\not\to Y$ is equivalent with a being semicausal and semilocalizable. Here, we provide a quantum channel definition of semicausal channel.
\begin{defbox}{Semicausal channel}
    \begin{definition}
        A quantum channel $\mM\in\cptp(AB\to AB')$ is $A\not\to B'$ \emph{semi-casual} if 
        \begin{equation}
            \Tr_A [\mM^{AB\to AB'}(\rho^{AB})] = \Tr_A [\mM^{AB\to AB'}(\cal{E}^{A\to A}\otimes\mI^B(\rho^{AB}))]
        \end{equation}
        for any $\mE\in\cptp(A\to A)$ and $\rho\in\fk{D}(AB)$. See Figure~\ref{fig:semicausal} for the diagram.
    \end{definition}
\end{defbox}

In quantum domain, a channel is $A\not\to B$ semicausal if and only if it is $A\not\to B$ no-signalling. The same statement still holds true in the restriction to the domain of classical channel. We provide a proof for classical channels in the following theorem.
\begin{propbox}{Classical $X\not\to Y$ semicausal channels and $X\not\to Y$ no-signalling}
    \begin{theorem}
        A classical channel $\cal{M}^{XY\to XY'}$ is $X\not\to Y$ semicausal if and only if it is $X\not\to Y$ no-signalling.
    \end{theorem}
\end{propbox}
\begin{proof}
    In the only-if direction, $\mM^{XY\to XY'}$ is $X\not\to Y$ semicausal. 
    We want to shows that
    \begin{equation*}
        \sum_{x'\in[m]} \mu_{x'y'\vert xy} = r_{y'\vert y}
    \end{equation*}
    where $\mu_{x'y'\vert xy}$ is an element of a transition matrix $M$ associated with the channel $\mM^{XY\to XY'}$.
    Suppose $\mN^{XY\to Y'} = \Tr_X \circ \mM^{XY\to XY'}$. Semicausal condition can be written as
    \begin{equation}\label{eq:semicausal}
        \cal{N}(\mE\otimes\mI) = \cal{N}
    \end{equation}
    for any classical channel $\cal{E}$.
    A transition matrix $N$ associated with a channel $\mN^{XY\to Y'}$ has their entries given by
    \begin{equation*}
        \nu_{y'\vert xy} = \sum_{x'\in[n]} \mu_{x'y'\vert xy}.
    \end{equation*}
    Suppose $\cal{E}$ has a transition matrix $E = \pqty{\eta_{vw}}$. The semicausal condition in Equation~\eqref{eq:semicausal} can be written as 
    \begin{equation*}
        \pqty{\sum_{y',x,y} \nu_{y'\vert xy} \*{e}_{y'} \*{e}_{x,y}^T}\pqty{\sum_{v,w} \eta_{v\vert w} \*{e}_{v}\*{e}_{w} ^T\otimes I} = \pqty{\sum_{y',x,y} \nu_{y'\vert xy} \*{e}_{y'} \*{e}_{x,y}^T}
    \end{equation*}
    where $ \*{e}_{x,y} = \*{e}_x \otimes \*{e}_y$. Consider the left side of equality
    \begin{align*}
        \pqty{\sum_{y',x,y} \nu_{y'\vert xy} \*{e}_{y'} \*{e}_{x,y}^T}\pqty{\sum_{v,w} \eta_{v\vert w} \*{e}_{v}\*{e}_{w} ^T\otimes I} &= \sum_{y',x,y,w} \nu_{y'\vert xy} \eta_{x\vert w} \*{e}_{y'}\*{e}_{w,y} ^T\\
        &= \sum_{y',y,w} \pqty{\sum_{x\in[n]} \nu_{y'\vert xy} \eta_{x\vert w}} \*{e}_{y'}\*{e}_{w,y} ^T
    \end{align*}
    That is for any $E$, $y$, and $y'$
    \begin{align*}
        \sum_{x} \nu_{y'\vert x y} \eta_{x\vert w} = \nu_{y'\vert w y} 
    \end{align*}
    Picking $\eta_{x\vert w} = \delta_{xx_0}$ for all $x\in[n]$ results in 
    \begin{equation*}
        \nu_{y'\vert x_0 y} = \nu_{y'\vert w y},
    \end{equation*}
    implying that $\nu_{y'\vert xy}$ is a constant in $x$. Then we define $r_{y'\vert y} := \nu_{y'\vert xy}$. Therefore,
    \begin{equation*}
        \nu_{y'\vert xy} = \sum_{x'\in[n]} \mu_{x'y'\vert xy} = r_{y'\vert y}.
    \end{equation*}
    
    Conversely, if the channel is $X\not\to Y$ no-signalling, we have that 
    \begin{equation}\label{eq:no-signal}
        \nu_{y'\vert xy} = \sum_{x'\in[n]} \mu_{x'y'\vert xy} = r_{y'\vert y}.
    \end{equation}
    A composition of channels $\mN(\mE\otimes \mI)$ is associated with a transitioin matrix
    \begin{align*}
        N(E\otimes I) = \pqty{\sum_{x,y,y'} \nu_{y'\vert xy} \*{e}_{y'}\*{e}_{x,y}^T}\pqty{\sum_{v,w\in[n]}\eta_{v\vert w} \*{e}_{v}\*{e}_{w} ^T\otimes \sum_{z\in[m]} \*{e}_{z} \*{e}_{z} ^T }.
    \end{align*}
    Apply the property of $X\not\to Y$ no-signalling (Equation~\eqref{eq:no-signal}) leads to
    \begin{align*}
        N(E\otimes I) &= \pqty{\sum_{x,y,y'} r_{y'\vert y} \*{e}_{y'}\*{e}_{x,y}^T}\pqty{\sum_{v,w}\eta_{v\vert w} \*{e}_{v}\*{e}_{w} ^T \otimes I} \\
        &=\sum_{y',y,w}  r_{y'\vert y}  \pqty{\sum_{x\in[n]}\eta_{x\vert w}} \*{e}_{y'}\*{e}_{w,y} ^T \\ 
        &=\sum_{y',y,w}  r_{y'\vert y}  \*{e}_{y'}\*{e}_{w,y} ^T  = N
    \end{align*}
    showing $X\not\to Y$ semicausality.
\end{proof}

Moreover, $X\not\to Y$ non-signalling is realizable with a semilocalizable operation, an operation which decomposable into local operations on $X$ and $Y$ with one-way communication from $Y$ to $X$~\cite{ESW2002}. Here we present a quantum channel definition.
\begin{figure}[h]
    \centering
    \includegraphics[width=\textwidth]{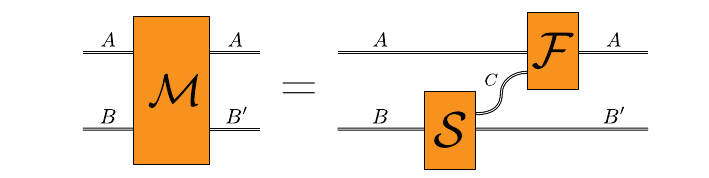}
    \caption{Diagram visualization of a semilocalizable map $\cal{M}$.}
    \label{fig:semilocalizable}
\end{figure}
\begin{defbox}{Semilocalizable channel}
    \begin{definition}
        A quantum channel $\cal{M}\in\cptp(AB\to AB')$ is \emph{semilocalizable} with a communication $B\to A$ if it can be decomposed into 
        \begin{equation*}
            \cal{M} = \pqty{\cal{F}^{AC\to A} \otimes \id^{B'}}\circ \pqty{\id^{A} \otimes \cal{S}^{B\to B' C}}
        \end{equation*}
        for some $\cal{F}\in\cptp(AC\to A)$ and $\cal{S}\in\cptp(B\to B' C)$. See Figure~\ref{fig:semilocalizable} for diagram.
    \end{definition}
\end{defbox}
This definition of channel gives a specific construction of a semicausal channel and consequently, our conditionally mixing operation. In the classical case, any classical channel has their corresponding transition matrix. We define the classical systems $X,Y,$ and $Z$ in the place of $A,B,$ and $C$ respectively. We denotes transition matrix with the above channel $\cal{M}$ to be $M$ and $F$ and $S$ for their corresponding decomposition of the transition matrix
\begin{equation*}
    M = (F^{XZ\to X} \otimes I^{Y'}) \circ (I^X \otimes S^{Y\to YZ'}).
\end{equation*}
Suppose that $Z$ has dimension $\ell$. We denotes each distinct states of $Z$ with $j\in[\ell]$, we can then write $S^{Y\to YZ}$,
\begin{equation*}
    S^{Y\to YZ} = \sum_{j\in[\ell]} S^{Y\to Y} _j \otimes \*{e} ^{Z} _j
\end{equation*}
where $S_j ^{Y\to Y}$ is a substochastic matrix and $\sum_{j\in[n]} S_j ^{Y\to Y}$ is a substochastic matrix. Similarly, we can write $F$ in terms of its conditioning on $j\in[\ell]$
\begin{equation*}
    F^{XZ \to X} = \sum_{j\in[\ell]} F_j ^{X\to X} \otimes (\*{e}_j ^Z)^T,
\end{equation*}
where $F_j ^{X \to X}$ is a stochastic matrix. Puting above equations together, we can equivalently write the classical semilocalizable channel $M$ with communication from $Y\to X$ as 
\begin{equation}\label{eq:cl_semilocal}
    M = \sum_{j\in[\ell]} F^{X\to X} _j \otimes S^{Y \to Y} _j.
\end{equation}

\begin{propbox}{}
    \begin{lemma}\label{lm:semilocal-semicausal}
        A classical channel is $X\not\to Y$ no-signalling if and only if it is semilocalizable with a communication from $Y\to X$.
    \end{lemma}
\end{propbox}
\begin{proof}
    In the only-if direction, suppose $M = \pqty{\mu_{x'y'\vert xy}}$ is a transition matirx associated with a classical $X\not\to Y$ no-signalling. By being $X\not\to Y$ no-signalling, we have that 
    \begin{equation*}
        \sum_{x'\in[n]} \mu_{x'y'\vert xy} = r_{y'\vert y}.
    \end{equation*}
    Define 
    \begin{equation*}
        t^{(y,y')}_{x'\vert x} = \frac{\mu_{x'y'\vert xy}}{r_{y'\vert y}}.
    \end{equation*}
    We have that $M$ can be put in the form 
    \begin{align}
        M &= \sum_{\substack{x,x'\in[n]\\ y\in[m]\\y'\in[m']}} \mu_{x'y'\vert xy} \*{e}_{x'} \*{e}_{x} ^T\otimes \*{e}_{y'} \*{e}_{y} ^T= \sum_{y,y'} r_{y'\vert y} \sum_{x,x'} t^{(y,y')}_{x'\vert x} \*{e}_{x'} \*{e}_{x} ^T\otimes \*{e}_{y'} \*{e}_{y} ^T\\
        &= \sum_{y,y'} T^{(y,y')} \otimes r_{y'\vert y}~\*{e}_{y'} \*{e}_{y} ^T, \label{eq:M-T-r-yy}
    \end{align}
    where $T^{(y,y')} := \sum_{x,x'} t^{(y,y')}_{x'\vert x} \*{e}_{x'} \*{e}_{x} ^T\in\stoch(m',m)$. This shows that $M$ is a semilocalizable with a communication from $Y \to X$. To put it clearly, $S^{(y,y')}=r_{y'\vert y}~\*{e}_{y'} \*{e}_{y} ^T$ is the substochastic matrix.

    Conversly, if $M$ is semilocalizable, then it has a form as in Equation~\eqref{eq:cl_semilocal}. Consider taking its elements $M = \pqty{\mu_{x'y'\vert xy} : x,x'\in[n],  y\in[m], y'\in[m']}$
    \begin{align*}
        \mu_{x'y'\vert xy} &= \*{e}_{x'} ^T \otimes \*{e}_{y'} ^T \pqty{\sum_{j\in[\ell]} F_j \otimes S_j}  \*{e}_{x} \otimes \*{e}_{y}\\
        &= \sum_{j\in[\ell]} (\*{e}_{x'} ^T  F_j \*{e}_x) (\*{e}_{y'} ^T S_j  \*{e}_{y} ).
    \end{align*}
    Since $F_j$ is stochastic, we have that the sum over $x'$ result in
    \begin{equation*}
        \sum_{x' \in[n]} \mu_{x'y'\vert xy} = \sum_{x' \in[n]} (\*{e}_{x'} ^T  F_j \*{e}_x)=(\*{e}_{y'} ^T S_j  \*{e}_{y} ) = \*{e}_{y'} ^T S_j  \*{e}_{y}
    \end{equation*}
    independent of $x$ showing that $M$ is $X\not\to Y$ no-signalling.
\end{proof}

The identification of $X\not\to Y$ no-signalling with semilocalizable channel characterize the set of conditional mixing operations with a specific schema of transformation on a joint probability vector. With this result we are ready to state the constructive characterization of conditional majorization.

\subsection{Constructive characterization}
\begin{figure}
    \centering
    \includegraphics[height = 4cm]{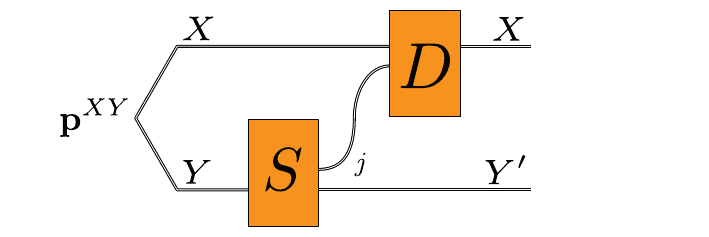}
    \caption{A diagram of conditionally doubly stochastic map, where $S = \sum_{j\in[\ell]} S_j ^{Y\to Y'} \otimes \ketbra{j}{j}$ and $D = \sum_{j\in[\ell]} D_j ^{X\to X} \otimes \ketbra{j}{j}$.}
    \label{fig:conditionally-doubly-stochastic}
\end{figure}

Suppose there are a correlated source $\p^{XY}$ and two agents: Alice and Bob. Alice has access to the system $X$ and a receiving end of the communication from Bob, and Bob has access to the system $Y$. Alice and Bob cooperate to implement a map $M$ which has to following structure. For Bob, he can read what $Y$ is, apply data processing to his side and ask Alice to apply doubly-stochastic matrix on Alice side. The doubly-stochastic operation can be conditioned on Bob's information of $Y$. However, Bob cannot receive or request Alice to send him information regarding $X$. After the application of this map, Bob end up having the same or higher level uncertainty regarding the system $X$. This is because Bob ask Alice to randomly permute outcome on $X$ with a doubly-stochastic matrix, and Alice never have told or hinted Bob on the state of $X$. 
These constriants give a concrete mathematical structure of the set of mixing operations, which we shall call \emph{conditionally doubly stochastic}, as follows.
\begin{defbox}{Conditionally doubly stochastic (CDS)}
    \begin{definition}
        A stochastic matrix $M\in\stoch(nm',nm)$ is said to be a \emph{conditionally doubly stochastic} (CDS) conditioned on $Y$ if it can be written in the form 
        \begin{equation}\label{eq:CMO-expression}
            M  = \sum_{j\in[\ell]} D_j^{X\to X} \otimes S_j ^{Y\to Y'}
        \end{equation}
        where $D_j$ is a doubly stochastic matrix and $S_j$ is a sub-stochastic matrix such that $\sum_{j\in[\ell]} S_j \in \stoch(m',m)$. See Figure~\ref{fig:conditionally-doubly-stochastic} for the diagram.
    \end{definition}
\end{defbox}
Notice that a conditionally doubly stochastic as expressed in Equation~\eqref{eq:CMO-expression} is indicative of being semilocalizable. We saw previously that the no-signalling condition from conditionally mixing operation is equivalent with semilocal operation. Indeed, with the condition of preserving uniform vector, the operation on $X$ is restricted to be doubly stochastic.

\begin{propbox}{Conditionally mixing operations and conditionally doubly stochastic}
    \begin{theorem}\label{th:cmo-cds}
        A stochastic matrix $M\in\stoch(nm',nm)$ is conditionally mixing operation if and only if it is conditionally doubly-stochastic.
    \end{theorem}
\end{propbox}
\begin{proof}
In the if direction, suppose $M\in\CDS(nm',nm)$. We want to show that $M$ satisfies both axioms of CMO. Since $\CDS(nm',nm)$ is a subset of semilocalizable operations, it is $X\not\to Y$ no-signalling. For the second axiom, $M$ preserves uniform distribution, since each $D_j$ is doubly-stochastic.

Conversly, suppose $M$ is conditionally mixing operation, we have that $M$ is semilocalizable and can be expressed as in~\eqref{eq:M-T-r-yy}. Left to show is that $T$ is doubly stochastic. By the second axiom of CMO, we have that $M$ preserve uniform distribution on $X$, this makes 
\begin{align*}
    M\pqty{\u^X\otimes \e^Y _y} &= \u^X \otimes \q^{Y'} _y \quad \exists\q_{y} \in\prob(m')\\
    &= \sum_{y'\in[n']} q_{y' \vert y} \u^{X}\otimes \e^{Y'} _{y'}
\end{align*}
Equate to previous form of $M$ 
\begin{equation*}
    \sum_{y'\in[n]} r_{y'\vert y} T^{(y,y')}\*{u}^X \otimes \e^{Y'}_{y'} = \sum_{y'\in[n]} q_{y' \vert y} \u^{X}\otimes \e^{Y'} _{y'}.
\end{equation*}
Taking a dot product with $\e^{Y'} _{y'}$ on both sides of the equation results in 
\begin{equation}
    r_{y'\vert y} T^{(y,y')}\*{u}^X = q_{y'\vert y} \u^{X}.
\end{equation}
Since $T$ is a stochastic matrix, then taking a dot product with $\*{1}$ to the right on both side $r_{y'\vert y}=q_{y'\vert y}$. Consequently, $T^{(y,y')}\u = \u$, equivalently $T^{(y,y')}$ is doubly stochastic for all $y$ and $y'$.
\end{proof}

\subsubsection{Examples of conditionally mixing operations}
For example, suppose we are given a certain $\p^{XY} = \sum_{y\in[m]} \p_y ^X \otimes \*{e}_y ^Y$. Suppose 
$$M=\sum_{j\in[m]} \Pi_j ^{X\to X} \otimes S_j ^{Y\to Y}\in\stoch(nm,nm),$$
where $S_j = \*{e}_j \e^T _j$ and the matrix $\Pi_j$ is a permutation matrix such that $\Pi_j \p_j = \p_j ^\da$ for any $j\in[m]$. The matrix $M$ leaves the state on $Y$ unchanged and apply $\Pi_y$ to rearrange $\p^X _y$ in to non-increasing order $\p^\da _y$ . Since permutations are invertible, we have that $M$ is invertible with the inverse
\begin{equation*}
    M^{-1} = \sum_{j\in[m]} \pqty{\Pi_j ^{-1}}^{X\to X}\otimes S_j ^{Y\to Y}.
\end{equation*}
This example proves that with conditionally doubly stochastic, one can permute the probability distribution on $X$ conditioned on $Y$ in non-increasing order. That is, we have the following lemma.
\begin{propbox}{}
    \begin{lemma}
        For any joint probability vector $\p^{XY}$, suppose $\p^{XY} = \sum_{y\in[m]} \p^X _y \otimes \*{e}_y ^Y$. The probability vector $\p^{XY}$ is reversibly convertible to $\tilde{\p}^{XY}$ by a conditionally doubly stochastic matrices where $\tilde{\p}^{XY} = \sum_{y\in[m]} (\p^\da _y)^{X} \otimes \*{e}_y$.
    \end{lemma}
\end{propbox}

The second example is a conditional mixing operation in the form, as appeared in Equation~\eqref{eq:in-proof-cmo-isometric-form},
\begin{equation*}
M = \sum_{y,y'} D_{(y,y')} \otimes r_{y'\vert y} \*{e}_{y'} \*{e}_{y} ^T,
\end{equation*}
where $D_{(y,y')}\in\dstoch(n,n)$ and $R = (r_{y'\vert y})\in\stoch(m',m)$. Notice that $\sum_{y,y'} r_{y'\vert y} \*{e}_{y'} \*{e}_{y} ^T \in \stoch(m',m)$. Given that $y$ is received on system $Y$, this map maps it to $y'$ and apply the map $D_{(y,y')}$ on system $X$ with probability $r_{y'|y}$. In general, any conditionally doubly stochastic can be put in this form.
\begin{propbox}{}
\begin{lemma}\label{lm:isometric-form-cmo}
    A matrix $M$ is a conditionally doubly stochastic if and only if there exists a set of doubly stochastic matrices $\qty{D_{(y,y')}: D_{(y,y')} \in \dstoch(n,n), y \in[m], y'\in[m']}$ and $R = (r_{y'\vert y})\in\stoch(m',m)$ such that 
    \begin{equation}\label{eq:constructive-Dreeyy}
        M = \sum_{y,y'} D_{(y,y')} \otimes r_{y'\vert y} \*{e}_{y'} \*{e}_{y} ^T
    \end{equation} 
\end{lemma}
\end{propbox}
\begin{proof}
    In the if direction, since $D_{(y,y')}$ is doubly stochastic and $\sum_{y',y} r_{y'\vert y} \*{e}_{y'}\*{e}_{y} ^T \in \stoch(m',m)$, then $M$ is conditionally doubly stochastic.
    
    Conversely, suppose $M \in\CDS(nm',nm)$. It is conditionally mixing operation. As in the proof Theorem~\ref{th:cmo-cds}, any CMO can be put in the form as in~\eqref{eq:in-proof-cmo-isometric-form} with $T$ being a doubly stochastic matrix. Identifying $T^{(y,y')}$ with $D_{(y,y')}$ completes the proof.
\end{proof}

With these two lemmas, convertibility via conditional mixing operation is equivalently stated with a unnormalized conditioned probability vector on the system $X$: $\p^X _y$ and $\q^X _{y'}$. The conditioned probability vector are defined to be such that
\begin{equation*}
    \p^{XY} = \sum_{y\in[m]} \p^{X} _y \otimes \e^{Y} _{y}, \qquad \q^{XY'} = \sum_{y'\in[m']} \q^{X} _{y'} \otimes \e^{Y'} _{y'}.
\end{equation*}
Explicitly, the elements of the vector is given by
\begin{equation}\label{eq:conditional-unnormalized}
    \p^{X} _y = \sum_{x\in[n]} p_{xy} \*{e}_x ^{X} = \pqty{I\otimes \e^T _{y}} \p^{XY}, \qquad \q^{X} _{y'} = \sum_{x\in[n]} q_{xy'} \*{e}_{x} ^{X} = \pqty{I\otimes \e^T _{y'}} \q^{XY'}.
\end{equation}
The equivalent statement is in the following lemma.
\begin{propbox}{Conditional mixing operation on the conditional}
    \begin{lemma}\label{lm:cmo-conditional}
        Suppose $\p^{XY} \in\prob(nm)$ and $\q^{XY'}\in\prob(nm')$. There is a conditional mixing operation $M$ such that $\p^{XY} = M \q^{XY'}$ if and only if there exists a set of doubly stochastic matrices $\qty{D_{(y,y')}: D_{(y,y')} \in \dstoch(n,n), y \in[m], y'\in[m']}$ and $R = (r_{y'\vert y})\in\stoch(m',m)$ such that for any $y'\in[m]$
        \begin{equation}
            \q_{y'} = \sum_{y\in[m]} r_{y'\vert y} D^{(y,y')} \p_y.
        \end{equation}
        where $\q_{y'}$ and $\p_{y}$ are defined as in Equation~\eqref{eq:conditional-unnormalized}.
    \end{lemma}
\end{propbox}
\begin{proof}
    First, consider the only-if direction. Suppose $\p^{XY} = M \q^{XY'}$ where $M$ is a conditional mixing operation. Then, by the Lemma~\ref{lm:isometric-form-cmo},
    \begin{equation}\label{eq:in-proof-cmo-isometric-form}
        M = \sum_{\substack{y\in[m]\\y'\in[m']}} D_{(y,y')} \otimes r_{y'\vert y} \*{e}_{y'} \*{e}_{y}^T.
    \end{equation}
    Since $\q_{y'} = \pqty{I \otimes \e^T _{y'}} \q^{XY'}$, we have that 
    \begin{align*}
        \q^{Y'} _{y'} &= \pqty{I \otimes \e^T _{y'}}\pqty{\sum_{\substack{y\in[m]\\z\in[m']}} D_{(y,z')} \otimes r_{z\vert y} \*{e}_{z} \*{e}_{y}^T}\pqty{ \sum_{w\in[m]} \p^{X} _w \otimes \e^{Y} _{w}}\\
        &= \sum_{y} r_{y'\vert y} D_{(y,y')} \p^{X} _y.
    \end{align*}
    Conversely, if there exists such $D$ and $R$, then $\p^{XY} = M \q^{XY'}$ where $M$ defined as in the Equation~\eqref{eq:in-proof-cmo-isometric-form}.
\end{proof}

The theorem above stated thtat one can turn a mixture of conditionals of $\p^{XY}$ to the conditional of $\q^{XY'}$ via doubly stochastic matrices. Would this implies majorization relations between the conditional as well? As it turns out, there are majorization relations between the conditionals but only once we put the conditional of $\p^{XY}$ in the non-increasing order.
\begin{propbox}{Convex sum characterization}
    \begin{theorem}\label{th:convex_sum_characterization}
        Suppose $\p ^{XY} \in \prob(nm)$, $\q^{XY'} \in \prob(nm')$, $\abs{X} = n$, $\abs{Y} = m$, and $\abs{Y'} = m$. There exists a conditional mixing operation $M$ such that $\q^{XY'} = M \p^{XY}$ if and only if for any $y\in[n]$ there exists a convex combination by the convex parameters $r_{y'\vert y}$ such that
        \begin{equation}\label{eq:condi-maj-convex-sum-char}
            \sum_{y\in[n]} r_{y'\vert y} \p^\da _y \succ \q_{y'} 
        \end{equation} 
        where $\q_{y'}$ and $\p_{y}$ are defined as in Equation~\eqref{eq:conditional-unnormalized}.
    \end{theorem}
\end{propbox}
\begin{proof}
    Suppose that there is $M\in\CMO(nm',nm)$ such that $\q^{XY'} = M \p^{XY}$. Equivalently, by the Lemma~\ref{lm:cmo-conditional}, we have that 
    \begin{equation*}
        \sum_{y\in[m]} r_{y'\vert y} D_{(y', y)} \p^X _y = \q^X _{y'} 
    \end{equation*}
    Recall that, for each $y$ and $y'$, $\p^X _y \succ D_{(y', y)} \p^X _y$. For a vector in non-increasing order,
    \begin{equation*}
        \p^\da _{y} \sim \p^X _y \succ D_{(y', y)} \Pi_y \p^\da _y  
    \end{equation*}
    where for each $y$, the permutation matrix $\Pi_y$ is such that $\Pi_y \p^\da _y = \p^{X} _y$. Since the composition $D_{(y,y')}\Pi_y$ is doubly-stochastic for any $y,y'$ and the majorization relation above, we have that for all $k\in[n]$
    \begin{equation*}
        \knorm{\vctr_{y'\vert y} \p^\da _y}{k} \geq \knorm{\vctr_{y'\vert y} D_{y',y} \Pi_y \p^\da _y}{k}.
    \end{equation*}
    Notice that summing over $y\in[n]$ on the right side of the inequality is equal to $\knorm{\q_{y'} ^X}{k}$. We have the following chain of inequalities for any $k\in[n]$
    \begin{align*}
        \knorm{\q^{X} _{y'}}{k} &= \knorm{\sum_{y\in[m]} r_{y'\vert y} D_{y',y} \Pi_y \p^\da _y}{k}\\ 
        &\leq \knorm{\sum_{y\in[m]} r_{y'\vert y} \pqty{D_{y',y} \Pi_y \p^\da _y}^\da}{k} = \sum_{y\in[m]} r_{y'\vert y} \knorm{D_{y',y} \Pi_y \p^\da _y}{k}\\
        \magenta{\pqty{\p^\da _{y} \succ D_{(y', y)} \Pi_y \p^\da _y} \rightarrow} &\leq \sum_{y\in[m]} r_{y'\vert y} \knorm{ \p^\da _y}{k} = \knorm{\sum_{y\in[m]} r_{y'\vert y} \p_y ^\da}{k}.
    \end{align*}
    We have that $\knorm{\q^{X} _{y'}}{k} \leq \knorm{\sum_{y\in[m]} r_{y'\vert y} \p_y ^\da}{k}$ for all $k\in[n]$, which is equivalent with the Equation~\eqref{eq:condi-maj-convex-sum-char}.

    Conversely, suppose that the Equation~\eqref{eq:condi-maj-convex-sum-char} holds. We want to show that there is a conditionally doubly stochastic $M$ such that $\q^{XY'} = M \p^{XY}$. The Equation~\eqref{eq:condi-maj-convex-sum-char}, can be equivalently stated with some doubly stochastic matrix $D_{y'}$ as 
    \begin{equation*}
        \sum_{y\in[n]} r_{y'\vert y} D_{y'} \p^\da _y = \q_{y'} \quad \forall y' \in [m'].
    \end{equation*}
    We define $\Pi_y ^{-1}$ such that $\Pi_y ^{-1} \p^X _y = \p^{\da} _y$, this is leading to 
    \begin{equation*}
        \q_{y'} = \sum_{y\in[n]} r_{y'\vert y} D_{y'} \p^\da _y = \sum_{y\in[n]} r_{y'\vert y} D_{y'} \Pi_y ^{-1} \p^X _y \quad \forall y'\in[m'],
    \end{equation*}
    which is equivalent to the convertibility in the Lemma~\ref{lm:cmo-conditional}.
\end{proof}

This lemma leads to the characterization of conditional majorization with a feasibility problem.
\begin{propbox}{Characterization from linear programming}
\begin{theorem}
    Let $\p^{XY}$ and $\q^{XY'}$, then $\p^{XY} \succ_{X} \q^{XY'}$ if and only if for all $S\in\stoch_\leq (m,m')$ we have that 
    \begin{equation}\label{eq:characterization-cond-maj-lin-prog}
        \sum_{y\in[n]} \max_{w\in[m']} \s_w ^\da \cdot \p_y ^\da \geq \sum_{w\in[n']} \s_w ^\da \cdot \q_{w} ^\da
    \end{equation}
    where $\s_w$ is the $w^\rm{th}$ column of $S$.
\end{theorem}
\end{propbox}
\begin{proof}
Suppose $\p^{XY} \succ_{X} \q^{XY'}$, we want to show that it is equivalent with the Equation~\eqref{eq:characterization-cond-maj-lin-prog}. The idea is to rewrite the Theorem~\ref{th:convex_sum_characterization} into a linear programming and then apply the Farkas lemma. For simplicity, we will denote in this proof $(\p^{X} _y)^\da = \p_y$ and $(\q_w ^{X})^\da = \q_w$

From the Theorem~\ref{th:convex_sum_characterization}, the conditional majorization relation is equivalent with the existence of a stochastic matrix $R = \pqty{r_{y'\vert y}}_{y',y}$ such that it satisfies 
\begin{equation}\label{eq:conditional-L-maj}
\sum_{y\in[m]} r_{y'\vert y} L \p_y ^X \geq L \q_{y'} ^X,
\end{equation} 
where $L$ is a lower triangular matrix with non-zero elements equal to 1.
Denote each row of $R$ by $\vctr^T _{y'}$ for $y'\in[m']$. 
% That is 
% \begin{equation*}
%     R = \pmqty{ \vctr^T _1 \\  \vctr^T _2 \\ \vdots \\ \vctr^T _{m'}}.
% \end{equation*}
Define a concatenation of $\vctr_{y'}$ as $\vctr$,
\begin{equation*}
    \vctr= \pmqty{ \vctr_1 \\  \vctr_2 \\ \vdots \\ \vctr_{m'}}\in\bb{R}^{mm'} _+.
\end{equation*}
Define $P = \pmqty{ \p_1 & \p_2 & \p_3 & \ldots & \p_m}$. The condition in Equation~\eqref{eq:conditional-L-maj}that $R$ must satisfy can be written as
\begin{equation}\label{eq:conditional-linear-prog}
    \sum_{y\in[m]} r_{y'\vert y} L\p_y = LP\vctr_{y'} \geq L\q_{y'}\quad \forall y'\in[m']
\end{equation}
In addition, because $R$ is stochastic, $\vctr$ must satisfy 
\begin{equation}
    \sum_{y'\in[m']} \vctr_{y'} = \*{1}^{(m)},
\end{equation}
where $\*{1}$ is a vector containing only ones $1$ of dimension $1$. However, if there exists a $\vctr$ which sum of less than $\*{1}^{(m)}$ instead of equality, then there exists $\vctr'$ such that it satisfy the equality and the condition in~\eqref{eq:conditional-linear-prog}. Notice that,  
\begin{align*}
\sum_{y'\in[m']} \vctr_{y'} &= \sum_{y'\in[m]} \sum_{y\in[m]} r_{y'\vert y}  \*{e}_y = \sum_{y\in[m]} \sum_{y'\in[m]}  r_{y'\vert y}  \*{e}_y 
\end{align*}
If we have inequality for any $y\in[m]$
\begin{align*}
    \sum_{y'\in[m]}  r_{y'\vert y}   &\leq 1,
\end{align*}
then we have that there is $\alpha_y \in (0,1]$ such that 
\begin{equation}
    \sum_{y'\in[m]}  \frac{1}{\alpha_y} r_{y'\vert y}   = 1.
\end{equation}
The subscript $y$ in $\alpha_y$ signifies that $\alpha_y$ need not to be identical among all $y\in[m]$. By redefining $\vctr= (r_{y'\vert y})$ to $\vctr' = (\frac{r_{y'\vert y}}{\alpha})$, we have a that $\vctr' \geq \vctr$ and 
\begin{equation}
    LP\vctr' _{y'} \geq LP\vctr_{y'} \geq L\q_{y'}
\end{equation}
because $LP$ is a matrix with all entries being positive. 
% Now, the existence of the stochastic matrix $R$ can be put in terms of feasibility problem 
% \begin{align}
%     \rm{Is} \{ &\*{r}: LP\*{r}_{y'}\geq L\*{q}_{y'} \forall y'\in[m],
%         &\sum_{y'\in[m']} = \

%     \}
% \end{align}
With these notations, the existence of the stochastic matrix $R$ is equivalent with non-emptiness of the feasible set 
\begin{equation*}
    \qty{\vctr: M\vctr\leq \b, \vctr\geq 0},
\end{equation*}
where $M$ is defined as 
\begin{equation}
    M = \pmqty{-LP & 0 & 0 & \ldots & 0\\ 0 & -LP & 0 &\ldots & 0\\ \vdots & \vdots & \ddots & & \vdots\\ 0 & 0 & 0 & \ddots & 0 \\ 0 & 0 & 0 & \ldots & -LP \\ I_m & I_m & I_m & \ldots & I_m}
\end{equation}
and $\b$ is defined as 
\begin{equation}
    \b = \pmqty{-L \q_1 & -L \q_2 & -L \q_3 & \ldots & -L \q_{m'} & \*{1}^{(m)}} ^T .
\end{equation}
From Farkas lemma, the feasible set is non-empty if and only if for any $\vctv \in \bb{R}^{nm' + m} _+$, we have that $\vctv^T M \geq 0$ implies $\vctv\cdot \b \geq 0$. Suppose that $\vctv^T = \pmqty{\vctv_1 ^T& \vctv_2 ^T& \vctv_3 ^T& \ldots& \vctv_{m'} ^T & \t^T }$, where $\vctv_i \in \bb{R}^{m'}$ for all $i\in[m']$ and $\t \in \bb{R}^{m}$. The condition $\vctv^T M$ can be written as 
\begin{align*}
    &\vctv^T M \geq 0 \\ 
    &\iff \t^T  \geq \vctv_w ^T LP  \quad \forall w \in[m'] \\
    &\iff \t^T  \geq \max_{w\in[m]} \vctv_w ^T LP  \\
    &\iff t_y \geq \max_{w\in[m]} \vctv_w ^T L\p_y \quad \forall y \in[m].
\end{align*}
From $\vctv \cdot \b \geq 0$, equivalently write 
\begin{align*}
    &\sum_{w\in[m']} \vctv_w \cdot (-L\q_w) + \t \cdot \*{1} \geq 0 \\
    &\iff \sum_{y\in[m]} t_y \geq \sum_{w\in[m']} \vctv_w \cdot (L\q_w).
\end{align*}
The statement: for all $i \in [m']$, any $v_i\in\bb{R}^{n} _+$ and $\*{t}\in\bb{R}^{m}$, $t_y \geq \max_{w\in[m]} \vctv_w ^T L\p_y$ for all $y \in[m]$ implies $\sum_{y\in[m]} t_y \geq \sum_{w\in[m']} \vctv_w \cdot (L\q_w)$; holds if and only if for all $\vctu = \pmqty{\vctu_1 ^T& \vctu_2 ^T& \vctu_3 ^T& \ldots& \vctu_{m'} ^T}^T \in\bb{R}^{nm'} _+$
\begin{equation}
    \sum_{y\in[m]} \max_{w\in[m']} \vctu^T _w L \p_y \geq \sum_{w\in[m']} \vctu_w ^T \cdot L \q_w.
\end{equation}
% \magenta{\huge\textbf{Never leave your laptop unattended}} Thank you!
Using $ \vctu^T _w L = U\vctu _w$ and define $\s_w := U\vctu _w$. Note that $\s_w ^\da = \s_w$. Therefore, the feasible stochastic matrix $R$ exists if and only if for any $\s \in \bb{R}^m _+$,
\begin{equation}\label{ineq:condition-major-linear-prog}
    \sum_{y\in[m]} \max_{w\in[m']} \s_w ^\da \cdot \p_y ^\da  \geq \sum_{w\in[m']} \s_w ^\da \cdot \q_w ^\da.
\end{equation}
To rephrase in terms of stochastic matrix $S$, take $\overline{s} = \max_{w} \sum_{x\in[m]} \*{e}_x ^T \cdot \s_w$, we have that 
\begin{equation*}
    S = \frac{1}{\overline{s}} \pmqty{ \s_1 & \s_2 & \ldots & \s_{m'}} \in \stoch_\leq (m,m'). 
\end{equation*}
Since $\frac{1}{\overline{s}}\s_w$ is a scaling of $\s_w$, the inequality in~\eqref{ineq:condition-major-linear-prog} holds for all $S \in \bb{R}_{+} ^{m,m'}$ if and only if it holds for $S \in \stoch_\leq (m,m')$.
\end{proof}
With this lemma, a conditional majorization can be characterize with an optimization task in each dot product with the conditional. But what would be the task that naturally rised when one think about an uncertainty when given an access to a correlated random variable. We will introduce and disscuss such the task in the next section. 
\subsection{Operational characterization}
A $k$-gambling game characterizes the probability vector majorization. Similarly, conditional majorization relation can be equivalently stated via the comparison of a family of gambling games.

\begin{figure}[!h]
    \centering
    \includegraphics[height=2.5cm]{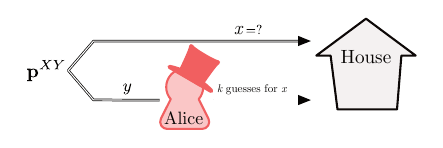}
    \caption{A diagram for a $k$-game in which a player is given $y$ from the joint probability vector $\p^{XY}$.}
    \label{fig:conditional-k-game}
\end{figure}

\begin{figure}[!h]
    \centering
    \includegraphics[height=2.5cm]{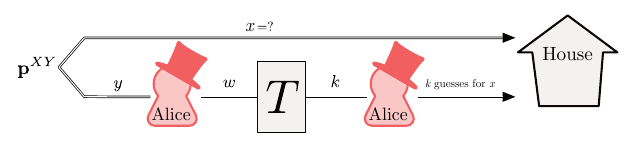}
    \caption{A diagram for a conditional $T$-game.}
    \label{fig:conditional-T-game}
\end{figure}

In the case of the probability vector majorization, we have an operation meaning for one probability vector $\p^{X}$ to be less uncertain than $\q^{XY'}$ if one has better chance winning with distribution $\p^{XY}$ in any $k$-games. In conditional majorization, we consider an uncertainty of $X$ when one have access to $Y$. That is a player in this family of gambling games must feature the utilization of information on $Y$ to make better guesses. The simplest game one could devise to respect the conditional majorization would be to let to player access $y$ in each $k$-game. That is the winning chance of player, instead of simply given by the Fan $k$-norm, one weight the Ky Fan $k$-norm with a probability of obtaining character $y$ on the system $Y$, we have winning chance
\begin{equation*}
    \sum_{y\in[m]} p_y \knorm{\p^{Y} _{\vert y}}{k}.
\end{equation*}
See Figure~\ref{fig:conditional-k-game} for a diagram.
In this family of games, each $k$ game specifies the number of guesses a player can make. However, one also could devise a bigger family of gambling games which includes the scenario where a player can influence a number of guesses they are given. To incorporate this, we model a game with a transition matrix $T$. Each time a player receives the value $y$, a player choose an action out of $\ell$ possibilities. The actions are labeled with $w$ which gives a probability on $k$ as $\*{t}_{\vert w} = \sum_{k\in[n]} t_{k\vert w} \*{e}_k$ and probability of losing on the spot $1-\sum_{k\in[n]} t_{k\vert w}$.
That is $T$ need not be a stochastic matrix. $T$ can be a sub-stochastic matrix $T\in\stoch_\leq (n,\ell)$. See Figure~\ref{fig:conditional-T-game} for the diagram.

The characterization of conditional majorization with a gambling game is first presented in~\cite{BGG2022}.

\begin{probbox}{Conditional $T$-gambling games}
    \textbf{Given:} a matrix $T = (t_{k\vert w})_{k\in[n],w\in[\ell]}$, a joint probability distribution $\p^{XY}$\\
    \textbf{Mechanism:} \\
    \begin{enumerate}
        \item draw $x$ and $y$ from the source with distribution $\p^{XY}$, give $y$ to the player.
        \item player pick $w$ and send $w$ to the house and $k$ is picked with probability $t_{k\vert w}$.
    \end{enumerate}
    \textbf{Goal:} gives correct value of $x$ within $k$ guesses.
\end{probbox}

To devise the best strategy, first consider a winning chance with $k$ guesses. Similarly to the $k$-game, the best guesses is to guess the first $k$ most likely outcomes conditioned on the received $y$. With $k$ guesses, the likelihood of a person winning a $k$ game is given by
\begin{equation*}
    \rm{Pr}_k (\p^{XY}) = \sum_{y\in[m]} p_y \sum_{x\in[k]} p_{x\vert y}^\da
\end{equation*}
where $p _{x\vert y}$ is the probability that $X=x$ given $Y=y$. If $y$ is given, and the player chose $w$ to be the input, the winning chance is 
\begin{equation*}
\sum_{k\in[n]} t_{k\vert w} \sum_{x\in[k]} p^\da _{x\vert y} = \sum_{x\in[n]} p^\da _{x\vert y} \sum_{k=x} ^{n} t_{k\vert w}
\end{equation*}
Define $s_{xw} = \sum_{k = x} ^{n} t_{k\vert w}.$ The above equation* becomes,
\begin{equation*}
\sum_{x\in[n]} p^\da _{x\vert y} \sum_{k=x} ^{n} t_{k\vert w} = \sum_{x\in[n]} s_{xw} p^\da _{x\vert y} 
\end{equation*}
Define $\s_w = \pqty{s_{1w},s_{2w},\ldots,s_{nw}}^T$ and $\p_y = p_y \pqty{p_{1\vert y}, p_{2\vert y}, \ldots, p_{n\vert y}}^T$. We have that
\begin{equation*}
    p_y \sum_{x\in[n]} p^\da _{x\vert y} \sum_{k=x} ^{n} t_{k\vert w} = \s_w \cdot \p_y ^\da.
\end{equation*}

The best strategy is to pick $w$ that maximize the probability of winning when $y$ is received. In a game, prior to receiving $y$, a player playing a conditional $T$-gambling game has a winning probability 
\begin{equation}
    \rm{Pr}_T (\p^{XY}) = \sum_{y\in[m]} \max_{w\in[\ell]} \s_w \cdot \p_y ^\da.
\end{equation}

This chance of winning $T$-gambling games characterize the majorization relation.
\begin{propbox}{Conditional majorization and the game of chance}
    \begin{theorem}\label{th:coincidence-cond-major-gameChance}
        Suppose $\p^{XY}\in\prob(nm)$ and $\q^{XY'}\in\prob(nm')$. The conditional majorization relation $\p^{XY} \succ_X \q^{XY'}$ holds if and only if for any $T-$gambling game where $T\in\stoch_{\leq}(n\ell)$, we have that $\rm{Pr}_T (\p^{XY}) \geq \rm{Pr}_T (\q^{XY})$.
    \end{theorem}
\end{propbox}
\begin{proof}
Suppose $\p^{XY}\succ \q^{XY'}$. By the characterization in Theorem~\ref{th:convex_sum_characterization}, we have that 
\begin{equation}
\sum_{y\in[m]} \frac{r_{y'\vert y}}{q_{y'}} p_y \p^{\da} _{\vert y} \succ \q_{\vert y'} \quad \forall y' \in [m']
\end{equation}
where $\p _{\vert y}$ is a probability vector such that $\p_{y} = p_y \p_{\vert y}$ and similarly for $\q_{\vert y'}$. A winning chance in a $T$-gambiling game is 
\begin{equation}
    \rm{Pr}_T (\p^{XY}) = \sum_{y\in[m]} p_y \max_{w\in[\ell]} \sum_{k\in[n]} t_{k\vert w} \knorm{\p_{\vert y}}{k}.
\end{equation}
Notice that the function $f:\prob(n)\to \bb{R}_+$ defined as 
\begin{equation*}
    f(\p^{X} _{\vert y}):= \max_{w\in[\ell]} \sum_{k\in[n]} t_{k\vert w} \knorm{\p_{\vert y}}{k}
\end{equation*}
is symmetric and convex. We rewrite the winning chance in the form 
\begin{equation}\label{eq:winning-chance-f-conditional-form}
    \rm{Pr}_T (\p^{XY}) = \sum_{y\in[m]} p_y f(\p^X _{\vert y}).
\end{equation}
Now, consider the winning chance with the probability vector $\q^{XY'}$
\begin{align*}
    \rm{Pr}_T (\q^{XY'}) &= \sum_{y'\in[m']} q_{y'} f(\q_{\vert y'} ^X) \\
    &= \sum_{y'\in[m']} q_{y'} f\pqty{\sum_{y\in[m']} \frac{r_{y'\vert y}}{q_{y'}} p_y \p^{\da} _{\vert y}} \\ 
    &\leq \sum_{y\in[m]} \sum_{y'\in[m']} r_{y'\vert y} p_y f\pqty{\p^{\da} _{\vert y}} \\ 
    &= \sum_{y\in[m]} p_y f\pqty{\p _{\vert y}} = \rm{Pr}_T (\p^{XY}).
\end{align*}

Conversely, suppose that we have the winning chance inequality 
\begin{equation*}
    \forall \ell\in[n]~\forall T\in\stoch_{\leq} (n,\ell) \quad \sum_{y\in[m]} \max_{w\in[\ell]} \s_w \cdot \p^\da _y \geq \sum_{y'\in[m']} \max_{w\in[\ell]} \s_w \cdot \q_{y'},
\end{equation*}
where $\s_w = \pqty{s_{1w},s_{2w},\ldots,s_{nw}}^T$ and $s_{xw} = \sum_{k = x} ^{n} t_{k\vert w}.$

For specific $\ell = m'$, we have the inequality as well. By removing the maximization over $m'$ on the right side of the inequality, we have a lessor quantity
\begin{equation}\label{ineq:proof-operational-conditional-01}
    \forall T \in\stoch_{\leq} (n,m') \quad \sum_{y\in[m]} \max_{w\in[m']} \s_w \cdot \p^\da _y \geq \sum_{y'\in[m']} \max_{w\in[m']} \s_w \cdot \q_{y'} \geq \sum_{y'\in[m']} \s_{y'} \cdot \q_{y'}.
\end{equation}
We want to show the implication of above inequality that for all $R\in\stoch_{\leq} ^\da (n,m')$
\begin{equation*}
    \sum_{y\in[m]} \max_{w\in[m']} \vctr_w  \cdot \p_y ^\da \geq \sum_{w\in[m']} \vctr_w \cdot \q_{w} ^\da 
\end{equation*}
where $\vctr_w$ is the $w$-th column of $R$. For any $R\in\stoch^\da _\leq (n,m')$, we have that there exists $T$ such that $R = UT$ where $U$ is an upper triangular matrix whose its non-zero elements are equal to one. We have that any column $\vctr_w$ of any $R\in\stoch^\da _\leq (n,m')$ can be written as 
\begin{equation*}
    \vctr_w = U \t_w = \sum_{x=1} ^{n} \sum_{k=x} ^{n}t_{k\vert w} \*{e}_x = \sum_{x=1} ^{n} s_x \*{e}_x = \s_w.
\end{equation*}
Since each element of $\vctr_w$ is a sum over $t_{k\vert w}$ and $\vctr_w$ is in non-increasing order, $T$ is a substochastic matrix. Notice that $\s_w$ is defined exactly the same as in the inequality~\eqref{ineq:proof-operational-conditional-01}. With the above identification of $\vctr_w$ with $\s_w$, we have the desired implication.
% To attain this, suppose $U\in\bb{R}^{n\times n}$ be an upper triangular matrix whose its non-zero elements are equal to one. This matrix $U$ maps $T\in\stoch_{\leq} (n,m')$ to $UT\in\bb{R}^{n\times m} _+$. Since this matrix is invertible any $\bb{R}^{n\times m} _+$ with properly rescaling, we have a bijection, denoted with $\tilde{U}$ from $\stoch_{\leq} (n,m')$ to $\stoch ^\da _{\leq} (n,m')$. That is we can write $\vctr_w$ as $\alpha\s_w$, where $\alpha$ is the rescaling factor. We have 
% \begin{align*}
%     \sum_{y\in[m]} \max_{w\in[m']} \vctr_w  \cdot \p_y ^\da = \sum_{y\in[m]} \max_{w\in[m']} alpha \s_w  \cdot \p_y ^\da 
% \end{align*}
\end{proof}

\subsection{Monotones of conditional majorization}

\begin{defbox}{Conditionally Schur-convex function}
\begin{definition}
    A function $f: \bigcup_{m,n\in\bb{N}} \prob(nm) \to \bb{R}$ is a \emph{conditionally Schur-convex function} if for any $\p^{XY} \in \prob(nm)$, any $\q^{XY'} \in\prob(nm')$, and any $n,m,m'\in \bb{N}$
    \begin{equation}
        \p^{XY} \succ_X \q^{XY'} \implies f(\p^{XY}) \geq f(\q^{XY}).
    \end{equation} 
\end{definition}
\end{defbox}

Notice that in the previous proof, we established that a function in the Equation~\eqref{eq:winning-chance-f-conditional-form} is convex. In showing that, we utilize only the characterization in Theorem~\ref{th:convex_sum_characterization} and the convexity of $f$. Therefore, we have the following lemma.
\begin{propbox}{$f$-conditionally Schur-convex}
    \begin{lemma}\label{lm:f-condition-schur-conv}
        Suppose a function $f:\bigcup_{n\in\mN} \prob(n) \to \bb{R}$ is symmetric and convex. Then, a function $g: \bigcup_{m,n\in\bb{N}} \prob(nm) \to \bb{R}$ defined as 
        \begin{equation}
            g(\p^{XY}) = \sum_{y\in[m]} p_y f(\p_{\vert y} ^{X})
        \end{equation}
        is conditionally Schur-convex. 
    \end{lemma}
\end{propbox}

This leads to a succinct definition of the one of the important conditionally Shur-convex functions, namely the conditional entropy.

\begin{defbox}{The classical conditional entropy}
    \begin{definition}
        The \emph{conditional entropy} of a classical system $X$ conditioned on a classical system $Y$ is defined as 
        \begin{align*}
            H(X\vert Y)_{\p^{XY}} &= \sum_{y\in[m]} p_y H(\p^X _{\vert y})
        \end{align*}
        where $H(\p^X _{\vert y})$ is a Shannon entropy, $\p^{XY}= (p_{x,y})_{x,y}$ is a joint probability vector, $p_y$ is a probability of $y\in[m]$ and $\p^X _{\vert y} = (\p_{x\vert y})_x$ is a probability vector associated the distribution of $x$ given $y$.
    \end{definition}
\end{defbox}
With the Lemma~\ref{lm:f-condition-schur-conv}, the classical conditional entropy is conditionally Schur-concave. Similarly to entropies and relative entropies, the classical conditional entropy is additive. To see this, consider a product of bipartite classical probability vector 
\begin{align*}
    \p^{XX'YY'} &= \vctq^{XY}\otimes \vctr^{X'Y'} = \pqty{\sum_{y\in[m]} q_y \q_{\vert y} ^{X} \otimes \*{e}_y ^{Y} }\otimes \pqty{\sum_{y'\in[m']} r_{y'} \*{r}_{\vert y'} ^{X'} \otimes \*{e}_{y'} ^{Y'}}\\ 
    &= \sum_{\substack{y\in[m]\\y'\in[m']}} q_y r_{y'} \*{q}_{\vert y} ^{X} \otimes \*{r}_{\vert y'} ^{X'} \otimes \*{e}_y ^{Y} \otimes \*{e}_{y'} ^{Y'}.
\end{align*}
The entropy is 
\begin{align*}
    H(XX'\vert YY')_{\p^{XX'YY'}} &= \sum_{y\in[m]} p_y H(\p^X _{\vert y}) = \sum_{\substack{y\in[m]\\y'\in[m']}} q_y r_{y'} H(\*{q}_{\vert y} ^{X} \otimes \*{r}_{\vert y'} ^{X'}) \\
    &=\sum_{\substack{y\in[m]\\y'\in[m']}} q_y r_{y'} H(\*{q}_{\vert y} ^{X}) +\sum_{\substack{w\in[m]\\w'\in[m']}} q_w r_{w'} H(\*{r}_{\vert w} ^{X'})\\
    &= \sum_{y\in[m]} q_y H(\*{q}_{\vert y} ^{X}) +\sum_{w'\in[m']}r_{w'} H(\*{r}_{\vert w'} ^{X'})\\
    &= H(X\vert Y)_{\*{q}^{XY}} + H(X'\vert Y')_{\*{r}^{X'Y'}}.
\end{align*}
where the third equality is from the additivity of Shannon entropy.

For example, if $\p^{XY} = \q^{X} \otimes \*{r}^{Y}$, then for every $y\in[m]$, $\p^{X} _{\vert y}$ is constant with $y$, indeed $\p^{X} _{\vert y} = \q^{X}$. We have 
\begin{equation*}
    H(X\vert Y)_{\p^{XY}}= \sum_{y\in[m]} p_y H(\p^X ) =  H(\q^{X}).
\end{equation*}
That is, the classical conditional entropy for two uncorrelated system $XY$ condition on $Y$ is a Shannon entropy of $X$.

%% Print bib in every chapter for ease of revision
% \bibliography{QuantumEntropy}
% \bibliographystyle{unsrt}
% \printbibliography

% \tableofcontents 
\setcounter{chapter}{2}
\chapter{Uncertainty of classical channels}
% \magenta{Add examples in multiple places}\\ 
%Assume that the readers are already introduced to the superchannel in the prelim chapter
This chapter presents our answer to one of the main thesis questions: what is uncertainty of classical channel? Given two classical channels of the same output system, $\cal{N}^{X\to Y}$ and $\cal{M}^{X'\to Y}$, we would like to define and compare the \emph{uncertainty} inherent in each channel. Specifically, the goal is to compare the uncertainty of the two classical channels given that we have all the access and control over the input of the channels. To do this, we propose a definition of a majorization relation on the set of classical channels. We follow the approach we set in the probability vector majorization, in which we defined mixing operations axiomatically and constructively. The operational approach gives us a definition of majorization in terms of the comparison between success probabilities in carrying out channel-related tasks.

This chapter begins with a discussion of classical channels and classical superchannels. Then, we discuss the definition of classical channel majorization with three conceptually distinct approaches, coincidence of the approaches, and basic properties. The chapter concludes with the characterization and connection to conditional majorization. 

\section{Classical channels and superchannels}
Given a classical channel $\cal{N}\in\cptp(X\to Y)$, there is an associated transition matrix $N$. The probability vector $\*{p}_x$ represents $x$-th column of the transition matrix $N$. And similarly, $\q_w$ represents the $w$-th column of the transition matrix $M$ associated with a classical channel $\cal{M}\in\cptp(X'\to Y')$.

\subsection{Classical superchannels}
Instead of concerning the whole class of quantum superchannels, we restrict to the superchannels that totally dephases an input channel and only output classical channels. Recall that a classical channel $\cal{N}^{X\to Y}$ is a channel that is invariant under the pre-composition and post-composition with totally dephasing channels $\cal{D}^X$ and $\cal{D}^Y$. Succinctly, we denote this operation as a totally dephasing superchannel $\Delta^{(X\to Y)}[\cal{N}]:= \cal{D}^Y\circ \cal{N}^{X\to Y} \circ \cal{D}^X$. Formally, we have the following definition of classical superchannel,
\begin{defbox}{Classical superchannels}
    \begin{definition}
        A superchannel $\Theta\in\super((X\to Y)\to (X'\to Y'))$ is a classical \emph{superchannel} if it is invariant under the conjugation with totally dephasing superchannel $\Delta^{\pqty{X\to Y}}$. That is
        \begin{equation}\label{eq:def-cl-super-ch}
            \Theta = \Delta^\pqty{X'\to Y'} \circ \Theta \circ \Delta^\pqty{X\to Y}.
        \end{equation}
    \end{definition}
\end{defbox}
Notice that we require the superchannel to dephase the input channels, as the requirement of outputting classical channel alone would give a more general superchannel called a \emph{maximally incoherent superchannel}~\cite{SCG2020}.

A classical superchannel can be realized as classical pre-processing and post-processing channels. To denote classical systems, we use the last letters of the English alphabet $X,$ $Y,$ $Z,$ and $W$.
\begin{propbox}{The realization theorem for classical superchannel}
\begin{theorem}\label{th:Cl-sup-realization}
    For any classical superchannel $\Theta\in\super((X\to Y)\to(X'\to Y'))$, there exists a classical system $Z$, a classical pre-processing channel $\cal{F}^{X' \to XZ}$, and a classical post-processing channel $\cal{E}^{YZ\to Y'}$ such that 
    \begin{equation}
    \Theta[\cal{N}] = \cal{E}^{{YZ\to Y'}}\circ \cal{N}^{X\to Y}\circ \cal{F}^{X' \to XZ}.
    \end{equation}
\end{theorem}
\end{propbox}
\begin{proof}
    From the realization Theorem~\ref{th:super-realization}, we have that there exist an isometry channel $\cal{V}\in\cptp(X'\to XE)$ and a channel $\cal{G}\in\cptp(YE\to Y')$,
    \begin{equation}
        \Theta[\cal{N}] = \cal{G}\circ \cal{N}\circ \cal{V}.
    \end{equation}
    We would like to put an emphasis here that the realization theorem does not require the system $E$ to be a classical system. The aim of this proof is to show that a classical channel is realizable with a classical system.
    
    Using Equation~\eqref{eq:def-cl-super-ch}, the realization of the superchannel can be equivalently written as 
    \begin{equation}
    \Theta [\cal{N}] = \cal{D}^{Y'}\circ \cal{G}^{YE\to Y'} \circ \pqty{\cal{D}^Y\otimes \cal{I}^Z}\circ \pqty{\cal{N}^{X\to Y} \otimes \cal{I}^Z}\circ\pqty{\cal{D}^X\otimes\cal{I}^Z}\circ\cal{V}^{X'\to XE}\circ\cal{D}^{X'}.
    \end{equation}
    The equation is expressed in a diagram in Figure~\ref{fig:cl-super-dephase}.
    \begin{figure}[h]
        \centering
        \includegraphics[height=3.6cm]{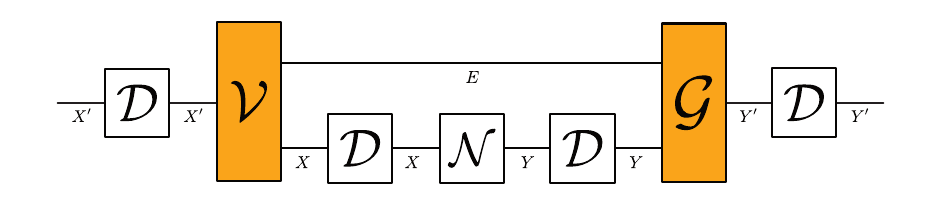}
        \caption{A realization of classical superchannel is invariant under composition with the dephasing channels.}
        \label{fig:cl-super-dephase}
    \end{figure}
    \\ Define these compositions with dephasing channels as new channels
    \begin{align*}
        \cal{P}^{X'\to XE} &:= \pqty{\cal{D}^X\otimes\cal{I}^E}\circ\cal{V}^{X'\to XE}\circ\cal{D}^{X'}\\
        \cal{Q}^{YE\to Y'} &:= \cal{D}^{Y'}\circ \cal{G}^{YE\to Y'} \circ \pqty{\cal{D}^Y\otimes \cal{I}^E}.
    \end{align*}
    Since $\cal{P}^{X'\to XE}$ is a classical-quantum channel, its output is a product state, denotes them by 
    \begin{equation}
        \cal{P}^{X'\to XE}(\*{e}_w ^{X'}) = \sum_{i} \rho_{iw}^{X}\otimes\eta_{iw}^{E}\in\fk{D}(XE)
    \end{equation}
    where the sum is over $i$ in some index set and $\e_w ^{X'} := \ketbra{w}{w}^{X'}$. Since all systems are finite dimensional, it can be rewritten as
    \begin{equation}
        \cal{P}^{X'\to XE}(\*{e}_w ^{X'}) = \sum_{x\in[m]} \*{e}_{x}^{X}\otimes \xi_{xw}^{E}\in\fk{D}(XE).
    \end{equation}
    Note that the matrices $\xi_{xw}^E$ are merely positive semi-definite matrices but not necessary density matrices.
    Consider $\Theta[\cal{N}](\*{e}_w ^{X'})$,
    \begin{align}
        \Theta[\cal{N}](\*{e}_w ^{X'}) &= \cal{Q}\circ \pqty{\cal{N}^{X\to Y}\otimes \cal{I}^E} \circ \cal{P} (\*{e}_w ^{X'})\\
        &= \sum_{x\in[m]} \cal{Q}\pqty{ \cal{N}^{X\to Y}(\*{e}_x ^X)\otimes\xi_{xw}^{E}}
    \end{align}
    Now, we define $\cal{Q}_{xw} ^{Y\to Y'}$ to be a currying of $\cal{Q}^{YE\to Y'}$,
    \begin{equation}
        \cal{Q}_{xw} ^{Y\to Y'} (\rho^Y):=\frac{1}{\Tr[\xi^E _{xw}]}~\cal{Q}^{YE\to Y'} (\rho^{Y}\otimes \xi_{xw}^{E}).
    \end{equation}
    If the trace $\Tr[\xi^E _{xw}]$ is zero, then define $\cal{Q}_{xw} ^{Y\to Y'} = \cal{I}_c$ a classical identity channel.
    This puts the above equation as 
    \begin{equation}
        \Theta[\cal{N}](\*{e}_w ^{X'}) = \sum_{x\in[m]} \cal{Q}^{YE \to Y'} (\cal{N}^{X\to Y}(\*{e}^X _x)\otimes \xi_{xw} ^E) = \sum_{x\in[m]} \Tr[\xi^E _{xw}]\cal{Q}^{Y\to Y'} _{xw} (\cal{N}^{X\to Y}(\*{e}_x ^X))
    \end{equation}
    Similarly, we define the pre-processing maps $\cal{P}_{xw}$ to be 
    \begin{equation}
    \cal{P}_{xw} ^{X' \to X} (\*{e}_v ^{X'})=\delta_{vw}\Tr[\xi_{xw}] \*{e}_x^X \quad \text{for any } x\in[m] \text{ and } w \in[m'].
    \end{equation}
    Notice that each pre-processing $\cal{P}_{xw}$ is merely a completely positive map but not necessary a channel. However, the sum of $\cal{P}_{xw} ^{X' \to X}$ is a channel, because
    \begin{align*}
        \sum_{\substack{x\in[m]\\w\in[m']}}\cal{P}_{xw} ^{X' \to X} (\*{e}_v) &= \sum_{\substack{x\in[m]\\w\in[m']}} \delta_{vw}\Tr[\xi_{xw}] \*{e}_x^X\\
        &=\sum_{x\in[m]} \Tr[\xi_{xv}]\*{e}_x ^X \in\prob(m).
    \end{align*}
    The inclusion is from $\cal{F}^{X'\to XE} (\*{e}_w ^{X'})= \sum_{x\in[m]} \*{e}^X _x\otimes \xi_{xw} ^{E} \in \fk{D}(XE)$ and $\*{e}^{X}_x\in\fk{D}(X)$.

    All together, we have that the realization of $\Theta$ can be written as 
    \begin{equation}
    \Theta[\cal{N}]= \sum_{\substack{x\in[m]\\w\in[m']}} \cal{Q}_{xw}\circ \cal{N}\circ\cal{P}_{xw}.
    \end{equation}
    Uncurrying by defining $Z$ to be a classical auxiliary space that its classical orthonormal basis contains all the pairs $(x,w)$,
    \begin{align*}
        \cal{F}^{X'\to XZ} (\rho^{X'}) &:= \sum_{\substack{x\in[m]\\w\in[m']}} \Tr[\cal{P}_{xw}(\rho)]\*{e}^X _x\otimes \*{e}^Z _{(x,w)},\\
        \cal{E}^{YZ \to Y'} (\sigma^{YZ}) &:= \sum_{\substack{x\in[m]\\w\in[m']}} \Tr_Z \bqty{\*{e}^{Z} _{(x,w)}\pqty{\cal{Q}^{Y\to Y'}_{xw}\otimes\cal{I}^Z (\sigma^{YZ})}}.
    \end{align*}
    Now, consider the composition with the state input $\*{e}^{X'} _{x'}$,
    \begin{align}
        \cal{E}\circ\cal{N}^{X\to Y}\otimes\cal{I}\circ\cal{F}(\*{e}^{X'}_{x'}) &= \cal{E}\circ\cal{N}^{X\to Y}\otimes\cal{I}^Z\pqty{\sum_{x,w} \Tr \bqty{\cal{P}_{xw}(\*{e}^{X'} _{x'})} \*{e}^X _x \otimes \*{e}^Z _{(x,w)}}\\
        &= \cal{E}\circ \cal{N}^{X\to Y}\otimes\cal{I}^{Z}\pqty{\sum_{x} \Tr[\xi_{xx'}]\*{e}^X _x \otimes \*{e}^Z _{(x,x')}}\\
        &= \sum_{x} \Tr[\xi_{xx'}] \cal{E}\pqty{\cal{N}^{X\to Y}(\*{e}^X _x) \otimes \*{e}^Z _{(x,x')}}\\
        \cal{E}\pqty{\cal{N}^{X\to Y}(\*{e}^X _x) \otimes \*{e}^Z _{(x,x')}} &= \sum_{\substack{u\in[m]\\ \vctv\in[m']}} \Tr_{Z} \bqty{\*{e}^Z _{(u,v)} \pqty{\cal{Q}_{uv}\otimes \cal{I}^Z}\pqty{\cal{N}(\*{e}^X _x)\otimes \*{e}^Z _\pqty{x,x'}}}\\
        &= \cal{Q}^{Y\to Y'} _{xx'} \pqty{\cal{N}(\*{e}^X _x)}\\
        \cal{E}\circ\cal{N}^{X\to Y}\otimes\cal{I}\circ\cal{F}(\*{e}^{X'}_{x'}) &=\sum_{x} \Tr[\xi_{xx'}] \cal{Q}^{Y\to Y'} _{xx'} \circ \cal{N}(\*{e}^X _x)\\
        &= \sum_{x,w} \cal{Q}^{Y\to Y'} _{xw} \cal{N}\circ\cal{P}^{X'\to X} _{xw}(\*{e}^{X'} _{x'}).\label{eq:Standard-form-in-realiz-proof}
    \end{align}
    The above equation holds for all $\*{e}^{X'} _{x'}$, then the superchannel $\Theta$ can be realized with the channel $\cal{E}$ and $\cal{F}$.  
\end{proof}

We call the form of superchannel in \eqref{eq:Standard-form-in-realiz-proof} the \emph{standard form} of a classical superchannel. The term standard this form exists and is unique for any classical superchannels.
\begin{propbox}{The standard form of a classical superchannel}
    \begin{corollary}\label{th:classical-supch-standardform}
        A classical superchannel $\Theta\in\rm{SUPER}((X\to Y) \to (X' \to Y'))$ can be realized as a sum of post-processing channels $\cal{E}_{x,w}\in\rm{CPTP}(Y\to Y')$ and the pre-processing maps $\cal{F}_{x,w}\in \rm{CP}(X'\to X)$ where $\cal{F}_{x,w} (\e_v ^{X'}) = \delta_{w,v} f(x\vert w) \e_x ^X$ for all $x\in[m]$ and $v\in[m']$.
        
        That is for any classical channel $\cal{N}\in \rm{CPTP}$ we have that 
        \begin{equation}
            \Theta[\cal{N}] = \sum_{x,w} \cal{E}_{x,w}^{Y\to Y'} \circ \cal{N}^{X\to Y}\circ \cal{F}_{x,w} ^{X'\to X}
        \end{equation}
    \end{corollary}
\end{propbox} 
The proof of this corollary can be taken directly from the proof of classical realization channel. We present an alternative proof using classical realization theorem and the \emph{currying} form of a classical superchannel.
\begin{lemma}[Currying form of the superchannel]
    A classical superchannel can be written in the currying form, 
    \begin{equation}
        \Theta[\cal{N}^{X\to Y}] = \sum_{z\in[k]} \cal{E}_z^{Y\to Y'}\circ \cal{N}^{X\to Y} \circ \cal{F}_z ^{X'\to X}
    \end{equation}
    where $k=\abs{Z}$ the dimension of reference system $Z$.
\end{lemma}
\begin{proof}
First, suppose a classical superchannel is realized as follows 
\begin{equation}
    \Theta[\cal{N}] = \cal{E}^{YZ \to Y'}\circ \cal{N}^{X\to Y}\otimes \cal{I}^Z \circ \cal{F}^{X'\to XZ}.
\end{equation}
We define $f(xz\vert x')$ to be such that $\cal{F}^{X'\to XZ}(\*{e}^{X'}_{x'})=\sum_{x,z} f(xz\vert x') \*{e}^X_{x}\otimes \*{e}^Z_{z}$. Now, consider an action of the channel $\Theta[\cal{N}^{X\to Y}]$ on an input $\*{e}^{X'}_{x'}$.
\begin{equation}
    \Theta[\cal{N}^{X\to Y}] (\*{e}^{X'}_{x'}) = \sum_{x,z} f(xz\vert x') \cal{E}(\cal{N}(\*{e}^X_x)\otimes \*{e}^Z_z)
\end{equation}
From the above equation, we define $\cal{E}_z^{Y\to Y'} (\*{p}^Y) = \cal{E}(\*{p}^Y\otimes\*{e}^Z_z)$.
\begin{align}
    \Theta[\cal{N}^{X\to Y}] (\*{e}^{X'}_{x'}) &= \sum_{x,z} f(xz\vert x') \cal{E}_z \circ \cal{N}(\*{e}^X_x)\\
    &=\sum_{z} \cal{E}_z^{Y\to Y'} \circ \cal{N}^{X\to Y}\pqty{\sum_x f(xz\vert x') \*{e}^X_x}\\
    &= \sum_{z} \cal{E}_z^{Y\to Y'} \circ \cal{N}^{X\to Y}\circ \cal{F}_z ^{X'\to X} (\*{e}^{X'}_{x'}).
\end{align}
That is, for all classical channel $\Theta$, we have that 
\begin{equation}
    \Theta[\cal{N}^{X\to Y}] = \sum_{z} \cal{E}_z^{Y\to Y'} \circ \cal{N}^{X\to Y}\circ \cal{F}_z ^{X'\to X}.
\end{equation}
\end{proof}

\begin{proof}[Proof of Corollary \ref{th:classical-supch-standardform}]
    From the currying form of the superchannel, we define
    \begin{align}
        \cal{E}_{xx'}^{Y\to Y} &:= \frac{1}{f(x\vert x')} \sum_{z\in[k]}f(xz\vert x') \cal{E}_z^{Y\to Y}\label{eq:stnd-cl-superch-post}\\
        \cal{F}_{xx'}^{Y\to Y} \pqty{\*{e}_w}&:= \delta_{wx'}f(x\vert x') \*{e}_x ^{X}\\
        f(x\vert x') &:= \sum_{z\in[k]} f(xz\vert x').
    \end{align}
    If $f(x\vert x')=0$, then we set $\cal{E}_{xx'}^{Y\to Y}:= \cal{I}_c$, a classical identity channel. Now, we have that 
    \begin{align}
        \sum_{\substack{x\in[m]\\x'\in[m']}} \cal{E}_{xx'} \circ \cal{N} \circ \cal{F}_{xx'} (\*{e}_w ^{X'}) &= \sum_{x\in[m]} f(x\vert w) \cal{E}_{xw} \circ \cal{N} (\*{e}^{X}_x)\label{eq:E.Nex-standardform}\\
        &= \sum_{\substack{x\in[m]\\ z\in[k]}} f(xz\vert w) \cal{E}_{z} \circ \cal{N} (\*{e}^{X} _x)\\
        &= \sum_{z\in[k]} \cal{E}_{z} \circ \cal{N} \circ \cal{F}_z (\*{e}^{X'}_w) \quad \forall w\in[m'],
    \end{align}
    showing that $\Theta$ can be written in the standard form.
\end{proof}
% \newpage

\begin{propbox}{Uniqueness of the standard form}
\begin{lemma}\label{lm:standard-form-unique}
    Suppose a classical superchannel $\Theta$ has the standard form $\Theta[\cal{N}] = \sum_{x,w} \cal{E}_{x,w} \circ \cal{N}\circ\cal{F}_{x,w}$. If there is another realization $\Theta[\mN] = \sum_{x,w} \tilde{\cal{E}}_{x,w} \circ \cal{N} \circ \tilde{\cal{F}}_{x,w}$ such that $\tilde{\cal{F}}_{x,w} (\e_{x'})=  \delta_{wx'} f(x\vert x') \e_x$. Then $\cal{F} = \tilde{\cal{F}}$ and $\cal{E} = \tilde{\cal{E}}$.
\end{lemma}
\end{propbox}
\begin{proof}
    Suppose $\cal{N}_{xy} (\rho) = \Tr \bqty{\ketbra{x}{x}\rho} \ketbra{y}{y}$. Consider,
    \begin{align}
        \Theta[\cal{N}_{xy}](\e_{x'}) &= \sum_{\tilde{x}{w}} \cal{E}_{\tilde{x}w} \circ \cal{N}_{xy} \cal{F}_{\tilde{x}w} (\e_{x'})\\
        &= \sum_{\tilde{x}} f(\tilde{x}\vert x') \cal{E}_{\tilde{x}w} \circ \Tr[\ketbra{x}{x}\e_{\tilde{x}}] \ketbra{y}{y} \\
        &= f(x\vert x') \cal{E}_{xx'} (\e_y).
    \end{align}
    Similarly, via another realization, we have 
    \begin{equation}
        \Theta[\cal{N}_{xy}](\e_{x'}) =\tilde{f}(x\vert x') \tilde{\cal{E}}_{xx'} (\e_y).
    \end{equation}
    Both realizations are of the same superchannel on the same channel, implying that
    \begin{align}
        f(x\vert x') \cal{E}_{xx'} (\e_y) &= \tilde{f}(x\vert x') \tilde{\cal{E}}_{xx'} (\e_y)
    \end{align}
    Since the post-processing $\cal{E}_{xx'}$ and $\tilde{\cal{E}}_{xx'}$ are channels, then taking trace on both sides will yield $f(x\vert x') = \tilde{f}(x\vert x')$ asserting $\mF = \tilde{\mF}$ and directly from the above equation, we have $\cal{\mE} = \tilde{\mE}$.
\end{proof}
% \newpage
\section{Classical channel majorization}
With the classical realization theorem and the standard form of classical superchannel, we are ready to define majorization relation between classical channels. From now on, when refering to the classical channels, a channel $\cal{N}$ is a classical channel from system $X$ to $Y$, and a classical channel $\cal{M}$ has input system $X'$ and output system $Y$. Their dimensions are denoted as follows
\begin{align*}
    &\cal{N}^{X\to Y} \quad \abs{X}=m \quad \abs{Y} = n\\
    &\cal{M}^{X'\to Y} \quad \abs{X'}=m' \quad.
\end{align*}
Notice that we require the output system to be the same. However, we later will show that we can compare channels with different output dimensions.

\subsection{Constructive approach}
\begin{figure}[h]
    \centering
    \includegraphics[width = .7\textwidth]{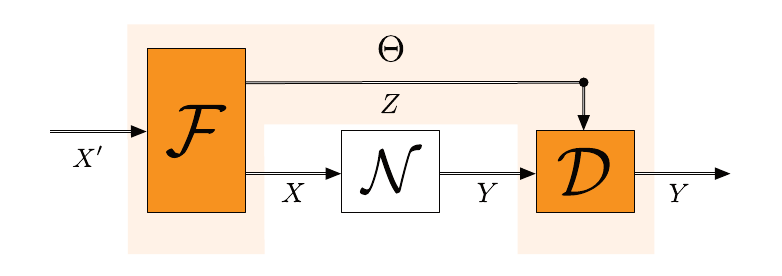}
    \caption{Random-permutation superchannel. The pre-processing channel $S$ maps an input on system $X'$ to $XZ$. The system $Z$ contains an instruction of which permutation $D_z$ to apply. This figure previously appeared in~\cite{GKNSY2025}.}
    \label{fig:constructive_random_perm}
\end{figure}
In the constructive approach, we will give a concrete description of the operations allows on the channel. As the goal is to conceptualize uncertainty of the outcome when given a control over input, the post-processing channel should not decrease the uncertainty of the outcome probability distribution. Such uncertainty non-decreasing operation is doubly-stochastic channel. Therefore, the only allowed operations are applications of doubly stochastic maps to the output of the channel $\cal{N}$ and conditioning which doubly stochastic maps to use based on the pre-processing of the input. We call such superchannel a random-permutation superchannel.
\begin{defbox}{Random-permutation superchannel}
    \begin{definition}
        A random-permutation superchannel $\Theta\in\rm{SUPER}((X\to Y)\to (X'\to Y))$ is a superchannel that can be realized with an auxiliary system $Z$ of dimension $k$ as
        \begin{equation}
            \Theta [\cal{N}] = \sum_{z\in[k]} \cal{D}_z ^{Y\to Y} \circ \cal{N}^{X\to Y} \circ \cal{F}^{X'\to X}_z
        \end{equation}
        where $\cal{D}_z$ is a doubly-stochastic channel and $\cal{F}_z$ is a completely positive trace non-increasing map on $X' \to X$. See figure~\ref{fig:constructive_random_perm} for the diagram.
    \end{definition}
\end{defbox}
For example, a pre-composition by a classical identity channel, $\cal{I}_C$, and a post-composition by a classical doubly-stochastic channel, $\cal{D}$, $\Theta[\cal{N}] = \cal{D}^{Y \to Y'}\circ\cal{N}\circ\cal{I}_C^{X' \to X}$ is a random permutation superchannel. This highlights the fact that, the superchannel need not permute the output of $\cal{N}$ randomly. A random-permutation superchannel could permute the output deterministically, as this superchannel $\Theta$ does. 

% \[
% \Qcircuit @C=1m @R=.7em {
%     \cw & \gate{\cal{I}} & \cw 
% }
% \]
% \begin{figure}[h]
    % \centering
    % \begin{center}
    %     \begin{quantikz}[wire types={c,c}]
    %         &\gate{\cal{I}}&\gate{\mathcal{N}}&\gate{\mathcal{D}}&\\
    %     \end{quantikz}
    % \end{center}
    % \caption{A simple random-permuation superchannel apply a doubly-stochastic channel on the output of the input channel $\cal{N}$.}
%   \end{figure}

A random-permutation superchannel might not be exclusively realizable with the doubly-stochastic post-processing. For example, consider a classical superchannel $\Theta\in\super((X \to Y)\to (X'\to Y))$ which realizable as 
\begin{equation}
    \Theta[\cal{N}] = \sum_{z\in[4]} \cal{E}_z \circ \cal{N} \circ \cal{F}_z,
\end{equation}
where $\cal{E}_z$ and $\cal{F}_z$ are defined as follows. For each $\cal{E}_z$, it has a corresponding transition matrix $E_z$ and for each $\mF_z$ has a corresponding matrix $F_z$. $E_1 = \pmqty{1 & 1 \\ 0 & 0}, E_2 = \pmqty{1 & 0 \\ 0 & 1}, E_3 = \pmqty{0 & 1 \\ 1 & 0},$ and $E_4 = \pmqty{0 & 0 \\ 1 & 1}$. For $\mF$, we define $F_{1} = \pmqty{1/2 & 0\\ 0 & 0}, F_{2} = \pmqty{0 & 0\\ 0 & 1}, F_{3} = \pmqty{0 & 0\\ 0 & 0}$ and $F_{4} = \pmqty{1/2 & 0\\ 0 & 0}$. However, this channel is realizable with doubly-stochastic post-processing.
\begin{equation}
    \Theta[\cal{N}] = \sum_{z\in[4]} \mD_z \circ \cal{N} \circ \mS_z
\end{equation}
where $D_1 = \frac{1}{2} \pmqty{1 & 1 \\ 1 & 1}$, $D_2 = \pmqty{1 & 0 \\ 0 & 1}, D_3 = D_4 = I$,  $S_1 = \pmqty{1 & 0 \\ 0 & 0}$, $S_2 = \pmqty{0 & 0 \\ 0 & 1}$, and $S_3 = S_4 =0$. To see that this realization give the same superchannel as the earlier realization, we compare transition matrices of the output channel denoted by $M$ and $\tilde{M}$ when the input channel is $\cal{N}$ associated with the transition matrix $N$.
In the first realization, we have
\begin{align}
    \sum_{z\in[4]} \cal{E}_z\circ \cal{N}\circ \cal{F}_z \leftrightarrow M &= \pmqty{1 & 1 \\ 0 & 0} N \pmqty{ 1/2 & 0 \\ 0 & 0} +\pmqty{1 & 0 \\ 0 & 1} N \pmqty{ 0 & 0 \\ 0 & 1} + \pmqty{0 & 0 \\ 1 & 1} N \pmqty{ 1/2 & 0 \\ 0 & 0} \\
    &= \pmqty{\overbrace{N_{11} + N_{21}}^{=1} & \overbrace{N_{12} + N_{22}}^{=1} \\ 0 & 0}\pmqty{1/2 & 0 \\ 0 & 0} +\pmqty{ 0 & N_{12}\\ 0 & N_{22}} \nonumber \\
    &\qquad + \pmqty{0 & 0\\ N_{11} + N_{21} & N_{12} + N_{22}} \pmqty{1/2 & 0 \\ 0 & 0}\\
    &=\pmqty{1 & 1 \\ 1 & 1}\pmqty{1/2 & 0 \\ 0 & 0} + \pmqty{0 &N_{12} \\ 0 & N_{22}} = \pmqty{1/2 & N_{12}\\ 1/2 & N_{22}}.
\end{align}
In the second realization,
\begin{align}
\sum_{z\in[4]} \mD_z \circ \cal{N} \mS \leftrightarrow \tilde{M} &= \frac{1}{2}\pmqty{ 1 & 1 \\ 1 & 1} N \pmqty{ 1 & 0 \\ 0 &0 } + \pmqty{1 & 0 \\ 0 & 1} N \pmqty{0 & 0 \\ 0 & 1} \\
&= \frac{1}{2} \pmqty{ 1 & 1 \\ 1 & 1} \pmqty{ 1 & 0 \\ 0& 0} + N \pmqty{ 0 & 0 \\ 0 & 1} \\
&=  \pmqty{1/2 & N_{12} \\ 1/2 & N_{22}}.
\end{align}
We have $M = \tilde{M}$ for any classical channel $\cal{N}$. This means both realizations maps a classical channel to a same classical channel, showing that both realizations are of the same superchannel.

We see that a random-permutation superchannel may not reveal its true nature in some realizations. A question is then if there is a streamline process to check if the superchannel is random-permutation or not. The answer to this question is yes. Looking back to the later realization of the random-permutation superchannel, it is exactly in the standard form when set $(x,w) = (\lfloor z/2\rfloor, z~\ssf{mod}~2)$. Indeed, we have the following theorem.
\begin{propbox}{The standard form of random-permutation superchannel}
    \begin{theorem}\label{th:random-perm-standard-form-superchannel}
        A superchannel $\Theta\in\super((X\to Y) \to (X' \to Y))$ is a random-permutation superchannel if and only if in the standard form it has only doubly-stochastic post-processing.
    \end{theorem}
\end{propbox}
\begin{proof}
    If the standard form has only doubly-stochastic post-processing channels, then it is trivially random-permutation. 
    
    To prove this theorem in the only-if direction, suppose the superchannel is random-permutation. Then, we have that its post-processing $\cal{D}_z$ are all doubly-stochastic. As the post-processing channel of $\cal{E}_{xx'}$ is defined as in the Equation~\eqref{eq:stnd-cl-superch-post} with a convex combination of $\cal{D}_z$, we have that $\mE_{xx'}$ are doubly-stochastic as well. Using the uniqueness lemma of the standard form (Lemma~\ref{lm:standard-form-unique}), this proves the theorem.
\end{proof}

With the random-permutation superchannel, we define the majorization of superchannel as follows 
\begin{defbox}{Constructive definition of channel majorization}
    \begin{definition}
        Suppose $\cal{N}^{X\to Y}$ and $\cal{M}^{X' \to Y}$ are classical channels. We say $\cal{N}$ majorize $\cal{M}$ constructively if there exists a random-permutation superchannel such that $\cal{M} = \Theta[\cal{N}]$.
    \end{definition}
\end{defbox}
Consider two classical channels $\cal{N}$ and $\cal{M}$ where any output of $\cal{M}$ can be permuted to simulate an output of $\cal{N}$. The act of permutation is itself a classical superchannel and indeed a random permutation one. This leads to the equivalence between the channel $\cal{N}$ and $\cal{M}$ since the permutation operation is invertible. In particular, a classical channel $\cal{N}$ is equivalent under constructive majorization to $\cal{N}^\da$ where all of its output vector is permuted to be in a non-increasing order. 

If any channel $\cal{N}$ that has its output vectors in non-increasing order, i.e. $\mN = \mN ^\da$. The constructive majorization of channels can be characterized as follows. Suppose $\cal{N}$ constructively majorizes $\cal{M}$. Given any output of $\cal{M}$, says $\*{q}$ one can always find the output of $\cal{N}$, denoted with $\*{p}$, such that $\*{p}\succ \*{q}$. Formally, we have the following equivalence. 
\begin{propbox}{Channel majorization and majorization of channel outputs}
    \begin{theorem}\label{th:convex-sum-outputs}
        For any classical channels $\cal{N}$ and $\cal{M}$, we have that $\cal{N}$ constructively majorizes $\cal{M}$ if and only if there exists a convex combination of $\p_x ^\da $'s such that 
        \begin{equation}\label{eq:convex-sum-outputs}
            \sum_{x\in [m]} s_{x\vert w} \*{p}_x ^\da \succ \*{q}_w
        \end{equation}
        where $s_{x\vert w} \geq 0$, $\sum_{x\in[m]} s_{x\vert w} = 1$, $\*{p}_x := \cal{N}(\*{e}_x)$ and $\*{q}_w := \cal{M}(\*{e}_w)$.
    \end{theorem}
\end{propbox}
\begin{remark}
    Columns of the transition matrix $N$ are $\qty{\p_x: x\in[m]}$, and of $M$ are $\qty{\q_w: w\in[m']}$.
\end{remark}
\begin{proof}
    Suppose, $\cal{N}^{X\to Y}$ constructively majorize $\cal{M}^{X'\to Y}$ and $\mN = \mN ^\da$ so that $\p_x = \p_x ^\da$. The superchannel $\Theta$ can be put in the standard form 
    \begin{equation}
        \Theta[\cal{N}] = \sum_{\substack{x\in[m]\\x'\in[m']}} \cal{E}_{xx'} \circ \cal{N} \circ \cal{F}_{xx'}.
    \end{equation}
    From the definition of $\cal{E}_{xx'}$ in Equation~\eqref{eq:stnd-cl-superch-post}, the superchannel $\Theta$ being random-permutation superchannel implies that $\cal{E}_{xx'}$ is a doubly-stochastic channel. Consider the definition of the vectors $\*{q}_{w}$,
    \begin{align}
        \*{q}_w = \Theta[\cal{N}](\*{e}_w) &= \sum_{\substack{x\in[m]\\x'\in[m']}}\cal{E}_{xx'} \circ \cal{N} \circ \cal{F}_{xx'}(\*{e}_w ^{X'})\\
        &=\sum_{x\in[m]} f(x\vert w) \cal{D}_{xw}\circ \cal{N}(\*{e}_x) = \sum_{x} f(x\vert w) \cal{D}_{xw}(\*{p}_x).
    \end{align}
    Since $\*{p} \succ D\*{p}$ for any $\*{p}\in\rm{Prob}(n)$ and $\p_x = \p_x ^\da$, we have that for any $k\in[n]$
    \begin{equation*}
        \knorm{\sum_{x\in[m]} f(x\vert w) \p_x}{k}\geq \sum_{x\in[m]} f(x\vert w) \knorm{D_{xw} \p_x}{k} \geq  \knorm{\sum_{x\in[m]} f(x\vert w) \cal{D}_{xw} \*{p}_x}{k} = \knorm{\*{q}_w}{k},
    \end{equation*}
    showing that $\sum_{x\in[m]} f(x\vert w) \p_x \succ \q_w$.

    On the converse direction, suppose that there exists a doubly stochastic matrix $S = \pqty{s_{x\vert w}}$ such that for all $w\in[m']$ we have $\sum_{x\in[m]} s(x\vert w) \*{p}_x \succ\*{q}_w$. This is equivalent to the existence of a classical doubly-stochastic channel $\cal{D}_w$ such that 
    \begin{equation}
        \cal{D}_w \pqty{\sum_{x\in[m]} s(x\vert w) \*{p}_x } = \*{q}_w
    \end{equation}
    We define 
    \begin{align}
        S(\*{e}_w ^{X'}) &\eqdef \sum_{x\in[m]} s(x\vert w) \*{e}_x ^X \otimes \*{e}_w ^Z,\\
        \cal{D} (\*{p}^Y \otimes \*{e}_w ^Z) &\eqdef \cal{D}_w(\*{p}^Y)\\
        \Theta[\cal{N}] &\eqdef \cal{D}\circ \cal{N}\circ \cal{S}.
    \end{align}
    The classical superchannel $\Theta$ is a random-permutation superchannel since $\cal{D}_z$ is doubly stochastic. From the above equation we have,
    \begin{equation}
    \Theta[\cal{N}](\*{e}_w ^{X'}) = \cal{D}_{w} \pqty{\sum_{x\in[m]} s(x\vert w)  \cal{N}(\*{e}_x)} =\*{q}_w.
    \end{equation}
    This holds for all $w\in[m']$ which implies that $\Theta[\cal{N}] = \cal{M}$.
\end{proof}
\begin{remark}
    It is important to note here that removing the non-increasing-order condition in Equation~\eqref{eq:convex-sum-outputs} will not yield a valid statement. This invalidity illustrates that, even if $\cal{N}\succ\cal{M}$, not every output of the majorized channel $\cal{M}$ can be majorized by the output of the majorizing channel $\cal{N}$, in contrast to what one might expect.
\end{remark}
\begin{example}
    This is an example of the above remark. Consider $N = \pmqty{\p_1 & \p_2}$ and $M = \pmqty{\q}$, where $\p_1 = \frac{1}{100} \pmqty{70 & 15 & 15}^T$, $\p_2 = \frac{1}{100} \pmqty{5 & 45 & 50}^T$ and $\q = \frac{1}{100} \pmqty{60 & 30 & 10}^T$. There is no convex combination of $\p_1$ and $\p_2$ such that it majorizes $\q$. However, there is a convex combination of $\p_1 ^\da$ and $\p_2 ^\da$ such that it majorizes $\q$. The convex combination is exactly $\p = \frac{1}{2}(\p_1 ^\da +\p_2 ^\da)$. 

    We will now show that there is no convex combination of $\p_1$ and $\p_2$ that majorizes $\q$. The convex combination can be parametrized by $s\in[0,1]$ as 
    \begin{equation*}
        \p:= (1-s)\p_1 + s\p_2 =\pmqty{(1-s)0.70 +s\times 0.05\\ (1-s)0.15+s\times 0.45 \\ (1-s)0.15+s\times 0.50} = \pmqty{0.75 - 0.65 s\\ 0.15 + 0.30 s \\ 0.15 + 0.35s}
    \end{equation*}
    For the first Ky Fan norm, $\knorm{\q}{1} =0.6$. For $\p$ to majorize $\q$, the element $0.75-0.65s$ need to be its biggest element with $s\in[0,1/65]$. The other elements are at most $0.5$ when $s=1$. The infeasibility is in the second Ky Fan norm, $\knorm{\q}{2} = 0.9$. Notice that $0.15 + 0.35s\geq 0.15 + 0.30s$. It follows that $\knorm{\p}{2} = 0.75 - 0.65 s + 0.15 + 0.30s = 0.85 -0.35 s$, which is lesser than $0.9$ for any non-negative value $s$.
\end{example}

\subsection{Axiomatic approach}
\begin{figure}[h]
    \centering
    \includegraphics[width = .7\textwidth]{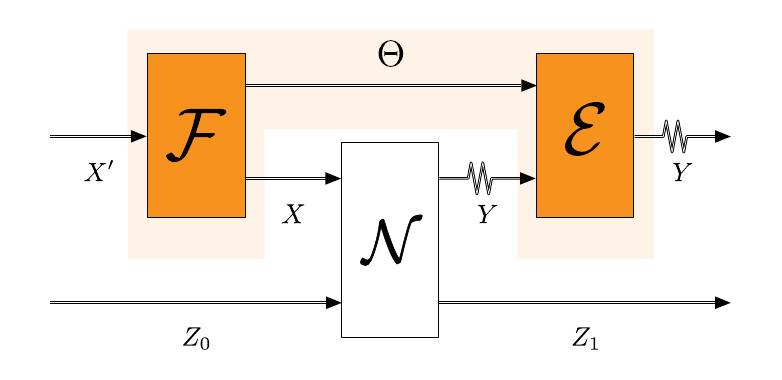}
    \caption{Completely uniformity preserving superchannel. In this figure, the channel $\mN$ is a conditionally uniform channel. The channel outputs a product state where the state on $Y$ is maximally mixed, denoted by a squiggly line. The channel $\Theta[\mN]$ is conditionally uniform as the output state on $Y'$ is maximally mixed. Previously appeared in \cite{GKNSY2025}.}
    \label{fig:axiomatic_diag}
\end{figure}
In the axiomatic approach, the maximally randomizing channel $\cal{R}$ is the nosiest channel and the mixing operation must preserve the nosiest channel. However, there is a classical superchannel which preserve $\cal{R}^{X\to Y}$ but does not preserve a \emph{classical conditionally uniform} channel $\cal{N}^{XZ_0 \to YZ_1}$, which is defined as follows.
\begin{defbox}{Conditionally uniform channels}
\begin{definition}
    A classical channel $\cal{N}^{XZ_0 \to YZ_1}$ is said to be a conditionally uniform classical channel if for all $\*{p}\in\rm{Prob}(nk)$ where $k:= \abs{Z_0}$ we have that 
    \begin{equation}
        \cal{N}^{XZ_0 \to YZ_1}(\*{p}) = \*{u}^Y \otimes \*{q}^{Z_1}
    \end{equation}
    See Figure~\ref{fig:axiomatic_diag} for a diagram.
\end{definition} 
\end{defbox}
\noindent If a classical superchannel fails to preserve this class of channels, then the output of the channel $\Theta[\cal{N}]$ can have less uncertainty than $\cal{N}$, by creating correlation between system $Y$ and $Z_1$. To illustrate, suppose $\cal{M}^{XZ_0 \to Y'Z_1} = \Theta[\cal{N}]$, if $\cal{N}$ is conditionally uniform channel but $\cal{M}$ is not, then the outputs of the channel $\cal{M}$, $Y'$ and $Z_1$, are not independent and hence can have lesser uncertainty than that of $\cal{N}$.
% To illustrate, suppose $\cal{N}^{XZ_0 \to YZ_1} (\p^{XZ_0}) =\u^{Y} \otimes \p^{Z_1}$ and $\Theta[\mN](\p^{XZ_0}) = \q^{Y}\otimes \p^{Z_1}$. We have that $\q^{Y} \otimes \p^{Z_1} \succ \u^{Y} \otimes \p^{Z_1}$, making the resultant channel $\Theta[\cal{N}]$ less uncertain than the original channel $\cal{N}$.

This class of superchannel is a strict subset of the class of uniformity preserving superchannel. There exists a superchannel which is uniformity preserving but does not completely preserve uniformity. Presenting in the standard form of superchannel, an example of such superchannel is as follows. Consider a superchannel $\Theta\in\super((X\to Y)\to (X'\to Y'))$ where $\abs{X'} =1$ and $\abs{X}= \abs{Y}= \abs{Y'} =2$.
Preprocessing channels $\cal{F}_{x,w}$ and processing channel $\cal{E}_{x,w}$ have corresponding transition matrices
\begin{align}
    E_{1,1} &= \begin{bmatrix}1 & 1 \\ 0 & 0\end{bmatrix},
    &E_{2,1}  = \begin{bmatrix}0 & 0 \\ 1 & 1\end{bmatrix},\\
    F_{1,1} &= \frac{1}{2} \begin{bmatrix}1 \\ 0\end{bmatrix},
    &F_{2,1} = \frac{1}{2} \begin{bmatrix}0 \\ 1\end{bmatrix}.
\end{align}
The indices $x$ and $w$ belong to $x\in\{1,2\}$ and $w\in\{1\}$. This superchannel is uniformity preserving. Suppose $\mN^{X\to Y}$ is a uniform channel and its corresponding transition matrix is $N$. We have 
\begin{align}
    \Theta[\mN^{X\to Y}] &= \sum_{x\in[2]} E_{x,1} N  F_{x,1}\\
    &= \begin{bmatrix}1 & 1 \\ 0 & 0\end{bmatrix}\times\frac{1}{2} \begin{bmatrix}1 & 1 \\ 1 & 1\end{bmatrix}\times\frac{1}{2} \begin{bmatrix}1 \\ 0\end{bmatrix} + \begin{bmatrix}0 & 0 \\ 1 & 1\end{bmatrix}\times\frac{1}{2} \begin{bmatrix}1 & 1 \\ 1 & 1\end{bmatrix}\times\frac{1}{2} \begin{bmatrix}0 \\ 1\end{bmatrix}\\
    &= \frac{1}{2} \begin{bmatrix}1 \\ 1\end{bmatrix}.
\end{align}
This shows that $\Theta$ is uniformity preserving. Next, we will show that the superchannel is not completely uniformity preserving. Suppose $\mN^{XZ\to YW}$ is a conditionally uniform channel with a corresponding transition matrix $N$
\begin{equation*}
    N= \*{u}^{Y} \otimes I^{XZ\to W} = \frac{1}{2} \begin{bmatrix}1 & 0 \\ 0 & 1 \\ 1 & 0 \\ 0 & 1\end{bmatrix}.
\end{equation*}
Denote a transition matrix corresponding to a marginal channel $\mN^{XZ\to W}$ with $\overline{N}$.
The action of the superchannel on this channel is $\Theta\otimes\id^{(Z\to W)}[\mN^{XZ\to YW}]$. Suppose its corresponding transition matrix is $M$
\begin{align}
M &= \sum_{x,w} \mathcal{E}_{x,w}[\u^Y]\otimes \mN^{XZ\to W'}\circ (\mathcal{F}_{x,w}^{X'\to X} \otimes \mathcal{I}^{Z\to Z}) \\
&=E_{1,1} \u^Y \otimes \overline{N}(F_{1,1}\otimes I^Z)+ E_{2,1} \u^Y \otimes \overline{N} (F_{2,1}\otimes I^Z)\\
&=  \e_1 \otimes \begin{bmatrix}1 & 0 \\ 0 & 1\end{bmatrix} \left( \frac{1}{2} \begin{bmatrix}1 \\ 0 \end{bmatrix} \otimes 1 \right) + \e_2 \otimes \begin{bmatrix}1 & 0 \\ 0 & 1\end{bmatrix} \left( \frac{1}{2} \begin{bmatrix}0 \\ 1\end{bmatrix}\otimes 1\right) \\
&=\e_1 \otimes\frac{1}{2} \begin{bmatrix}1 \\ 0\end{bmatrix} + \e_2\otimes\frac{1}{2} \begin{bmatrix}0 \\ 1\end{bmatrix} = \frac{1}{2} \begin{bmatrix}1 \\ 0\\ 0\\ 0\end{bmatrix} + \frac{1}{2} \begin{bmatrix}0 \\ 0\\ 0\\ 1\end{bmatrix} = \frac{1}{2} \begin{bmatrix}1 \\ 0\\ 0\\ 1\end{bmatrix}.
\end{align}
This is not a conditionally uniform channel, since the conditionally uniform channel is in the form $\u^{(2)}\otimes (p_1, p_2)^T = \frac{1}{2} (p_1, p_2,p_1, p_2)^T$.

To have a well-defined ordering and an uncertainty comparison relation, we require that $\Theta$ preserve conditionally uniform classical channels as well. We call this kind of superchannel \emph{completely uniformity-preserving} classical superchannel. This leads to the following definition of majorization.
\begin{defbox}{Axiomatic definition of channel majorization}
    \begin{definition}
        Given two classical channels $\cal{N}^{X\to Y}$ and $\cal{M}^{X'\to Y}$, we have that $\cal{N}$ axiomatically majorize $\cal{M}$ if there exists a classical superchannel $\Theta^{(X\to Y)\to (X'\to Y)}$ such that $\cal{M} = \Theta[\cal{N}]$ and for any conditionally uniform classical channel $\cal{E}$, we have that $\Theta[\cal{E}]$ is also a conditionally uniform classical channel.
    \end{definition}
\end{defbox}

With requirement of uniformity preserving in complete sense, the two approaches of majorization coincide.
\begin{propbox}{Equivalence of constructive and axiomatic approaches}
    \begin{theorem}\label{th:equiv_const-axiom}
        A classical superchannel is completely uniformity-preserving if and only if it is a random-permutation superchannel.     
    \end{theorem}
\end{propbox}
\begin{proof}
    First, a random-permutation superchannel is completely uniformity-preserving. Suppose $\Theta$ is random-permutation superchannel, then we can realize $\Theta$ as
    \begin{equation}
    \Theta [\cal{N}] = \sum_{z\in[k]} \cal{D}^{Y\to Y} \circ \cal{N}^{X\to Y} \circ \cal{S}^{X' \to XZ}.
    \end{equation} 
    Suppose $\cal{N}^{XZ_0 \to YZ_1}$ is a conditionally uniform classical, the output channel from the superchannel $\Theta$ is given by
    \begin{align}
        \Theta^{(X\to Y)\to(X'\to Y)}\otimes {1}^{(Z_0 \to Z_1)} [\cal{N}^{XZ_0 \to YZ_1}] &= \sum_{z\in[k]} (\cal{D}^{Y\to Y}\otimes I^{Z_1}) \circ \cal{N}^{XZ_0 \to YZ_1} \circ (\cal{S}^{X' \to XZ}\otimes {I}^{Z_0})
    \end{align}
    where $I^{Z_0}$ and $I^{Z_1}$ denote classical identity channels on $Z_0$ and $Z_1$ respectively.

    We will shortly denote the left side of the equation by $\cal{M}$, to see that this is a conditionally uniform, consider any classical input $\*{e}_{x'} ^{X'}\otimes \*{e}_w ^{Z_0}$,
    \begin{align}
        \cal{M}(\*{e}_{x'} ^{X'}\otimes \*{e}_w ^{Z_0}) &= \sum_{z\in[k]} (\cal{D}^{Y\to Y}\otimes I^{Z_1}) \circ \cal{N}^{XZ_0 \to YZ_1} \circ (\cal{S}^{X' \to XZ} \*{e}_{x'} ^{X'} \otimes {I}^{Z_0}\*{e}_w ^{Z_0})\\
        &= \sum_{z\in[k]} (\cal{D}^{Y\to Y}\otimes I^{Z_1}) \circ \cal{N}^{XZ_0 \to YZ_1} (\*{s}_{x'} \otimes \*{e}_w)\\
        &= \sum_{z\in[k]} \cal{D}^{Y \to Y}(\*{u}^Y) \otimes I^{Z_1}(\*{q}^{Z_1}) = \*{u}^Y\otimes \*{q}^{Z_1}.
    \end{align}
    This is for any $x'\in[m']$ and $w\in[k_1]$. Therefore, $\cal{M}$ is conditionally uniform classical channel.

    In the converse direction, a completely uniformity-preserving is a random-permutation superchannel. Suppose that $\Theta$ is completely uniformity-preserving classical superchannel, we consider its action on the channel 
    \begin{equation}
        \cal{N}^{X\to YZ_1} = \*{u}^{Y}\otimes \cal{D}^{Z}\circ \cal{V}^{X\to Z_1}\circ\cal{D}^{X}
    \end{equation}
    where $\cal{V}^{X\to Z_1}$ is an isometry channel and $\cal{D}^{Z}$ and $\cal{D}^{X}$ are totally dephasing channels. This is equivalent to 
    \begin{equation}
    \cal{N}^{X\to YZ_1} (\*{e}_x ^X)= \*{u}^Y \otimes \*{e}_x^{Z_1} \quad \forall x\in[m].
    \end{equation}
    Now, consider $\cal{M}^{X'\to YZ_1} \pqty{\*{e}_x ^X}$ where $\cal{M}^{X'\to YZ_1} = \Theta^{(X\to Y)\to(X' \to Y)}[\cal{N}^{X\to Y Z_1}]$,
    \begin{align}
        \cal{M}^{X'\to YZ_1} \pqty{\*{e}_{x'} ^{X'}} &= \cal{E}^{YZ \to Y} \circ \cal{N}^{X\to Y} \circ \cal{F}^{X'\to XZ}(\*{e}_{x'} ^{X'})\\
        \*{u}^Y \otimes \*{q}^{Z_1}&= \sum_{\substack{x\in[m]\\z\in[k]}} f(xz\vert x')\cal{E}^{YZ\to Y}\pqty{\cal{N}^{X\to Y}(\*{e}^X_x)\otimes \*{e}^Z _z} \\
        &= \sum_{\substack{x\in[m]\\z\in[k]}} f(xz\vert x')\cal{E}^{YZ\to Y}\pqty{\*{u}^Y\otimes \*{e}^{Z_1}_x \otimes \*{e}^Z _z} \\
        &= \sum_{\substack{x\in[m]\\z\in[k]}} f(xz\vert x')\cal{E}^{YZ\to Y}\pqty{\*{u}^Y \otimes \*{e}^Z _z}\otimes \*{e}^{Z_1}_x \\
        \pqty{\*{q}^{Z_1} \cdot \*{e}^{Z_1}_x } \*{u}^Y &= \sum_{z\in[k]}f(xz\vert x')\cal{E}^{YZ\to Y}\pqty{\*{u}^Y \otimes \*{e}^Z _z}\\
        \frac{\*{q}^{Z_1} \cdot \*{e}^{Z_1}_x }{f(x\vert x')} \*{u}^Y &= \frac{1}{f(x\vert x')} \sum_{z\in[k]}f(xz\vert x')\cal{E}^{YZ\to Y}\pqty{\*{u}^Y \otimes \*{e}^Z _z}= \cal{E}^{Y\to Y} _{xx'}\pqty{\*{u}^Y}
    \end{align}
    Since $\cal{E}_{xx'}$ is a channel, we have that $\*{q}^{Z_1} \cdot \*{e}^{Z_1}_x /f(x\vert x') = 1$. This shows that post-processing channels in the standard form are doubly-stochastic channels. This superchannel is a random-permutation superchannel.
    That is a superchannel can be realized as 
    \begin{equation}
        \Theta\bqty{\cal{N}^{X\to Y}} = \sum_{\substack{x\in[m]\\w\in[m']}} \cal{E}^{Y\to Y}_{x,w} \circ \cal{N}^{X\to Y}\circ \cal{F}^{X'\to X}_{x,w},
    \end{equation}
    where $\cal{E}^{Y\to Y}$ is doubly-stochastic.
\end{proof}

\subsubsection{A sufficient condition}
Instead of proving property of uniformity preserving of all possible conditionally uniform channels, it is sufficient to show the property by showing that on a particular channel the superchannel preserve its marginal uniformity.
\begin{propbox}{Classical completely uniformity-preserving checking channel}
    \begin{theorem}
        $\Theta$ is completely uniformity preserving if and only if $\Theta[\mN^{X\to YZ}]$ is conditionally uniform channel where $\cal{N}^{X\to YZ}(\*{e}^X_x) = \*{u}^Y \otimes \*{e}^{Z_1}_x$.
    \end{theorem}
\end{propbox}
\begin{proof}
The only-if direction is trivial. Suppose that $\Theta[\mN^{X\to YZ}]$ is a conditionally uniform channels.
First consider an input $\e_{x'} ^{X'}$ of the channel $\Theta[\mN]$,
\begin{align}
    \Theta[\cal{N}](\*{e}^{X'}_{x'}) &= \sum_{z\in[k]} \cal{E}_z\circ \cal{N}\circ \cal{F}_z(\*{e}^{X'}_{x'})\\
    &= \sum_{\substack{x\in[m]\\z\in[k]}} f(xz\vert x)\cal{E}_{z}\circ\cal{N}(\*{e}^X_x) = \sum_{\substack{x\in[m]\\z\in[k]}} f(xz\vert x)\cal{E}_{z}(\*{u}^Y)\otimes \*{e}^{Z_1} _x
\end{align}
The left side of the equality is a conditionally uniform channel,
\begin{align}\label{eq:marginally-checking-form}
    \*{u}^Y \otimes \*{r}^{Z_1}_{x'}= \sum_{\substack{x\in[m]\\z\in[k]}} f(xz\vert x)\cal{E}_{z}(\*{u}^Y)\otimes \*{e}^{Z_1} _x
\end{align}

Suppose $\cal{M}^{XZ_0 \to YZ_1}$ is another classical conditionally uniform channel, then its transformation by $\Theta$ is given by 
\begin{align}
    \Theta[\cal{M}](\*{e}_{x'}^{X'}) &= \sum_{x,z}f(xz\vert x') \cal{E}_z (\*{u}^Y)\otimes \*{q}_x ^{Z_1}\\
    &= \sum_{x,z,w}f(xz\vert x') \cal{E}_z (\*{u}^Y)\otimes q_{w}\*{e}_w ^{Z_1}\\
    &= \sum_{w} q_{w} \pqty{\sum_{x,z} f(xz\vert x') \cal{E}_z (\*{u}^Y)\otimes \*{e}_w ^{Z_1}}
\end{align}
The parenthesis terms on the right side of the last line is identical to the right side of the Equation~\eqref{eq:marginally-checking-form}. The convex sum of $\*{u}^Y\otimes \*{r}^{Z_1}_{x'}$ over $x'\in[m']$ is equal to $\*{u}^Y \otimes \*{q}^{Z_1}$ for some $\*{q}^{Z_1}$, concluding the proof.
\end{proof}

\subsection{Operational approach}
In an operational approach, the channel majorization is defined by a game of chance played with a channel $\cal{N}$. The game is identical to that previously discuss in \cite{BGG2022}. This game is a generalization of the game of chance called $k$-game introduced in the section of probability vector majorization. In a $k$-game with probability vector, a player aware of the probability distribution associated with probability vector $\p$ and given $k$ guesses of a sample drawn from the distribution. 

Suppose we want to play a similar gambling game with a channel. A classical channel $\cal{N}$ receive an input $\e^{X} _x$ for some $x\in[m]$ and output a sample drawn from a probability distribution associated with the probability vector $\p_x := \cal{N}(\e_x ^{X})$. That is, we can play a $k$-game with a probability vector $\p_x$, an output of the classical channel. 

Unlike a probability vector, the player can choose an input to the channel, consequently influencing what probability distribution a sample of output to be drawn from. Moreover, the optimal input of the channel depends on whether the player knows the value of $k$. For example, if a player knows $k$ prior to giving the input, the best strategy for the player is to pick an input that has the highest $k^\text{th}$ Ky Fan norm, similar to the $k$-game with a probability vector. If a player knows nothing about $k$ prior to giving an input ($k$ is uniformly distributed), then the best strategy is to pick the input that result in $\p_x$ with the highest average $k^\text{th}$ Ky Fan norm over all possible $k\in[n]$.

\begin{figure}
    \centering
    \includegraphics[width = .7\textwidth]{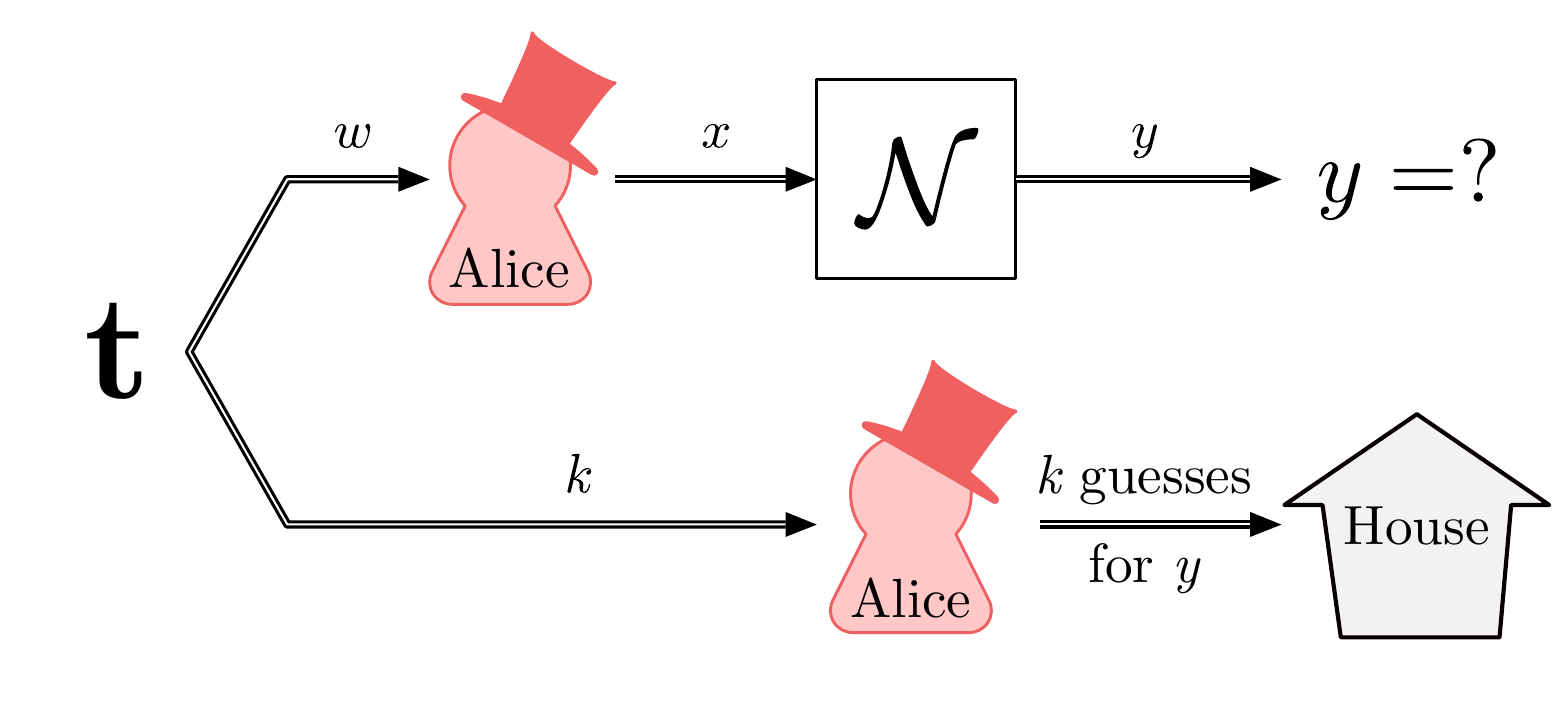}
    \caption{The diagram of how a $\t$-game is played. A player, named Alice, is given a full knowledge of a joint probability distribution $\t$ and a transition matrix $N$ associated with the classical channel $\mN$. Once the game start, she will be made aware of the hint $w$ and then give an input $x$. To win the game, she has to provide a correct guess of the value $y$, the output of the channel, within $k$ guesses.  This figure previously appeared in \cite{GKNSY2024}.}
    \label{fig:t-games}
\end{figure}

How much knowledge of $k$ that the player know prior to giving an input need not be between these two extremes: certainly know $k$ and complete ignorant. Some partial information could be sent to the player prior to give an input via a number $w\in[\ell]$. For example, $w=0$ represents $k$ being an even number and $w=1$ represent $k$ being an odd number. In general, we can represent a full spectrum of prior knowledge of $k$ by a joint probability vector $\*{t}$ between the random variables $k$, the number of guesses, and $w$, the value player learned before giving an input. Therefore, we coined the term $\t$-game to denote a gambling game with a classical channel when the prior knowledge of $k$ is described by the joint probability $\t$. A diagram illustration for this family of games can be found in Figure~\ref{fig:t-games}.

% A player is given the knowledge of transition matrix $N$ corresponding to the channel $\cal{N}$. They have a complete control over the input. The goal is to choose an input and correctly guess an outcome of a channel within $k$ guesses. The number of guesses $k$ is randomly picked. 

% How much knowledge of $k$ that the player know prior to giving an input depending on the joint probability vector $\*{t}$ between the random variables $k$ and $w$, where $w\in[\ell]$ is the value player fully knows before giving an input. 
To give more examples, suppose a player knows nothing about $k$ prior to giving an input to the channel. The joint conditional probability distribution is described by a vector $\t = \t^W\otimes \u^K$ which represent the most uncertainty of $k$ and independence of $w$ and $k$. On the other hand, if a player gain perfect knowledge of $k$ prior to giving the input, then the probability vector is $\t = \sum_{k\in[n]} t_k \e_k ^{W} \otimes \e_k ^{K}$ where $\sum_{k\in[n]} t_k = 1$ and $0\leq t_k\leq 1$.

The best strategy for any $t$-games is as follows. After a player receives a value of $w$, they choose an input $\e_x ^X$ such that it maximizes a convex sum of the $k^\text{th}$ Ky Fan norm of $\p_x$ weighted by the conditional probability $\qty{t_{k\vert w}: k\in[n]}$.
% In the case of perfect knowledge of $k$ prior to giving the input, the best strategy for the player is to pick an input that has the highest $k^\text{th}$ Ky Fan norm. 
% In this case, the winning chance, denoted by $\rm{Pr}_{\*{t}}$, is given by
% \begin{equation}
%     \rm{Pr}_{\*{t}} (\cal{N}) = \sum_{k\in[n]} t_k \max_{x\in[m]} \knorm{\p_x}{k}
% \end{equation}
% where $\p_x$ is the $x^{\rm{th}}$ column of the transition matrix $N$. 
% On the other hand, if the player is completely ignorant of $k$, picking an input $x$ that maximize the convex sum of $\knorm{\p_x}{k}$ over $x\in[m]$ weight by the conditional probability of $k$ on $w$. 
That is, a winning chance of a $\t$-game is expressible as 
\begin{equation}
    \rm{Pr}_{\*{t}} (\cal{N}) =\sum_{w \in[\ell]} t_{w} \max _{x \in[m]} \sum_{k \in[n]} t_{k\vert w}\knorm{\*{p}_x}{k}.
\end{equation}

We operationally define majorization as the comparison of the winning chance. 
\begin{defbox}{Operational definition of majorization}
    \begin{definition}
        For any classical channels $\cal{N}^{X\to Y}$ and $\cal{M}^{X'\to Y}$, we say $\cal{N}$ majorize $\cal{M}$ operationally if 
        \begin{equation}
            \rm{Pr}_{\*{t}} (\cal{N}) \geq \rm{Pr}_{\*{t}} (\cal{M}) \quad \forall \*{t}\in\rm{Prob}(n\ell)
        \end{equation}
    \end{definition}
\end{defbox}

In this approach, we can extend the winning chance to the concept of predictability. We define the \emph{predictability} of a classical channel via a function $\mathbf{P}:\bb{R}^n \to \bb{R}$ as follows.
\begin{defbox}{Predictability of a classical channel}
    \begin{definition}
        The predictability of a classical channel $\cal{N}$ with the vector $\*{s}$ is defined as
        \begin{equation}
            \predict_\cal{N} (\*{s}) = \max_{x\in[m]} \*{s}\cdot\*{p}_x ^\da,
        \end{equation}
        where $\*{p}_x ^\da$ is the $x$-th column of the transition matrix associated with $\cal{N}$ and is rearranged in the non-increasing order.
    \end{definition}
\end{defbox}
The function $\predict_\cal{N}\pqty{\s}$ captures the notion of predictability whenever $\*{s}\in\prob^\da (n)$. Specifically, we have the following equivalent statement for operational majorization.
\begin{corollary}\label{cor:predict-majorization}
For any classical channels $\cal{N}^{X\to Y}$ and $\cal{M}^{X'\to Y}$, we have that $\cal{N}\succ\cal{M}$ operationally if and only if 
\begin{equation}\label{eq:predictability}
    \predict_\cal{N} (\*{s})\geq \predict_\cal{M} (\*{s})\quad\forall \*{s}\in\prob^\da(n).
\end{equation}
\end{corollary}
\begin{proof}
    To prove that operational majorization implies the inequality of predictability functions, consider $\t \in\prob(n)$, which is the scenario where player know nothing prior to giving the input, $\ell = 1$. From $\rm{Pr}_\*{t} (\cal{N}) = \max_{x\in[m]} \sum_{k} t_k \knorm{\p_x}{k} = \max_{x\in[m]} \t\cdot L\p^\da _x$, the channel $\mN$ operationally majorize $\mM$ implies
    \begin{equation}
        \max_{x\in[m]} U\t\cdot \p_x ^\da \geq \max_{x'\in[m']} U\t\cdot \q_{x'} ^\da \quad\forall \t\in\prob(n),
    \end{equation}
    where $L$ is a lower triangular matrix with all non-zero elements equal to one and $U=L^T$. By defining $\s := \frac{U\t}{\norm{U\*{t}}_1}$, we have that the Equation~\eqref{eq:predictability} holds for all $\s\in\prob^\da(n)$.

    Conversely, the winning chance of a $\t$-game can be written as a sum of $\rm{Pr}_\cal{N} (\s_w)$ where $\s_w = U\t_w$ and $\t_w = \sum_{k\in[n]} \t_{k\vert w} \e_{k}$.
    % \begin{align}
    %     \rm{Pr}_\*{t}(\cal{N}) &= \max_{x\in[m]} \sum_{k\in[n]} t_{k} \knorm{\*{p}_x}{k}=\max_{x\in[m]} U\t\cdot\p_x = \predict_{\cal{N}}(U\t)
    % \end{align}
    \begin{align}
        \rm{Pr}_\*{t}(\cal{N}) &= \sum_{w\in[\ell]} t_w \max_{x\in[m]} \sum_{k\in[n]} t_{k\vert w} \knorm{\*{p}_x}{k}\\
        &= \sum_{w\in[\ell]} \max_{x\in[m]} \sum_{k\in[n]} t_{kw} \knorm{\*{p}_x}{k}\\
        % \sum_{k\in[n]}  t_{kw} \knorm{\*{p}_x}{k} &= \*{t}_{w}\cdot L\*{p}_x = U\*{t}_w\cdot \*{p}_x\\
        % \rm{Pr}_\*{t}(\cal{N}) &= \sum_{w\in[\ell]} \max_{x\in[m]} U\*{t}_w \cdot \*{p}_x 
        &= \sum_{w\in[\ell]} \predict_{\cal{N}}(\s_w)
    \end{align}
    If we have that the Equation~\eqref{eq:predictability} holds for all $\s\in\prob(n)$, then we have that
    \begin{align}
        \rm{Pr}_{\*{t}}(\cal{N}) = \sum_{w\in[\ell]} \predict_\cal{N} (\s_w) \geq \sum_{w\in[\ell]} \predict_\cal{M} (\s_w) = \rm{Pr}_{\*{t}}(\cal{M}),
    \end{align}
    showing the converse direction of the proof.
    % Firstly, note that the winning chance can be phrased in terms of the predictability function.
    
    % Since $\t\in\prob(n)$, $U\t$ is a non-negative vector in non-increasing order. With normalization, define $\s := \frac{1}{\knorm{U\t}{n}} U\t \in \prob^\da(n)$.
    % In fact, given any $\s\in\prob^\da (n)$, one can define $\t\in\prob(n)$ such that $\frac{1}{\knorm{U\t}{n}} U\t = \s$ since $U$ is invertible.
    
    % Now consider the inequality, we have that 
    % \begin{align}
    %     \predict_\cal{N} (\*{s})\geq \predict_\cal{M} (\*{s})\quad\forall \*{s}\in\prob^\da(n) \quad &\iff \quad \predict_\cal{N} (U\t)\geq \predict_\cal{M} (U\t)\quad\forall \*{t}\in\prob(n)\\
    %     &\iff \quad  \max_{x\in[m]} U\t\cdot\p^\da _x \geq \max_{w\in[m']} U\t\cdot\q^\da _w \quad \forall \t\in\prob(n)\\
    %     &\iff \quad  \rm{Pr}_{\*{t}} (\cal{N}) \geq \rm{Pr}_{\*{t}} (\cal{M}) \quad \forall \t\in\prob(n),
    % \end{align}
    % proving the corollary.
\end{proof}

\subsubsection{A sufficient condition}
This previous corollary implies that it is sufficient to consider only the $\*{t}$-games where prior to giving an input, the player is completely clueless about the number of guesses allowed, that is $\*{t}\in\prob(n)$. This family of $\*{t}$-games gives a sufficient condition to conclude the inequality in winning chance of any $\t$-games where $\t\in\prob(n\ell)$.
\begin{corollary}
    Suppose $\cal{N}^{X\to Y}$ and $\cal{M}^{X'\to Y}$ are classical channels. We have that 
    \begin{equation}
    \forall \t \in \prob(n\ell)\quad \pr_\t (\mN) \geq \pr_\t (\mM) \quad \iff \quad \forall \t \in \prob(n)\quad \pr_\t (\mN) \geq \pr_\t (\mM).
    \end{equation}
    % In word, the winning chance of any $\t$-games where $\t\in\prob(n\ell)$ with the channel $\cal{N}$ is greater than or equal to that of $\cal{M}$ if and only if for any $\t\in\prob(n)$ the chance of winning $\t$-game with $\cal{N}$ is greater than or equal to $\cal{M}$.
\end{corollary}
\begin{proof}
    The $\implies$ direction is trivial by setting $\ell = 1$. We left to show the converse implication.

    Using the previous corollary, we want to show that the winning chance of $\t$-games for all $\t\in\prob(n)$ is enough to yield the predictability inequality $\predict_{\cal{N}} (\s) \geq \predict_{\cal{M}} (\s)$ for any $\s \in\prob^\da (n)$. The winning chance can be rephrased as the predictability as follows $\predict_{\cal{N}} (U\t) = \frac{1}{\norm{Ut}_1} \pr_{\t} (\cal{N})$ where $U$ is an upper triangular matrix with non-zero elements equal to one. Since $U$ is invertible, there is a one-to-one correspondence between $\t\in\prob(n)$ and $\s\in\prob(n)$. If $\Pr_\t (\cal{N})\geq \Pr_\t (\cal{M})$ for all $\t\in\prob(n)$, then we have $\predict_{\mN} (\s) \geq \predict_{\mM} (\s)$ for all $\s\in\prob^{\da}(n)$ as well.
\end{proof}

% \magenta{Even simpler, it is enough to consider all $k$-games played with a channel to conclude inequality for all $\*{t}$-game.}
% \begin{corollary}
%     Suppose $\cal{N}^{X\to Y}$ and $\cal{M}^{X'\to Y}$ are classical channels. We have a majorization relation $\cal{N}\succ\cal{M}$ if and only if for all $\*{t} = \*{e}_k \in\prob(n)$ for any $i\in[n]$ we have 
%     \begin{equation*}
%         \pr_\*{t} (\mN) \geq \pr_\*{t} (\mM).
%     \end{equation*}
% \end{corollary}
% \begin{proof}
%     Suppose the channel $\cal{N}$ majorizes the channel $\cal{M}$, then we have the desire inequality as $\*{t} = \*{e}_k \in\prob(n)$ is a special case of $\t \in \prob(n\ell)$. 

%     Conversely, suppose the inequalties of winning chance hold for any $\*{t} = \*{e}_k \in\prob(n)$. We have that 
%     \begin{equation*}
%         \max_{x\in[m]} \knorm{\*{p}_x}{k} \geq \max_{x'\in[m']} \knorm{\*{q}_{x'}}{k}.
%     \end{equation*}
%     Consider any $\*{t}\in\prob(n)$ and the following chain of inequalities
%     \begin{equation*}
%         \max_{x\in[m]} \sum_{k \in[n]} t_k \knorm{\*{p}_x}{k} 
%     \end{equation*}
% \end{proof}

The operational approach is coincided with the previous two approaches.
\begin{propbox}{Equivalence between axiomatic and operational approaches}
    \begin{theorem}\label{th:equiv_axiom-operational}
        Suppose $\cal{N}^{X\to Y}$ and $\cal{M}^{X'\to Y}$ are two classical channels. There exists a completely uniformity-preserving superchannel $\Theta$ such that $\cal{M} = \Theta[\cal{N}]$ if and only if 
        \begin{equation}
            \rm{Pr}_{\*{t}} (\cal{N}) \geq \rm{Pr}_{\*{t}} (\cal{M}) \quad \forall \*{t}\in\rm{Prob}(n).
        \end{equation}
    \end{theorem}
\end{propbox}
\begin{proof}
By the Corollary~\ref{cor:predict-majorization}, $\mN$ majorizing $\mM$ operationally is equivalent to having
\begin{equation}\label{eq:predictability-02}
    \max_{x \in [m]} \s\cdot\p_x ^\da \geq \max_{x' \in [m']} \s\cdot\q_{x'} ^\da,
\end{equation}
for any $\s\in\prob^{\da}(n)$, where $\p_x := \mN(\e_x)$ for each $x \in [m]$ and $\q_{x'} := \mM(\e_{x'})$ for each $x' \in [m']$. Furthermore, in Theorem~\ref{th:convex-sum-outputs}, there exists $\Theta$ such that $\cal{M} =\Theta [\cal{N}]$ if and only if there exists $S\in\stoch(n,n)$, $S=(S_{x\vert w})$ such that 
\begin{equation}\label{eq:op-constr-proof-conv-output}
    \forall w\in[m']\quad \sum_{x\in[m]} s_{x\vert w} \p_x ^\da \succ \q_w.
\end{equation}
The strategy is to phrase this in terms of feasibility problem and use Farkas' lemma to show equivalence to the Inequality~\ref{eq:predictability-02}. 

Suppose $\p_x=\p_x^\da$ and $\q_w=\q_w^\da$ so that the Equation~\ref{eq:op-constr-proof-conv-output} can be written as 
\begin{align}
    \sum_{x\in[m]} s_{x\vert w} L\p_x \geq L \q_w \quad \text{(element-wise)}
\end{align}
or equivalently,
\begin{equation}\label{eq:pv-LNsw}
LN\s_w \geq L\q_w
\end{equation}
where $\s_w$ is the $w^\text{th}$ column of the stochastic matrix $S$. $\s_w$ has to have its elements sum to $1$. That is $\*{1}\cdot\s_w = 1$. Notice that if we have $\*{1}\cdot \s < 1$ then we can multiply by the number that is greater 1 to get equality and not violate the Inequality~\ref{eq:pv-LNsw}. Therefore, we can rewrite the condition in the Theorem~\ref{th:convex-sum-outputs} as 
\begin{equation}
    \bmqty{-LN\\ \*{1}^T _m} \s_w \leq \bmqty{-L\q_w\\ 1}.
\end{equation}
From Farkas' lemma, such $\s$ exists if and only if for all $\t\in\bb{R}^n$ and $\lambda\in\bb{R}$, we define $\vctr:= \t \oplus \lambda$ and have the following implication,
\begin{equation}
\vctr^T \bmqty{-LN\\ \*{1}^T _m} \geq 0 \implies \vctr^T\bmqty{-L\q_w\\ 1} \geq 0
\end{equation}
The premise is equivalent to 
\begin{equation}
\lambda \geq \t^T LN = (U\t)^T N.
\end{equation}
Then the least $\lambda$ that satisfy the Inequality is $\lambda_{\textrm{min}} = \max_{x\in[m]} U\t\cdot \p_x.$ The implication is equivalent to $\lambda \geq U\t\cdot\q_w$. Define $\s = U\t$ and consider the smallest $\lambda$ that the premise holds, we have that 
\begin{equation}
    \max_{x\in[m]} \s\cdot \p_x \geq \s\cdot \q_w.
\end{equation}
Since this holds for all $w$, then it holds for the maximum $\s\cdot\q_w$ as well.
% First, to prove that $\cal{M}=\Theta[\cal{N}]$ implies the Inequality~\ref{eq:predictability-02}. Consider
% \begin{align}
% \q_{x'} ^{Y} = \cal{M}(\*{e}_{x'} ^{X'}) &= \Theta\bqty{\cal{N}}(\*{e}_{x'} ^{X'}) \\
% \magenta{\pqty{\textrm{Equation~\ref{eq:E.Nex-standardform}}}\rightarrow} &= \sum_{x\in[m]} f(x\vert x') \cal{E}_{xx'} \circ \cal{N} (\*{e}^{X}_x)
% \end{align}
% Now, consider the dot product with any $\s\in\prob^\da(n)$, which can be written as $\s=U\t$ for some $\t\in\bb{R}^n _{\geq 0}$
% \begin{align}
%     \s\cdot\q_{x'} &= U\t \cdot \sum_{x\in[m]} f(x\vert x') \cal{E}_{xx'} (\p_x ^Y) \\
%     &=\t \cdot \sum_{x\in[m]} f(x\vert x') L \cal{E}_{xx'} (\p_x ^Y)\\
%     \magenta{\pqty{\cal{E}_{xx'}\text{ is doubly stochastic}}\rightarrow}&\leq \t\cdot \sum_{x\in[m]}f(x\vert x') L \p_x ^Y\\
%     &=\s \sum_{x\in[m]} f(x\vert x') \p^Y _x \leq \max_{x\in[m]} \s \cdot \p.
% \end{align}
% This is true for all $\s\cdot\q_{x'}$ implying the Inequality~\ref{eq:predictability-02}.
% \red{complete this}
\end{proof}
\subsection{Equivalence of the three approaches}
From the Theorem~\ref{th:equiv_const-axiom} and the Theorem~\ref{th:equiv_axiom-operational}, we see that all three approaches results in the same ordering of classical channels. This way, we can establish the majorization preorder on the domain of classical channel. From now on, the majorization of classical channels will be simply referred as majorization without any adverb: constructively, axiomatically, or operationally. Symbolically, the symbol $\succ$ represents majorization of classical channels.
\begin{table}[!h]
    \centering
    \begin{tabular}{|l|l|}
        \hline
        Approach     & Definition of majorization                          \\ \hline
        Constructive & Convertibility via random permutation superchannel  \\
        Axiomatic    & Convertibility via completely uniformity preserving \\
        Operational  & Comparison of winning chance in any $\*{t}$-games   \\ \hline
    \end{tabular}
    \caption{Definitions of classical channel majorization in each approach.}
\end{table}

In hindsight, the coincidence makes an intuitive sense. The constructive approach prescribes a construction which randomly permutes the outcome, which preserves the uniform distribution and reduce the predictability of the outcome. The axiomatic approach enforces that the uniform distribution must be preserved similarly to the axiomatic characterization of the standard probability majorization. Lastly, the operational approach is an extension of $k$-games which characterize the standard probability majorization operationally.

\subsection{Comparison of classical channels across all output dimensions}
In previous sections, we consider only channels which its output has the same dimension. However, we can compare two channels that output states in different dimensions. Suppose that $\cal{N}^{X\to Y}$ and $\cal{M}^{X'\to Y'}$ are classical channels and $\dim{Y}<\dim{Y'}$. The channel majorization relation is defined as a majorization relation between $\cal{M}^{X'\to Y'}$ and $\cal{V}^{Y\to Y'}\circ\cal{N}^{X\to Y}$. That is 
\begin{align}
\cal{N}\succ\cal{M} &:\iff \cal{V}^{Y\to Y'}\circ \cal{N}^{X\to Y} \succ \cal{M}^{X' \to Y'}\\
\cal{M}\succ\cal{N} &:\iff \cal{M}^{X' \to Y'} \succ \cal{V}^{Y\to Y'}\circ \cal{N}^{X\to Y}
\end{align}
In terms of transition matrix associated with $N$, the dimensions of two transition matrices can be matched by adding zero into the smaller dimensions, similarly to the case of probability vectors.

\section{Properties of channel majorization}
In this section, we explore basic properties of the newly proposed majorization: its optimal elements, the preordering given by the majorization, and the recipe for picking a unique representative of an equivalence class given by this preordering.
\subsection{Minimal and maximal elements}
Consider the set of all  $\bigcup_X \rm{CPTP}(X\to Y)$, the most uncertain channel in this domain is a maximally randomizing channel $\cal{R}\in\cptp(1 \to Y)$, and the most certain are channels $\cal{N}_y \in\cptp(1\to Y)$ that produce a standard basis vector $\*{e}_{y}^Y$ for a certain $y\in[n]$.
\begin{proposition}
    For any classical channel $\cal{M}^{X\to Y}$, we have that 
    \begin{enumerate}
        \item $\cal{M}^{X\to Y} \succ \cal{R}^{1 \to Y}$.
        \item $\cal{N}_y ^{1\to Y} \succ \cal{M}^{X\to Y}$
    \end{enumerate}
\end{proposition}
\begin{proof}
For $\cal{N}\succ\cal{R}$, we have $\mR = \Theta [\cal{N}]$ where $\Theta[\cal{N}] = \Tr_Y[\cal{N}(\e_1 ^X)]~\u^{Y}$. This superchannel is completely uniformity preserving, as 
$$\Theta\otimes\id[\cal{N}^{XW\to YZ}] (\p^{W})= \Tr_Y [\cal{N}^{XW\to YZ} (\e_1 ^{X} \otimes \p^{W})]\otimes \u^{Y} = \cal{N}^{XW \to Z} (\e_1 ^X \otimes \p^{W})\otimes \u^Y.$$ 
Notice that $\mR^{X' \to Y}$ for any $X'$ also the minimal as the standard form of the maximally randomizing channel $\cal{R}^{X' \to Y}$ is the preparation channel preparing $\u^Y$ from the trivial state $1$, $\mR^{1\to Y}$.

For $\mN^{1\to Y} _y \succ \cal{N}^{X\to Y}$, suppose $\p_x$ is the $x$-th column of a transition matrix corresponding to $\cal{N}$. We have that for any $\s = \pmqty{s_1 & s_2 & \ldots & s_n}^T\in\prob^\da (n)$
\begin{equation}
\s \cdot \e _y ^\da =\s \cdot \e _1 = s_1 \geq \max_{x\in[m]} \s\cdot \p_x.
\end{equation}
By the Corollary~\ref{cor:predict-majorization}, it follows that $\mN_y ^{1\to Y} \succ \mN ^{X\to Y}$.
\end{proof}

\subsection{Classical channel majorization is a preorder}
\begin{lemma}
    A classical channel majorization ($\succ$) is a preorder on the set of classical channels. 
\end{lemma}
\begin{proof}
    % To show that $\succ$ is a preorder, it has to satisfy
    % \begin{enumerate}
    %     \item (Reflexivity) For any $\cal{N}\in\clch$, we have that $\mN\succ\mN$.
    %     \item (Transitivity) For any $\cal{M},\cal{N},\cal{P}\in\clch$ if $\mM\succ\mN$ and $\mN\succ\mP$ then $\mM\succ\mP$.
    % \end{enumerate}
    To show the reflexivity of $\succ$, notice that an identity superchannel $1^{(X\to Y)}$ is a completely uniformity-preserving superchannel. Therefore, any $\cal{N}^{X\to Y}$ is convertible to itself via an application of the identity superchannel.

    On the transitivity of $\succ$, we will show that the composition of completely uniformity-preserving superchannels is also a uniformity preserving. Suppose a classical superchannel $\Theta \in \super((X_0\to Y )\to (X_1\to Y))$, a classical superchannel $\Upsilon\in\super((X_1\to Y )\to (X_2\to Y))$, and a conditionally uniform channel $\cal{M}^{X_0 W \to Y Z}$. We have that $\cal{N} := \Theta\otimes 1^{(W\to Z)} [\cal{M}]$ is a conditionally uniform channel and $\Upsilon\otimes \otimes 1^{(W\to Z)} [\cal{N}]$ is also a conditionally uniform channel.

    Suppose $\cal{N}\succ\cal{M}$ and $\cal{M}\succ\cal{P}$. Suppose $\cal{M} = \Theta[\cal{N}]$ and $\cal{P} = \Upsilon [\cal{M}]$. We have that $\Upsilon \circ \Theta$ is a completely uniformity-preserving superchannel and therefore $\cal{P} = \Upsilon \circ \Theta [\cal{N}]$ and $\cal{N}\succ\cal{P}$.
\end{proof}

\subsection{Replacement channel: reduction to probability majorization}
One can look at a quantum state as a replacement channel that map the trivial state, $1$, to the quantum state. For this reason, one would expect that the state majorization relation is reflected in the channel majorization as well. By definition, a replacement channel is a channel that always output the same quantum state, says $\sigma \in \fk{D}(B)$, for any input $\rho\in\fk{D}(A)$. The replacement channel $\cal{N}_\sigma \in \rm{CPTP}(A\to B)$ is defined as
\begin{equation}
    \cal{N}_\sigma(\rho) = \Tr[\rho]\sigma.
\end{equation}
In the classical domain, a classical replacement channel takes classical state $\rho\in\fk{D}(X)$ and another classical state output $\sigma\in\fk{D}(Y)$. As we use probability vectors to represent classical states, one can express a classical replacement channel as a transition matrix with all columns being an associated vector of $\sigma$, here we call it $\*{q}$. We have the following equivalence
\begin{lemma}\label{lm:reduce-to-majorization}
    Suppose $\cal{N}\in\rm{CPTP}(X\to Y)$ and $\cal{M}\in\rm{CPTP}(X\to Y')$ be classical replacement channels replacing with $\*{p}$ and $\*{q}$ respectively. We have that $\cal{N}\succ \cal{M}$ if and only if $\*{p} \succ \*{q}$.
\end{lemma}

\begin{proof}
    Suppose $\p\succ\q$; then we have $\q = D\p$. To show that $\cal{N}\succ\cal{M}$, define $\Theta$ such that 
    \begin{equation}
    \Theta[\cal{F}] = \cal{D}\circ\cal{F}.
    \end{equation}
    This shows that $\Theta[\cal{N}] = \cal{D}\circ\cal{N} = \cal{M}$.

    On the converse direction, suppose $\cal{N} \succ\cal{M}$. We have that there is $\Theta$ superchannel such that $\cal{M}= \Theta[\cal{N}]$. Writing this superchannel in its standard form results in 
    \begin{equation}
        \Theta[\cal{N}] = \sum_{x,x'}\cal{E}_{x,x'}\circ \cal{N}\circ \cal{F}_{x,x'}
    \end{equation}
    Then input any $\*{e}_w$ into the channel,
    \begin{align}
        \sum_{x,x'}\cal{E}_{x,x'}\circ \cal{N}\circ\cal{F}_{x,x'}(\*{e}_w) &=  \sum_{x,x'} f(x\vert w)\cal{E}_{x,w}\circ\cal{N}^{X\to Y}(\*{e}^X).\\
        &= \sum_{x,x'} f(x\vert w)\cal{E}_{x,w}(\p) = D\p
    \end{align}
\end{proof}

\subsection{The standard form of channel}
In the domain of probability vectors, two equivalent probability vectors are equal if both vectors have their elements in non-increasing order. Similarly, there is a way to turn majorization preorder on the classical channels into the partial order by restricting the set of classical channels to only those in the \emph{standard form}.

Suppose we have a classical channel $\cal{N}$. If there is another classical channel $\mM$ that is equivalent with the channel $\mN$, then we say that it is in the same equivalence class, but need not be the same channel. The goal of defining the standard form is to pick one unique representative of any equivalent class of channels. We define the standard form for a classical channel as follows.
\begin{defbox}{The standard form of classical channel}
\begin{definition}\label{def:std-form}
    A classical channel $\cal{N}$ with the associated transition matrix $N = \pqty{\p_1 \ldots\p_n}$ is in the standard form, if the followings are true for any $x\in[m]$.
    \begin{enumerate}
        \item Non-increasing order: $\p_x = \p^\da _x$.
        \item Incomparable columns: $\p_x$ is not majorized by a convex sum of other columns.
        \item Lexicographical order: If $x\neq m$ and $\ell$ is the lowest index such that $(\p_x)_\ell \neq (\p_{x+1})_\ell$, then $(\p_x)_\ell > (\p_{x+1})_\ell$.
    \end{enumerate}
\end{definition}
\end{defbox}

To turn a classical channel $\cal{N}$ into its standard form, one obtains the transition matrix $N_s$ associated with the standard form of the channel $\cal{N}$, denoted with $\cal{N}_s$, by the following steps.
\begin{enumerate}
    \item Replace each and every column $\p_x$ with their own in non-increasing ordered vector $\p_x ^\da$.
    \item Remove, one at a time, a column that is majorized by a convex sum of other columns.
    \item Rearrange the columns so that for each $\p_x$ and $\p_{x+1}$, the first $\ell-1$ elements are equal while the $\ell^\rm{th}$ elements of $\p_x$ is greater than that of $\p_{x+1}$.
\end{enumerate}
The above procedure turns a classical channel into its standard form. This process produces a unique and equivalent classical channel to the original one.
\begin{lemma}
    Suppose $\cal{N}^{X\to Y}$ is a classical channel. Its standard form, $\cal{N}_s ^{X' \to Y}$, is equivalent to $\cal{N}^{X\to Y}$.
\end{lemma}
\begin{proof}
    We use the Theorem~\ref{th:convex-sum-outputs} in both directions. 
    
    Consider the direction $\cal{N}_s\succ\cal{N}$. Any $\p_x$ for $x\in[m]$ is either in $\qty{\cal{N}_s(\e_x): x\in[m']}$ or is majorized by a convex sum of columns of $N_s$. Therefore, $\cal{N}_s\succ \cal{N}$.
    
    Conversely, consider the direction $\cal{N}\succ\cal{N}_s$. Since, no new column is added to $N_s$, any $\q_w = \cal{N}_s (\e_w)$ is $\q = \p^\da$ for some $\p \in\cal{N}(\prob(m))$. This concludes the proof. 
\end{proof}

Once the channel is in the standard form, any two equivalents classical channels have to be equal. That is a preorder of channel majorization turns into a partial order on the set of classical channels in the standard form.
\begin{propbox}{Channel majorization is a partial order on the set of channels in the standard form}
    \begin{theorem}
        Suppose $\cal{N}^{X\to Y}$ and $\cal{M}^{X'\to Y}$ are classical channels in the standard form. We have that $\cal{N}^{X\to Y} \sim \cal{M}^{X'\to Y}$ if and only if $\cal{N}^{X\to Y} = \cal{M}^{X'\to Y}$.
    \end{theorem}
\end{propbox}
\begin{proof}
    The if direction is trivial, we will prove the only-if direction.
    First, the majorization conditions are equivalent with
    \begin{align}
        \cal{N}\succ\cal{M} &\iff \exists S = (s_{x\vert w})\in\stoch(m,m') \quad \sum_{x} s_{x\vert w} \p_x \succ \q_w \quad \forall w\in[m']\\
        &\iff LNS \geq LM \label{eq:prf-std-form-eq-01}\\ 
        \cal{M}\succ\cal{N} &\iff \exists T = (t_{w\vert x})\in\stoch(m',m) \quad \sum_{w} t_{w\vert x} \q_w \succ \p_x \quad \forall w\in[m']\\
        &\iff LMT \geq LN \label{eq:prf-std-form-eq-02}
    \end{align}
    Apply $T$ on the right in the Inequality~\ref{eq:prf-std-form-eq-01}, and $S$ to the right in the Inequality~\ref{eq:prf-std-form-eq-02}, and define $R = ST$ and $R' = TS$. Using the transitive property, we have
    \begin{align}
        LNR \geq LMT \geq LN  \quad \text{and} \quad LMR' \geq LNS \geq LM.
    \end{align}
    That is
    \begin{equation}
        LN\vctr_x \geq L\p_x \quad \forall x\in[m] \quad \text{and} \quad LM\vctr_w \geq L\q_w \quad \forall w\in[m']
    \end{equation}
    where $\vctr_x = (r_{x' \vert x})$ and $\vctr_w = (r' _{w' \vert w})$ are the $x^\rm{th}$ column of $R$ and the $w^\rm{th}$ column of $R'$ respectively.
    If $r_{x\vert x} \neq 1$, then 
    \begin{equation}
        \sum_{x' \neq x} r_{x'\vert x} L\p_{x'} \geq (1-r_{x\vert x}) L\p_x,
    \end{equation}
    which implies the existence of convex combination of other columns that majorizing the $x^\rm{th}$ column,  therefore $r_{x'\vert x} = \delta_{xx'}$ and similarly with $R'$, $r'_{w'\vert w} = \delta_{ww'}$. With the consideration of $ST = R = I_m$ and $TS = R' = I_{m'}$, we have that $S$ and $T$ is inverse of one another. The only invertible stochastic matrices are permutation matrices, which are square. From the following inequality, we have the equality
    \begin{equation}
        LNS \geq LM \text{ and } LM = LMTS \geq LNS \implies LNS = LM \implies NS = M.
    \end{equation}
    This implies that $N$ and $M$ are identical up to column permutation, however being in the standard form impose a unique way of arranging columns. Therefore, $\cal{N} = \cal{M}$.
% \red{New idea, check with people} \\
% $\cal{N}\sim\cal{M}$ implies that the convex hulls $\frak{N} := \conv\pqty{\qty{\Pi\p_x:\Pi\in\perm(n),x\in[m]}}$\\ and $\frak{M} := \conv\pqty{\qty{\Pi\q_w:\Pi\in\perm(n),w\in[m']}}$ coincide. Since from the Definition~\ref{def:std-form}, every columns is in non-increasing order and there is no columns that is majorized by another columns, therefore $\qty{\p_w}_{x\in[m]} = \qty{\q_w}_{w\in[m']}$ are sets of extreme points and particularly $m = m'$. The only freedom left is the arrangement of columns which is determined by number $3$ of the definition.
\end{proof}

% \subsection{Monoid}
% A set of classical channels in the standard form together with a tensor product can be recognise as an ordered commutative monoid. The definition of ordered commutative monoid is given as follows.
% \begin{definition}
% An ordered commutative monoid is an ordered set equipped with a binary relation $\otimes$ that satisfies the following.
% \begin{enumerate}
%     \item The binary relation is associative and commutative 
%     \begin{equation}
%         \mM\otimes(\mN \otimes \mP) = (\mM\otimes \mN) \otimes \mP \quad \text{and} \quad \mM\otimes\mN = \mN\otimes\mM
%     \end{equation}
%     \item Identity $\mM\otimes I = \mM$
%     \item Addition respect the ordering $\cal{M}\succ\cal{N} \implies \cal{M}\otimes \cal{P} \succ \cal{N}\otimes \cal{P}$
% \end{enumerate}
% \end{definition}

% \begin{theorem}
%     A set of all classical channels, $\clch$, equipped with the Kronecker tensor product $\otimes$, is an ordered commutative monoid.
% \end{theorem}
% \begin{proof}
%     \magenta{Decide what to do with this}
% \end{proof}
\section{Characterizations of classical channel majorization}
In this section, we provide additional characterizations of the classical channel majorization. This is in addition to the characterization we previously provided in the Theorem~\ref{th:convex-sum-outputs} and in the corollary~\ref{cor:predict-majorization}.

\subsection{Majorization of outputs}
One interesting characterization of channel majorization is through the set of its outputs, which is a convex hull of the columns of its corresponding transition matrix. The image of all $\p\in\prob(m)$ by $\cal{N}$ is a convex set. This is because the channel $\cal{N}$ is linear and $\p$ can be put in a convex sum of vectors in the standard basis,
\begin{equation}
    \cal{N}(\p) = \cal{N}(\sum_{x\in[m]} p_x \e_{x}) = \sum_{x\in[m]}  p_x \cal{N}(\e_{x})
\end{equation}
Now we can see that the image is a convex hull of columns of $\cal{N}$,
\begin{align}
    \cal{N}\pqty{\prob(m)} &=\qty{\cal{N}(\p):\p\in\prob(n)} = \qty{\sum_{x\in[m]}  p_x \cal{N}(\e_{x}) : \p = \sum_{x\in[m]} p_x \e_x \in\prob(m)}\\
    &=\conv\qty{\cal{N}(\e_x) : x\in[m]}.
\end{align}

Using the Theorem~\ref{th:convex-sum-outputs} and the standard form of channels, we have the following characterization classical channel majorization. 
\begin{propbox}{Majorization of outputs of channels}
    \begin{corollary}\label{cor:convex-set-maj}
        Given two classical channels, $\cal{N}$ and $\cal{M}$, in their standard form, we have $\cal{N}\succ\cal{M}$ if and only if for any elements $\q \in \cal{M}(\prob(m'))$, there exists an element $\p\in\cal{N}(\prob(m))$ such that $\p\succ\q$. 
    \end{corollary}
\end{propbox}
\begin{proof}
The proposition is directly followed from the Theorem~\ref{th:convex-sum-outputs}. In the if direction, we have that all any columns of $M$ can be majorized by a convex combination of columns of $N$ showing that, $\mN \succ \mM$.

On the only-if direction, suppose $\mN \succ \mM$. We have that any column of $M$, $\q_w$, is majorized by a convex combination of columns of $N$. Denote the convex combination with $\vctr_w = \sum_{x\in[m]} s_{x\vert w} \p_x$. Any $\q\in\mM(\prob(m'))$, is a convex combination of $\q_w$, i.e. $\q = \sum_{w\in[m']} t_w \q_w$. Since $\vctr_w \succ \q_w$ and $\p_x, \vctr_w,$ and $\q_w$ are in non-increasing order, we have the following chain of inequalities. For any $k\in[n]$, 
\begin{equation*}
\knorm{\sum_{w\in[m']} t_w \vctr_w }{k} = \sum_{w\in[m']} t_w \knorm{\vctr_w}{k} \geq \sum_{w\in[m']} t_w \knorm{\q_w}{k} = \knorm{\q}{k},
\end{equation*}
showing that $\vctr:= \sum_{w\in[m']} t_w \vctr_w \succ \q$. The vector $\vctr$ is in the image $\mN\pqty{\prob(m)}$ as it is a convex combination of $\p_x$ with the convex coefficients $\qty{\sum_{w\in[m']} t_w s_{x\vert w}: x\in[m]}$.

% By recognizing $\sum_{x\in[m]} s_{x\vert w} \p_x = \sum_{x\in[m]} p_x \cal{N}(\*{e}) \in \cal{N}(\prob(m))$ and that any elements $\q\in\cal{M}(\prob(m'))$ can be written as a convex sum of $\q_{w} := \cal{M}(\e_w)$, we have the following equivalent statements
% \begin{align}
% \cal{N}\succ\cal{M} &\iff \sum_{x\in[m]} s_{x\vert w} \p_x \succ \q_w \quad\forall w\in[m]\\
% &\iff \sum_{x\in[m]} s_{x\vert w} L \p_x ^\da \geq L \q_w ^\da \quad \forall w\in[m] \label{eq:proof-conv-set-output-1}
% \end{align}
% Then taking the convex sum of the above inequality will not change the inequality,
% \begin{equation}
% \sum_{w\in[m']} t_w \sum_{x\in[m]} s_{x\vert w} L \p_x \geq \sum_{w\in[m']} t_w L \q_w \quad \forall \t = (t_w) \in\prob(m').
% \end{equation}
% Equivalently,
% \begin{equation}
%     \sum_{w\in[m']} t_w \sum_{x\in[m]} s_{x\vert w} \p_x \succ \sum_{w\in[m']} t_w \q_w \quad \forall \t = (t_w) \in\prob(m'),
% \end{equation}
% proving the Corollary.
\end{proof}

\subsection{Set containment}
In the case of classical probability vector, $\p\succ\q$ if and only if $\q$ is in the convex hull of all permutations of $\p$. Similarly, in the case of classical channels, we have the following characterization.
\begin{propbox}{Set-containment characterization}
    \begin{theorem}
        Suppose $\cal{N}$ and $\cal{M}$ are classical channels, we have that $\cal{N}\succ\cal{M}$ if and only if
        \begin{equation}
            \conv\qty{\Pi\q_w:\Pi\in\perm(n),w\in[m']}\subseteq \conv\qty{\Pi\p_x:\Pi\in\perm(n),x\in[m]}
        \end{equation}
    \end{theorem}
\end{propbox}
\begin{proof}
    Suppose the set containment is true. For any $w\in[m']$, we have that any $\q_w$ can be written as a convex sum of $\p_x$'s and their permutations,
    \begin{equation*}
        \q_w = \sum_{\substack{x\in[m]\\ \vctv\in[n!]}} s_{xv} \Pi_v \p_x,
    \end{equation*}
    where $s_{xv}$'s are the convex coefficients for each $\Pi_v \p_x$. Any $\p_x$ can be turned into $\p_x ^\da$ by applying an appropriate permutation matrix, denote such permutation for each $\p_x$ with $S_x$,
    \begin{align*}
        \q_w &= \sum_{\substack{x\in[m]\\ v\in[n!]}} s_{xv} \Pi_v S_{x}^{-1} S_{x}\p_x \\
    \magenta{\pqty{P_{xv}:=\Pi_v S_{x}^{-1}} \rightarrow} &= \sum_{\substack{x\in[m] \\ v\in[n!]}} s_{xv} P_{xv} \p_x ^\da = \sum_{x\in[m]} \pqty{\sum_{v\in[n!]} s_{xv}P_{xv}}\p_x ^\da = \sum_{x\in[m]} t_x D_x \p_x ^\da,
    \end{align*}
    where $t_x D_x := \sum_{v\in[n!]} s_{xv} P_{xv}$ and $D_x \in \rm{DSTOCH}(n)$. Now, consider the following chain of inequalities of the Ky Fan norms, for any $k\in[n]$
    \begin{equation*}
        \knorm{\q_w}{k} = \knorm{\sum_{x\in[m]} t_x D_x \p_x ^\da}{k} \leq \sum_{x\in[m]} t_x \knorm{D_x \p_x}{k} \leq \sum_{x\in[m]} t_x \knorm{\p_x}{k} = \knorm{\sum_{x\in[m]} t_x \p_x ^\da}{k},
    \end{equation*}
    showing that $\sum_{x\in[m]} t_x \p_x^\da \succ \q_w$. By Theorem~\ref{th:convex-sum-outputs}, we have that $\cal{N}\succ\cal{M}$.

    Conversely, in the only-if direction, we suppose that $\mN \succ \mM$.\\ For any $\q\in\conv\qty{\Pi \q_w: \Pi\in\perm(n), w\in[m']}$, it can be written as 
    \begin{equation*}
    \q = \sum_{\substack{u\in[n!]\\w\in[m]}} t_{uw} \Pi_u \q_w
    \end{equation*}
    Since $\mN\succ \mM$, then any $w\in[m']$ there is majorized by a convex combination of $\p_x$'s. This leads to 
    \begin{equation*}
        \q = \sum_{\substack{u\in[n!]\\w\in[m]}} t_{uw} \Pi_u D_w \sum_{x\in[m]} s_{x\vert w} \p_x.
    \end{equation*}
    Any doubly stochastic matrix can be written as a convex combination of permutation matrices. Suppose the combination for each $D_w$ is $D_w = \sum_{v\in[n!]} r_{v\vert w} \Pi_v.$ We have that
    \begin{equation*}
        \q = \sum_{\substack{u,v\\w,x}} t_{uw} s_{x\vert w} r_{v\vert w} \underbrace{\Pi_u \Pi_v}_{=: P_{uv}} \p_x = \sum_{\substack{u,v\\w,x}} t_{uw}s_{x\vert w} r_{v\vert w} P_{uv} \p_x,
    \end{equation*}
    showing that $\q$ is a convex combination of $\p_x$'s and their permutations.
\end{proof}

\subsection{Characterization in low dimensional cases}
For any classical channel with one dimensional output, we have the trivial classical states. Recall that the standard form requires no column of the transition matrix to be majorized by a convex sum of another columns. If a channel has two-dimensional output, its standard form is a probability vector, because all two-dimensional vectors are comparable by the majorization relation. 
% The case of three and higher dimensions output could have a nice property and characterization but we have not yet to reveal it.

On the side of the input dimension, the input dimensions represent the number of choices of the output. For a one-dimensional output, we have probability vector, the channel majorization relation reduces to probability vector majorization by the Lemma~\ref{lm:reduce-to-majorization}. For the higher dimensions, a dimension of the input can be inflated by adding columns that is in the convex hull of the pre-existing columns. The next following characterization theorems require classical channels to be in the standard form to avoid this elusive notion of the input dimension.

Suppose a classical channel $\cal{N}^{X\to Y}$ has two-dimensional input $m=2$ and another classical channel $\cal{M}^{X'\to Y}$ has $m'$-dimensional input . Suppose the channel $\cal{N}$ has columns $\p_1$ and $\p_2$. By Theorem~\ref{th:convex-sum-outputs}, we have that
\begin{equation}
    \cal{N}\succ \cal{M} \iff \forall w\in[m'] ~\exists t_w\in[0,1] \quad t_w \p_1 + (1-t_w) \p_2 \succ \q_w.
\end{equation}
Since we require $\cal{N}$ and $\cal{M}$ to be in the standard form, we have $\p_1$ and $\p_2$ are in non-increasing order. By the majorization relation, we have that their Ky Fan norms satisfy
\begin{equation}
    \knorm{t_w \p_1 + (1-t_w) \p_2}{k} = t_w \knorm{\p_1}{k} + (1-t_w) \knorm{\p_2}{k} \geq \knorm{\q_w}{k}.
\end{equation}
That is 
\begin{equation}
    t_w (\knorm{\p_1}{k} - \knorm{\p_2}{k}) \geq \knorm{\q_w}{k} - \knorm{\p_2}{k}.
\end{equation}
This equality gives enough condition to check existence of $t_w$. The idea is to divide both side by $\knorm{\p_1}{k} - \knorm{\p_2}{k}$. We categorize the each index $k$ into one of three sets $\fk{I}_+,\fk{I}_0,\fk{I}_-$ where the denominator is positive, zero, and negative respectively. That is 
\begin{align}
    \fk{I}_+ &:= \qty{k: \knorm{\p_1}{k} - \knorm{\p_2}{k} > 0}, \\
    \fk{I}_0 &:= \qty{k: \knorm{\p_1}{k} - \knorm{\p_2}{k} = 0}, \\
    \fk{I}_- &:= \qty{k: \knorm{\p_1}{k} - \knorm{\p_2}{k} < 0}.
\end{align}
Notice that $k\in \fk{I}_0$, we have $\knorm{\p_2}{k} \geq \knorm{\q_w}{k}$.

For $k\in \fk{I}_+$, dividing both side of the inequality by $\knorm{\p_1}{k} - \knorm{\p_2}{k}$ yields
\begin{align}
    t_w &\geq \frac{\knorm{\q_w}{k} - \knorm{\p_2}{k}}{\knorm{\p_1}{k} - \knorm{\p_2}{k}} \quad \forall k\in[n] \\
    \iff t_w &\geq \max_{k\in\fk{I}_+}~\frac{\knorm{\q_w}{k} - \knorm{\p_2}{k}}{\knorm{\p_1}{k} - \knorm{\p_2}{k}}.
\end{align}
Similarly in $k\in \fk{I}_-$, with converse inequality
\begin{align}
    t_w &\leq \min_{k\in\fk{I}_-}~\frac{\knorm{\q_w}{k} - \knorm{\p_2}{k}}{\knorm{\p_1}{k} - \knorm{\p_2}{k}}.
\end{align}
Combine with $t_w\in[0,1]$ we have 
\begin{align}
    \max\qty{0,\max_{k\in\fk{I}_+}~\frac{\knorm{\q_w}{k} - \knorm{\p_2}{k}}{\knorm{\p_1}{k} - \knorm{\p_2}{k}}} \leq t_w \leq \min \qty{1, \min_{k\in\fk{I}_-}~\frac{\knorm{\q_w}{k} - \knorm{\p_2}{k}}{\knorm{\p_1}{k} - \knorm{\p_2}{k}}}
\end{align} 
We define the leftmost quantity to be $\mu_w$ and the rightmost $\nu_w$. We have the following theorem.
\begin{propbox}{Characterization theorem for channels with two dimensional input}
    \begin{theorem}
        For any $\cal{N}^{X\to Y}$ a classical channel in the standard form with two-dimensional input and any classical channel $\mM$, we have that $\cal{N}\succ \cal{M}$ if and only if the followings hold.
        \begin{enumerate}
            \item For all $w\in[m']$, $\nu_w \geq \mu_w$.
            \item For all $k\in \fk{I}_0$, $\knorm{\p_2}{k} \geq \knorm{\q_w}{k}$.
        \end{enumerate} 
        $\nu_w$, $\mu_w$, and $\fk{I}_0$ are as previously defined.
    \end{theorem}
\end{propbox}

\subsection{Connection to conditional majorization}
We proposed a channel majorization to capture uncertainty of the output given full control of the input. This notion is closely related with conditional majorization. Suppose $\p^{XY}$ and $\q^{X'Y}$ are two joint probability vectors. The conditional majorization relation compares uncertainty of two probability distribution, given that we have access to the information on the systems $X$ and $X'$~\cite{GGH+2018}.

The classical channel naturally gives rise to a joint probability vector but not uniquely. Suppose Alice is an agent on the sending side of a classical channel $\cal{N}^{X\to Y}$ and Bob is at the receiving end. If Alice randomly gives an input from distribution $\p^X = (p_x)$, then the joint probability between the input and the output is $\p^{XY} = \sum_{x} p_x \e_x \otimes \cal{N}(\e_x)$. Notice that the joint probability vector $\p^{XY}$ is dependent on the choice of input distribution $\p^{X}$, leading to multiple joint probability vectors associated with a classical channel.

Recall from the theorem~$\ref{th:convex-sum-outputs}$ that a classical channels $\cal{N}$ in its standard form majorizes another classical channel $\cal{M}$ if we can find an input $\p$ which give an output which majorize $\q_w := \cal{M}(\e_w)$ for each of any $w\in[m']$. This motivates the following theorem.
\begin{propbox}{Characterization with conditional majorization}
\begin{theorem}
    Suppose $\cal{N}^{X\to Y}$ and $\cal{M}^{X' \to Y}$ are classical channels and $\cal{N}^{X\to Y}$ is in the standard form. We have that $\cal{N}\succ\cal{M}$ if and only if there exists $\p = (p_x)$ and $\q = (q_w)$ such that
    \begin{equation}\label{eq:th-conditional-majorization-channel-majorization}
    \sum_{x\in[m]} p_x \e_x ^{\tilde{X}} \otimes \cal{N}(\e_x) \succ_Y \sum_{w\in[m']} q_w \e_w^{\tilde{X}'} \otimes\cal{M}(\e_w).
    \end{equation}
\end{theorem}
\end{propbox}
\begin{proof}
    Define $\cal{N}(\e_x) = \p_x$ and $\cal{M}(\e_w) = \q_w$. The Equation~\eqref{eq:th-conditional-majorization-channel-majorization} is equivalently written as  
    \begin{equation}
        \sum_{x\in[m]} p_x \e_x ^{\tilde{X}} \otimes \p_x \succ_Y \sum_{w\in[m']} q_w \e^{\tilde{X}'} _w \otimes \q_w.
    \end{equation}
    Using a characterization of conditional majorization with the existence of $R = (r_{w\vert x}) \in \stoch(m',m)$ such that for any $w\in[m']$ 
    \begin{equation}
        \sum_{x\in[m]} r_{w\vert x} p_x \p_{x} \succ q_w \q_w.
    \end{equation} 
    For all $w\in[m']$, we have that
    \begin{equation}
        \sum_{x\in[m]} \frac{r_{w\vert x} p_x}{q_w} \p_{x} ^\da =\sum_{x\in[m]} \frac{r_{w\vert x} p_x}{q_w} \p_{x} \succ \q_w,
    \end{equation} 
    showing that $\cal{N}\succ\cal{M}$ by the characterization in theorem \ref{th:convex-sum-outputs}.
\end{proof}

%%%% Reference %
%%%% for editing purpose only %%%
% \bibliography{QuantumEntropy}
% \bibliographystyle{unsrt}
% \printbibliography
%%%%%%%%%%%%%%%%%%%%%%%%%%%%%%%%%

% \tableofcontents
\setcounter{chapter}{3}
\chapter{Measures of classical channel uncertainty}
With the foundation of channel majorization laid in previous chapter, we can provide a general definition of \emph{a measure of classical channel uncertainty}. Instead of taking the preorder, we can define a \emph{monotone} of the preorder and compare the mapping to the real number. Eventually, we want the monotone to represent an extensive physical quantity, hence requiring an additivity into the monotone, which form a class of monotones called \emph{entropy} of classical channels.
\section{Monotones of classical channel majorization}
The classical channel majorzation relation is a preorder in the domain of classical channels. Similarly to the case of probability vectors and quantum states, the monotone of the classical channel majorization is a map that preserve the majorization relation between any two classical channels.
\begin{defbox}{Classical channel majorization monotones}
    \begin{definition}
        A function $g:\mathsf{ClassicalChannels} \to \mbR$ is said to be a channel majorization monotone if for every classical channel $\mN\in\cptp(X\to Y)$ and $\mM\in\cptp(X'\to Y')$
        \begin{equation}
            \mN\succ \mM \implies g(\mN)\geq g(\mM).
        \end{equation}
        If $-g$ is a channel majorization monotone, then $g$ is called channel majorization antitone.
    \end{definition}
\end{defbox}
\begin{remark}
    The implication is one way, because the $\geq$ relation on $\bb{R}$ is a total order while $\succ$ on $\ssf{ClassicalChannels}$ is merely a preorder. 
\end{remark}

As we have seen in the previous chapter, the winning chance for a $\t$ game $\pr$ and the predictability function $\predict_{\cal{N}}$ are examples of the classical channel monotones. However, we are interested in a monotone that quantifies an extensive property of classical channel, which require an additional property.

Similarly to the case of states, if $\bb{H}$ is an antitone and it is additive under tensor product, then it is called an entropy. We write $\bb{H}(Y\vert X)_\cal{N}$ to represent an entropy of the channel $\cal{N}^{X\to Y}$. The $(Y\vert X)$ part signifies conditioning $Y$ on $X$.
\begin{defbox}{Classical Channel entropy}
    \begin{definition}
        A function $\bb{H}:\ssf{Classical Channels}\to \bb{R}$ is an entropy of a classical channel if for every two classical channels $\mN\in\cptp(X\to Y)$ and $\mM\in\cptp(X'\to Y')$ we have that
        \begin{enumerate}
            \item $\bb{H}$ is an antitone of channel majorization.
            \begin{equation}
                \cal{N}\succ\cal{M} \implies \bb{H}(Y\vert X)_{\cal{N}} \leq \bb{H}(Y'\vert X')_{\cal{M}}
            \end{equation}
            \item $\bb{H}$ is additive under parallel composition $\otimes$
            \begin{equation}
                \bb{H}(YY'|XX')_{\cal{N}\otimes\cal{M}}=\bb{H}(Y|X)_{\cal{N}}+\bb{H}(Y'|X')_\cal{M}
            \end{equation}
        \end{enumerate}
    \end{definition}
\end{defbox}
The additivity property ensures that the entropy is a measure of uncertainty that reflects amount of uncertainty inherent in a classical channel. With additivity, a bipartite channel $\cal{P}^{XX' \to YY'}$ that is composed of two local operations, $\cal{P} = \cal{N}^{X\to Y}\otimes\cal{M}^{X' \to Y'}$, has its entropy equal to the sum of the entropies of its parts. This allows entropy, as a measure of uncertainty, to reflect a well-defined physical quantity related to a classical channel.

Since the majorization of classical channel reduces to the majorization of probability vectors when the channels are preparation channels, a monotone of channel majorization is Schur-convex on the domain of probability vectors. This also implies that any $\bb{H}$ entropy of a classical channel $\cal{N}^{X\to Y}$ is bounded below by 0 and is bounded above by $\log (n)$.

\section{Extensions of the known entropy to channel domains}

There are multiple monotones (Schur-convex function) and entropies on the domain of probability vectors. We take advantage of the existing monotones by extending it to be the monotone of channel majorization. 
\begin{definition}
    Suppose $A\supseteq B$. A function $f:A\to \bb{R}$ is an \emph{extension} of function $g:B\to \bb{R}$ to the domain $A$ if for all $x\in B$, we have $f(x) = g(x)$.
\end{definition}

%I would like to rephrase this back in term of resource monotones.
It has been shown in~\cite{GT2020} that with a given resource monotone $M_1$ define on a domain $\mathfrak{R}_1$. One can extend the monotone onto a domain $\mathfrak{R}\supset\mathfrak{R}_1$ by the optimal extensions, namely \emph{maximal} extension and \emph{minimal} extension. These two extensions give a bound for any monotone extending $M_1$ is bounded above and below by the optimal extensions.  Similarly, given that we have an entropy function for a probability vector $\bb{H}$, we can apply minimal and maximal extensions to define a monotone on the domain of classical channels.
In particular, the maximal extension $\Hmax(Y|X)_\cal{N}$ and minimal extension $\Hmin(Y|X)_\cal{N}$ of entropy are defined as 
\begin{align}
    \Hmax(Y|X)_\cal{N} &= \inf_{\substack{{\q\in\prob(m)}\\ m\in\bb{N}}} \qty{\bb{H}(\q):\cal{N}\succ\q}.\label{eq:max-extension}\\
    \Hmin(Y|X)_\cal{N} &= \sup_{\substack{{\q\in\prob(m)}\\ m\in\bb{N}}} \qty{\bb{H}(\q):\q\succ\cal{N}}\label{eq:min-extension}
\end{align}
In terms of uncertainty associated with a classical channel $\cal{N}$, the minimum extension is an entropy of the most uncertain probability vector that is more certain than the channel $\cal{N}$. On another end, the maximum extension is an entropy of the most predictable probability vector that is not more predictable than the channel $\cal{N}$. Moreover, any other extension $\bb{H}'$ to the domain of classical channel is bounded above and below by the maximal extension $\overline{\bb{H}}$ and minimal extensions $\underline{\bb{H}}$. 
\begin{propbox}{Boundedness of entropy extensions}
    \begin{theorem}\label{th:optimal-bound-extensions}
        Given a classical channel entropy $\bb{H}$. The functions $\overline{\bb{H}}$ and $\underline{\bb{H}}$ defined as in equations~\eqref{eq:max-extension} and~\eqref{eq:min-extension} respectively are channel majorization monotones. Moreover, if $\bb{H}'$ is an extension of $\bb{H}$ onto the classical channel domain, then it is bounded by $\underline{\bb{H}} \leq \bb{H}'\leq \overline{\bb{H}}$.
    \end{theorem}
\end{propbox}
\begin{proof}
    First, monotonicity. For the minimal extension, suppose that two classical channel $\cal{N}^{X\to Y}\succ \cal{M}^{X'\to Y'}$. Suppose further that $\q^{Y} \succ \cal{N}^{X\to Y}$. By transitivity of majorization preorder, we have that $\q^Y \succ \cal{M}^{X'\to Y'}$ as well. This show that 
    \begin{equation*}
        \qty{\q : \cal{N} \succ \q} \subseteq \qty{\q: \mM \succ\q}.
    \end{equation*}
    The supremum of the image of $\qty{\q : \cal{N} \succ \q}$ by $\bb{H}$ is greater than the supremum of the image of $\qty{\q : \cal{M} \succ \q}$ by $\bb{H}$, i.e.
    \begin{equation*}
        \underline{\bb{H}} (\cal{N}) = \sup_{\substack{{\q\in\prob(m)}\\ m\in\bb{N}}} \qty{\q:\cal{N}\succ\q} \leq \sup_{\substack{{\q\in\prob(m)}\\ m\in\bb{N}}} \qty{\q: \mM \succ\q} = \underline{\bb{H}} (\cal{M}),
    \end{equation*}
    showing that $\underline{\bb{H}}(Y|X)_\cal{N} \leq \underline{\bb{H}}(Y'|X')_\cal{M}$.

    Similarly, with the monotonicity of the maximal extension. Suppose $\cal{N}^{X\to Y} \succ \cal{M}^{X'\to Y'}$. Suppose $\cal{M}^{X\to Y}\succ\q^{Y}$. By transitivity, we have that, $\mN\succ\q$. Following the similar line of logic from above, we have 
    \begin{align*}
        \qty{\q : \q\succ\cal{N}} &\supseteq \qty{\q: \q\succ\mM}\\ 
        \implies  \quad \inf_{\substack{{\q\in\prob(m)}\\ m\in\bb{N}}} \qty{H(\q):\q\succ\cal{N}} &\leq \inf_{\substack{{\q\in\prob(m)}\\ m\in\bb{N}}} \qty{H(\q) : \q\succ\mM},
    \end{align*}
    showing that $\overline{\bb{H}}(Y|X)_\cal{N} \leq \overline{\bb{H}}(Y'|X')_\cal{M}$.

    For the optimality, since $\bb{H}'$ is a monotone, we have that for any $\mN \in \clch$, $\p\in\prob(m)$, for any $m\in\bb{N}$ such that $\p\succ\mN$
    \begin{equation}
        \bb{H}' (Y \vert X)_\cal{N} \leq \bb{H} (\p).
    \end{equation}
    Then,
    \begin{equation}
        \bb{H}' (Y \vert X)_\cal{N} \leq \inf_{\substack{{\q\in\prob(m)}\\ m\in\bb{N}}} \qty{\bb{H} (\q) :\q\succ\mN} = \overline{\bb{H}} (Y\vert X)_\mN
    \end{equation}
    For $\mN \in \clch$, $\p\in\prob(m)$, for any $m\in\bb{N}$ such that $\mN\succ\p$, we have 
    \begin{equation}
        \bb{H}' (Y \vert X)_\cal{N} \geq \bb{H} (\p).
    \end{equation}
    Then, we complete the proof, 
    \begin{equation}
        \bb{H}' (Y \vert X)_\cal{N} \geq \sup_{\substack{{\q\in\prob(m)}\\ m\in\bb{N}}} \qty{\bb{H} (\q) :\mN\succ\q} = \underline{\bb{H}} (Y\vert X)_\mN.
    \end{equation}
\end{proof}

We will apply the technique in the upcoming sections. Since we have a characterization of channel majorization in terms of convex set of probability vectors, the extensions can be equivalently defined by optimal bound of a set probability vectors. Specifically, for a probability vector $\q$ to majorize a channel $\mN$, it needs to majorize all outputs of $\mN$, denoted with $\mN(\prob(n))$.

\subsection{Optimal bounds of probability vector majorization}
\begin{defbox}{The optimal bounds}
    \begin{definition}    
        Suppose $A = \qty{\p_1, \p_2, \ldots, \p_n}$ is a set of probability vectors. A vector $\check{\p}$ is the optimal upper bound of $A$  if the followings hold.
        \begin{enumerate}
            \item $\check{\*{p}}\succ \p_i$ for any $i\in[n]$.
            \item Any probability vector $\q$ such that $\q\succ\p_i$ for all $i\in[n]$, we have that $\q\succ\vee\p$.
        \end{enumerate}
        Similarly, a probability vector $\hat{\p}$ is the optimal lower bound of $A$ if the following holds.
        \begin{enumerate}
            \item $\p_i \succ \hat{\p}$ for any $i\in[n]$.
            \item Any probability vector $\q$ such that $\p_i\succ\q$ for all $i\in[n]$, we have that $\wedge\p\succ\q$.
        \end{enumerate}
    \end{definition}
\end{defbox}
Do the optimal upper bound and lower bound always exists? Yes, to see this, we show the existence of the optimal lower bound first. Once the optimal lower bound is established, the optimal upper bound can be given as an optimal lower bound of all upper bound.  
\begin{proposition}
    Given a set of probability vectors $A$, a probability vector $\hat{\p} = (\hat{p}_1 , \hat{p}_2 , \ldots , \hat{p}_n)^T$ defined as 
    \begin{equation}
        \hat{p}_k := \inf_{x\in[m]} \knorm{\p_x}{k} - \inf_{y\in[m]} \knorm{\p_y}{k-1}
    \end{equation}
    is an optimal lower bound of $A$. For $k=1$, $\knorm{\p}{0} =0$.
\end{proposition}
\begin{proof}
    Suppose a probability vector $\p_x\in A$ form a pair $\pqty{\p_x,\vctu^{(n)}}$ which a lower Lorenz curve $r_x (t)$. By monotonicity of the lower Lorenz curve, we have that if $\knorm{\p_x}{k} > \knorm{\p_w}{k}$, then $r_x (t) < r_w (t)$ at least on the interval $[\knorm{\p_w}{k}, \knorm{\p_x}{k}]$.  Then, the lower Lorenz curve of $\pqty{\hat{\p},\vctu^{(n)}}$ is $\hat{r}(t) = \sup \qty{ r(t) = \rm{LLC}(\p,\*{u}^{(n)}): \p \in A }$. Since each $r (t)$ is convex, the maximization over all possible $r(t)$ is also convex.
    % Notice that $\hat{\p}$ is a probability vector define from finite intersection of all lower Lorenz curve of $\qty{\p,\vctu^{(n)}}$. That is $\fk{T}(\hat{\p},\*{u}^{(n)}) = \bigcap_{x\in[m]} \fk{T}(\p_x,\*{u}^{(n)})$. Since a testing region is convex, its finite intersection and its lower Lorenz curve are convex. 
    This means that $\hat{\p}$ is a probability vector in non-increasing order. Consequently, the Ky Fan $k$-norm of $\hat{\p}$ is given simply by 
    \begin{equation}
        \knorm{\hat{\p}}{k} = \inf_{x\in[m]} \knorm{\p_x}{k},
    \end{equation}
    which is the highest possible Ky Fan norm, otherwise some $\p\in A$ does not majorize  the vector.
\end{proof}

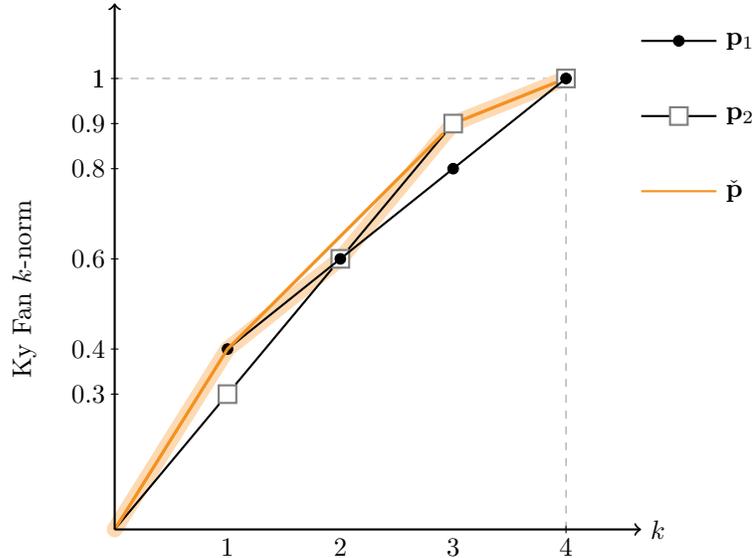
\begin{figure}[h]
    \centering
    \begin{tikzpicture}[squarednode/.style={rectangle, draw=gray, fill=white, thick, minimum size=1pt},]
        %Frame for 1x1 square
        \draw [color = black!40, dashed] (0,6) -- (6,6);
        \draw [color = black!40, dashed] (6,0) -- (6,6);

        %BurntOrange
        \draw [line width = 6pt, line cap = round, color = BurntOrange!30] (0,0) -- (1.5,2.4) -- (3,3.6) -- (4.5,5.4) -- (6,6);
        
        %BurntOrange
        %p_2
        \node[squarednode, centered] (a) at (1.5,1.8) {};
        \node[squarednode, centered] (b) at (3,3.6) {};
        \node[squarednode, centered] (c) at (4.5,5.4) {};
        \node[squarednode, centered] (d) at (6,6) {};
        \draw [thick, color = black] (0,0) -- (a) -- (b) -- (c)--(d);
        
        %p_1
        \draw [thick, color = black] (0,0) -- (1.5,2.4) -- (3,3.6) -- (4.5,4.8)--(6,6);
        \filldraw [color=black] (1.5,2.4) circle (2pt);
        \filldraw [color=black] (3,3.6) circle (2pt);
        \filldraw [color=black] (4.5,4.8) circle (2pt);
        \filldraw [color=black] (6,6) circle (2pt);
        
        %p-tilde
        \draw [very thick, color = BurntOrange] (0,0) -- (1.5,2.4) -- (3,3.9) -- (c)--(d);
        
        %Legend
        \draw [thick, color = black] (7,6.5)--(8,6.5);
        \filldraw[color=black] (7.5,6.5) circle (2pt);
        \node[right] at (8,6.5) {$\p_1$};

        \draw [thick, color = black] (7,5.5)--(8,5.5);
        \node[squarednode, centered] at (7.5,5.5) {};
        \node[right] at (8,5.5) {$\p_2$};

        \draw [thick, color = BurntOrange] (7,4.5)--(8,4.5);
        \node[right] at (8,4.5) {$\check{\p}$};

        % y-axis 
        \node [rotate = 90] at (-1.2,3.2) {Ky Fan $k$-norm};

        \node[left] at (0,1.8) {$0.3$};
        \draw [color = black] (-0.05,1.8) -- (0.05,1.8);
        \node[left] at (0,2.4) {$0.4$};
        \draw [color = black] (-0.05, 2.4) -- (0.05, 2.4);
        \node[left] at (0,3.6) {$0.6$};
        \draw [color = black] (-0.05, 3.6) -- (0.05, 3.6);
        \node[left] at (0,4.8) {$0.8$};
        \draw [color = black] (-0.05, 4.8) -- (0.05, 4.8);
        \node[left] at (0,5.4) {$0.9$};
        \draw [color = black] (-0.05, 5.4) -- (0.05, 5.4);
        \node[left] at (0,6) {$1$};

        \draw [color = black] (6,-0.05) -- (6,0.05);
        \draw [color = black] (-0.05,6) -- (0.05,6);
        \node[below] at (1.5,0) {$1$};
        \node[below] at (3,0) {$2$};
        \node[below] at (4.5,0) {$3$};
        \node[below] at (6,0) {$4$};
        \node[left] at (0,6) {$1$};
        \draw [thick, ->] (0,0) -- (7,0);
        \node[right] at (7,0) {$k$};
        \draw [thick, ->] (0,0) -- (0,7);
    \end{tikzpicture}
    \caption{The Ky Fan $k$-norm of $\p_1$, $\p_2$ and the optimal upper bound $\check{\p}$.}
    \label{fig:KyFanNorm-not-the-optimal-upper-bound-example}
\end{figure}

\remark{The opposite is not true for the optimal upper bound. A probability vector $\tilde{\p}$ whose elements defined to be 
\begin{equation*}
    \tilde{p}_k := \max_{x\in[m]} \knorm{\p_x}{k} - \max_{y\in[m]} \knorm{\p_y}{k-1}
\end{equation*}
is not be necessary an optimal upper bound. For example, consider a set of $\p_1$ and $\p_2$ where
\begin{align*}
    \p_1 = \pqty{ 0.4, 0.2, 0.2, 0.2} ^T \quad\rm{and}\quad \p_2 = \pqty{ 0.3, 0.3, 0.3, 0.1} ^T.
\end{align*}
This set of probability vector gives $\tilde{\p} = \frac{1}{10} \pqty{4,2,3,1}$, which is not in non-increasing order. The probability vector $\check{\p} = \frac{1}{10} \pqty{4,2.5,2.5,1}$ is not majorizing $\tilde{\p}$, $\check{\p}\not\succ\tilde{\p}$, and it is the optimal upper bound.  See Figure~\ref{fig:KyFanNorm-not-the-optimal-upper-bound-example} for illustration.
}
In the previous example, we claimed that $\check{\p}$ prescribed above is the optimal upper bound. One can give a geometric reasoning from the Figure~\ref{fig:KyFanNorm-not-the-optimal-upper-bound-example}. The vector $\check{\p}$ has all Ky Fan $k$-norm equal to that of $\p_1$ or $\p_2$ except at $k=2$ where it is strictly higher. This qualifies the curve to define an upper bound of the set $\qty{\p_1,\p_2}$. On one hand, nudging the point $(2,0.65)$ a down will disqualify the drawn line to be a plot of a Ky Fan norm, as it break non-increasing ordering of the slope. Equivalently, it breaks the concavity of the curve. On the other hand, if the point $(2,0.65)$ is nudged up, then the probability vector defined by the nudged point cannot be majorized by the given upper bound $\check{\p}$. 

Relating this back to the extensions of majorization monotones, we have that an optimal upper bound for the image of $\mN$ has the entropy equal to the maximal extension entropy of $\mN$, i.e.
\begin{equation}\label{eq:minmimal-ext-sup}
    \bb{H}\pqty{\check{\p}} = \underline{\bb{H}}\pqty{Y\vert X}_{\mN}.
\end{equation}
To see this, consider any $\q\in\prob(m)$ such that $\q\succ\p$ for all $\p\in\mN(\prob(n))$. We have that $\q\succ \check{\p}$ by the definition of $\check{\p}$ being an optimal upper bound. By monotonicity of an entropy with respect to the majorization ordering, the inequality $\bb{H}(\check{\p}) \geq \bb{H}(\q)$ holds. Moreover, it is enough to check that $\check{p}$ majorizes all $\p_x = \mN\qty(\ketbra{x}{x})$ to conclude majorization over the entire image, because $\check{\p}\succ\p_i$ and $\check{\p}\succ\p_j$ implies $\check{\p}\succ\alpha\p_i+(1-\alpha) \p_j$ for all $\alpha \in[0,1]$.

On the other hand, the optimal lower bound for the image of $\mN$ does not necessary yield the minimal extension entropy. This is because the majorization condition $\cal{N}\succ\q$ does not require $\q$ to be majorized by all elements in the image of $\mN$.

\subsection{Entropies from extensions}
These extensions are only guaranteed to be a monotone and reduced to entropy on state but not necessary are additive. On some class of function, we find that its extension is additive. A quasiconcave entropy, e.g. R\'enyi entropies, minimally extends to an entropy function.
\begin{propbox}{Maximal extension of a R\'enyi entropy}
    \begin{theorem}\label{th:minClCh}
        Suppose $\cal{N}$ is a classical channel, $\bb{H}$ is a quasiconcave classical entropy, then the maximal extension of the entropy is 
        \begin{equation}
            \Hmax (Y|X)_\cal{N} = \min_{x\in[m]} \bb{H}(\p_x)
        \end{equation}
        where $\p_x\eqdef\cal{N}(\*{e}_x)$ and $\Hmax$ is a classical channel entropy.
    \end{theorem}
\end{propbox}
\begin{remark}
    The entropy function defined as above is called minimal entropy output. It is first purposed by~\cite{KR2001}.
\end{remark}
\begin{proof}
    By definition the maximal extension is $\overline{H} (Y\vert X)_{\cal{N}} = \inf \qty{\bb{H}(\q): \cal{N}\succ\q}$. A probability $\q$ satisfies $\cal{N}\succ\q$ if and only if $\exists \s = (s_x)\in \prob(m)~\text{s.t.}~\sum_{x\in[m]} s_x \p_x \succ \q$. In terms of the entropy function, 
    \begin{equation}
        \bb{H}\pqty{\sum_{x\in[m]} s_x \p_x} \leq \bb{H}(\q)
    \end{equation}
    If $\bb{H}$ is quasiconcave, then any point in the convex hull of $\conv\qty{\p_x}$ is greater than the minimum of $\p_x$, that is
    \begin{equation}
        \min_{x\in[m]} \bb{H}(\p_x) \leq \bb{H}\pqty{\sum_{x\in[m]} s_x \p_x} \leq \bb{H}(\q).
    \end{equation}
    This shows that the entropy of $\q$ is at least the minimal of the $\cal{N}$'s outputs. To show that this value is attainable, pick $x_0$ to be the argument of the minimum $\bb{H}(\p_{x_0})$, then define $\s = \e_{x_0}$ shows that $\cal{N}\succ\p_{x_0}$.
    
    Left to show is the additivity of the function, consider two channels $\cal{N}^{X\to Y}$ and $\cal{M}^{X'\to Y'}$. Set $m = \abs{X}, m'=\abs{X'}, Z = X\otimes Y$, its max-entropy is 
    \begin{align}
        \Hmax(YY' \vert XX') _{\cal{N}\otimes \cal{M}} &= \min_{z\in[mm']} \bb{H} (\cal{N}\otimes \cal{M}(\*{e}^Z _{z})) \\ 
        &= \min_{\substack{x\in[m]\\ x'\in[m']}} \bb{H} (\cal{N}(\e^X _x)\otimes \cal{M}(\e^{X'} _{x'})) \\
        &= \min_{x\in[m]}\bb{H} (\cal{N}(\e^X _x)) + \min_{x' \in[m']}\bb{H}(\cal{M}(\e^{X'} _{x'})) \\
        &= \Hmax(Y\vert X)_{\cal{N}} +\Hmax(Y'\vert X')_{\cal{M}}.
    \end{align}
    This shows than a maximum extension of a quasiconcave entropy (e.g. R\'enyi entropies) is an entropy for classical channels.
\end{proof}

For example, the maximal extension of the min entropy is
\begin{align}
    \overline{H}_\rm{min} (Y\vert X)_\cal{N} &= \min_{x\in[m]}  - \log( \max_{w\in[n]} \qty{(\p_x)_w} ) \\
    &= -\log\max_{\substack{x\in[m]\\ w\in[n]}}  N_{wx} \label{eq:ob-H-min}
\end{align}
If $\cal{N}$ is in the standard form, then the transition matrix element $N_{wx}$ that is the biggest is $N_{11}$.
\begin{equation*}
    \overline{H}_\rm{min} (Y\vert X)_\cal{N} = -\log (N_{11}).
\end{equation*}
The maximal extension of the max entropy is 
\begin{align}
    \overline{H}_\rm{max} (Y\vert X)_\cal{N} = \log  \min_{x\in[m]}  \abs{\rm{supp}(\p_x)} \label{eq:ob-H-max}
\end{align}

In fact, there is no other way to extend max-entropy or min-entropy. Any function that extends $H_\rm{min}$ or $H_\rm{max}$ to the domain of classical channel has the expression equivalent to that in Equation~\eqref{eq:ob-H-min} or Equation~\eqref{eq:ob-H-max}.
\begin{propbox}{Uniqueness of min-entropy and max-entropies extensions}
\begin{theorem}
    The extensions of max-entropy and min-entropy to the domain of classical channels are unique.
\end{theorem}
\end{propbox}
\begin{proof}
    For the min-entropy, the minimal extension of the min-entropy is 
    \begin{align*}
        \underline{H}_\rm{min} (Y\vert X)_\cal{N} &= \sup \qty{- \log  \pqty{\max_{x\in[m]} q_x}: \q\succ\cal{N}}
    \end{align*}
    Suppose $\cal{N}$ is in the standard form. Define $k_0 = \lfloor \frac{1}{N_{11}} \rfloor$, define a probability vector 
    \begin{equation*}
        \q = ( \underbrace{N_{11}, \ldots ,N_{11}}_{k_0~\rm{elements}}, 1-k_0 N_{11})^T.
    \end{equation*}
    We will show that $\q\succ\cal{N}$. Since $\p_x$ is in non-increasing order and $\cal{N}$ is in the standard form, we have $N_{11} \geq N_{1j}$ for any $j\in[n]$, $\p_x ^\da$ for all $x\in[n]$, and for all $k\leq k_0$,
    \begin{equation*}
        \knorm{\p_x}{k} = \sum_{i=1} ^{k} p_{x,i} \leq \sum_{i=1} ^{k} N_{11} = \knorm{\q}{k}.
    \end{equation*}
    For $k> k_0$, $\knorm{\q}{k} = 1$ by definition. Therefore, $\q\succ\p_x$ for all $x\in[n]$ and, consequently, $\q\succ\cal{N}$.
    Going back to the minimal extension of the min-entropy, we have an inequality
    \begin{equation*}
        \underline{H}_\rm{min} (Y\vert X)_\cal{N} \geq -\log(N_{11}) = \overline{H}_\rm{min} (Y\vert X)_\cal{N}.
    \end{equation*}
    From the theorem, Theorem~\ref{th:optimal-bound-extensions}, we have that the maximal and minimal extensions are identical
    \begin{equation*}
        \underline{H}_\rm{min} (Y\vert X)_\cal{N} = \overline{H}_\rm{min} (Y\vert X)_\cal{N} = -\log(N_{11}) .
    \end{equation*}

    For the max-entropy, the minimal extension of the max-entropy is 
    \begin{align*}
        \underline{H}_\rm{max} (Y\vert X)_\cal{N} &= \sup \qty{\log  \pqty{\abs{\rm{supp}(\q)}}: \q\succ\cal{N}}
    \end{align*}
    We will give an element $\q$ of the above set which has support equal to the minimum support of $\p_x$ over all $x\in[n]$. Suppose the minimum support of $\p_x$ is $k_0$. Recall that the Ky Fan norm is increasing with $k$. If there is $\q$ such that $\knorm{\q}{1} = \max_{x\in[n]} \knorm{\p_x}{k_0-1}$, then $\knorm{\q}{k} \geq \knorm{\q}{1} \geq \knorm{\p_x}{k'}$ for all $k,k'\in[k_0-1]$.  
    
    Suppose $\knorm{\p_z}{k_0-1} =\max_{x\in[n]} \knorm{\p_x}{k_0-1}$, define $\q = (\knorm{\p_z}{k_0 -1}, q_2, \ldots, q_{k_0}, \ldots, q_n)$ where $q_i = (1-q_1)/(k_0 -1)$ for all $i\in[2: k_0]$ and $q_j = 0$ for $j> k_0$. This $\q$ is a probability vector in non-increasing order, has a cardinality of the support equal to $k_0$ and $\q \succ\p_x$ for all $x\in[n]$. That is, we have an inequality 
    \begin{equation*}
        \underline{H}_\rm{max} (Y\vert X)_\cal{N} \geq \log \abs{\rm{supp}(\q)} = \log \min_{x\in[m]} \abs{\rm{supp}(\p_x)} = \overline{H}_\rm{max} (Y\vert X)_\cal{N}.
    \end{equation*}
    From Theorem~\ref{th:optimal-bound-extensions}, we have that $\underline{H}_\rm{max}\leq \overline{H}_\rm{max}$. Therefore, extension of the max-entropy is unique, i.e
    \begin{equation*}
        \underline{H}_\rm{max} (Y\vert X)_\cal{N} = \overline{H}_\rm{max} (Y\vert X)_\cal{N} = \log \min_{x\in[m]} \abs{\rm{supp}(\p_x)}.
    \end{equation*}
    % If $\q\succ\p$, then $\rm{supp}(\q)\leq \rm{supp}(\p)$. We have $\q\succ\cal{N}$ if and only if $\q\succ\p_x$ for all $x\in[n]$. Therefore, if $\q\succ\cal{N}$ then $\rm{supp}(\q)\leq \min_{x\in[n]} \rm{supp}(\p_x)$. That is $\underline{H}_\rm{max} (\p) \geq \log \qty{\min_{x\in[n]} \rm{supp}(\p_x)}$
\end{proof}

\subsection{Regularization and uniqueness of extensions}
\begin{defbox}{Regularization of a classical channel majorization monotone}
    \begin{definition}
        Suppose $\bb{H}:\clch\to\bb{R}$ is a classical channel majorization monotone. Its regularization is defined as 
        \begin{equation}
            \bb{H}^{\rm{reg}} (Y\vert X)_\cal{N} := \lim_{k\to\infty} \frac{1}{n} \bb{H}(Y^{k}\vert X^{k})_{\cal{N}^{\otimes k}}
        \end{equation}
    \end{definition}
\end{defbox}
\begin{remark}
    The limit in above definition is not guaranteed to exist in general. In a resource theory that admit tensor product structure, the optimal extensions of monotone  are sub-additive and super-additive. By Fekete's subadditive lemma, the limit exists.
\end{remark}

If we apply the regularization to inequality in the Theorem~\ref{th:optimal-bound-extensions}, we have that 
\begin{equation*}
    \Hmin^\rm{reg}(Y\vert X )_\cal{N} \leq \bb{H} ^\rm{reg}(Y\vert X )_\cal{N} \leq \Hmax ^\rm{reg}(Y\vert X )_\cal{N}.
\end{equation*}
If $\bb{H}$ is an entropy, i.e. it is additive, then the regularized $\bb{H} ^\rm{reg}(Y\vert X )_\cal{N}$ is exactly the entropy $\bb{H}(Y\vert X)_{\cal{N}}$. That is we have, for $\bb{H}$ a channel entropy, 
\begin{equation}\label{ineq:reg-upper-lower}
    \Hmin^\rm{reg}(Y\vert X )_\cal{N} \leq \bb{H}(Y\vert X )_\cal{N} \leq \Hmax ^\rm{reg}(Y\vert X )_\cal{N}.
\end{equation}
This simple result will help us show the uniqueness of a channel entropy that extends Shannon entropy. In particular, the extension in Theorem~\ref{th:minClCh} provide a unique extension of the Shannon entropy to the classical channel majorization.
\begin{propbox}{Uniqueness of the channel entropy extending Shannon entropy}
    \begin{theorem}\label{th:unique-channel-shannon-entropy}
        Suppose $\bb{H}$ be a classical channel entropy. If $\bb{H}$ extends Shannon entropy, then it is the maximal extension, i.e.
        \begin{align*}
            \bb{H}\pqty{Y\vert X}_\cal{N} &= \underline{H}^\rm{reg}(Y\vert X)_\cal{N} = \overline{H}(Y\vert X)_\cal{N} = \min_{x\in[m]} H(\p_x).
        \end{align*}
    \end{theorem}
\end{propbox}
\begin{remark}
    The statement of this is weaker than the statement with min-entropy and max-entropy. It does not state the uniqueness for a monotone that extends Shannon entropy but is not additive.
\end{remark}
\begin{remark}[2]
    The proof of this theorem follows the same spirit as the uniqueness of a channel divergence extending KL-divergence in~\cite{GW2021}.
\end{remark}
We are working toward the proof of this theorem, which requires the following lemma. 
\begin{lemma}\label{lm:typical-majorization}
    Suppose $\p\in\prob(n)$, and $0<\varepsilon, \delta < 1$. There exists an integer $m$ such that for all integers $k\geq m$, 
    \begin{equation*}
        \*{r}_{k} = \pqty{\delta, \underbrace{2^{-k(H(\p) - \varepsilon)},\ldots,2^{-k(H(\p) - \varepsilon)}}_{c_k -\rm{times}},s_k} \succ \p^{\otimes k},
    \end{equation*}
    where $c_k = \lfloor (1-\delta) 2^{k (H(\p) - \varepsilon)}\rfloor$ and $s_k$ is a remainder of the division $(1-\delta) /2^{- k (H(\p) - \varepsilon)}$.
\end{lemma}
\begin{proof}
    For any probability vector $\p\in\prob(n)$, its tensor product $\p^{\otimes k}$ describes a probability of obtaining a sequence $\*{X}= X_1, X_2, \ldots, X_k$ where each $X_i$ is drawn from the same distribution $\p$ independently. The vector $\p^{\otimes k}$ is equivalent with any permutation, so we have 
    \begin{equation*}
        \p^{\otimes k} \sim \bigoplus_{\*{x} \in [n]^k} p_{\*{x}}
    \end{equation*}
    where $\*{x}$ is a sequence of the values of $X_1, \ldots, X_k$ and $p_{\*{x}}$ is a probability of the sequence $\*{x}$ drawn.

    Next, we merge all the entries that are not $\varepsilon$-typical into an element. This results in 
    \begin{equation}\label{eq:typical-majorization}
        \p^{\otimes k} \sim \bigoplus_{\*{x} \in [n]^k} p_{\*{x}} \prec \rm{P}(\*{X}\not\in A^{(k)} _\varepsilon) \oplus \bigoplus_{\*{x}\in A^{(k)} _\varepsilon} p_{\*{x}} 
    \end{equation}
    where $A^{(k)} _\varepsilon$ denotes a set of all $\varepsilon$-typical sequences of length $k$. The majorization relation above can be seen quickly by applying finite number of $T$-transform mapping the elements $\rm{P}(\*{X}\not\in A^{(k)} _\varepsilon)$ back to the original in $\p^{\otimes k}$.
    %%%%% ARGUMENT USING T_TRANSFORM WOULD BE EASIER... T maps the RHS back to LHS by unsum the terms 
    % The majorization relation follows from the fact that for all $p_{\*{x}}\in A^{(k)} _\varepsilon$ it is bounded in the interval, $[2^{-k(H(\p)+\varepsilon)}, 2^{-k(H(\p)-\varepsilon)}]$. If $p_\x \leq 2^{-k(H(\p)-\varepsilon)}$ for all $\*{x} \in [n]^k$, then the RHS of the Equation~\eqref{eq:typical-majorization} sum muliple small elements 
    
    We will use the transitivity of majorization relation to show that the right side of the Equation~\eqref{eq:typical-majorization} is majorized by $\*{r}_k$. First, by the application of the weak law of large number on i.i.d. sequence (see Theorem~\ref{th:typical-seq-prob}), for any $\delta > 0$, there exists a $k_0 \in\bb{N}$ such that for all $k \geq k_0$, we have
    \begin{equation*}
        \rm{P}(\*{X}\in A^{(k)} _\varepsilon) > 1 - \delta,
    \end{equation*}
    or equivalently
    \begin{equation*}
        \rm{P}(\*{X}\not\in A^{(k)} _\varepsilon) < \delta.
    \end{equation*}
    For any $\x \in A^{(k)} _\varepsilon$, the probability of drawing the sequence $\x$ is bounded, 
    \begin{equation*}
        p_\*{x} \leq 2^{-k(H(\p)-\varepsilon)}.
    \end{equation*}
    % Since $s_k$ is defined to be the remainder of the division $(1-\delta) / 2^{-k (H(\p) - \varepsilon)}$, we have that 
    % \begin{equation*}
    %     s_k \leq 2^{-k(H(\p)-\varepsilon)}
    % \end{equation*}
    Two equations above give a sufficient condition for 
    \begin{equation*}
        \*{r}_k \succ \rm{P}(\*{X}\not\in A^{(k)} _\varepsilon) \oplus \bigoplus_{\*{x}\in A^{(k)} _\varepsilon} p_{\*{x}} \succ \p^{\otimes k}.
    \end{equation*}
\end{proof}

Now, we are ready to proof the uniqueness theorem.
\begin{proof}[Proof of Theorem~\ref{th:unique-channel-shannon-entropy}]
    The minimal extension of the Shannon entropy $\underline{H}(Y^k \vert X^k)_{\cal{N}^{\otimes k}}$ is the entropy of $\check{\p}$ the optimal upper bound of a set of probability vectors $\qty{\bigotimes_{x\in\x} \p_x\quad \forall \x \in [m]^k}$. Such bound can be assumed to have the same dimension, i.e. $\check{\p}\in\prob(mk)$. We have
    \begin{align*}
        \underline{H}(Y^k \vert X^k)_{\cal{N}^{\otimes k}} &= \sup \qty{H(\q): \q \succ \bigotimes_{x\in\x} \p_x\quad \forall \x \in [m]^k, \q \in\prob(mk)} \\ 
        &= \sup \qty{H(\q): \q \succ \bigotimes_{x\in[m]} \p_x ^{\otimes kt_x}\quad \forall \t \in \bb{T}_{m,k},\q \in\prob(mk)},
    \end{align*}
    where $\bb{T}_{m,k}$ is a set of all types $\*{t}$ of length-$k$ sequence with possible $m$ alphabets.
    % From the previous lemma, we have that $H(\bigvee_{x\in[m]} p_x \otimes s_x) \leq H(\bigvee_{x\in[m]} p_x) + \log(\ell)$ where $s_x\in \prob(\ell)$.
    % \begin{align*}
    %     \underline{H}(Y^k \vert X^k)_{\cal{N}^{\otimes k}} &\geq \sup \qty{ H(\q): \q\succ \bigotimes_{x\in[m]} \p_x ^{\otimes kt_x} \otimes \bigotimes_{x\in[m]} \p_x ^{\otimes \kappa}} - \log(n^{\kappa})\\ 
    %     &= \sup\qty{H(\q): \q\succ \bigotimes_{x\in[m]} \p^{\otimes kt_x + \kappa}} - \kappa \log(n).
    % \end{align*}
    % Define $k_x = k t_x + \kappa$.
    
    Define $k_x = k t_x$. Define for all $x\in[m]$ and $\varepsilon,\delta \in(0,1)$, 
    \begin{equation*}
        \overline{\*{p}}^{(k_x)} _x := \pqty{\delta, r_2, \ldots, r_{c_x}, s'_x},
    \end{equation*}
    where $c_x = \lfloor (1-\delta) 2^{k_x (H(\p_x))}\rfloor $, for all $i\in[2:c_x]$, we define $r_i =2^{-k_x (H(\p_x))-\varepsilon}$, and $s' _x = 1 - \delta - c_x 2^{-k_x (H(\p_x))}$. We use Lemma~\ref{lm:typical-majorization}, $\p^{\otimes k_x} _x \prec \overline{\*{p}}^{(k_x)} _x$
    Furthermore, suppose $H(\*{p}_z) = \min_{x\in[m]} H(\*{p}_x)$. We have that 
    \begin{equation*}
        \*{v}_x := \pqty{\delta, v_2, \ldots, v_{d_x}, s _x}
    \end{equation*}
    where $d_x = \lfloor (1-\delta) 2^{k_x (H(\p_z)-\varepsilon)}\rfloor $, for all $i\in[2:d_x]$, we define $v_i =2^{-k_x (H(\p_z)-\varepsilon)}$, and $s _x = 1 - \delta - d_x 2^{-k_x (H(\p_z)-\varepsilon)}$. We have $\*{v}_x\succ \overline{\p}_x ^{(k_x)} \succ \p^{\otimes k_x} _x$. That is  
    \begin{equation*}
        \underline{H}(Y^k \vert X^k)_{\cal{N}} \geq H(\*{v}_1 \otimes \*{v}_2 \otimes \ldots \otimes \*{v}_m).
        % - \kappa \log(n).
    \end{equation*}
    Using additiviy of Shannon entropy, 
    \begin{equation*}
        H(\*{v}_1 \otimes \*{v}_2 \otimes \ldots \otimes \*{v}_m) = \sum_{x\in[m]} H(\*{v}_x) = -m\delta \log(\delta) - \sum_{x\in[m]} s_x \log(s_x) + d_x 2^{-k_x (H(\*{p}_z) - \varepsilon)} k_x (H(\p_z) - \varepsilon).
    \end{equation*}
    Notice that, at each $k$ and each $x$, 
    \begin{equation*}
        d_x = \left\lfloor \frac{1-\delta}{2^{-k_x (H(\*{p}) - \varepsilon)}} \right\rfloor 
        = \frac{1-\delta }{2^{-k_x (H(\vctp_z) - \varepsilon)}} - \frac{g^{(k)} _x}{ 2^{-k_x (H(\vctp_z) - \varepsilon)} }
    \end{equation*}
    for some $g^{(k)} _x  \in [0,2^{-k_x (H(\vctp_z) - \varepsilon)})$. Taking the limit $k \to \infty$, $2^{-k_x (H(\vctp_z) - \varepsilon)} \to 0$ and $g^{(k)} \to 0$. We have 
    \begin{equation*}
        \lim_{k\to \infty} d_x 2^{-k_x (H(\p_z)-\varepsilon)} = \lim_{k\to \infty}\left\lfloor \frac{1-\delta}{2^{-k_x (H(\vctp_z) - \varepsilon)}} \right\rfloor 2^{-k_x (H(\vctp_z) - \varepsilon)}  = 1 - \delta.
    \end{equation*}

    We have 
    \begin{align*}
        \lim_{k\to \infty} \frac{1}{k} \sum_{x\in[m]} s_x \log s_x &= \lim_{k\to\infty} \frac{1}{k}  \sum_{x\in[m]} \pqty{1-\delta - d_x 2^{-k_x (H(\p_z)-\varepsilon)}} \log \pqty{1-\delta - d_x 2^{-k_x (H(\p_z)-\varepsilon)}} \\
        &= \lim_{k\to \infty} \frac{g^{(k)} _x}{k}   \log(g^{(k)} _x) = 0
    \end{align*}
    
    Now, we are ready to apply these results into the regularized maximally extended entropy,
    \begin{align*}
        \underline{H}^\rm{reg}(Y\vert X)_{\cal{N}} &= \lim_{k\to \infty} \frac{1}{k}\underline{H}(Y^k \vert X^k)_{\cal{N}^{\otimes k}}\\ 
        &\geq \lim_{k\to \infty} \frac{1}{k} H(\*{v}_1 \otimes \ldots \*{v}_m) \\
        % + \frac{\kappa}{k} \log(n) \\ 
        &= \lim_{k\to \infty} \frac{k_x}{k} \pqty{H(\p_z) - \varepsilon} d_x 2^{-k_x (H(\*{p}_z) - \varepsilon)} \\ 
        &= \lim_{k\to \infty} \frac{k_x}{k} \pqty{H(\p_z) - \varepsilon}(1-\delta - g^{(k)} _x) \\ 
        &= \sum_{x\in[m]} t_x (H(\p_z) - \varepsilon)(1-\delta) \\
        &= (H(\p_z) - \varepsilon)(1-\delta).
    \end{align*}
    Since the above equations hold for all $\epsilon, \delta \in (0,1)$, we have an inequality
    \begin{equation}\label{ineq:proof-lower-reg-geq-upper}
        \underline{H}^\rm{reg}(Y\vert X)_{\cal{N}} \geq H(\p_z) = \max_{x\in[n]} H(\p_x) = \overline{H}(Y\vert X)_\cal{N}.
    \end{equation}

    The bound in the inequality~\eqref{ineq:reg-upper-lower}, leads to 
    \begin{equation*}
        \underline{H}^\rm{reg}(Y\vert X)_\cal{N}\leq \bb{H}(Y\vert X)_{\cal{N}} \leq \overline{H}^\rm{reg}(Y\vert X)_{\cal{N}}= \overline{H}(Y\vert X)_\cal{N} = \max_{x\in[n]} H(\p_x).
    \end{equation*}
    Notice that $\overline{H}^\rm{reg}(Y\vert X)_{\cal{N}}= \overline{H}(Y\vert X)_\cal{N}$ because the maximal extension of Shannon entropy is additive. Together with the inequality~\eqref{ineq:proof-lower-reg-geq-upper}, we have 
    \begin{equation}
        \bb{H}\pqty{Y\vert X}_\cal{N} = \underline{H}^\rm{reg}(Y\vert X)_\cal{N} = \overline{H}(Y\vert X)_\cal{N} = \min_{x\in[m]} H(\p_x).
    \end{equation}
    
\end{proof}

\section{Channel entropy from divergences}
Other than extending entropy from the known classical state entropy, we can define a classical channel entropy from the divergence of a quantum channel. The divergence-defined channel entropy relates to the degree of indistinguishability of classical channels. In this section, we present two previously proposed divergence-defined channel entropies and their relations to our axiomatic definition of channel entropy.

\subsection{Divergence from the maximally randomizing channel}
This class of channel entropies appears in multiple literature, to the best of my knowledge Gour and Wilde~\cite{GW2021} are the first proposed this.

Given ${D}$ is a divergence of a pair of classical states. Define a channel divergence $D$ as 
\begin{equation*}
    D(\cal{N}\lVert \cal{M}) = \sup_{\*{q}\in\prob(m)} D(\cal{N}(\*{p})\lVert \cal{M}(\*{p})).
\end{equation*}
A channel entropy of $\cal{N}^{X\to Y}$ is defined from a distance from a maximally randomizing channel, 
\begin{equation*}
    H(\cal{N}) = \log_2 (\abs{Y}) - D(\cal{N}\lVert \cal{R}).
\end{equation*}
There are two important properties: being an antitone of channel majorization and is additive (if $D$ is additive). The first property follows from a completely uniformity-preserving superchannel is uniformity-preserving as well and that $D$ satisfies DPI, we can arrive at $H(\cal{N})\leq H(\Theta[\cal{N}])$.

\subsection{Divergence from a bipartite quantum channel}
In~\cite{DGP2024}, the author purposed a set of axiom for a function to be called conditional entropy of a quantum channel. Their purpose entropy is called \emph{generalized conditional entropy}. The conditional entropy of a channel $\cal{N}^{A'B' \to AB}$ is defined as 
\begin{equation}\label{eq:uniform-divergence-entropy}
    \bb{H}(A\vert B)_\cal{N} = \log \abs{A} - \inf_{\cal{M}\in\cptp(B'\to B)} \bb{D}(\cal{N}\lVert\cal{R}^{A'\to A} \otimes \cal{M}^{B' \to B}). 
\end{equation}
This definition of entropy is shown to reduce to an entropy of quantum channel when the input channel is a product bipartite channel, e.g. $\cal{N}^{A'B' \to AB} = \cal{N}_1 ^{A'\to A} \otimes \cal{N}_2 ^{B' \to B}$. From the example, 
\begin{equation*}
    \bb{H}(A\vert B)_\cal{N} = \bb{H}(\cal{N}_1 ^{A' \to A}) = \log\abs{A} - \cal{D}(\cal{N}\vert \cal{R}).
\end{equation*}

This definition of conditional entropy is monotone under an action of a completely uniformity-preserving superchannel. Suppose that $\cal{N},\cal{M}\in\cptp(A'B'\to AB)$, $\Theta$ is a superchannel, $\Theta \in \super((A\to B))$, and $\Theta\otimes \id [\cal{N}]= \cal{M}$. Consider the divergence 
\begin{align*}
    \inf_{\cal{P}\in\cptp(B'\to B)} \bb{D}(\cal{M}\Vert \cal{R} \otimes \cal{P}) &= \inf_{\cal{P}\in\cptp(B'\to B)} \bb{D}(\Theta\otimes\id[\cal{N}]\Vert \Theta\otimes\id[\cal{R}\otimes\cal{P}]) \\
    &\leq \inf_{\cal{P}\in\cptp(B'\to B)} \bb{D}(\cal{N}\Vert \cal{R}\otimes\cal{P}).
\end{align*}
With this inequality, we have monotonicity, for any $\cal{N}\in\cptp(A'B'\to AB)$ and $\Theta$ completely uniformity-preserving,
\begin{equation*}
    \bb{H}(A\vert B)_\cal{N} \leq \bb{H}(A\vert B)_{\Theta[\cal{N}]}.
\end{equation*}
Notice that with above derivation of the inequality, we require only that $\Theta$ preserve a maximally randomizing channel $\cal{R}$. Therefore, this definition of quantum channel conditional entropy reduces to the entropy of channel defined on the first section.

With all of these, we can define an entropy of a unipartite channel $\cal{N}^{A' \to A}$ to be a conditional entropy of a product of the channel with any channel in $\cptp(B' \to B)$. It satisfies monotonicity and additivity.

In the same work~\cite{DGP2024}, the author define a non-signalling quantum conditional entropy to be a divergence of a channel $\cal{N}^{A'B' \to AB}$ from $\cal{R}^{A\to A}\circ \cal{N}^{A'B' \to AB}$. Instead, we can take the infimum on the second argument, we define a channel entropy to be 
\begin{equation}\label{eq:conditionally-uniform-divergence-entropy}
    \bb{H}(\cal{N}) = \log(A) - \inf_{\substack{\cal{P}\in\cptp(A'B'\to B)\\ B,B'}} \bb{D}(\cal{N}^{A' \to A} \otimes \id^{B'\to B} \Vert \*{u}^{A} \otimes \cal{P}^{A'B' \to B}).
\end{equation}
This definition of a quantum channel entropy from a quantum channel relative entropy satisfies monotonicity under completely uniformity preserving and is additive. The proof is similar to that of the generalized conditional entropy being a channel entropy. 

We have no positive or negative result if the function coincide with the entropy defined by the divergence from maximally randomizing channel. However, specifically for the Umegaki relative entropy, we have that the channel entropy as defined in the Equation~\eqref{eq:uniform-divergence-entropy}, and in the Equation~\eqref{eq:conditionally-uniform-divergence-entropy} are coincided. This is due to a theorem in~\cite{DGP2024}. 
\begin{proposition}
    For a quantum channel $\mathcal{N}^{A'B' \to AB}$, its von Neumann conditional entropy $H(A \vert B)_\cal{N}$ (as defined in Equation~\eqref{eq:uniform-divergence-entropy}) satisfies the following identity,
    \begin{equation*}
        H(A\vert B)_\cal{N} = \log(A) - \inf_{\substack{\cal{P}\in\cptp(A'B'\to B)\\ B,B'}} D(\cal{N}^{A' \to A} \otimes \id^{B'\to B} \Vert \*{u}^{A} \otimes \cal{P}^{A'B' \to B})
    \end{equation*}
    if and only if $\cal{N}^{A'B' \to AB}$ is non-signalling from $A' \to B$. Here $D$ is Umegaki relative entropy defined as 
    \begin{equation*}
        D(\rho\lVert \sigma) = \Tr [\rho\log(\rho)] - \Tr [\rho\log(\sigma)].
    \end{equation*}
\end{proposition}
The proof of this statement use an identity 
\begin{equation*}
    D(\rho^{AB} \Vert I^A \otimes \sigma^B) = D(\rho^{AB} \Vert I^{A} \otimes \rho^{B}) + D(\rho^{B} \Vert \sigma^{B}),
\end{equation*}
which is specific to the Umegaki relative entropy.

Suppose the proposition is true. Since $\cal{N}^{A' \to A}\otimes \id^{B \to B}$ is non-signalling, then the entropy as defined in Equation~\eqref{eq:uniform-divergence-entropy} coincide with that as defined in Equation~\eqref{eq:conditionally-uniform-divergence-entropy}.

\section{Channel entropy from Choi isomorphism}
In \cite{CBZ2019,CHLZ2022}, the authors purposed a channel entropy that can be defined from the Choi matrix from Choi--Jamiołkowski isomorphism. They showed that this function satisfies all properties put forward from~\cite{Gour2019}, which is similar to ours but replace the monotonicity with completely uniformity-preserving with monotonicity under random-unitary superchannel, a superchannel which realizable with a unitary pre-processing channel and a post-processing channel and has no auxiliary system. However, this difference leads to incompatibility with our definition as a random-unitary superchannel is a strict subset of completely uniformity-preserving.

Suppose that $\cal{N}\in\cptp(X\to Y)$. We define $J_\cal{N}\in\mathfrak{L}(XY)$ to be its Choi matrix and $\hat{J}_\cal{N}$ to be the normalized Choi matrix. A function $f$ of $N$ is defined by
\begin{equation*}
    f(\cal{N}) = H\pqty{\hat{J}_\cal{N}} - \log(\abs{X})
\end{equation*}
where $H$ is a von Neumann entropy. The function $f$ is purposed to be an entropy function. The key property that the author showed in their work is that $f$ is antitone with respect to a random-unitary superchannel, which is a superchannel $\Theta$ of the form 
\begin{equation*}
    \Theta[\cal{N}^{A\to B}] = \sum_{x\in[\ell]} p_x \cal{U}_x ^{B\to B'} \circ \cal{N}^{A\to B} \circ \cal{V}_x ^{A' \to A}
\end{equation*}
where $\cal{U}_x$ and $\cal{V}_x$ are unitary channels, $\sum_{x\in[\ell]} p_x = 1$, and $p_x \geq 0$.

This function is antitone with respect to a random-unitary superchannel. To see this, consider 
\begin{align*}
    f(\Theta[\cal{N}]) &= H(\hat{J}_{\Theta[\cal{N}]}) - \log(\abs{X}) \\
                       &= H(\Delta_\Theta (\hat{J}_\cal{N})) - \log(\abs{X}) \\ 
                       \textrm{($\Delta_\Theta$ is unital)}\rightarrow&\geq H(\hat{J}_\cal{N}) - \log(\abs{X}) = f(\cal{N}).
\end{align*}
% Since a Choi isomorphism of a product channel is a product of the Choi of each channel, we have additivity. 
However, this function is not antitone with respect to the completely-uniformity preserving.

For Shannon entropy, we know that its entropy extension to the domain of classical channel is unique and is a minimum entropy output. However, this purposed formulation of entropy doesn't suggest that it would coincide with the minimum entropy output, even when restrict the domain to classical channels. It turns out that this function is not a quantum channel entropy. To see this, we consider a classical channel counter example. Suppose 
\begin{equation*}
    N = \pmqty{1 \\ 0} \quad M = \pmqty{1 & 1/2 \\ 0 & 1/2}
\end{equation*}
Since $\cal{N} \sim \cal{M}$, any classical channel entropy $H$ must have the two channel agree on $H(\cal{N}) = H(\cal{M})$. The entropy of a Choi matrix does not satisfy this property. The entropy of both channel is $0$ as measure by any channel entropies. For the channel $\cal{M}$, its Choi is 
\begin{align*}
    \hat{J}_\cal{M} &= \frac{1}{2} \ketbra{1}{1}^X \otimes\ketbra{1}{1}^Y + \frac{1}{2} \ketbra{2}{2}^X \otimes  \frac{1}{2}(\ketbra{1}{1}^Y +\ketbra{2}{2}^Y) = \pmqty{\dmat{1/2,0,1/4,1/4}}.
\end{align*}
A von Neumann entropy for this state is 
\begin{equation*}
    H(\hat{J}_\cal{M}) = -\frac{1}{2} \log\pqty{1/2} - \frac{1}{2} \log\pqty{1/4} = \frac{3}{2}\log(2).
\end{equation*}
This gives the function $f$ on $\cal{N}$ to be 
\begin{equation*}
    f(\cal{N}) = \frac{3}{2} \log(2) - \log(2) = \frac{1}{2}\log(2) = \frac{1}{2}\neq 0.
\end{equation*}

To see why we have this missed matching of the value of two equivalent channels, consider the proof that $\Delta_\Theta$ is unital channel if $\Theta$ is a random-unitary channel.
\begin{lemma}
    If $\Theta\in\super((A\to B))$ is random-unitary, then the induced map on Choi matrix $\Delta_\Theta \in \cptp(AB)$ is unital.
\end{lemma}
\begin{proof}
    For the maximally randomizing channel $\cal{R}\in\cptp(A\to B)$, we have that
    \begin{equation*}
        \Theta[\cal{R}^{A\to B}] = \cal{R}^{A\to B}.
    \end{equation*}
    The Choi matrix is 
    \begin{align*}
        J[\Theta[\cal{R}]] &= J[\cal{R}] = \*{u}^{AB}\\
        J[\Theta[\cal{R}]] = \Delta_\Theta (J[\cal{R}]) &=\Delta_\Theta (\*{u}^{AB})  = \*{u}^{AB},
    \end{align*}
    showing that $\Delta_\Theta$ is unital.
\end{proof}
This leads to the answer of why we see missed match in entropy value between channel-majorization equivalent channels. The induced map $\Delta_\Theta$ is not necessary unital if $\Theta$ is completely uniformity-preserving. 
From the counter example, the induced map of a superchannel transforming $\cal{N}$ to $\cal{M}$ cannot be unital, since it is a mapping between two different dimensions of physics systems.

% \section{Monotones from Channel divergence}
% One can define classical channel entropy by from channel divergence,
% \begin{equation}
%     \Hmax (Y|X)_\cal{N} = \log\abs{Y} - \bb{D}(\cal{N} \lVert \cal{R})
% \end{equation}
% where the channel divergence is defined as
% \begin{equation*}
%     \bb{D}(\cal{N} \lVert \cal{M}) = \max_{\p\in\prob(n)} \bb{D}(\cal{N}(\p) \lVert \cal{M}(\p))
% \end{equation*}
% In fact, the divergence defined this way is the maximal extension of the monotones.

%%% Print bib in every chapter for ease of revision
% \bibliography{QuantumEntropy}
% \bibliographystyle{unsrt}
% \printbibliography

\chapter{Conclusion and outlook}
\section*{Conclusion}
In this thesis, we give a definite notion of classical channel uncertainty. We propose that the exact notion of uncertainty inherent in the channel can be characterized from three perspectives. From the construction perspective, uncertainties of two classical channels are compared by an ability to reconstruct the more uncertain one with random permutations of the outcomes. In the axiomatic approach, the uncertainty of a classical channel is proposed to be preserved or increased under transformations by a specific type of superchannel, namely completely uniformity-preserving superchannel. The operational task we proposed to characterize uncertainty is a family of $\t$-games, where a player strategically sends an input through a classical channel to best predict the outcome of the channel. These three conceptually distinct propositions of uncertainty of classical channels are proven, in this thesis, to induce the same ordering between classical channels, establishing a unique and conceptually as well as operationally meaningful preorder reflecting uncertainty inherent in a classical channel.

With the firm foundation of majorization, we approach defining the most general class of functions evaluating uncertainty inherent in classical channels, namely monotones. We require that a monotone must behave monotonically with the majorization relation. With the additional requirement of additivity under the tensor product of channels, we derive the definition of classical channel \emph{entropy}. We give examples of such functions by extending the existing classical state entropies into the domain of classical channels. Specifically, we show that the extensions of min-entropy and max-entropy to the classical channel domain are entropies. Moreover, with Shannon entropy, the extension that results in an entropy is uniquely identified with the maximal extension, namely the min-entropy output. In addition, the axiomatic definition of classical channel entropy disqualifies the entropy of Choi of a channel from being a valid classical channel entropy.

In complement to majorization of classical channels, we prescribe the standard form as a unique representative of classical channels in each equivalence class. We characterized the majorization relation in terms of majorization of the outputs of channels, convex set containment, predictability function, optimization problems, and, most importantly, conditional majorization. These characterizations offer perspectives and insights into the uncertainty of classical channels.

\section*{Outlook}
One of the natural questions to ask is the extension to the quantum domain. Starting with the notion of majorization itself, would this notion of majorization and entropy of classical channel extend to the quantum domain? The answer is presented in~\cite{GKNSY2025} and the answer was partially affirmative; the axiomatic and constructive extend naturally, while the game of chance does not have a direct analogue in the quantum regime. This leaves the door open for further study of the operational approach to characterize quantum channel uncertainty. On entropy, quantum channel entropy can be defined as an additive monotone on quantum channel majorization. It is interesting to see if the extensions of entropy or classical channel entropy to the domain of quantum channel are unique.

Even on the classical channel domain, this thesis answered many questions and opened many new ones. To start with, what information-theoretic task that the channel majorization play a role in? We see in this thesis that the family of $t$-games is directly tied to the majorization of classical channels. Specifying the game could enable to simulate an informational theoretic protocol such as in cryptographic and communication tasks. 

The foundation of classical channel majorization seems to suggest a path to \emph{conditional channel entropy}. Suppose there is a classical bipartite conditionally uniform channel $\cal{P}^{XX' \to YY'} (\*{p}^{XX'}) = \*{u}^{Y} \otimes \*{q}^{Y'}$ for any $\*{p}\in\prob(\abs{XX'})$. A completely uniform-preserving superchannel ensures that, even when accessing information on the output of the channel $Y'$, the \emph{conditional} probability distribution on $Y$ remains maximally mixed. The further analysis in this venue could yield foundational, insightful results.

For the measure of channel uncertainty, it is shown in the thesis that the extension of min-entropy and max-entropy is unique and results in classical channel entropies. The questions remain in the operational interpretation of which one could extend from the work of~\cite{KRS2009}. The thesis also showed that any classical entropy that extends Shannon entropy can be nothing other than the maximal extension. This is a strong statement, but it still leaves room to wonder if there is a non-additive monotone that extends Shannon entropy but is not the maximal extension. For other entropies, would the extension of them be unique as well?

%%%%%%%%%%%%%%%%%%%%%%%%%%%%%%%%%%%%%%%%%%%%%%%%%%%%%%%%%%%%%%%%%%%%%%%%
%                                                                     %%
% Bibliography                                                        %%
%                                                                     %%
%%%%%%%%%%%%%%%%%%%%%%%%%%%%%%%%%%%%%%%%%%%%%%%%%%%%%%%%%%%%%%%%%%%%%%%%
%  \bibliography{QuantumEntropy}
%  \bibliographystyle{unsrt}
\printbibliography

@Misc{DGP2024,
  author        = {Siddhartha Das and Kaumudibikash Goswami and Vivek Pandey},
  title         = {Conditional entropy and information of quantum processes},
  archiveprefix = {arXiv},
  date          = {2024},
  eprint        = {2410.01740},
  eprintclass   = {quant-ph},
  eprinttype    = {arxiv},
}

@Article{CBZ2019,
  author       = {Czartowski, Jakub and Braun, Daniel and \.{Z}yczkowski, Karol},
  title        = {Trade-off relations for operation entropy of complementary quantum channels},
  number       = {05},
  pages        = {1950046},
  volume       = {17},
  date         = {2019},
  doi          = {10.1142/s0219749919500461},
  journaltitle = {Int. J. Quantum Inf.},
}

@Article{Stinespring1955, 
  title       ={Positive functions on $C^*$-algebras}, 
  volume      ={6}, 
  DOI         ={10.1090/s0002-9939-1955-0069403-4}, 
  number      ={2}, 
  journaltitle={Proceedings of the American Mathematical Society}, 
  publisher   ={American Mathematical Society (AMS)}, 
  author      ={Stinespring, W. Forrest}, 
  date        ={1955},
  pages       ={211–6} 
}

@Article{KR2001,
  author       = {King, C. and Ruskai, M.B.},
  title        = {Minimal entropy of states emerging from noisy quantum channels},
  number       = {1},
  pages        = {192--209},
  volume       = {47},
  date         = {2001},
  doi          = {10.1109/18.904522},
  journaltitle = {IEEE Trans. Inform. Theory},
}

@Article{CHLZ2022,
  author       = {Chu, Yanjun and Huang, Fang and Li, Ming-Xiao and Zheng, Zhu-Jun},
  title        = {An entropy function of a quantum channel},
  issn         = {1573-1332},
  number       = {1},
  pages        = {27},
  volume       = {22},
  date         = {2022-12},
  doi          = {10.1007/s11128-022-03778-1},
  journaltitle = {Quantum Inf Process},
}

@Book{Billingsley2012,
  author    = {Billingsley, Patrick.},
  publisher = {Wiley},
  title     = {Probability and measure},
  edition   = {Anniversary ed.},
  isbn      = {9781118122372},
  series    = {Wiley series in probability and statistics},
  booktitle = {Probability and measure},
  date      = {2012},
  language  = {eng},
  lccn      = {2012382323},
  location  = {Hoboken, N.J},
}

@Book{Durrett2019,
  author    = {Durrett, Richard},
  publisher = {Cambridge University Press},
  title     = {Probability : Theory and Examples},
  edition   = {Fifth edition.},
  isbn      = {1-108-59263-5},
  series    = {Cambridge series in statistical and probabilistic mathematics ; 49},
  booktitle = {Probability : Theory and Examples},
  date      = {2019},
  language  = {eng},
  location  = {Cambridge, England},
}

@Article{ESW2002,
  author       = {T. Eggeling and D. Schlingemann and R. F. Werner},
  title        = {Semicausal operations are semilocalizable},
  number       = {6},
  pages        = {782},
  volume       = {57},
  date         = {2002-03},
  doi          = {10.1209/epl/i2002-00579-4},
  journaltitle = {Europhys. Lett.},
}

@Article{BBPS1996,
  author       = {Bennett, Charles H. and Bernstein, Herbert J. and Popescu, Sandu and Schumacher, Benjamin},
  title        = {Concentrating partial entanglement by local operations},
  pages        = {2046--2052},
  volume       = {53},
  date         = {1996-04},
  doi          = {10.1103/PhysRevA.53.2046},
  issue        = {4},
  journaltitle = {Phys. Rev. A},
  numpages     = {0},
  publisher    = {American Physical Society},
}

@Article{Vidal2000,
  author       = {Guifré Vidal},
  title        = {Entanglement monotones},
  number       = {2-3},
  pages        = {355--376},
  volume       = {47},
  date         = {2000},
  doi          = {10.1080/09500340008244048},
  journaltitle = {J. Mod. Opt.},
  publisher    = {Taylor \& Francis},
}

@Misc{Aberg2006,
  author        = {Johan Aberg},
  title         = {Quantifying Superposition},
  archiveprefix = {arXiv},
  date          = {2006},
  eprint        = {quant-ph/0612146},
  eprintclass   = {quant-ph},
  eprinttype    = {arxiv},
}

@Article{WY2015,
  author       = {Andreas J. Winter and Dong Yang},
  title        = {Operational Resource Theory of Coherence.},
  pages        = {120404},
  volume       = {116 12},
  date         = {2015},
  journaltitle = {Phys. Rev. Lett.},
  url          = {https://api.semanticscholar.org/CorpusID:15630647},
}

@Article{Wyner1974,
  author       = {Wyner, A.},
  title        = {Recent results in the Shannon theory},
  number       = {1},
  pages        = {2--10},
  volume       = {20},
  date         = {1974},
  doi          = {10.1109/tit.1974.1055171},
  journaltitle = {IEEE Trans. Inform. Theory},
}

@Article{DHW2008,
  author       = {Devetak, Igor and Harrow, Aram W. and Winter, Andreas J.},
  title        = {A Resource Framework for Quantum Shannon Theory},
  number       = {10},
  pages        = {4587--4618},
  volume       = {54},
  date         = {2008},
  doi          = {10.1109/tit.2008.928980},
  journaltitle = {IEEE Trans. Inform. Theory},
}

@Article{GMVRN2015,
  author       = {Gilad Gour and Markus P. Müller and Varun Narasimhachar and Robert W. Spekkens and Nicole {Yunger Halpern}},
  title        = {The resource theory of informational nonequilibrium in thermodynamics},
  issn         = {0370-1573},
  note         = {The resource theory of informational nonequilibrium in thermodynamics},
  pages        = {1--58},
  volume       = {583},
  date         = {2015},
  doi          = {10.1016/j.physrep.2015.04.003},
  journaltitle = {Phys. Rep.},
  url          = {https://www.sciencedirect.com/science/article/pii/S037015731500229X},
}

@Article{SKWGB2018,
  author       = {Streltsov, A. and Kampermann, H. and Wölk, S. and Gessner, M. and Bruß, D.},
  title        = {Maximal coherence and the resource theory of purity},
  pages        = {053058},
  volume       = {20},
  date         = {2018},
  doi          = {10.1088/1367-2630/aac484},
  issue        = {5},
  journaltitle = {New J Phys},
}

@Article{ZMCFV2017,
  author       = {Zhu, Huangjun and Ma, Zhihao and Cao, Zhu and Fei, Shao-Ming and Vedral, Vlatko},
  title        = {Operational one-to-one mapping between coherence and entanglement measures},
  pages        = {032316},
  volume       = {96},
  date         = {2017-09},
  doi          = {10.1103/PhysRevA.96.032316},
  issue        = {3},
  journaltitle = {Phys. Rev. A},
  numpages     = {13},
  publisher    = {American Physical Society},
}

@Article{CG2016,
  author       = {Chitambar, Eric and Gour, Gilad},
  title        = {Comparison of incoherent operations and measures of coherence},
  pages        = {052336},
  volume       = {94},
  date         = {2016-11},
  doi          = {10.1103/PhysRevA.94.052336},
  issue        = {5},
  journaltitle = {Phys. Rev. A},
  numpages     = {19},
  publisher    = {American Physical Society},
}

@Article{DBG2015,
  author       = {Du, Shuanping and Bai, Zhaofang and Guo, Yu},
  title        = {Conditions for coherence transformations under incoherent operations},
  pages        = {052120},
  volume       = {91},
  date         = {2015-05},
  doi          = {10.1103/PhysRevA.91.052120},
  issue        = {5},
  journaltitle = {Phys. Rev. A},
  numpages     = {5},
  publisher    = {American Physical Society},
}

@Article{BHNOW2015,
  author       = {Fernando Brandão and Michał Horodecki and Nelly Ng and Jonathan Oppenheim and Stephanie Wehner},
  title        = {The second laws of quantum thermodynamics},
  number       = {11},
  pages        = {3275--3279},
  volume       = {112},
  date         = {2015},
  doi          = {10.1073/pnas.1411728112},
  journaltitle = {Proc. Natl. Acad. Sci.},
}

@Article{BBCJ+1993,
  author       = {Bennett, Charles H. and Brassard, Gilles and Cr\'epeau, Claude and Jozsa, Richard and Peres, Asher and Wootters, William K.},
  title        = {Teleporting an unknown quantum state via dual classical and Einstein-Podolsky-Rosen channels},
  pages        = {1895--1899},
  volume       = {70},
  date         = {1993-03},
  doi          = {10.1103/PhysRevLett.70.1895},
  issue        = {13},
  journaltitle = {Phys. Rev. Lett.},
  numpages     = {0},
  publisher    = {American Physical Society},
}

@Article{BW1992,
  author       = {Bennett, Charles H. and Wiesner, Stephen J.},
  title        = {Communication via one- and two-particle operators on Einstein-Podolsky-Rosen states},
  pages        = {2881--2884},
  volume       = {69},
  date         = {1992-11},
  doi          = {10.1103/PhysRevLett.69.2881},
  issue        = {20},
  journaltitle = {Phys. Rev. Lett.},
  numpages     = {0},
  publisher    = {American Physical Society},
}

@Article{LBL2020,
  author       = {Li, Lu and Bu, Kaifeng and Liu, Zi-Wen},
  title        = {Quantifying the resource content of quantum channels: An operational approach},
  pages        = {022335},
  volume       = {101},
  date         = {2020-02},
  doi          = {10.1103/PhysRevA.101.022335},
  issue        = {2},
  journaltitle = {Phys. Rev. A},
  numpages     = {8},
  publisher    = {American Physical Society},
}

@Article{Neumann1927,
  author       = {Neumann, J. von},
  title        = {Thermodynamik quantenmechanischer Gesamtheiten},
  pages        = {273--291},
  volume       = {1927},
  date         = {1927},
  journaltitle = {Nachr. Ges. Wiss. Gottingen, Math.-Phys. Kl.},
  language     = {ger},
  url          = {http://eudml.org/doc/59231},
}

@InBook{Petz2001,
  author    = {Petz, D{\'e}nes},
  editor    = {R{\'e}dei, Mikl{\'o}s and St{\"o}ltzner, Michael},
  pages     = {83--96},
  publisher = {Springer Netherlands},
  title     = {Entropy, von Neumann and the von Neumann Entropy},
  isbn      = {978-94-017-2012-0},
  booktitle = {John von Neumann and the Foundations of Quantum Physics},
  date      = {2001},
  doi       = {10.1007/978-94-017-2012-0_7},
  location  = {Dordrecht},
}

@Article{Yuan2019,
  author       = {Yuan, Xiao},
  title        = {Hypothesis testing and entropies of quantum channels},
  pages        = {032317},
  volume       = {99},
  date         = {2019-03},
  doi          = {10.1103/PhysRevA.99.032317},
  issue        = {3},
  journaltitle = {Phys. Rev. A},
  numpages     = {8},
  publisher    = {American Physical Society},
}

@Article{RR2011,
  author       = {Renes, Joseph M. and Renner, Renato},
  title        = {Noisy Channel Coding via Privacy Amplification and Information Reconciliation},
  number       = {11},
  pages        = {7377--7385},
  volume       = {57},
  date         = {2011},
  doi          = {10.1109/tit.2011.2162226},
  journaltitle = {IEEE Trans. Inform. Theory},
}

@Article{HW2010,
  author       = {Hsieh, Min-Hsiu and Wilde, Mark M.},
  title        = {Entanglement-Assisted Communication of Classical and Quantum Information},
  number       = {9},
  pages        = {4682--4704},
  volume       = {56},
  date         = {2010},
  doi          = {10.1109/tit.2010.2053903},
  journaltitle = {IEEE Trans. Inform. Theory},
}

@Article{Devetak2005,
  author       = {Devetak, I.},
  title        = {The private classical capacity and quantum capacity of a quantum channel},
  number       = {1},
  pages        = {44--55},
  volume       = {51},
  date         = {2005},
  doi          = {10.1109/tit.2004.839515},
  journaltitle = {IEEE Trans. Inform. Theory},
}

@Article{RT2021b,
  author       = {Regula, Bartosz and Takagi, Ryuji},
  title        = {One-Shot Manipulation of Dynamical Quantum Resources},
  pages        = {060402},
  volume       = {127},
  date         = {2021-08},
  doi          = {10.1103/PhysRevLett.127.060402},
  issue        = {6},
  journaltitle = {Phys. Rev. Lett.},
  numpages     = {9},
  publisher    = {American Physical Society},
}

@Article{TSP2020,
  author       = {Theurer, Thomas and Satyajit, Saipriya and Plenio, Martin B.},
  title        = {Quantifying Dynamical Coherence with Dynamical Entanglement},
  pages        = {130401},
  volume       = {125},
  date         = {2020-09},
  doi          = {10.1103/PhysRevLett.125.130401},
  issue        = {13},
  journaltitle = {Phys. Rev. Lett.},
  numpages     = {7},
  publisher    = {American Physical Society},
}

@Article{HWXLG2022,
  author       = {Ren-Dong He and Kang-Da Wu and Guo-Yong Xiang and Chuan-Feng Li and Guang-Can Guo},
  title        = {Experimental quantification of dynamical coherence via entangling two qubits},
  number       = {7},
  pages        = {10346--10353},
  volume       = {30},
  date         = {2022-03},
  doi          = {10.1364/oe.453504},
  journaltitle = {Opt. Express},
  publisher    = {Optica Publishing Group},
  url          = {https://opg.optica.org/oe/abstract.cfm?URI=oe-30-7-10346},
}

@Article{GS2021,
  author       = {Gour, Gilad and Scandolo, Carlo Maria},
  title        = {Entanglement of a bipartite channel},
  number       = {6},
  volume       = {103},
  date         = {2021-06},
  doi          = {10.1103/physreva.103.062422},
  journaltitle = {Phys. Rev. A},
  publisher    = {American Physical Society (APS)},
}

@Article{KRS2009,
  author       = {Konig, Robert and Renner, Renato and Schaffner, Christian},
  title        = {The Operational Meaning of Min- and Max-Entropy},
  number       = {9},
  pages        = {4337--4347},
  volume       = {55},
  date         = {2009},
  doi          = {10.1109/tit.2009.2025545},
  journaltitle = {IEEE Trans. Inform. Theory},
}

@Book{Khinchin1957,
  author    = {Khinchin, A Ya},
  publisher = {New York: Dover},
  title     = {Mathematical foundations of information theory},
  date      = {1957},
}

@Article{AFN1974,
  author       = {Aczél, J. and Forte, B. and Ng, C. T.},
  title        = {Why the Shannon and Hartley entropies are ‘natural’},
  number       = {1},
  pages        = {131--146},
  volume       = {6},
  date         = {1974},
  doi          = {10.2307/1426210},
  journaltitle = {Adv. Appl. Probab.},
}

@Article{Renyi1961,
  author       = {Alfr{\'e}d R{\'e}nyi},
  title        = {On Measures of Entropy and Information},
  pages        = {547--561},
  journaltitle = {Berkeley Symp. on Math. Statist. and Prob.},
  date         = {1961},
  url          = {https://projecteuclid.org/proceedings/berkeley-symposium-on-mathematical-statistics-and-probability/Proceedings-of-the-Fourth-Berkeley-Symposium-on-Mathematical-Statistics-and/Chapter/On-Measures-of-Entropy-and-Information/bsmsp/1200512181?tab=ArticleFirstPage},
}

@Article{NPVV1992,
  author       = {K.K. Nambiar and Pramod K. Varma and Vandana Saroch},
  title        = {An axiomatic definition of Shannon's entropy},
  issn         = {0893-9659},
  number       = {4},
  pages        = {45--46},
  volume       = {5},
  date         = {1992},
  doi          = {10.1016/0893-9659(92)90084-M},
  journaltitle = {Appl. Math. Lett.},
  url          = {https://www.sciencedirect.com/science/article/pii/089396599290084M},
}

@Article{Shannon1948,
  author       = {Shannon, C. E.},
  title        = {A mathematical theory of communication},
  number       = {3},
  pages        = {379--423},
  volume       = {27},
  date         = {1948},
  doi          = {10.1002/j.1538-7305.1948.tb01338.x},
  journaltitle = {The Bell System Tech. J.},
}

@Article{Hasselbarth1986,
  author       = {Hässelbarth, Werner},
  title        = {The incompleteness of {Rényi} entropies},
  issn         = {1432-2234},
  number       = {2},
  pages        = {119--121},
  volume       = {70},
  date         = {1986-08},
  doi          = {10.1007/bf00532209},
  journaltitle = {Theor. Chim. Acta},
}

@Article{KL1951,
  author       = {S. Kullback and R. A. Leibler},
  title        = {{On Information and Sufficiency}},
  number       = {1},
  pages        = {79--86},
  volume       = {22},
  date         = {1951},
  doi          = {10.1214/aoms/1177729694},
  journaltitle = {The Annals of Mathe. Statist.},
  publisher    = {Institute of Mathematical Statistics},
}

@Article{SCG2020,
  author       = {Saxena, Gaurav and Chitambar, Eric and Gour, Gilad},
  title        = {Dynamical resource theory of quantum coherence},
  pages        = {023298},
  volume       = {2},
  date         = {2020-06},
  doi          = {10.1103/PhysRevResearch.2.023298},
  issue        = {2},
  journaltitle = {Phys. Rev. Res.},
  numpages     = {27},
  publisher    = {American Physical Society},
}

@Article{VSGC2022,
  author       = {Vempati, Mahathi and Shah, Saumya and Ganguly, Nirman and Chakrabarty, Indranil},
  title        = {A-unital {O}perations and {Q}uantum {C}onditional {E}ntropy},
  issn         = {2521-327x},
  pages        = {641},
  volume       = {6},
  date         = {2022-02},
  doi          = {10.22331/q-2022-02-02-641},
  journaltitle = {{Quantum}},
  publisher    = {{Verein zur F{\"{o}}rderung des Open Access Publizierens in den Quantenwissenschaften}},
}

@book{Kraus1983,
author = {Kraus, Karl and Böhm, Arno and Dollard, J. D. (John Day) and Wootters, W. H.},
address = {Berlin},
booktitle = {States, effects, and operations : fundamental notions of quantum theory : lectures in mathematical physics at the University of Texas at Austin},
isbn = {3540127321},
language = {eng},
lccn = {83016906},
publisher = {Springer-Verlag},
series = {Lecture notes in physics ; 190},
title = {States, effects, and operations : fundamental notions of quantum theory : lectures in mathematical physics at the University of Texas at Austin },
year = {1983},
}

@Misc{GKNSY2024,
  author      = {Gilad Gour and Doyeong Kim and Takla Nateeboon and Guy Shemesh and Goni Yoeli},
  title       = {Inevitable Negativity: Additivity Commands Negative Quantum Channel Entropy},
  date        = {2024},
  eprint      = {2406.13823},
  eprintclass = {quant-ph},
  eprinttype  = {arXiv},
}

@article{GKNSY2025,
  title = {Inevitable negativity: Additivity commands negative quantum channel entropy},
  author = {Gour, Gilad and Kim, Doyeong and Nateeboon, Takla and Shemesh, Guy and Yoeli, Goni},
  journal = {Phys. Rev. A},
  volume = {111},
  issue = {5},
  pages = {052424},
  numpages = {12},
  publisher = {American Physical Society},
  date = {2025-05},
  doi = {10.1103/PhysRevA.111.052424},
  url = {https://link.aps.org/doi/10.1103/PhysRevA.111.052424}
}

@Article{GW2021,
  author       = {Gilad Gour and Mark M. Wilde},
  title        = {Entropy of a quantum channel},
  issn         = {2643-1564},
  pages        = {023096},
  volume       = {3},
  date         = {2021-05},
  doi          = {10.1103/PhysRevResearch.3.023096},
  issue        = {2},
  journaltitle = {Phys. Rev. Res.},
  numpages     = {26},
  publisher    = {American Physical Society},
}

@Book{HLP1934,
  author    = {Hardy, G. H. and Littlewood, J. E. and P\'{o}lya, George},
  publisher = {Cambridge University Press},
  title     = {Inequalities},
  date      = {1934},
}

@Article{Birkhoff1946,
  author       = {Birkhoff, Garrett},
  title        = {Three observations on linear algebra},
  pages        = {147--151},
  volume       = {5},
  date         = {1946},
  journaltitle = {Univ. Nac. Tacuman, Rev. Ser. A},
}

@Article{Nielsen1999,
  author       = {Nielsen, Michael A.},
  title        = {Conditions for a Class of Entanglement Transformations},
  pages        = {436--439},
  volume       = {83},
  date         = {1999-07},
  doi          = {10.1103/PhysRevLett.83.436},
  issue        = {2},
  journaltitle = {Phys. Rev. Lett.},
  numpages     = {0},
  publisher    = {American Physical Society},
}

@Article{CDP2008,
  author       = {Chiribella, Giulio and D'Ariano, G Mauro and Perinotti, Paolo},
  title        = {Transforming quantum operations: Quantum supermaps},
  number       = {3},
  pages        = {30004},
  volume       = {83},
  date         = {2008},
  journaltitle = {Europhys. Lett.},
  publisher    = {IOP Publishing},
}

@Book{MOA2011,
  author    = {Marshall, Albert W. and Olkin, Ingram and Arnold, Barry C.},
  publisher = {Springer},
  title     = {Inequalities: Theory of Majorization and Its Applications},
  edition   = {2},
  series    = {Springer Series in Statistics},
  date      = {2011},
  doi       = {10.1007/978-0-387-68276-1},
}

@Article{CMW2016,
  author       = {Cooney, Tom and Mosonyi, Mil{\'a}n and Wilde, Mark M},
  title        = {Strong converse exponents for a quantum channel discrimination problem and quantum-feedback-assisted communication},
  number       = {3},
  pages        = {797--829},
  volume       = {344},
  date         = {2016},
  journaltitle = {Commun. Math. Phys.},
  publisher    = {Springer},
}

@Article{FGG2013,
  author       = {Friedland, Shmuel and Gheorghiu, Vlad and Gour, Gilad},
  title        = {Universal Uncertainty Relations},
  pages        = {230401},
  volume       = {111},
  date         = {2013-12},
  doi          = {10.1103/PhysRevLett.111.230401},
  issue        = {23},
  journaltitle = {Phys. Rev. Lett.},
  numpages     = {5},
  publisher    = {American Physical Society},
}

@Article{Gour2019,
  author       = {Gour, Gilad},
  title        = {Comparison of quantum channels by superchannels},
  number       = {9},
  pages        = {5880--5904},
  volume       = {65},
  date         = {2019},
  doi          = {10.1109/tit.2019.2907989},
  journaltitle = {IEEE Trans. Inform. Theory},
}

@Article{GT2020,
  author       = {Gour, Gilad and Tomamichel, Marco},
  title        = {Optimal extensions of resource measures and their applications},
  pages        = {062401},
  volume       = {102},
  date         = {2020-12},
  doi          = {10.1103/PhysRevA.102.062401},
  issue        = {6},
  journaltitle = {Phys. Rev. A},
  numpages     = {13},
  publisher    = {American Physical Society},
}

@Article{GS2020a,
  author       = {Gour, Gilad and Scandolo, Carlo Maria},
  title        = {Dynamical Entanglement},
  pages        = {180505},
  volume       = {125},
  date         = {2020-10},
  doi          = {10.1103/PhysRevLett.125.180505},
  issue        = {18},
  journaltitle = {Phys. Rev. Lett.},
  numpages     = {6},
  publisher    = {American Physical Society},
}

@Misc{GS2020b,
  author      = {Gilad Gour and Carlo Maria Scandolo},
  title       = {Dynamical Resources},
  date        = {2020},
  eprint      = {2101.01552},
  eprintclass = {quant-ph},
  eprinttype  = {arXiv},
}

@Article{LY2020,
  author       = {Liu, Yunchao and Yuan, Xiao},
  title        = {Operational resource theory of quantum channels},
  pages        = {012035},
  volume       = {2},
  date         = {2020-02},
  doi          = {10.1103/PhysRevResearch.2.012035},
  issue        = {1},
  journaltitle = {Phys. Rev. Res.},
  numpages     = {6},
  publisher    = {American Physical Society},
}

@Misc{BGWG2021,
  author      = {Sarah Brandsen and Isabelle J. Geng and Mark M. Wilde and Gilad Gour},
  title       = {Quantum conditional entropy from information-theoretic principles},
  date        = {2021},
  eprint      = {2110.15330},
  eprintclass = {quant-ph},
  eprinttype  = {arXiv},
}

@Article{GT2021,
  author       = {Gour, Gilad and Tomamichel, Marco},
  title        = {Entropy and Relative Entropy From Information-Theoretic Principles},
  number       = {10},
  pages        = {6313--6327},
  volume       = {67},
  date         = {2021},
  doi          = {10.1109/tit.2021.3078337},
  journaltitle = {IEEE Trans. Inform. Theory},
}

@Article{BGG2022,
  author       = {Brandsen, Sarah and Geng, Isabelle Jianing and Gour, Gilad},
  title        = {What is entropy? A perspective from games of chance},
  pages        = {024117},
  volume       = {105},
  date         = {2022-02},
  doi          = {10.1103/PhysRevE.105.024117},
  issue        = {2},
  journaltitle = {Phys. Rev. E},
  numpages     = {5},
  publisher    = {American Physical Society},
}

@book{Gour2024a,
  author    = {Gour, Gilad},
  booktitle = {Quantum resource theories},
  isbn      = {1-009-56090-5},
  language  = {eng},
  publisher = {Cambridge University Press},
  title     = {Quantum resource theories },
  year      = {2025},
}

@Article{Renes2016,
  author       = {Renes, Joseph M.},
  title        = {{Relative submajorization and its use in quantum resource theories}},
  issn         = {0022-2488},
  number       = {12},
  pages        = {122202},
  volume       = {57},
  date         = {2016-12},
  doi          = {10.1063/1.4972295},
  journaltitle = {J. Math. Phys.},
}

@Article{Partovi2011,
  author       = {Partovi, M. Hossein},
  title        = {Majorization formulation of uncertainty in quantum mechanics},
  pages        = {052117},
  volume       = {84},
  date         = {2011-11},
  doi          = {10.1103/PhysRevA.84.052117},
  issue        = {5},
  journaltitle = {Phys. Rev. A},
  numpages     = {10},
  publisher    = {American Physical Society},
}

@Article{PRZ2013,
  author       = {Zbigniew Puchała and Łukasz Rudnicki and Karol Życzkowski},
  title        = {Majorization entropic uncertainty relations*},
  number       = {27},
  pages        = {272002},
  volume       = {46},
  date         = {2013-06},
  doi          = {10.1088/1751-8113/46/27/272002},
  journaltitle = {J. Phys. A: Math. Theor.},
  publisher    = {IOP Publishing},
}

@Article{RT2021,
  author        = {Regula, Bartosz and Takagi, Ryuji},
  title         = {Fundamental limitations on distillation of quantum channel resources},
  year          = {2021},
  number        = {1},
  pages         = {4411},
  volume        = {12},
  date          = {2021-07-20},
  doi           = {10.1038/s41467-021-24699-0},
  id            = {Regula2021},
  isbn          = {2041-1723},
  journaltitle  = {Nat. Commun.},
}

@Article{HS2018,
  author       = {Matty J Hoban and Ana Belén Sainz},
  title        = {A channel-based framework for steering, non-locality and beyond},
  number       = {5},
  pages        = {053048},
  volume       = {20},
  date         = {2018-05},
  doi          = {10.1088/1367-2630/aabea8},
  journaltitle = {New J. Phys.},
  publisher    = {IOP Publishing},
}

@Article{NV2001,
  author       = {Michael A. Nielsen and Guifr{\'e} Vidal},
  title        = {Majorization and the interconversion of bipartite states},
  pages        = {76--93},
  volume       = {1},
  date         = {2001},
  journaltitle = {Quantum Inf. Comput.},
  url          = {https://api.semanticscholar.org/CorpusID:4654708},
}

@Article{GGH+2018,
  author       = {Gour, Gilad and Grudka, Andrzej and Horodecki, Micha\l{} and K\l{}obus, Waldemar and \L{}odyga, Justyna and Narasimhachar, Varun},
  title        = {Conditional uncertainty principle},
  number       = {4},
  pages        = {042130},
  volume       = {97},
  date         = {2018-04},
  doi          = {10.1103/PhysRevA.97.042130},
  issue        = {4},
  journaltitle = {Phys. Rev. A},
  numpages     = {14},
  publisher    = {American Physical Society},
}

@Article{HO2013,
  author        = {Horodecki, Micha{\l} and Oppenheim, Jonathan},
  title         = {Fundamental limitations for quantum and nanoscale thermodynamics},
  year          = {2013},
  number        = {1},
  pages         = {2059},
  volume        = {4},
  date          = {2013-06-26},
  doi           = {10.1038/ncomms3059},
  id            = {Horodecki2013},
  isbn          = {2041-1723},
  journaltitle  = {Nat. Commun.},
  publisher     = {Nature Publishing Group UK London},
}
%%%%%%%%%%%%%%%%%%%%%%%%%%%%%%%%%%%%%%%%%%%%%%%%%%%%%%%%%%%%%%%%%%%%%%%%
%%                                                                    %%
%% Appendicies                                                        %%
%%                                                                    %%
%%%%%%%%%%%%%%%%%%%%%%%%%%%%%%%%%%%%%%%%%%%%%%%%%%%%%%%%%%%%%%%%%%%%%%%%
\appendix

% \chapter{Misc.}

\chapter{Miscellaneous}

\section{Probability theory and Shannon entropy}
We present necessary terminologies and theorems from probability theory. The weak law will be stated. The detailed study of these can be found on the standard probability theory textbooks, e.g. Durrett's \emph{Probability: Theory and Examples}~\cite{Durrett2019} and Billingsley's \emph{Probability and Measure}~\cite{Billingsley2012}.

A framework of probability theory begins with probability space, which is a measure space.
\begin{definition}
    A probability space is a triplet $\pqty{\Omega, \cal{F}, P}$ of a set sample space $\Omega$ with its $\sigma$-algebra $\cal{F}$ and the probability measure $P:\cal{F} \to \bb{R}_+$.
\end{definition}

A $\sigma$-algebra is a collection of sets that close under complement and countable unions. Elements of the $\sigma$-algebra are called \emph{events}. A probability measure $P:\cal{F}\to \bb{R}_+$ is a measure, which satisfies 
\begin{enumerate}
    \item for any $A\in\cal{F}$, we have $P(A)> P(\emptyset) = 0$, and 
    \item for any countable collections of disjoint sets $\qty{A_i : i \in \bb{N}} \subseteq \cal{F}$, we have that 
    \begin{equation*}
        P\pqty{\bigcup_{i\in\bb{N} A_i}} = \sum_{i\in\bb{N}} P(A_i)
    \end{equation*}
\end{enumerate}

Central to probability theory is a notion of random variable, which actually is a function rather than a variable.
\begin{definition}
    Given a probability space $(\Omega, \cal{F}, P)$. A function $X:\Omega \to \bb{R}$ is a \emph{random variable} if $X^{-1}(\cal{B}(A)) \in \cal{F}$, where $A\in\mathfrak{B}(\bb{R})$. The measure-theoretic term is $X$ is $P$-measurable.
\end{definition}
\begin{remark}
    $\mathfrak{B}(\bb{R})$, Borel $\sigma$-algebra, is the smallest $\sigma$-algebra containing all open sets.
\end{remark}
A random variable is commonly denoted with $X$, $Y$, and $Z$. In this thesis, $X$, $Y$, $Z$, and $W$ denote random variables, labels of classical systems, and Hilbert spaces corresponding to the physical system of the same letter.

% A random variable comes with a probability distribution, which is defined to be a measure of its preimages.
% \begin{definition}[]

% \end{definition}

\subsection{Law of large number and Shannon entropy}
Laws of large number are main tools and central to the study of probability theory. The theorem lies on the notion of independence and the convergence in probability.

\begin{definition}[Independence]
    Events $A,B\in\cal{F}$ are independent if $P(A\cap B) =P(A)P(B)$. Random variables $X,Y$ are independent if $P(X\in C, Y \in D) =P(X\in C)P(Y\in D)$ for any $C,D\in\cal{B}(\bb{R})$. 
\end{definition}
\begin{remark}
    $P(X\in C)$ is defined as the probability of an event $\qty{\omega: X(\omega) \in C}$.
\end{remark}

\begin{definition}[Independent and identically distributed]
    A sequence of random variables $\*{X} = X_1, X_2, X_3, \ldots$ is independent and identically distributed (i.i.d.) if for all $X_i$ are independent and they have the same distribution.
\end{definition}
    
\begin{definition}
    We say $X_n$ converges to $X$ \emph{in probability} and write $X_n\xrightarrow{P} X$ if for any $\varepsilon > 0$ we have that 
    \begin{equation*}
        \lim_{n\to\infty} P(\abs{X_n - X} > \varepsilon) =0.
    \end{equation*}
\end{definition}
\begin{remark}
    In measure theoretic terms, convergence in probability is a convergence in measure, where the measure is a probability measure.
\end{remark}

\begin{theorem}[The weak law of large number]
    Suppose $X_i$---i.i.d. with $\bb{E}[X_i] = \mu$ and $\bb{E}[\abs{X_i}]<\infty$. Then, $S_n /n\to \mu$ in probability.
\end{theorem}

\subsection{Typicality}
We will present the Shannon's theorem. The theorem concern an i.i.d. source $\p$ which $n$ random alphabets ${X_i: i\in[m]}$ are drawn from. That is, for any $k\in[m]$, $X_i = k$ with probability $p_k$. The theorem says that the probability of drawing the drawn sequence is equally likely, with a probability given by $2^{-H(\p)}$.

A probability of an observed event of a random vector $\*{X}(\omega) = \pqty{X_1 (\omega), X_2 (\omega), \ldots, X_n (\omega)}$, is
\begin{equation*}
    \pi(\omega) = \prod_{i\in[n]} p\pqty{X_i}
\end{equation*}
where $p\pqty{X_i}=p_{X_i (\omega)}$ to simplify our expression. Note that a function $p\pqty{X_i}:\Omega\to[0,1]$ and $\pi:\Omega \to [0,1]$ are random variables.

\begin{theorem}[Asymtotic equipartition property]
    \label{th:aep}
    Suppose $X_1, X_2, \ldots $ is a sequence of independently and identically distributed with distribution $\p = \pqty{p_1, p_2, \ldots, p_m}^T \in \prob(m)$, then 
    \begin{equation*}
        \frac{1}{n} \log \pqty{\pi(\omega)} \xrightarrow{P} - \sum_{x\in[m]} p_x \log(p_x) 
    \end{equation*}
\end{theorem}

\begin{proof}
Since $X_i$ are independent, we have that 
\begin{align*}
    \log(\pi(\omega)) &= \log\qty(\prod_{i\in[n]} p(X_i)) = \sum_{i\in[n]} \log(p(X_i)).
\end{align*}
By the weak law of large number, 
\begin{align*}
\lim_{n\to \infty} -\frac{1}{n} \sum_{i\in[n]} \log(p(X_i)) &= - \bb{E} \bqty{\log(p(X_1))} \quad\rm{in}~P\\
% \magenta{\pqty{X_i \text{ are i.i.d.}}\rightarrow} & \\ 
&= -\sum_{x\in[m]} p_x \log(p_x) \quad\rm{in}~P.
\end{align*}
\end{proof}
Later, the quantity $-\sum_{x\in[m]} p_x \log(p_x)$ will be referred to as a Shannon entropy, 
\begin{equation*}
H(\p) = - \sum_{x\in[n]} p_x \log (p_x).
\end{equation*}
That is with a long enough sequence, any sequences that occur, occur with probability 
\begin{equation*}
\pi(\omega) \approx 2^{-H(\p)}.
\end{equation*}

We define an $\varepsilon$-\emph{typical sequence} to be sequence such that its probability of occuring is $\varepsilon$-close to $2^{-H(\p)}$.
\begin{definition}[Typical sequences]
    A sequence $\*{x}\in[m]^n$ of length $n$ is $\varepsilon$-typical if 
    \begin{equation*}
        2^{-(H(\p)+\varepsilon)} \leq P\pqty{\*{X} = \*{x}} \leq 2^{-(H(\p)-\varepsilon)}.
    \end{equation*}
    A set of all $\varepsilon$-typical sequences of length $n$ is denoted with $A^{(n)} _{\varepsilon}$.
\end{definition}
One consequence of the asymptotic equipartition theorem is that as $n$ get larger, $\*{X} \in A^{(n)} _\varepsilon$ with probability approaching $1$. 
\begin{theorem}\label{th:typical-seq-prob}
    Suppose $\*{X} = X_1, X_2, \ldots$ is a sequence of i.i.d. random variables. For any $\varepsilon > 0$, there exists $m\in\bb{N}$ such that for all $n>m$,
        $P(\*{X}\in A^{(n)} _\varepsilon ) > 1 - \varepsilon$.
\end{theorem}
That is given we observe a long enough sequence of i.i.d. random variables, the sequence is typical with probability close to $1$.
\begin{proof}
    Recall that, a sequence $\*{X}(\omega)$ is in $A^{(n)} _\zeta$ if $2^{-H(\p) - \zeta} < \pi(\omega) < 2^{-H(\p) + \zeta}$. The inequality can be put equivalently as 
    \begin{equation*}
    2^{-H(\p)}(2^{-\zeta} - 1) \leq \pi(\omega) - 2^{-H(\p)} \leq 2^{-H(\p)}(2^{\zeta} - 1).
    \end{equation*}
    The set of elements $\omega\in\Omega$ satisfying above expression is a superset of the set of $\omega$ satisfying 
    \begin{equation*}
        \abs{\pi(\omega) - 2^{-H(\p)}} \leq 2^{-H(\p)}(1 - 2^{-\zeta}) =: \delta_0 (\zeta),
    \end{equation*}
    because $0<1-2^{-\zeta} < 2^{\zeta}-1$ for all $\zeta >0$. By theorem~\ref{th:aep}, a random variable $\pi(\omega)$ converges to $2^{-H(\p)}$ in probability. That is, for any $\delta > 0$ and $\varepsilon \in(0,1)$, there exists a natural number $m$ such that for all $n>m$,
    \begin{equation*}
        P\pqty{\abs{\pi(\omega) - 2^{-H(\p)}} < \delta} > 1 - \varepsilon.
    \end{equation*}
    Since $\delta_0 (\varepsilon) > 0$, we have that 
    \begin{equation*}
        P\qty(\*{X}(\omega)\in A^{n} _\varepsilon) \geq P\pqty{\abs{\pi(\omega) - 2^{-H(\p)}} < \delta_0 (\varepsilon)}  > 1- \varepsilon.
    \end{equation*}
\end{proof}

\section{Linear programming}
One of the main techniques we exploit multiple times in this thesis is Linear programming. Here,we provide a succinct introduction to linear programming, its dual, weak duality, the Farkas lemma, and the strong duality.

\begin{definition}
    A linear program is an optimization problem of the form 
    \begin{align*}
        \textrm{max } &\*{c} \cdot \*{x} \\
        \textrm{subject to } &A \*{x} \leq \*{b} \\
        &\*{x} \geq 0
    \end{align*}
    where $x\in\bb{R}^n, $ 
\end{definition}

\begin{definition}
    Given a linear program, its feasibility set $\ssf{F}$ is a set of all $\*{x}$ that satisfy the constraints. A linear program with non-empty feasibility set is said to be \emph{feasible}.
\end{definition}

\begin{definition}
    A dual of a linear program is a minimization problem of the form
    \begin{align*}
        \textrm{min } &\*{b} \cdot \*{y} \\
        \textrm{subject to } &\*{y}^T A \geq \*{c}^T \\
        &\*{y} \geq 0
    \end{align*}
\end{definition}
The dual of a linear program is simply referred to as the dual program while its original program is called the primal program.

\begin{theorem}[Weak duality]
    The optimal value for the primal program is no greater than the optimal value of the dual program.
\end{theorem}
\begin{proof}
    Suppose that the feasibility set of the primal ($\ssf{F}_P$) and of the dual $\ssf{F}_D$ are not empty. Suppose that $\*{x}\in\ssf{F}_P$ and $\*{y} \in \ssf{F}_D$. We have that $\*{x},\*{y} \geq 0$, $\*{b}\geq A\*{x}$, and $\*{y}^T A \geq \*{c}$. We have that 
    \begin{align*}
        \*{y} \cdot \*{b} \geq \*{y} \cdot A\*{x} \geq \*{c}\cdot\*{x},
    \end{align*}
    showing the desired inequality.
\end{proof}

\begin{theorem}[Hyperplane separating theorem for a conic hull and a point]
    Suppose that $K\in\bb{R}^m$ is a conic hull of $\qty{\*{a}_1, \*{a}_2, \ldots, \*{a}_n: \*{a}_i \in\bb{R}^m}$ and $\*{b}$ is a point. The point $\*{b}$ is not in the cone $K$ if and only if there exists a plane described by $h(\*{x}) = \*{n} \cdot \*{x} = 0$ separating $K$ and $\*{b}$, i.e 
    \begin{equation*}
        \*{n} \cdot \*{b} < 0 \quad\rm{and}\quad \*{n}\cdot\*{a}_i \geq 0~\forall i\in[n].
    \end{equation*}
\end{theorem}
\begin{proof}
    Suppose $\*{b}$ is not in the cone $K$. We define $d$ to be a distance between the cone and the point, that is 
    \begin{equation*}
        d = \inf \qty{ \norm{\*{b}-\*{k}} ~\vert~ \*{k} \in K}.
    \end{equation*}
    Consider a closed ball ${B}$ of radius $\norm{\*{b}}$ centered at $\*{b}$. This ball has non-empty intersection with $K$, at least it contains the origin. Any points $k$ outside this ball is further away from $\*{b}$ than the points in the intersection. That is $d = \min \qty{ \norm{\*{b}-\*{k}} ~\vert~ \*{k} \in K\cap {B}}$ the replacement of infimum with a minimum is due to existence of $\*{k}$ yielding infimum value in $K\cap B\subset K$.

    Suppose $\knot$ is a point in $K$ such that $d = \norm{\*{b} - \knot}$. Define $\*{n}' = \*{b}- \*{k}_0$. We will show first that any $\*{k} \in K$ satisfies $(\*{b} - \knot)\cdot \*{k} \geq 0$. Since $\*{b} - \knot$ is perpendicular to $\knot$, we have that $(\*{b} - \knot)\cdot \*{k}  = (\*{b} - \knot)\cdot \pqty{\*{k} - \knot}$. This gives us a nice geometric visualization (see fig~\ref{fig:geometric-Hyperplan-Cone-dot}).
    \begin{figure}
        \centering
        \begin{tikzpicture}
            \filldraw[gray] (2,-1) circle (2pt);
            \filldraw[gray] (0,2) circle (2pt);
            \node[below] at (2,-1) {$\*{k}$};
            \node[left] at (0,2) {$\*{b}$};

            \filldraw[black] (0,0) circle (2pt);
            \node[left] at (0, 0) {$\knot$ };
            \draw[->,thick] (0,0) -- (2,-1);
            \draw[->,thick] (0,0) -- (0,2);
            \draw (0,0.7) arc (90:-27:0.7);
            \node[above right] at (0.6,0.3) {$\theta$};
        \end{tikzpicture}
        \caption{A geometric visualization of $\*{b} - \knot$ and $\*{k} - \knot$.}
        \label{fig:geometric-Hyperplan-Cone-dot}
    \end{figure}

    Suppose on contrary that $(\*{b} - \knot)\cdot \pqty{\*{k} - \knot} > 0$. This leads to $\norm{\*{b} - \knot}\times \norm{\*{k} - \knot}\times \cos \theta > 0$, where $\theta$ is an angle between the two vectors. That is the point $\*{k}$ is strictly closer to $\*{b}$ than $\knot$. Therefore, $(\*{b} - \knot)\cdot \pqty{\*{k} - \knot} \leq 0$.

    For the inequality with the point $\*{b}$, consider $(\*{b} - \knot) \cdot \*{b} = (\*{b} - \*{k}) \cdot (\*{b} -\knot) = \norm{\*{b} - \knot} ^2 > 0$, due to $\*{b} \neq \knot$. 
    To get the direction of inequalities in the theorem statement, we define $\*{n} = -\*{n}'$.
\end{proof}

\begin{corollary}[Geometric Farkas lemma]
    Suppose $\*{a}_1, \ldots, \*{a}_n, \*{b}\in\bb{R}^m$. Exactly one of the following is true:
    \begin{enumerate}
        \item A vector $\*{b}$ is in a conic hull of $\qty{\*{a}_1, \ldots, \*{a}_n}$.
        \item There is $\*{n}$ such that $\*{n}\cdot \*{a}_i \geq 0$ and $\*{n}\cdot\*{b} < 0$ for all $i\in[n]$.
    \end{enumerate}
\end{corollary}
\begin{proof}
    Notice that $\*{b}$ is either inside or outside the cone. It is enough to show that $2$ is equivalent with $\*{b}$ being outside the cone. Denote the cone by $K$. Suppose that $\*{b}$ is outside $K$, then there exists a hyperplane separating $\*{b}$ and the cone $K$. The normal vector $\*{n}$ is then given by the plane. Conversely, if there is a plane separating between the point and the cone, then they are disjoint.
\end{proof}

\begin{theorem}[Farkas lemma]
    Suppose $A\in\bb{R}^{m\times n}$, $\*{b} \in \bb{R}^m$, and $\*{x}\in\bb{R}^n$. The feasibility set $\qty{ \*{x} ~\vert~ A\*{x} = \*{b}, \*{x}\geq 0}$ is non-empty if and only if for all $\*{y}\in\bb{R}^m$, we have that $\*{y}^T A \geq 0^T$ implies $\*{y}^T \*{b}\geq 0$.
\end{theorem}
\begin{proof}
    We denote columns of $A$ by $A = (\*{a}_1, \*{a}_2, \ldots ,\*{a}_n)$. The non-emptiness of the feasibility set is equivalent to 
    \begin{equation*}
        \sum_{i\in[n]} x_i \*{a}_i = \*{b} \quad\rm{and}\quad x_i \geq 0\quad \textrm{for any } i \in[n].
    \end{equation*}
    This show that the feasibility set is not empty if and only if $\*{b}$ is in the conic hull of $\qty{\*{a}_1,\ldots, \*{a}_n}$.  
    
    The point $\*{b}$ is outside the cone if and only if there exists $\*{y}\in\bb{R}^m$ such that $\*{y}\cdot \*{a}_i \geq 0 $ for any $i\in[n]$ and $\*{y}\cdot \*{b} < 0$. The point $\*{b}$ being inside the cone is equivalent with the statement: 
    \begin{center}
        for any $\*{y}\in\bb{R}^m$, the statement $\*{y}\cdot \*{a}_i \geq 0 $ for all $i\in[n]$ is not true or $\*{y}\cdot \*{b} \geq 0$. 
    \end{center}
    We turn this or-statement into a conditional, we have the statement: 
    \begin{center}
        for any $\*{y}\in\bb{R}^m$, $\*{y} \cdot \*{a}_i \geq 0$ for all $i\in[n]$ implies $\*{y}\cdot \*{b} \geq 0$.
    \end{center}
    Compactly, for all $\*{y}\in\bb{R}^m$, we have that $\*{y}^T A \geq 0^T$ implies $\*{y}^T \*{b}\geq 0$.
\end{proof}

This leads to the variation of Farkas lemma which we use throughout the thesis.
\begin{propbox}{Farkas lemma with inequality}
    \begin{corollary}
        Using a notation from previous theorem, we have that a feasibility set
        \begin{equation*}
            \qty{\*{x} ~\vert~ A\*{x} \geq \*{b} ~\text{and}~ \*{x} \geq 0}
        \end{equation*}
        is non-empty if and only if for all $\*{y}\in\bb{R}^m _+$, we have $\*{y}^T A \geq 0^T$ implies $\*{y}^T \*{b}\geq 0$.
    \end{corollary}
\end{propbox}
\begin{proof}
    Define $A' = \pqty{A ~\vert~ I_m}$. There is a non-negative solution of $\*{x}$ to $A\*{x} \leq \*{b}$ if and only if there is a non-negative solution $\*{x}'$ to $A' \*{x}' = \*{b}$. From the Farkas lemma, this is equivalent to $\*{y}\in\bb{R}^m _+$, we have $\*{y}^T A' \geq 0^T$ implies $\*{y}^T \*{b}\geq 0$. The expression $\*{y}^T A'\geq 0$ can be stated with $\*{y}^T A \geq 0$ and $\*{y}\geq 0$.
\end{proof}

Now, we can prove the strong duality result for linear programming.
\begin{propbox}{Strong duality}
    \begin{theorem}
        If a primal linear program is feasible, the optimal values for the primal and the dual programs are equal.
    \end{theorem}
\end{propbox}
\begin{proof}
    Suppose the primal linear program and the optimal value are $p^\star = \min \qty{\*{x}~\vert~A\*{x} \leq \*{b}, \*{x} \geq 0}$. The dual of this program is 
    \begin{align*}
        \textrm{min } & \*{y} \cdot \*{b} \\
        \textrm{subject to } &\*{y}^T A \geq \*{c} \\
        & \*{y} \geq 0.
    \end{align*}
    We define an augmented matrix, and a constraint vector 
    \begin{equation*}
        A ' = \pmqty {A \\ -\*{c}^T} \quad \*{b}_\varepsilon = \pmqty{ \*{b}\\ -p^\star - \varepsilon}.
    \end{equation*}
    Define a feasibility set by $A'$ and $\*{b}_\varepsilon$,
    \begin{align*}
        \ssf{F}_\varepsilon = \qty{\*{x} ~\vert~ A' \*{x} \leq \*{b}, ~\*{x} \geq 0}.
    \end{align*}  
    The condition $A' \*{x} \leq \*{b}$ is equivalent with  $A \*{x} \leq \*{b}$ and $\*{c}\cdot \*{x} \geq p^\star + \varepsilon$. Since $\pstar$ is the optimal value, the feasibility set $\ssf{F}_\varepsilon$ is empty unless $\varepsilon = 0$. 
    
    Consider the case where $\varepsilon > 0$. By Farkas lemma, the feasibility set $\ssf{F}_\varepsilon$ is empty if and only if there exists $\*{y} = \pmqty{\*{u} \\ z}$ where $\*{u}\in\bb{R}^m _+$ and $z \in \bb{R} _+$ such that $\*{y} ^T A' \geq \*{0}^T$ and $\*{y}\cdot \*{b}_\varepsilon < 0$. Notice that the condition $\*{y} ^T A' \geq \*{0}^T$ is independent of the value of $\varepsilon$. Equivalently, 
    \begin{equation*}
        \*{y}^T A' \geq \*{0}^T\quad\textrm{and}\quad \*{u}\cdot\*{b} < z(\pstar + \varepsilon).
    \end{equation*}
    This give non-emptiness of a feasibility set 
    \begin{equation*}
        \ssf{F}' _{\varepsilon, z} =  \qty{\*{u} ~\vert~ \*{u}^T A \geq z\*{c}^T ,~ \*{u}\cdot \*{b} \leq \pstar+ \varepsilon, ~\*{u} \geq 0} \neq \emptyset
    \end{equation*}

    For $\varepsilon = 0$, the feasibility set is not empty. In the case of $\varepsilon > 0$, there is such $\*{y}$ that make $\*{y}^T A' \geq \*{0}^T$, then the vector $\*{y}$ must satisfy
    \begin{align*}
        \*{y}\cdot\*{b} = \*{u}\cdot\*{b}-z(\pstar+\varepsilon) \geq 0.
    \end{align*}
    Equivalently, $\*{u}\cdot\*{b}\geq z(\pstar+\varepsilon)$. If $z=0$, $\*{u}\cdot\*{b} \geq 0$, but from the case of $\varepsilon > 0$, $\*{u}\cdot\*{b} < 0$. Therefore, the element ${z} > 0$ must be strictly positive. Since the set $\ssf{F}' _{\varepsilon, z}$ is non-empty for all $\varepsilon > 0$ and $z > 0$, a feasibility set 
    \begin{equation*}
        \ssf{G} _{\varepsilon} =  \qty{\*{v} ~\vert~ \*{v}^T A \geq \*{c},~ \*{v}\cdot \*{b} \leq \pstar + \varepsilon, ~\*{v} \geq 0}
    \end{equation*}
    is non-empty for any $\varepsilon > 0$. Because of $\*{v} \cdot \*{b} \leq \pstar + \varepsilon$ for all $\varepsilon > 0$, we have that $\*{v} \cdot \*{b} \leq \pstar$ and consequently that the dual problem is feasible with optimal value at most $\pstar$, that is $d^\star \leq p^\star$. From the weak duality, we have $d^\star \geq \pstar$. Therefore, $d^\star = \pstar$ showing the strong duality.
\end{proof}

  %\include{appendixb}

%%%%%%%%%%%%%%%%%%%%%%%%%%%%%%%%%%%%%%%%%%%%%%%%%%%%%%%%%%%%%%%%%%%%%%%%
%%                                                                    %%
%% End of document                                                    %%
%%                                                                    %%
%%%%%%%%%%%%%%%%%%%%%%%%%%%%%%%%%%%%%%%%%%%%%%%%%%%%%%%%%%%%%%%%%%%%%%%%
\end{document}